\documentclass[11pt,twoside]{report}

\usepackage{graphicx}

\usepackage{phd_thesis}

\usepackage{diakriti}

\usepackage{journals}

\usepackage{mynatbib}
\bibpunct[, ]{(}{)}{;}{a}{}{,}

\usepackage{stmaryrd}
\usepackage{amssymb}

\ifpdf
 \usepackage[pdftex, plainpages=false, colorlinks=true, pdfstartview=FitV,
             linkcolor=red, citecolor=red, urlcolor=blue, pagebackref]{hyperref}
\else
 \def\phantomsection{}
 \usepackage[dvipdf, breaklinks=true, plainpages=false, 
             pdfstartview=FitV, bookmarks=true, 
             pdfpagescrop={0 0 600 800},pagebackref]{hyperref}
\fi

\def\ads#1{\href{#1}{ADS}}

\usepackage{backrefx}
\renewcommand{\backrefpagesname}{}
\renewcommand{\backrefpagesnames}{}
\renewcommand{\backreflist}{,\space}
\renewcommand{\backrefformat}[1]{[#1]}
\renewcommand{\backrefnocite}{-}

\newcommand{\DD}{{\rm D}}
\newcommand{\dd}{{\rm d}}
\newcommand{\obs}{{\rm o}}
\newcommand{\mso}{{\rm ms}}
\newcommand{\init}{{\rm i}}

\newcommand{\tk}{\hat{t}}
\newcommand{\uk}{\hat{u}}
\newcommand{\muk}{\hat{\mu}}
\newcommand{\phik}{\hat{\varphi}}
\newcommand{\rhok}{\tilde{\rho}}
\newcommand{\Deltak}{\tilde{\Delta}}
\newcommand{\Ak}{\tilde{\cal A}}
\newcommand{\mat}[1]{\left( \begin{array}{cccc} #1 \end{array} \right)}
\newcommand{\fr}[2]{\displaystyle\frac{#1}{#2}}

\newcommand{\mbh}{M_{\bullet}}

\newcommand{\loc}{{\rm l}}
\renewcommand{\vec}[1]{\mbox{\protect\boldmath$#1$}}
\newcommand{\fekalfa}{{Fe~K$\alpha$} }
\newcommand{\fekbeta}{{Fe~K$\beta$} }

\begin{document}

\renewcommand{\thepage}{\roman{page}}
\pagestyle{empty}
\begin{center}
 
\vspace{1cm} 
{\Large Faculty of Mathematics and Physics \vspace{.3cm}

CHARLES UNIVERSITY in PRAGUE\vspace{.3cm}

{\large Institute of Theoretical Physics \\
\&\\ 
Astronomical Institute}}

\vspace{1.5cm}

\begin{figure}[hbtp]
\hspace*{5.1cm}
\includegraphics[width=4.5cm]{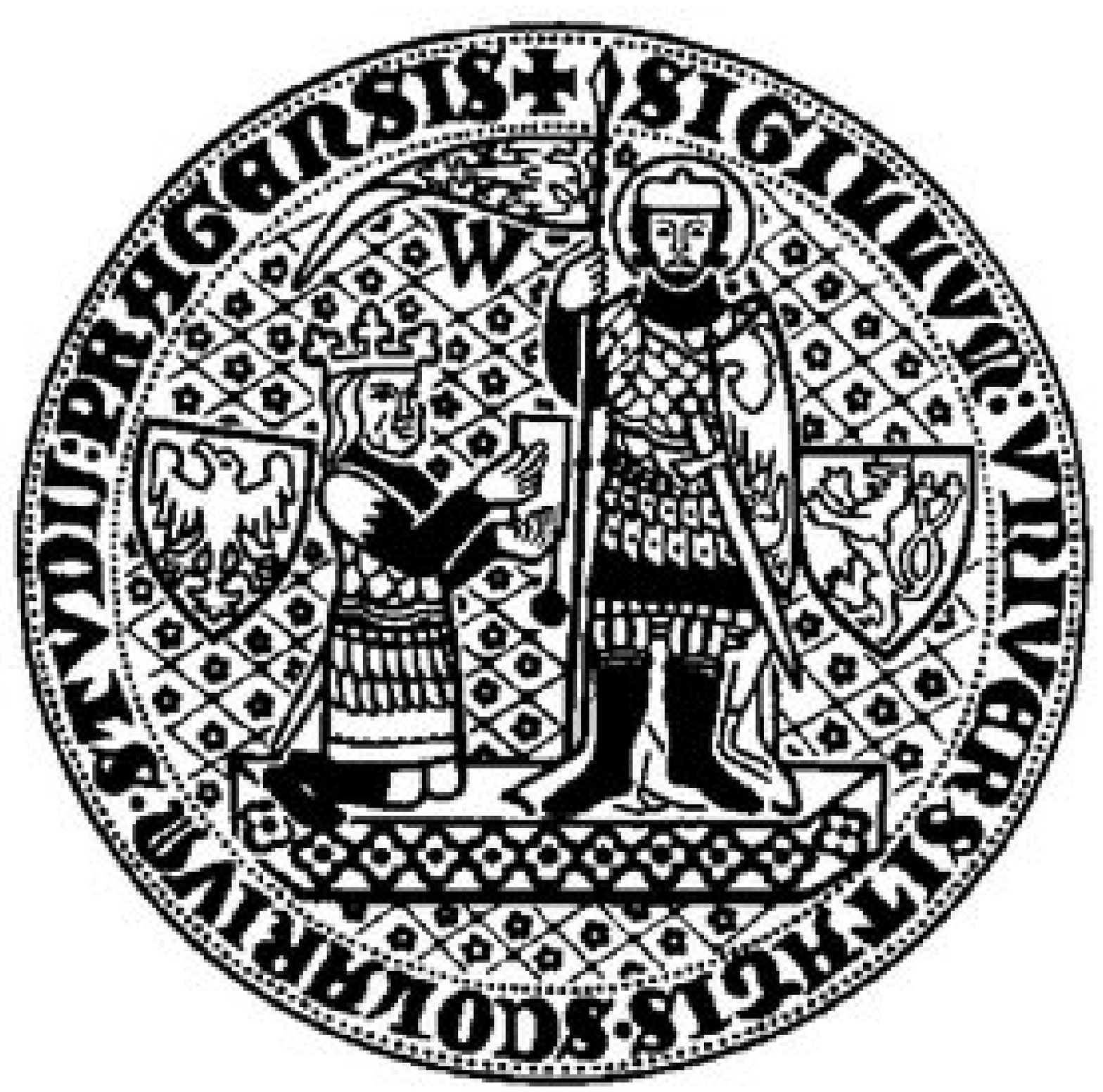}
\end{figure}

\vspace{1.5cm}

{\bf\LARGE Radiation of accretion discs in strong gravity}

\vspace{1cm}
{\large Thesis submitted for the degree of Doctor Philosophiae}

\vspace{1cm}
{\Large Michal Dov\v{c}iak}

\vfill
{\large Supervisor: doc.\ RNDr.\ Vladim\'{\i}r Karas, DrSc.}

\rule{140mm}{0.1mm}

\vspace{1.5cm}
{\large Prague, June 2004}
\end{center}


\clearpage
\pagestyle{empty}
\vspace*{\fill}
\hspace*{\fill}\parbox{6.1cm}
{\large
\noindent                                                                                                                                       
{\bf Referees:}\\[2mm]

Professor Annalisa Celotti \\
Astrophysics sector \\
International School for Advanced Studies in Trieste \\
Italy \\[2mm]

\noindent
Professor Andrew Fabian  \\
Institute of Astronomy \\
University of Cambridge \\
United Kingdom \\[2mm]

\noindent
Professor Zden\v{e}k Stuchl\'{\i}k\\
Institute of Physics \\
Faculty of Philosophy and Science \\
Silesian University at Opava \\
Czech Republic \\
}


\cleardoublepage
\pagestyle{empty}
\vspace*{\fill}
\section*{Acknowledgements}

  Firstly I would like to thank my supervisor, Vladim{\' \i}r Karas, for all of
his helpful advice, useful ideas and pertinent comments on this text and my
research from which it has germinated.
I wish to thank Andrea Martocchia and Giorgio Matt for discussing the lamp-post
model and polarization.
Many thanks go to my colleague, Ladislav \v{S}ubr,
with whom I had many long discussions and who initiated me into the great open
world of Linux. I also want to thank Adela Kawka for reading the manuscript and
filling in all the missing (mostly definite) articles.


  I thank everyone whom I worked with at my {\em Alma Mater}, Charles University
in Prague, and at the Institute of Astronomy in Ond\v{r}ejov for their
friendship. I gratefully acknowledge the support from the
Czech Science Foundation grants 202/02/0735 and 205/03/0902, and from the Charles
University grant GAUK 299/2004.

 Last but not least, I would like to thank my parents, who supported me during
the whole period of my PhD studies.


\cleardoublepage
\pagestyle{myheadings}
\markright{CONTENTS}
\tableofcontents
\addtocontents{toc}{\protect\thispagestyle{myheadings}}
\addtocontents{toc}{\vspace*{-\baselineskip}}

 \clearpage
 \thispagestyle{empty}

\cleardoublepage
\renewcommand{\thepage}{\arabic{page}}
\setcounter{page}{1}
\phantomsection\addcontentsline{toc}{chapter}{Preface}
\markboth{PREFACE}{PREFACE}
\chapter*{Preface}
 \thispagestyle{empty}
 Einstein's general relativity remains, after almost a century, our best
theory of gravitation. Its mathematical formalism and quantitative
predictions will most likely play a crucial role also in future, especially
on temporal and spatial scales that are relevant for astronomical purposes.
This concerns mainly the description of stellar-mass compact objects as well as
dark supermassive bodies in galactic nuclei.  In spite of many evident
successes, the general theory of relativity has not been tested yet with
sufficient accuracy in the regime of strong gravity. Nevertheless, partial
observational tests have been obtained, mainly by employing the methods of X-ray
spectroscopy. Some questions have been answered, while others have emerged.
In the near future, improved technology will allow to perform high-throughput
high-resolution spectroscopy and polarimetry in X-rays with capabilities
for detailed mapping of the close neighbourhood of compact objects. In this
way, physical parameters of the central compact bodies will be measured and
properties of accretion flows will be studied, including the innermost
regions near the black-hole horizon. The aim of the present thesis is to
contribute to the effort by developing an advanced computational tool for
X-ray spectro-polarimetry in a strong gravity regime.

We will turn our attention to nuclei of galaxies that harbour very
massive objects, attracting a gaseous environment from their imminent
surroundings. It appears beyond
all reasonable doubt that these are black holes. The case for black
holes residing in hearts of galaxies is not based on any single type of
evidence. Instead, it is a mosaic of various observations that strongly
supports the aforementioned interpretation. A similar situation arises on
a vastly different scale of masses, namely in compact binary
stars containing a stellar-mass accreting black hole as one of its
components.

This thesis concentrates on a fragment of the whole picture. Our main
goal is to discuss and improve the procedures and computational packages
that are appropriate for spectroscopical studies of accreting black
holes, including timing and polarimetry. The aim of this research
is to discriminate between objects
containing black holes from those containing other forms of compact
bodies and to assess the possibility of determining physical parameters
of these systems via the spectroscopy of accretion discs. We study the
effects of curved space-time on light rays and specific spectral
features which should arise when the gaseous
environment emits radiation while falling onto the black-hole horizon. We
thus develop an extended tool to fit X-ray data of real sources within a
full general relativistic context.

\vspace*{2\bigskipamount}
\begin{flushright}
Prague, June 2004
\end{flushright}

\clearpage
\phantomsection\addcontentsline{toc}{chapter}{Introduction}
\markboth{INTRODUCTION}{INTRODUCTION}
\chapter*{Introduction}
 \thispagestyle{empty}
 \enlargethispage*{\baselineskip}
It has been established observationally that significant amounts of
glowing gas is often present in the cores of galaxies. This is particularly
true in the case of very bright active galactic nuclei (AGN) and also in
Galactic black-hole candidates (BHC). In spite of the big scale difference
between the two types of cosmic objects, we can be rather confident that
the accretion of gas proceeds in disc-like configurations in both cases.
However, the sizes of the discs and their geometries are often quite uncertain.


\section*{On black holes and accretion discs}
\addcontentsline{toc}{section}{On black holes and accretion discs}


It is not so long ago that the general theory of relativity was brought
to life by Albert \cite{einstein1916}. Black holes belong to the first of
its offspring, the static ones being described as early as in 1916 by
Karl \citeauthor{schwarzschild1916}. These objects are characterized by
strong gravity and an imaginary horizon that enwraps a region from which nothing
can escape, not even the light. Black holes have their ancestors in the ideas of
John \cite{michell1784} and Pierre Simon Laplace (\citeyear{laplace1796,
laplace1799}), who defined similar objects more than one century before --
stars with such large masses that the escape velocity from their
surface is larger than the speed of light. But this would be a different story,
a Newtonian one. Soon the family of black holes grew bigger, several new types
were classified -- black holes with a non-zero
angular momentum \citep{Kerr1963}, with an electric charge or with a hypothetic
magnetic monopole \citep{reissner1916,nordstrom1918,newman1965}.

However, electrically charged non-rotating black holes are
not of direct astrophysical interest since, if the hole is immersed in a cloud
of ionized or partially ionized plasma, which is a typical situation in cosmic
environments and we assume it hereafter in this work, then any excess
charge becomes rapidly neutralized by selective accretion. Nevertheless, charged
black holes have a role as a model for the realistic black holes formed in a
spinning collapse, see \cite{wald1974,blandford1977,damour1978,damour1980,
price1986,karas1991}. The properties of black holes have been studied by
numerous authors in various contexts;\break see e.g.\ \cite{thorne1986,kormendy1995,
wald1998} for a
survey of the problems and the general discussion of astrophysical connotations
of black holes. As far as the Kerr (rotating) black holes are concerned, the
possibility to extract its
rotational energy either in the ergosphere by Penrose process
\citep{penrose1971} or by Blandford-Znajek process, when
the black hole is embedded in the electromagnetic field \citep{blandford1977},
are really intriguing.

The problem of the origin of black holes is very exciting as well.
Oppenheimer 
considered black holes as the last stage in the evolution of massive stars when
after having burnt all their fuel neither
the pressure of electrons \citep{chandrasekhar1931} nor the pressure of
neutrons \citep{oppenheimer1939, rhoades1974} can stop them from collapsing.
There are theories that
the black holes may be nearly as old as the universe itself. These primordial
black holes \citep{hawking1971} may have been the seeds for the formation of
galaxies.

From the astrophysical point of view the motion of matter and the propagation of
light near black holes has a special significance because the matter can emit
light that can be observed and thus we can deduce what is going on out there.
The emission
from the stellar-mass and supermassive black holes themselves is rather low
\citep{hawking1975}. Many authors investigated these issues and calculated
either the time-like and null geodesics in the black-hole space-times
\citep{sharp1979,chandrasekhar1983,bicak1993} or even the more complex
hydrodynamical models for the accreting matter or the outflowing wind. The
spherical accretion onto the black hole was firstly considered as early as in
\citeyear{bondi1944} by Bondi and Hoyle. The disc-like accretion
\citep{lynden-bell1969, pringle1972, novikov1973, shakura1973, shapiro1976}
is potentially much more efficient in converting the gravitational energy into
the thermal energy which can be radiated away. It is also a more realistic
configuration when we consider that the accreting matter has a non-zero angular
momentum.

The family of accretion discs is quite large. Let us just mention
the Shakura-Sunyaev or standard discs (SS), the advection dominated accretion
flow (ADAF), the convection dominated accretion flows (CDAF), the advection
dominated inflows-outflows solution (ADIOS), the Shapiro-Lightman-Eardley discs
(SLE) and the slim discs.
Their behaviour depends mainly on the density, pressure, viscosity and opacity
of the matter that they consist of as well as on the equation of state and their
accretion rate. For a basic review on accretion disc theories see e.g.\
\cite{kato1998} or \cite{frank2002}.

The interaction between an accretion disc and the matter above and below the
disc may also play an important role in the accretion process and therefore the
configuration of the black hole, accretion disc and corona seems to be more
appropriate \citep{paczynski1978,haardt1993}.

As can be seen from the multitude of the disc models, some of the processes
involved in the accretion are quite well understood. Nevertheless, there are
still many open issues, one of them being the origin of viscosity
in accretion discs. Turbulence and
magnetic fields may play an important role here. Another important phenomenon
is the existence of jets emerging from the vicinity of the central black hole,
which may be due to the interaction between the black hole and the accretion
disc. To tackle
these problems much more complex three dimensional numerical models of accretion
discs in curved space-time with magnetic field present have to be investigated
\citep[e.g.][]{hawley1995,nishikawa2001,villiers2003}.

If the disc is thick and/or dense, its own gravity, in particular farther away
from the central black hole,  cannot be neglected.
These self-gravitating discs were investigated by various
authors in various approximations, starting with \cite{ostriker1964},
\cite{bardeen1973}, \cite{fishbone1976}, \cite{paczynski1978a} and
\cite{karas1995}. A very useful and pertinent
review of this subject was recently written by Karas, Hur\'{e} \& 
Semer\'{a}k \citeyearpar{karas2004}.

\section*{Linking theory with observation}
\addcontentsline{toc}{section}{Linking theory with observation}

Last century was rich in discovering highly energetic and violent objects in
astronomy. In 1918 \citeauthor{curtis1918} discovered a long jet of material
coming out from elliptical galaxy M87. Later, galaxies with particularly
bright and compact nuclei were observed \citep{seyfert1943} and they were
classified as a new type of galaxies, the so-called Seyfert galaxies.
But truly amazing
discoveries came when radio astronomy fully developed. When quasars were
observed for the first time in 1960s, they were called radio stars
\citep{matthews1960, matthews1962} because their emission in the radio part
of the spectrum was enormous and they appeared like point sources. It was three
years later that \cite{schmidt1963} realized that these objects were not hosted
by our Galaxy but that they were rather distant extragalactic objects with
highly
redshifted spectral lines. Since then many strange radio sources with jets and
lobes of various shapes and sizes have been discovered, mostly associated with
optical extragalactic objects -- centres of distant galaxies. The luminosity,
size and variability of these sources indicate that highly energetic
processes were once at work there. Because of their huge activity they are
called active galactic nuclei.
In 1960s it was realized that the only reasonable way of explaining the
production of such large amounts of energy was to be sought in the gravitational
energy conversion by means of an accreting massive black hole
\citep{salpeter1964,zeldovich1964}.

A new branch of astronomy has emerged, in which many novel discoveries were made:
it was the X-ray astronomy and it developed quickly. The first observations were
made by Geiger counters carried by rockets in 1950s and balloons in 1960s and
later by specially designed X-ray astronomy satellites that have been orbiting
the Earth since 1970s.
The first Galactic X-ray source was discovered in 1962 -- a neutron star
Sco X-1 \citep{giacconi1962}. Soon more discoveries followed, X-ray binary
source Cygnus~X-1 \citep{bowyer1965} being one of them. The fact that this
object is a
strong X-ray emitter and that the optical and X-ray emission varies on very
short time scales (as short as one thousandth of a second) suggests that the
companion might be a black hole. Many Galactic objects (most of them binaries
containing neutron stars) and active galactic nuclei emitting X-rays have been
observed since then. It is believed that AGN
sources host a supermassive black hole with
a mass of approximately $10^6-10^8M_{\odot}$ and that some of the Galactic X-ray
binaries contain a black hole with the mass equal to several solar masses.

A breakthrough has come with the two most recent X-ray satellites, {\it
Chandra}\/ and {\it XMM-Newton}, which provide us with the most detailed
images and spectra with an unprecedented resolution.
Some of the observed objects exhibit redshifted, broad or narrow features in
their spectra, thus suggesting that the emission is coming from the very close
vicinity of a black hole.
But still, the data acquired by the instruments on board these satellites
are not as complete and detailed as would be necessary for an unambiguous
interpretation and comparison with the theoretical models. New
X-ray missions are being planned -- {\it Astro-E2}\/ (already being assembled),
{\it Constellation-X}\/ and {\it Xeus}.
Hopefully, with the data measured by the satellites involved in these
observations, it
will be possible not only to decide whether there is a black hole in the
observed system but also to determine its properties, its mass and angular
momentum, as well as the properties of the surrounding accretion disc.

\section*{This thesis}
\addcontentsline{toc}{section}{This thesis}

Nowadays both the theory describing accreting black-hole systems and the
instruments for observing such objects are quite advanced.
Computational tools for comparing acquired data with
theories have been developed to such an advanced state that we are able now
to find out properties of
particular observed systems on the one hand and check if our models are good
enough for describing these systems on the other. One of the tools used for
processing X-ray data, which also implements the possibility to fit the
data within a certain set of models,\footnote{Notice that, in accordance with
traditional terminology, we use
the term `model' in two slightly different connotations: first, it can
mean a physical model of an astronomical system or a process that is under
discussion; other possible meaning of the word is more technical, referring
to a computational representation, or a routine that is employed in order
to compare actual observational data with the theory.} is {\sc{}xspec}.
This X-ray spectral fitting
package contains various models for explaining the measured spectra but
it lacks fully general relativistic models.
One of the aims of this thesis has been to help to fill this gap and provide
new general
relativistic models that would be fast and flexible enough and would be
able to fit parameters describing a black hole, mainly its mass and angular
momentum. Although it is not possible to fit time resolved data within
{\sc{}xspec} and polarimetric data have not been available for X-ray sources yet
(though they may become available when future X-ray missions materialize), we
developed new models able to deal also with these issues.

Our objective is to model X-ray spectra from accretion discs
near compact objects. The observed spectrum depends not only on the local
radiation emitted from the disc but it is also affected by strong
gravitation on its way to the observer. Emission from different parts of the
disc may be either amplified or reduced by gravitational effects. We will focus
our attention on these effects in Chapter~\ref{chapter1}, where we deal with
calculations of six functions
that are needed for the integration of the total spectra of the accretion disc
(including time varying spectra and polarimetric information). These functions
include the Doppler and gravitational shifts, gravitational lensing, relative
time delay, two emission angles and change of the polarization angle.

The basic types of local emission coming from an accretion disc are summarized
in Chapter~\ref{chapter2}, where also the equations for the observed
flux and polarization at infinity are derived.

In Chapter~\ref{chapter3} we describe new {\sc ky} models that we have
developed and that can be used either
inside {\sc{}xspec} or as standalone programs for studying the non-stationary
emission from accretion discs or for studying polarimetry. These are
models for a general relativistic line and Compton reflection continuum,
general convolution models and a model for an orbiting spot.
We also compare these models with some of the models already present in
{\sc{}xspec}.


Some applications of the new models are summarized in
Chapter~\ref{chapter4}. Firstly, we employ the models to high-quality X-ray data
that are currently available for the Seyfert galaxy MCG--6-30-15.
Then we calculate the flux from X-ray illuminated orbiting spot
and polarization from an accretion disc illuminated from a primary source
located above the black hole.

There are several appendixes in this thesis.
There is a summary of the basic equations for Kerr space-time in
Appendix~\ref{kerr}.
Appendix~\ref{fits} describes the layout of data files used in this thesis.
A detailed description of the integration
routines that we have developed and that can be used inside the {\sc xspec}
framework for general
relativistic computations is included in Appendix~\ref{ide}.
In Appendix~\ref{atlas} an atlas of contour figures of all transfer
functions is shown.

\enlargethispage*{\baselineskip}
This thesis describes my original work except where references are given
to the results of other authors. Parts of the research were carried out in
collaboration and published in papers, as indicated in the text. In
particular, various features and usage of the newly developed code have
been described in \cite{dovciak2004} and \cite{dovciak2004b}.
Chapter~\ref{chapter4} is based
on papers  \cite{dovciak2004c};\break\cite{dovciak2004} and \cite{dovciak2004a}.
Another application to random emitting spots in the
accretion disc can be found in \cite{czerny2004} but it is not included in
the present thesis.



\chapter{Transfer functions for a thin disc in Kerr space-time}
 \chaptermark{Transfer functions for a thin disc \ldots}
 \label{chapter1}
 \thispagestyle{empty}
  \label{transfer_functions}

Photons emitted from an accretion disc are affected by gravitation in
several ways. They change their energy due to the gravitational and
Doppler shifts because photons are emitted by matter moving close to the source of
strong gravitation.
The cross-section of light tube changes as the photons propagate in the curved
space-time. This effect is strongest for high inclinations of the observer
when trajectories of photons coming from behind the black hole are bent very
much. Bending of light also influences the effective area from which photons
arrive due to different emission angles. Here, aberration caused by motion of
the disc matter plays its role as well. All three of these effects, the
$g$-factor, lensing and emission angle, influence the intensity of light that
the observer at infinity measures.

If the local emission from the disc is non-stationary one
must take into account the relative delay with which photons from different
parts of the disc arrive to the observer.
A part of the local emission may be due to the reflection of the light incident
on the disc from the corona above. In this case the disc radiation may be
partially polarized. The polarization vector changes as the light propagates
through the curved space-time. The change of polarization angle that occurs
between the emission of light at the disc and its reception by the observer at
infinity is needed for calculation of the overall polarization that the
observer measures. When calculating the local polarization we need to know the
geometry of incident and emitted light rays. Therefore we need to know also the
azimuthal emission angle of the emitted photons.
Thus six functions are necessary to transfer the local flux to the observer at
infinity. The exact definitions for all of them will be given in next
sections. For simplicity we will refer to them as transfer functions.

Various authors have computed radiation from matter moving around a black hole
in different approximations and in different parameter space.
De Felice, Nobili \& Calvani \citeyearpar{felice1974}
computed the effects of gravitational dragging on the electromagnetic radiation
emitted by particles moving on bound orbits around a Kerr black hole.
\mbox{\cite{cunningham1975}} studied the combined effects of the gravitational and
Doppler shifts together with the gravitational lensing effect on the X-ray
radiation from an accretion disc in the strong gravitational field of a black
hole. He introduced a concept of a transfer function where he includes all of
the above-mentioned effects into a single function that describes the overall
influence of the gravitational field on light rays emerging from the disc. He
also studied self-irradiation of the disc and its impact on the disc emission
(Cunningham \citeyear{cunningham1976}). Since then numerous authors have
investigated emission from disc (line, blackbody, reflected, etc.) in strong
gravity -- \cite{gerbal1981,asaoka1989};\break\cite{fabian1989,kojima1991,
laor1991,karas1992};\break\cite{viergutz1993};\hspace*{1em}\cite{bao1994};
\hspace*{1em}\cite{hameury1994};
\cite{zakharov1994,speith1995,bromley1997};\break\cite{martocchia2000,
cadez2003,beckwith2004}. In the situation when
full spectral resolution is not available, power spectra of light curves provide
us with partial information on the emission from accretion discs
\citep[see e.g.][]{abramowicz1991,schnittman2004}.
Other works also included the studies of net polarization from accretion discs,
here
we should mention\break\cite{connors1980,laor1990,chen1991};\break
\cite{matt1993b,matt1993,agol1997,bao1997};\break
\cite{ogura2000}. All of the above-mentioned authors used various
methods
for computing the transfer of photons from the accretion disc to the observer
far away from the central black hole. Some of them assumed the black hole to be
non-rotating, others performed their calculations for a rotating Kerr black
hole. \cite{karas1995} and \cite{usui1998} explored a substantially more
complex case when self-gravity of the disc is not neglected and photons move
in a space-time, the metric of which is itself derived from a numerical solution
of Einstein's equations. Similar approach is appropriate also for ray-tracing
in the field of a fast rotating neutron star, and in this way these techniques
are pertinent to the study of sources of quasi-periodic oscillations in X-ray
binaries.

Our numerical codes for computing the emission from accretion discs
(see Chapter~\ref{chapter3}) have to be fast, so
that they are suitable for fitting data. That is the reason for which
we have decided to pre-calculate the transfer functions and store them in the
form of tables in a
FITS\footnote{Flexible Image Transport System. See e.g.\ \cite{hanisch2001}
or \href{http://fits.gsfc.nasa.gov/}{\tt http://fits.gsfc.nasa.gov/} for
specifications.} file (see Appendix~\ref{fits}). We have chosen to
compute the transfer of photons from the infinity (represented by setting
the initial radius $r_{\rm i}=10^{11}$ in our numerical calculations) to the
disc numerically by
solving the equation of the geodesic. For this purpose the Bulirsch-Stoer method
of integration has been used (alternatively, the semi-analytical method of
elliptical integrals could be used, see e.g.\ \citealt{rauch1994}).
An optically thick and geometrically thin flat disc has been assumed. Only null
geodesics starting at the observer at infinity and ending at the equatorial
plane of the black hole, without crossing this plane, has been taken into
account (higher order images of the disc have not been computed).
The Kerr space-time has been assumed and the medium between the disc and the
observer has been supposed to be optically thin
for the wavelengths that we are interested in (X-rays in our case). Special Kerr
ingoing coordinates, which are non-singular on the
horizon and bring spatial infinity to a finite value (to zero),
have been used (see Appendix~\ref{kerr}). This has enabled us
to calculate the transfer functions with a high precision even very close to the
horizon of the black hole. By using these coordinates the
Boyer-Lindquist azimuthal coordinate $\varphi$ has been unfolded to the Kerr
coordinate $\varphi_{\rm K}$, which is more precise when interpolating between
the computed values on the disc. Some of the transfer functions depend on the
motion of the matter in the disc, which has been assumed to be Keplerian above
the marginally stable orbit $r_{\rm ms}$ and freely falling below it, where it
has the same energy and angular momentum as the matter which is on the marginally
stable orbit. All of the transfer functions
except lensing have been actually calculated analytically (see formulae in the
next sections) but with the numerical mapping of the impact parameters $\alpha$
and $\beta$ to the disc coordinates $r$ and $\varphi_{\rm K}$.
The lensing has been computed
by numerical integration of the equation of the geodesic deviation by the same
method as described above. The initial conditions for the numerical integration
are summarized in Appendix~\ref{initial}. In numerical computations,
approximately $10^5$
geodesics have been integrated. They have covered the disc from the horizon to
an outer radius of $r_{\rm out}\sim 1000$ in such a way that the region closest
to the horizon has been most densely covered. Thus we have obtained values of
transfer functions on a non-regular grid over the disc which have been
interpolated to a regular grid in $r$ and $\varphi_{\rm K}$ coordinates
using Delaunay triangulation.
Graphical representation of the transfer functions can be seen in
Appendix~\ref{atlas}.

In the following sections we summarize the equations used for the evaluation of
particular transfer functions over the disc. In this chapter as well as
everywhere else in the thesis we use units where
$G\mbh=c=1$ with $\mbh$ being the mass of the central black hole.

\section{Gravitational and Doppler shift}

The four-momentum of the photons emitted from the disc
is given by (see e.g.\ \citealt{carter1968} and \citealt{misner1973})
\begin{eqnarray}
\label{eq1}
p_{\rm e}^t & = & [a(l-a)+(r^2+a^2)(r^2+a^2-al)/\Delta]/r^2\, ,\\
p_{\rm e}^r & = & {\rm R}_{\rm sgn}\{(r^2+a^2-al)^2-
\Delta[(l-a)^2+q^2]\}^{1/2}/r^2\, ,\\
p_{\rm e}^{\theta} & = & -q/r^2\, ,\\
\label{eq4}
p_{\rm e}^{\varphi} & = & [l-a+a(r^2+a^2-al)/\Delta]/r^2\, ,
\end{eqnarray}
where $a$ is the dimensionless angular momentum of the black hole
($0\leq a \leq 1$).
In eqs.~(\ref{eq1})--(\ref{eq4}) we employed usual notation
(see Appendix~\ref{kerr} for the exact meaning of the quantities).

The combined gravitational and Doppler shift ($g$-factor) is defined as the
ratio of the energy of a photon received by an observer at infinity
to the local energy of the same photon when emitted from an accretion disc
\begin{equation}
\label{gfac}
g=\frac{\nu_{\rm o}}{\nu_{\rm e}}=\frac{{p_{{\rm o}\,t}}}{{p_{{\rm e}\,\mu}}\,
U^{\mu}}=-\frac{1}{{p_{{\rm e}\,\mu}}\,
U^{\mu}}\, .
\end{equation}
Here $\nu_{\rm o}$ and $\nu_{\rm e}$ denote the frequency of the observed and
emitted photons respectively, and $U^\mu$ is a four-velocity of the matter
in the disc.

\begin{figure}[tb]
\dummycaption\label{mm_gfac_angle}
\includegraphics[width=0.5\textwidth]{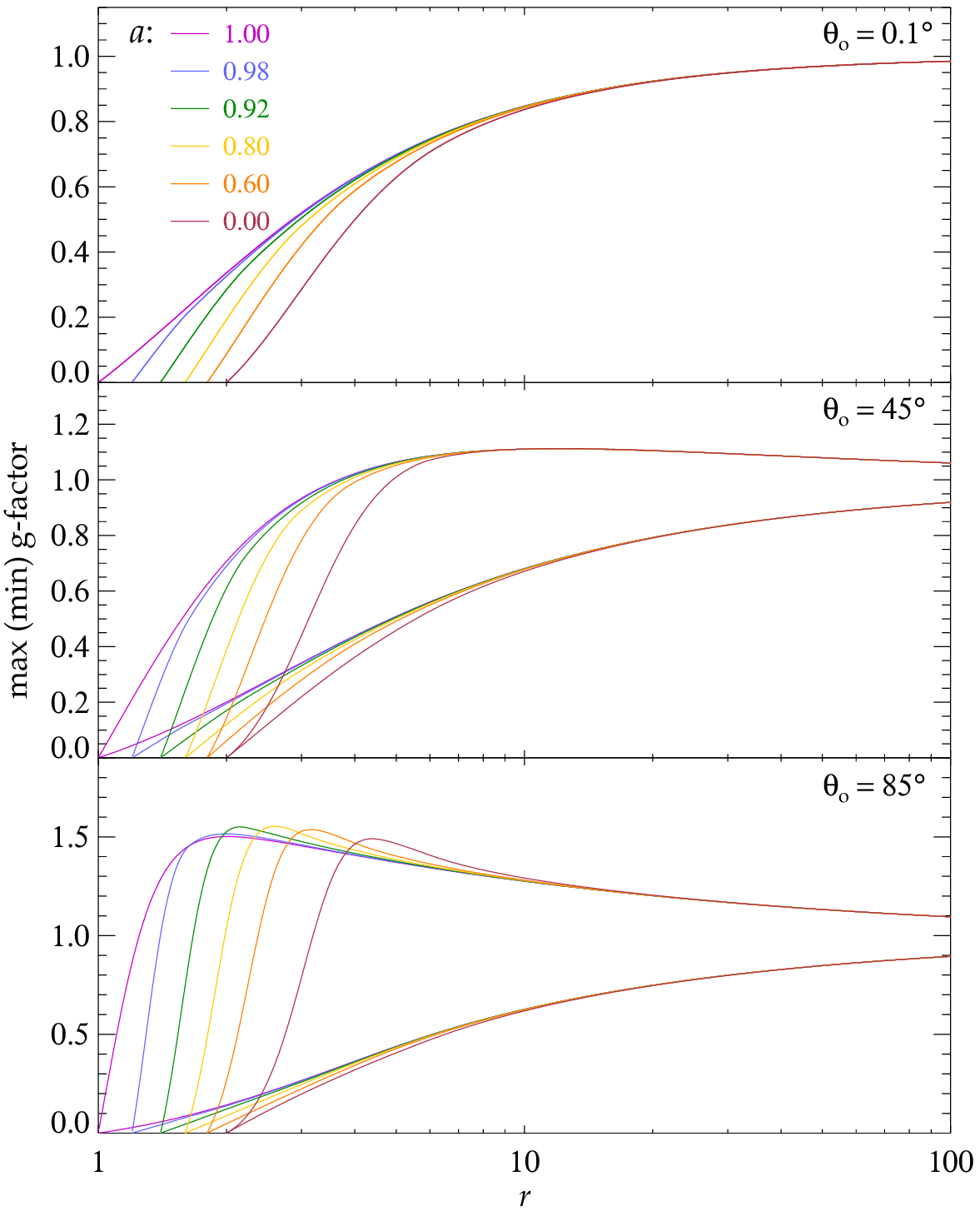}
\includegraphics[width=0.5\textwidth]{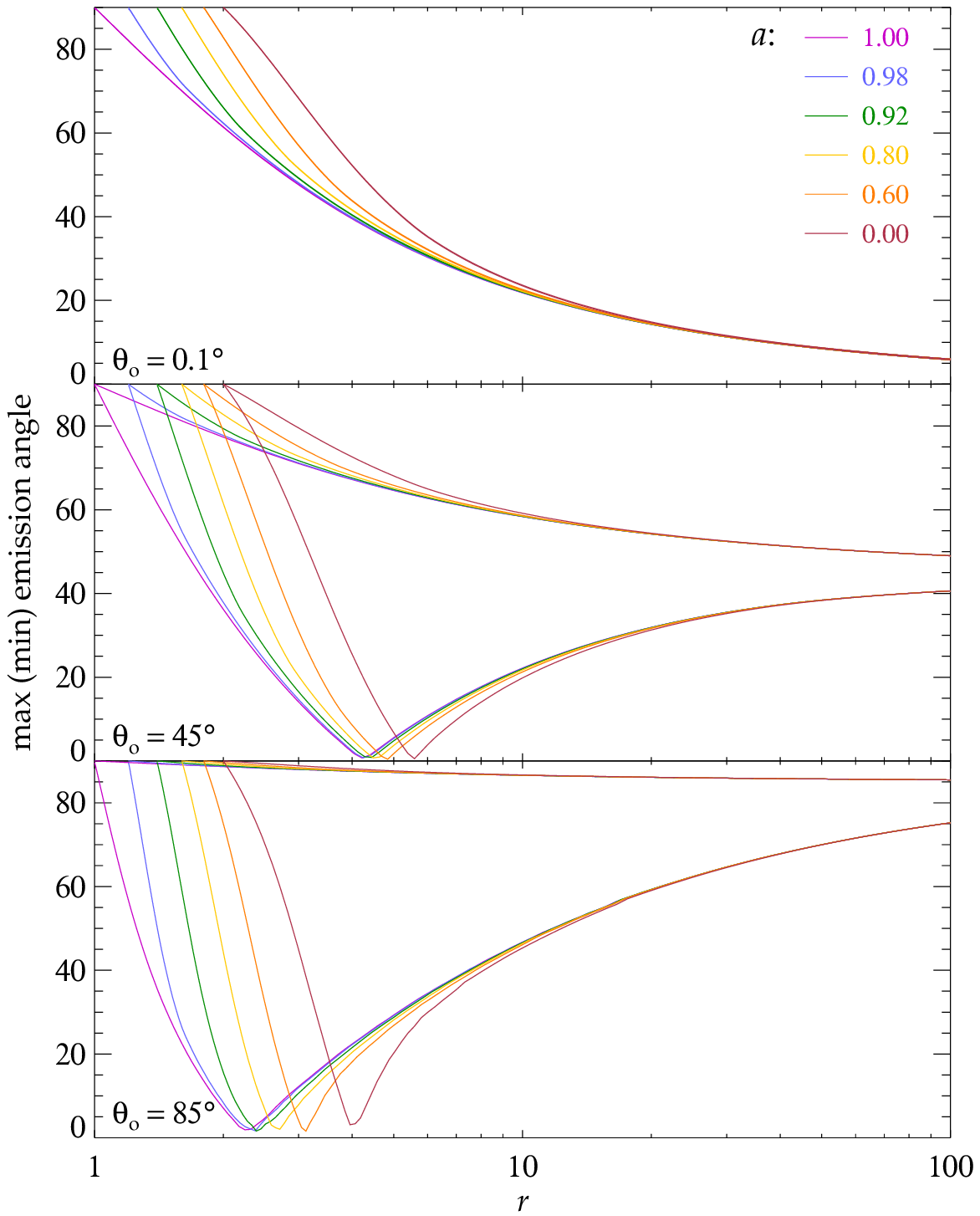}
\mycaption{Maximum and minimum values of energy shift (left) and emission angle
(right) of photons originating from different radii in the disc. The dependence
on the black-hole angular
momentum, $a$, is shown for three different inclination angles,
$\theta_{\rm o}=0.1^\circ,\,45^\circ$ and $85^\circ$.
Each curve corresponds to a fixed value of $a$ in the range $\langle0,1\rangle$,
encoded by line colours.}
\vspace*{1.3em}
\end{figure}

\begin{figure}[tb]
\vspace*{1.9em}
\dummycaption\label{mm_gfac_NN}
\includegraphics[width=0.5\textwidth]{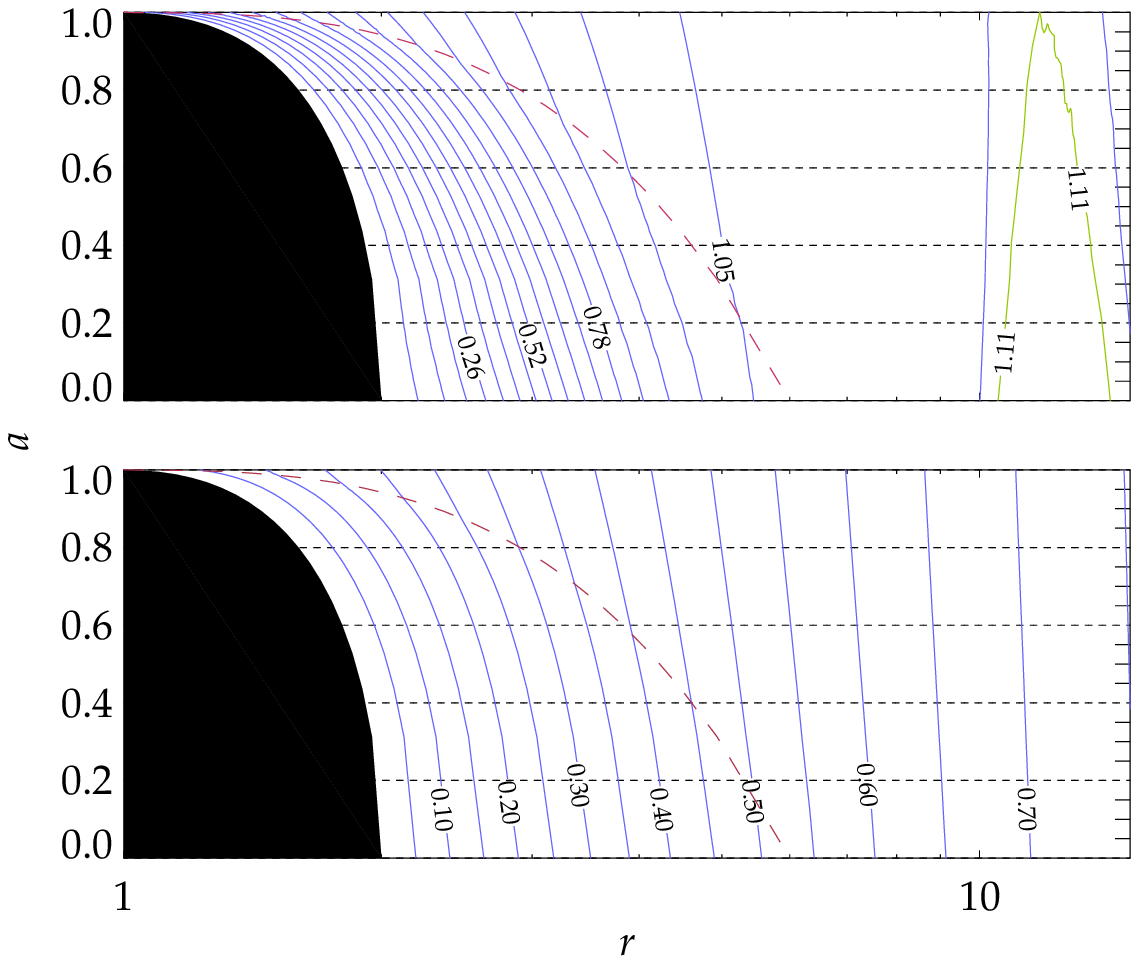}
\includegraphics[width=0.5\textwidth]{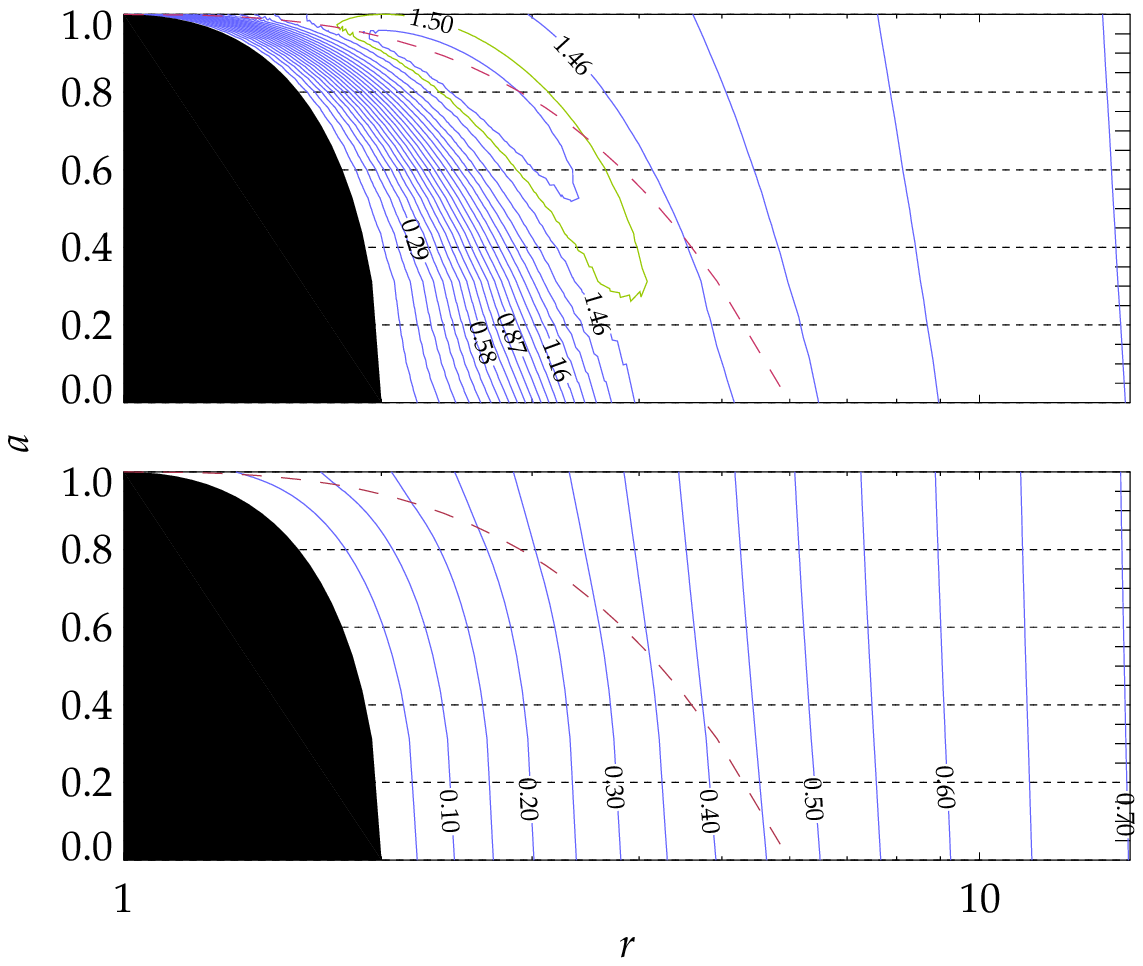}
\vspace*{0.5em}
\mycaption{Contour lines of extremal energy shift factors in the plane of the
rotation parameter $a$ versus radius $r$ for observer inclination
$\theta_{\rm o}=45^\circ$ (left) and $85^\circ$ (right). Maximum $g$-factor
is shown at the top, minimum $g$-factor at the bottom. Radius is in units of $GM/c^2$.
The horizon is shown in black. The curve $a(r)_{|r=r_{\rm{}ms}}$ (dashed) is also
plotted across the contour lines. Notice that, in the traditional
disc-line scheme, no radiation is supposed to originate from radii below marginally
stable orbit $r_{\rm ms}$. If this is the case then
one must assume that all photons originate outside the dashed curve
and so the effect of frame-dragging is further reduced.}
\vspace*{1.7em}
\end{figure}

  Contour graphs of the $g$-factor in Appendix~\ref{atlas} show that the effect of
rotation of the black hole is visible only in its vicinity
($r\lesssim10$). Near the horizon the $g$-factor decreases down to zero due to
the gravitational redshift, whereas far from the black hole the Doppler shift
prevails. Very far from the black hole, where matter of the disc rotates slowly,
the $g$-factor goes to unity.

Preliminary considerations on the line emission (its energy shift and its width)
can be based on the extremal values of the energy shift, $g_{+}$ and
$g_{-}$, which  photons experience when arriving at the observer's
location from different parts of the disc. This is particularly relevant
for some narrow lines whose redshift can be determined more accurately
than if the line is broad (e.g.\ \citealt{turner2002,guainazzi2003,
yaqoob2003}). Careful discussion of $g_{\pm}$ can, in principle,
circumvent the uncertainties which are introduced by the uncertain form of
the intrinsic emissivity of the disc and yet still constrain some of
parameters. Advantages of this technique were pointed out already by
\cite{cunningham1975} and it was further developed by \cite{pariev2001}.

Fig.~\ref{mm_gfac_angle} shows the extremal values of the
redshift factor for the observer inclinations
$\theta_{\rm{}o}=0.1^{\circ},45^{\circ}$ and
$85^{\circ}$. These were computed along circles with radius $r$
in the disc plane. Corresponding contour lines of constant
values of $g_{\pm}$ in the plane $a$ versus $r$ can be seen
in Fig.~\ref{mm_gfac_NN}.
In other words, radiation is supposed to originate from radius
$r$ in the disc, but it experiences a different redshift depending
on the polar angle. Contours of the redshift factor provide a very useful
and straightforward technique to determine the position of a flare or a
spot, provided that a narrow spectral line is produced
and measured with sufficient accuracy. The most direct
use of extremal $g$-values would be if one were able to
measure variations of the line profile from its lowest-energy
excursion (for $g_{-}$) to its highest-energy excursion
(for $g_{+}$), i.e.\ over a complete cycle.
Even partial information can help to constrain models (for example,
the count rate is expected to dominate the observed line at the
time that $g=g_{+}$, and so the high-energy peak is
easier to detect). While
it is very difficult to achieve sufficient precision on the highly
shifted and damped red wing of a broad line,
prospects for using narrow lines are indeed interesting,
provided that they originate close enough to the black hole and that
sufficient resolution is achieved both in the energy and time domains.

Fig.~\ref{mm_gfac_NN} also makes it clear that it is possible in
principle (but intricate in practice) to deduce the $a$-parameter value
from spectra. It can be seen from the redshift factor that
the dependence on $a$ is rather small and it quickly becomes
negligible if the light from the source is dominated by contributions
from $r\gtrsim10$.

\section{Emission angle}
Here we examine the angle of emission with respect to
the disc normal. We assume the local frame co-moving with the
medium of the disc. The cosine of the local emission angle is
\begin{equation}
\label{cosine}
\mu_{\rm e}=\cos{\delta_{\rm e}}=
-\displaystyle\frac{{p_{{\rm e}\,\alpha}\,n^{\alpha}}}
{{p_{{\rm e}\,\mu}\,U^{\mu}}}\, ,
\end{equation}
where $n^\alpha=-e_{(\theta)}^\alpha$ are components of the disc normal.

One can see from contours of $\mu_{\rm e}$ plotted in the disc plane
(Appendix~\ref{atlas}) and from graphs of its maximum and minimum values
(Fig.~\ref{mm_gfac_angle}) that gravitation influences visibly only those
light rays that pass close to the black hole on their way to the observer,
and especially for large inclination angles. Due to the effect of light
bending and aberration there exists a point on the disc where the emission
angle is zero, i.e.\ the local direction of the emission is perpendicular to the
disc surface. On the other hand, near the horizon only the light rays emitted
almost parallel to the disc plane can reach the observer.
Very far from the black hole, where matter rotates slowly, the
emission angle is determined mainly by special-relativistic aberration
and it gradually approaches the value of the observer inclination.

\begin{figure}[tb]
\dummycaption\label{mm_lens}
\includegraphics[width=0.5\textwidth]{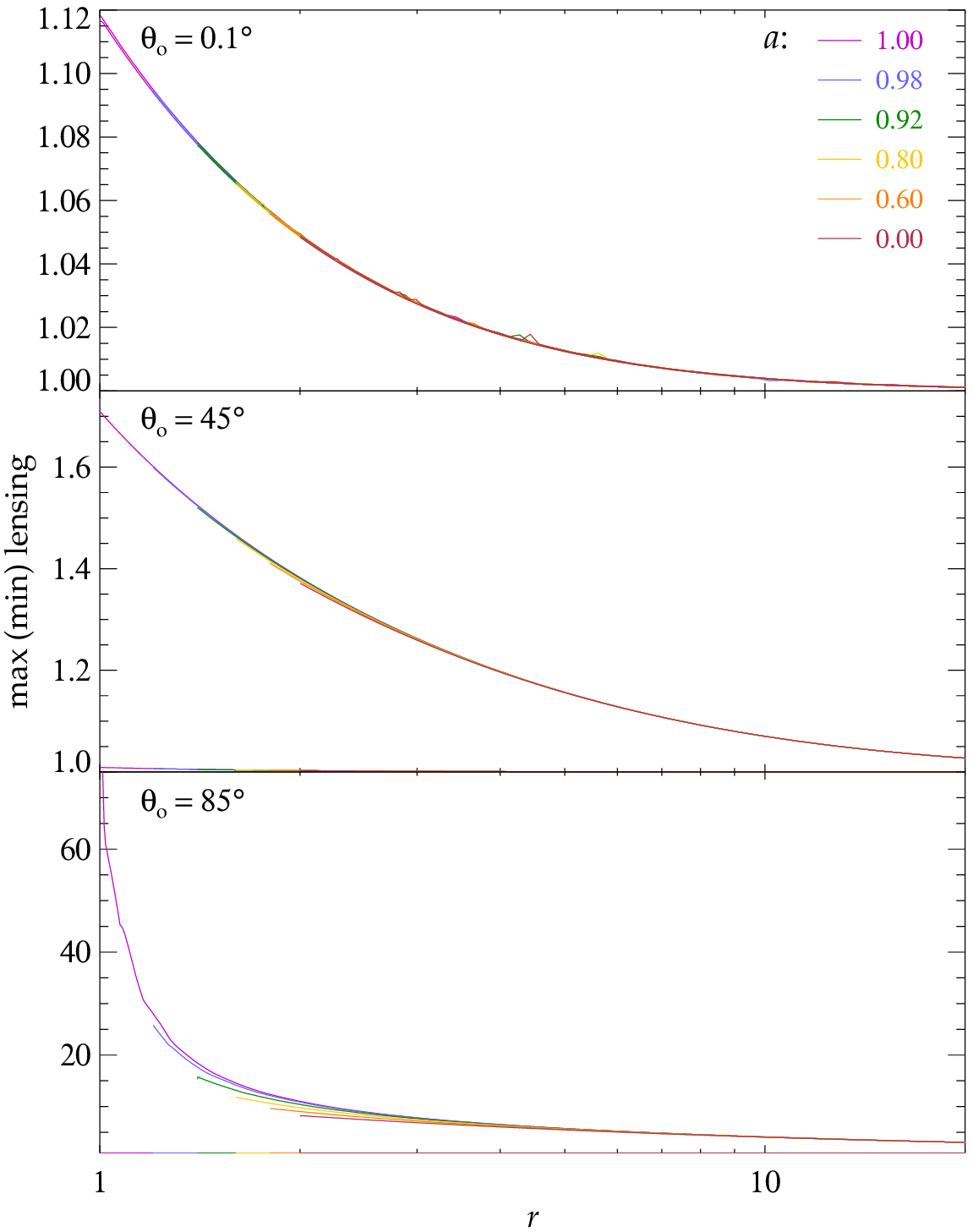}
\includegraphics[width=0.5\textwidth]{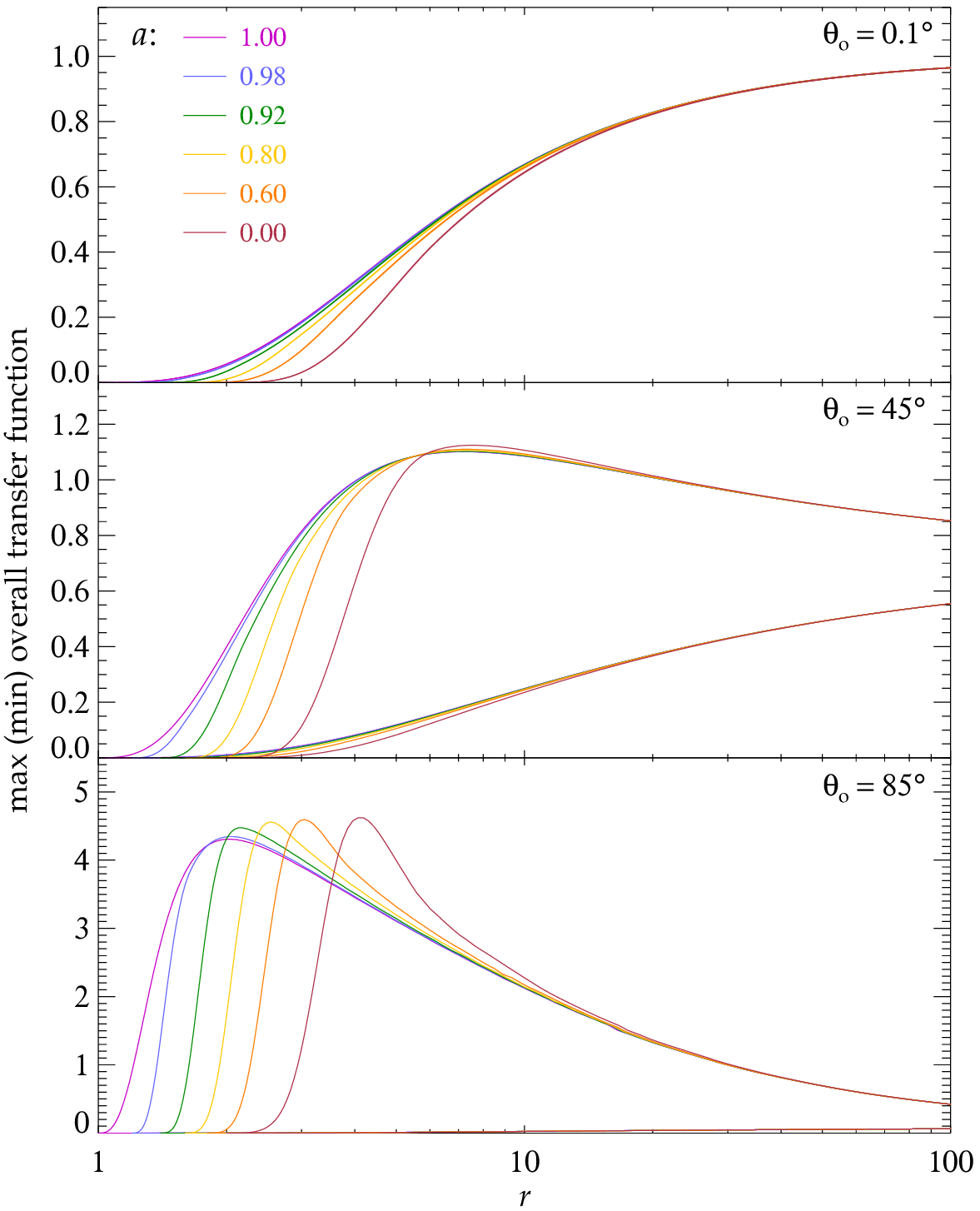}
\mycaption{Same as in Fig.~\ref{mm_gfac_angle} but for lensing (left) and
overall transfer function (right).}
\end{figure}

\section{Gravitational lensing}
We define lensing as the ratio of the cross-section ${\rm d}S_{\rm f}$ of the
light tube at infinity
to the cross-section ${\rm d}S_\perp$ of the same light tube at the disc:
\begin{equation}
\label{lensing}
l = \frac{{\rm d}S_{\rm f}}{{\rm d}S_\perp}=\frac{1}{\sqrt{\|Y_{\rm e1}\|^2
\|Y_{\rm e2}\|^2 - <Y_{\rm e1},Y_{\rm e2}>^2}}\, .
\end{equation}
The four-vectors $Y_{\rm e1}$ and $Y_{\rm e2}$ are transported
along the geodesic according to the equation of the geodesic deviation from
infinity where they are unit, space-like and perpendicular
to each other and to the four-momentum of light (for the exact definition of
these vectors very far from the black hole see Appendix~\ref{initial}).
In eq.~(\ref{lensing}) we have denoted the magnitude of a four-vector by
$\|\ \|$ and scalar product of two four-vectors by $<\ ,\ >$.
The cross section of light tube is constant for all observers
(see \citealt{schneider1992}).

Lensing can significantly amplify the emission from some parts of the disc
(located behind the black hole from the point of view of the observer and in
Kerr ingoing coordinates). This is true mainly for observers with large
inclination angles (see figures in Appendix~\ref{atlas} and Fig.~\ref{mm_lens}).

When one wants to estimate the total effect that gravitation has on the
intensity of light coming from different parts of the disc, one has to take into
account all the three effects -- the \mbox{$g$-factor}, the emission angle and
the lensing. These effects combine into a single function defined by
(see~eq.~(\ref{emission}))
\begin{equation}
\label{F}
F = g^2\mu_{\rm e}\,l\, .
\end{equation}
We call it the overall transfer function.
See contour graphs in Appendix~\ref{atlas} and
plots in Fig.~\ref{mm_lens} for the total gravitational amplification of photon
flux emitted from a Keplerian accretion disc.

\begin{figure}[tb]
 \begin{center}
  \dummycaption\label{mm_delay}
  \includegraphics[width=0.5\textwidth]{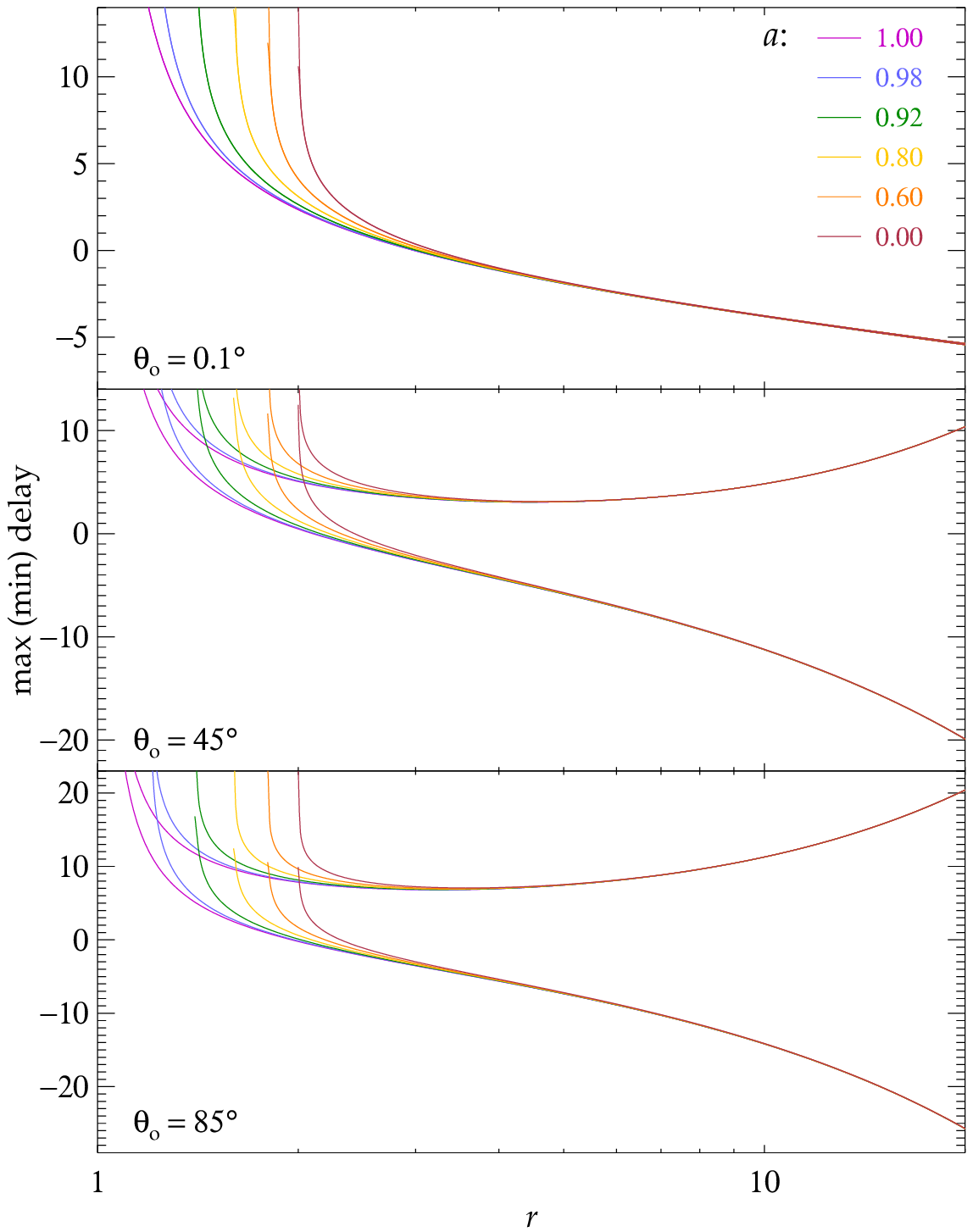}
  \mycaption{Same as in Fig.~\ref{mm_gfac_angle} but for maximum and minimum
	relative time delay.}
 \end{center}
\end{figure}

\section{Relative time delay}
The relative time delay $\Delta t$ is the Boyer-Lindquist time which elapses
between the emission of a photon from the disc and its reception by an observer
(plus a certain constant so that the
delay is finite close to the black hole but not too close). We integrate
the equation of the geodesic in Kerr ingoing coordinates and thus we calculate
the delay in the Kerr ingoing time coordinate $\Delta t_{\rm K}$. The
Boyer-Lindquist time can be obtained from $\Delta t_{\rm K}$
using the following equation:

\begin{figure}[tbh]
 \begin{center}
  \dummycaption\label{pol_angle}
  \begin{tabular}{c@{\quad}l}
   \parbox[t]{5cm}{\vspace*{1cm}\includegraphics[height=5cm]{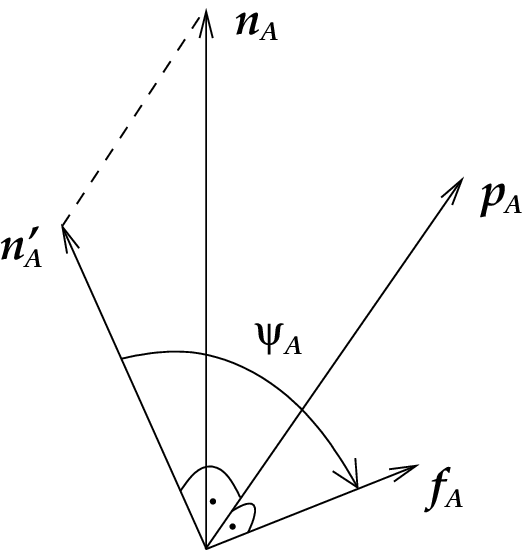}} &
   \parbox[t]{9cm}{\vspace*{-0.3cm}
     \begin{itemize} \itemsep -2pt
       \item [\rm(i)] Let three-vectors $\vec{p_A}$, $\vec{n_A}$, $\vec{n'_A}$
and $\vec{f_A}$ be the momentum of a photon, normal to
the disc, projection of the normal to the plane perpendicular to the momentum
and a vector which is parallelly transported along the geodesic (as four-vector),
respectively;
       \item [\rm (ii)] let $\Psi_A$ be an angle between $\vec{n'_A}$
and $\vec{f_A}$;
       \item [\rm (iii)] let the quantities in (i) and (ii) be evaluated at
the disc for $A=1$ with respect to the local rest frame co-moving with the
disc, and at infinity for \hbox{$A=2$} with respect to the stationary observer
at the same light geodesic;
       \item [\rm (iv)] then the change of polarization angle is defined
as $\Psi=\Psi_2-\Psi_1$.
     \end{itemize}}
  \end{tabular}
  \mycaption{Definition of the change of polarization angle $\Psi$.}
 \end{center}
\end{figure}

\begin{equation}
{\rm d}t = {\rm d}t_{\rm K}-\left [ 1+ \frac{2r}{(r-r_+)(r-r_-)}\right ]
{\rm d}r\, ,
\end{equation}
with $r_{\pm}=1\pm\sqrt{1-a^2}$ being inner ($-$) and outer ($+$) horizon of the
black hole.
Then we integrate the above equation and define the time delay as
\begin{myeqnarray}
\label{delay1}
\Delta t & = & \Delta t_{\rm K}-\left [r+\frac{2}{r_+ - r_-}\,
\ln{\frac{r-r_+}{r-r_-}}+\ln{[(r-r_+)(r-r_-)]}\right ]
& & {\rm for}\ \ a < 1\, ,\hspace*{2em}\\
\label{delay2}
\Delta t & = & \Delta t_{\rm K}-\left [r-\frac{2}{r-1}+2\ln{(r-1)}\right ]
& & {\rm for}\ \ a = 1\, .
\end{myeqnarray}
There is a minus sign in front of the brackets because the direction
of integration is from infinity to the disc.

For contour graphs of the relative time delay see figures in
Appendix~\ref{atlas} and for plots of maximum and minimum delay at a constant
radius see Fig.~\ref{mm_delay}.

\section{Change of polarization angle}
Various physical effects can influence polarization of light as it propagates
towards an observer. Here we examine only the influence of the gravitational
field represented by the vacuum Kerr space-time.
The change of the polarization angle is defined as the angle by which a vector
parallelly transported along the light geodesic rotates with respect to some
chosen frame. We define it in this
way because in vacuum the polarization vector is parallelly transported along
the light geodesic. This angle depends on the choice of the local frame at
the disc and at infinity. See Fig.~\ref{pol_angle} for an exact definition.

The change of polarization angle is \citep[see][]{
connors1977,connors1980}
\begin{equation}
\label{polar}
\tan{\Psi}=\frac{Y}{X}\, ,
\end{equation}
where
\begin{eqnarray}
X & = & -(\alpha-a\sin{\theta_{\rm o}})\kappa_1-\beta\kappa_2\, ,\\
Y & = & \phantom{-}(\alpha-a\sin{\theta_{\rm o}})\kappa_2-\beta\kappa_1\, ,
\end{eqnarray}
with $\kappa_1$ and $\kappa_2$ being components of the complex constant
of motion $\kappa_{\rm pw}$ (see Walker \& Penrose \citeyear{walker1970})
\begin{eqnarray}
\kappa_1 & = & arp_{\rm e}^\theta f^t-r\,[a\,p_{\rm e}^t-(r^2+a^2)\,
p_{\rm e}^\varphi]f^\theta-r(r^2+a^2)\,p_{\rm e}^\theta f^\varphi\, ,\\
\kappa_2 & = & -r\,p_{\rm e}^rf^t+r\,[p_{\rm e}^t-a\,p_{\rm e}^\varphi]f^r+
arp_{\rm e}^rf^\varphi\, .
\end{eqnarray}
Here the polarization vector $f^\mu$ is a four-vector corresponding to the
three-vector
$\vec{f_1}$ from Fig.~\ref{pol_angle} which is chosen in such a way that it
is a unit vector parallel with $\vec{n'_1}$ (i.e.\ $\Psi_1=0$)
\begin{equation}
f^\mu = \frac{n^\mu-\mu_{\rm e}\left( g\,p_{\rm e}^\mu-U^\mu\right)}
{\sqrt{1-\mu_{\rm e}^2}}\, .
\end{equation}
For contour graphs of the change of the polarization angle see figures in
Appendix~\ref{atlas}.

\section{Azimuthal emission angle}
\label{azimuth_angle}
We define the azimuthal emission angle as the angle between the projection of
the three-momentum of the emitted photon into the equatorial plane (in the local
rest frame co-moving with the disc) and the radial tetrad vector:
\begin{equation}
\label{azim_angle}
\Phi_{\rm e}=\arctan\frac{p_{\rm e}^\alpha\,e_{(\varphi)\,\alpha}}
{p_{\rm e}^\mu\,e_{(r)\,\mu}}=\arctan\left( g\,\frac{\Delta}{r}\,
\frac{-p_{\rm e}^t\,U^\varphi+p_{\rm e}^\varphi\,U^t}{g\,p_{\rm e}^r-U^r}\right)\, .
\end{equation}
The explicit formula for this angle is not necessary for computations of
light curves and spectral profiles, but it appears in discussions of
polarimetry. This is also the reason why we have computed $\Phi_{\rm{}e}$.
Like the quantities discussed previously, the resulting values of the azimuthal
angle depend on the adopted geometry of the source and the rotation law of
the medium. They are determined by mutual interplay of special- and
general-relativistic effects. Therefore, we remind the reader that our
analysis applies to geometrically thin Keplerian discs residing in the
equatorial plane, although generalization to more complicated situations
should be fairly straightforward. For contour graphs of the change of the
azimuthal emission angle see figures in Appendix~\ref{atlas}.


\chapter{Radiation of accretion discs in strong gravity}
 \thispagestyle{empty}
 \label{chapter2}

There is now plausible evidence that the emission in some
active galactic nuclei and some Galactic X-ray binary black-hole
candidates originates, at least in part, from an accretion
disc in a strong gravitational field. A lively debate is aimed at
addressing the question of what the spectral line profiles and the
associated continuum can tell us about the  central black hole, and
whether they can be used to constrain parameters of the accretion disc
in a nearby zone, about ten gravitational radii or less from the
centre. For a recent review of AGNs see \cite{fabian2000,
reynolds2003}, and references cited therein. For BHCs see
\cite{miller2002a,mcclintock2003} and references
therein. In several sources there is indication of iron K$\alpha$ line
emission from within the last stable orbit of a Schwarzschild black hole,
e.g.\ in the Seyfert galaxy MCG--6-30-15 \break \citep[see][]{iwasawa1996,
fabian2000,wilms2001,martocchia2002a} or from the region near above
the marginally stable orbit, as in the case of the X-ray transient source
XTE J1650--500 \citep{miniutti2004} that has been identified with a
Galactic black-hole candidate. In other cases the emission
appears to arise farther from
the black hole (e.g.\ in the microquasar GRS 1915+105, see
\citealt{martocchia2002b}; for AGNs see \citealt{yaqoob2004} and references
therein). Often, the results from X-ray line spectroscopy are
inconclusive, especially in the case of low spectral resolution data.
For example, a spinning  black hole is allowed but not required by the
line model of the microquasar V4641 Sgr \citep{miller2002b}. The
debate still remains open, but  there are good prospects for future
X-ray astronomy missions to be able to use the iron K$\alpha$ line to probe
the space-time in the vicinity of a black hole, and in particular to
measure the angular momentum, or the spin, associated with the metric.

One may also be able to study the `plunge region'
(about which very little is known), between the event horizon and the
last stable orbit, and to determine if any appreciable contribution to
the iron K$\alpha$ line emission originates from there
\citep{reynolds1997,krolik2002}. In addition to the Fe~K
lines, there is some evidence for relativistic soft X-ray emission
lines due to the Ly$\alpha$ transitions of oxygen, nitrogen, and
carbon \citep{mason2003}, although the observational support
for this interpretation is still controversial \citep{lee2001}.

\begin{table}[tbh]
\begin{center}
\begin{footnotesize}
\dummycaption\label{tab:models}
\begin{tabular}{lcccccc}
\hline
\multicolumn{1}{c}{Model}
 & \multicolumn{6}{c}{Effects that are taken into account\rule[-1.5ex]{0mm}{4.5ex}} \\
\cline{2-7}
 & \rule[-3.5ex]{0mm}{8ex}\parbox{20mm}{\footnotesize Energy shift/ Lensing effect}
 & {\footnotesize $a$}
 & \parbox{26mm}{\centering\footnotesize Non-axisymmetric emission region}
 & \parbox{20.5mm}{\centering\footnotesize Emission from plunge region}
 & {\footnotesize Timing}
 & {\footnotesize Polarization} \\
\hline
{\sc{}diskline}   \rule{0mm}{3ex} &  yes/no~    & $0$~
& no & no & no & no \\
{\sc{}laor}                       &  yes/yes    & $0.998$~
& no & no & no & no \\
{\sc{}kerrspec}                   &  yes/yes    & $\langle0,1\rangle^{\dag}$
& yes~ & no & no & no \\
{\sc{}ky}                         &  yes/yes    & $\langle0,1\rangle$
& yes$^{\ddag}$ & yes & yes & yes \\
\hline
\end{tabular}
\end{footnotesize}
\vspace*{2mm}\par{}
{\parbox{0.9\textwidth}{\footnotesize $^{\dag}$~The value of the dimensionless
$a$ parameter is kept frozen.\\
$^{\ddag}$~A one-dimensional version is available for the case of an
axisymmetric disc. In this axisymmetric mode, {\sc{}ky} still allows $a$ and
other relevant parameters to be fitted (in which case the computational speed of
{\sc{}ky} is then comparable to {\sc{}laor}). The results can be more accurate
than those obtained with other routines because of the ability to tune the grid
resolution.}}
\mycaption{Basic features of the new model in comparison with other
black-hole disc-line models. For references see \cite{fabian1989},
\cite{laor1991}, \cite{martocchia2000} and \cite{dovciak2004b},
respectively.}
\end{center}
\end{table}

With the greatly enhanced spectral resolution and throughput of
future X-ray astronomy missions, the need arises
for realistic theoretical models of the disc emission
and computational tools that are powerful enough to
deal with complex models and to
allow actual fitting of theoretical models to observational
data. It is worth noting that some of the current data
have been used to address the issue of distinguishing
between different space-time metrics around a black hole,
however, the current models available for fitting X-ray data are
subject to various restrictions.

In this chapter and the next one we describe a generalized scheme and a code
which can be
used with the standard X-ray spectral fitting package {\sc{}xspec}
\citep{arnaud1996}. We have in mind general relativity models for black-hole
accretion discs. Apart from a better numerical resolution, the principal
innovations compared to the currently available schemes
(see Tab.~\ref{tab:models}) are that the new
model allows one to  (i)~fit for the black-hole spin, (ii)~study the
emission from  the plunge region, and (iii)~specify a more general form
of emissivity  as a function of the polar coordinates in the disc plane
(both for the line and for the continuum). Furthermore, it is also
possible to (iv)~study time variability of the observed signal and
(v)~compute Stokes parameters of a polarized signal. Items
(i)--(iii) are immediately applicable to current data and modelling,
while the last two mentioned features are still mainly of theoretical
interest at the present time. Time-resolved analysis and polarimetry of
accretion discs are directed towards future applications when the necessary
resolution and the ability to do polarimetry are available in X-rays.
Thus our code has the advantage that it can be used with time-resolved
data for reverberation\break studies of relativistic accretion discs
\citep{stella1990,reynolds1999,ruszkowski2000,goyder2004}.
Also polarimetric analysis can be performed, and this will be extremely
useful because it can add very specific information on
strong-gravitational field effects \citep{connors1980,matt1993,bao1997}.
Theoretical spectra with
temporal and  polarimetric information can be analysed with  the current
version of our code and such analysis should provide tighter constraints
on future models than is currently possible.

\section{Basic spectral components of X-ray sources}
There are several components in the spectra of X-ray sources. Not all of them
are always present and some of them are more prominent in certain objects than
the others.

\begin{figure}[tbh]
 \begin{center}
  \dummycaption\label{compton_reflection}
  \begin{tabular}{cc}
   a) \hspace{5.8cm} & b) \hspace{5.8cm} \\[-3mm]
   \includegraphics[height=3.5cm]{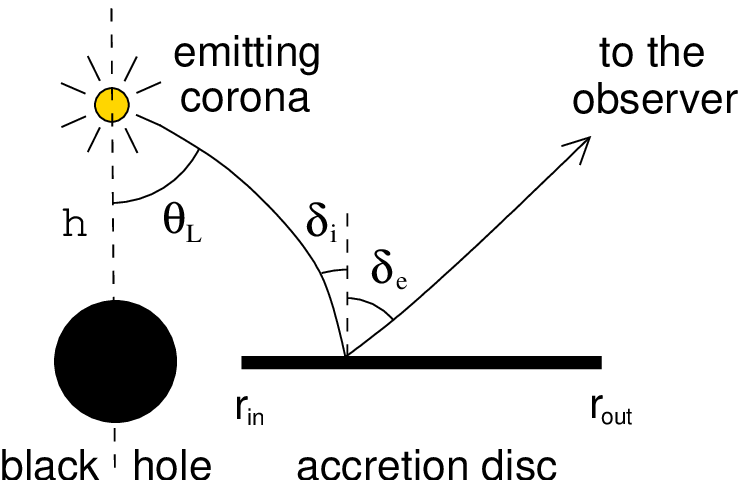} \hspace{2mm}
   & \includegraphics[height=3.5cm]{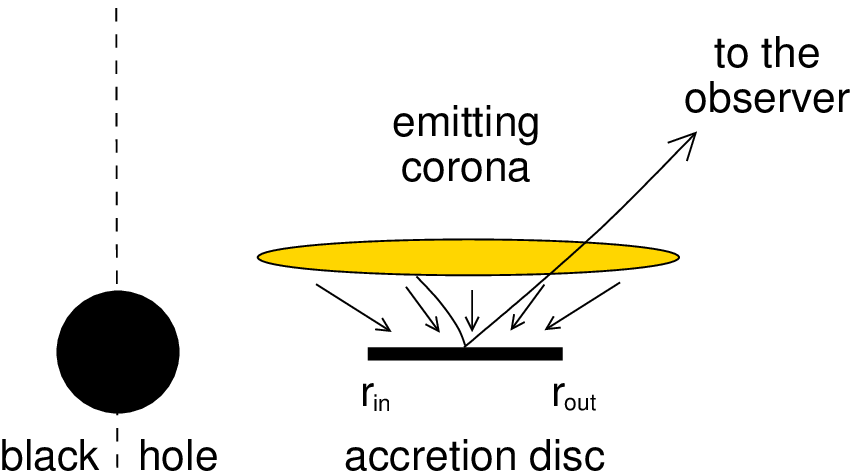}
  \end{tabular}
  \mycaption{Reflection models: a) lamp-post model; b) diffuse corona model.}
 \end{center}
\end{figure}

One of the characteristic spectral features that is almost always present is a
power-law component. It is assumed that this feature results from inverse
Compton scattering of thermal photons in a hot corona above the accretion disc.
Usually two different configurations are considered. The first one assumes a
patch of hot corona placed on the rotational axis of the central black hole
at some height above it (lamp-post model). It is usually supposed to be an
isotropic and
point-like source of stationary primary power-law emission. The second
configuration is a diffuse optically thin hot corona near above the accretion
disc. For simple sketches of both configurations see
Fig.~\ref{compton_reflection}.
In both cases the shape of this primary power-law continuum (in the rest
frame of the corona) is not affected by the relativistic effects acting on
photons during their journey to the observer at infinity. These
effects change only the normalization of the spectra.
This component extends up to a cut-off energy of some tenths or a few hundreds
keV.

Often, mainly in the spectra of active galactic nuclei, additional continuum
emission, a ``hump'', is added to the primary component. It is assumed that
this is due to the reflection of the primary emission from the illuminated disc.
The shape of the reflected continuum in the local rest frame co-moving with
the accretion disc depends mainly on photoelectric absorption and Compton
scattering of photons hitting the disc. The local emission is then smeared by
relativistic effects -- this
concerns mainly its sharp features (e.g.\ iron edge). The shape of the observed
spectra also depends on the illumination of the disc. This differs for the
lamp-post model and the diffuse corona model. In the former case the radial
dependence of illumination is determined by the height at which the patch of
corona is placed. In the latter case the illumination of the disc depends on
the emissivity of the diffuse corona near above the disc which may have quite
a complicated radial dependence (but usually is assumed to be decreasing with
radius as a power law).

Spectral lines are an important feature observed in X-ray spectra. We
assume that the origin of lines is the same as in the previous case --
the illumination of the cold
disc by primary (power-law) emission and reflection, in this case, by
fluorescence. Originally
narrow spectral lines are blurred by relativistic effects and thus they become
broad, their width being as large as several keV in several sources. The most
prominent examples are the iron lines K$\alpha$ and K$\beta$.

\begin{figure}[tbh]
 \begin{center}
  \dummycaption\label{denomination}
  \begin{tabular}{lll}
   a) \hspace{4.3cm} & b) \hspace{4.3cm} & c) \hspace{4.3cm} \\
   \hspace*{-1.65mm}\includegraphics[height=3cm]{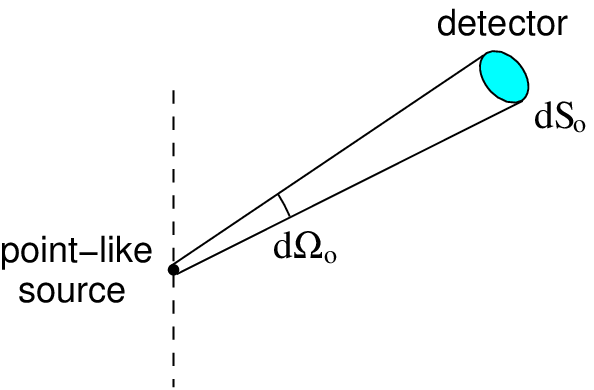} \hspace{2mm}
   & \includegraphics[height=3cm]{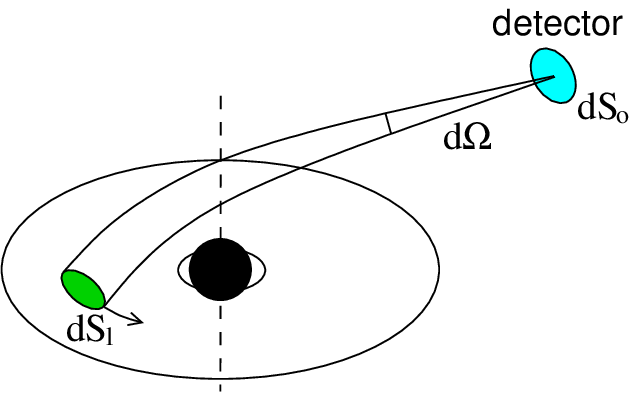} \hspace{2mm}
   & \includegraphics[height=3cm]{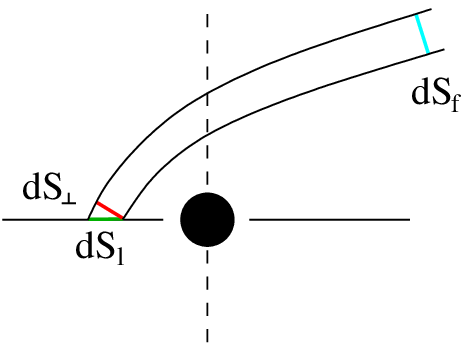}
  \end{tabular}
  \mycaption{Denomination of various elements of solid angles and areas
  defined in the text: a) the light source appears to the observer to
  be point-like;
  b) the light rays received by the detector are  coming from different parts
  of the disc (closer view of the disc than in previous figure);
  c) area of a light tube changes as the light rays travel close to the black
  hole (the disc is edge on).}
 \end{center}
\end{figure}

In the present thesis we are concerned with the spectral components described
above but here, we should not omit to mention other two important components --
the thermal emission and the warm absorber.
The thermal emission is more noticeable in the spectra of X-ray binaries with
the black-hole candidate as the central object. This is due to the fact that in
these sources the temperature of the accretion disc is much higher and therefore
the black body emission extends as far as the soft X-ray energy band. This type
of emission manifests itself in the spectra of active galactic nuclei as a
``big blue bump'' in the optical and UV frequencies.

If there is ionized matter, a warm absorber, in the line of sight of the
observer then the absorption spectral features are present in the observed
spectra. These can include edges or resonant absorption lines due to oxygen and
other ions.

\section{Photon flux from an accretion disc}
Properties of radiation are described in terms of photon numbers. The
source appears as a point-like object for a distant observer, so that
the observer measures the flux entering the solid angle ${\rm d}
\Omega_{\rm o}$, which is associated with the detector area
${\rm d}S_{\rm o}{\equiv}D^2\,{\rm d}\Omega_{\rm o}$
(see Fig.~\ref{denomination}a). This relation
defines the distance $D$ between the observer and the source.
We denote the total photon flux received by a detector,
\begin{equation}
\label{flux}
N^{S}_{\rm{}o}(E)\equiv\frac{{\rm d}n(E)}{{\rm d}t\,{\rm d}S_{\rm o}}
={\int}{\rm d}\Omega\,N_{\loc}(E/g)\,g^2\,,
\end{equation}
where
\begin{equation}
N_{\loc}(E_{\loc})\equiv
\frac{{\rm d}n_{\loc}(E_{\loc})}{{\rm d}\tau\,{\rm d}S_{\loc}\,
{\rm d}\Omega_{\loc}}
\end{equation}
is a local photon flux emitted from the surface of the disc, ${\rm
d}n(E)$ is the number of photons with energy in the interval
${\langle}E,E+{\rm d}E\,\rangle$ and $g=E/E_{\loc}$ is the redshift
factor. The local flux, $N_\loc(E_\loc)$, may vary over the disc as well
as in time, and it can also depend on the local emission angle. This
dependency is emphasized explicitly only in the final formula
(\ref{emission}), otherwise it is omitted for brevity.

The emission arriving at the detector within the solid angle ${{\rm d}\Omega}$
(see Fig.~\ref{denomination}b) originates from the proper area ${\rm
d}S_{\loc}$ on the disc (as measured in the rest frame co-moving with
the disc). Hence, in our computations we want to integrate the flux
contributions over a fine mesh on the disc surface. To achieve this aim, we
adjust eq.~(\ref{flux}) to the form
\begin{equation}
\label{N_o^S}
N^S_{\rm o}(E)=\frac{1}{D^2}\int {\rm d}S\,\frac{D^2{\rm d}\Omega}{{\rm d}S}\,
N_{\loc}(E/g)\,g^2=\frac{1}{D^2}\int {\rm d}S\,
\frac{{\rm d}S_{\perp}}{{\rm d}S}\,\frac{{\rm d}S_{\rm f}}{{\rm d}S_{\perp}}\,
N_{\loc}(E/g)\,g^2\, .
\end{equation}
Here ${{\rm{}d}S_{\rm f}}$ stands for an element of area perpendicular to light
rays corresponding to the solid angle ${\rm d}\Omega$ at a distance $D$,
${{\rm{}d}S_{\perp}}$ is the proper area measured in the local frame
of the disc and perpendicular to the
rays, and ${\rm d}S$ is the coordinate area for integration.
We integrate in a two-dimensional slice of a four-dimensional space-time, which
is specified by coordinates $\theta=\pi/2$ and $t=t_{\rm o}-\Delta t$ with
$\Delta t$ being a time delay with which photons from different parts of the
disc (that lies in the equatorial plane) arrive to the observer (at the same
coordinate time $t_{\rm o}$). Therefore, let us define the coordinate area by
(we employ coordinates $t',\,\theta,\,r,\,\varphi$
with $t'=t-\Delta t$ and $\Delta t=\Delta t(r,\theta,\varphi)$)
\begin{equation}
\label{dS}
{\rm d}S\equiv|{\rm d}^2{\!S_{t'}}^{\theta}|=\left|\frac{\partial x^\mu}
{\partial t'}{\rm d}^2{\!S_{\mu}}^{\theta}\right|=|{\rm d}^2{\!S_t}^{\theta}|=
|g^{\theta\mu}{\rm d}^2\!S_{t\mu}|\, .
\end{equation}
We define the tensor ${\rm d}^2\!S_{\alpha\beta}$ by two four-vector elements
${\rm d}x_1^\mu\equiv({\rm d}t_1,{\rm d}r,0,0)$ and
${\rm d}x_2^\mu\equiv({\rm d}t_2,0,0,{\rm d}\varphi)$ and by
Levi-Civita tensor
$\varepsilon_{\alpha\beta\gamma\delta}$. The time components of these vectors,
${\rm d}t_1$ and ${\rm d}t_2$, are such that the vectors ${\rm d}x_1^\mu$ and
${\rm d}x_2^\mu$ lie in the tangent
space to the above defined space-time slice. Then we obtain
\begin{equation}
\label{dS2}
{\rm d}S=|g^{\theta\theta}\varepsilon_{t\theta\alpha\beta}\,{\rm d}x_1^{[\alpha}
{\rm d}x_2^{\beta]}|=g^{\theta\theta}\sqrt{-\|g_{\mu\nu}\|}\,{\rm d}r\,{\rm
d}\varphi={\rm d}r\,{\rm d}\varphi\, ,
\end{equation}
where $g_{\mu\nu}$ is the metric tensor and
$\|g_{\mu\nu}\|$
is the determinant of the metric. The proper area, ${\rm d}S_\perp$,
perpendicular to the light ray can be expressed covariantly in the following
way:
\begin{equation}
\label{dSperp}
{\rm d}S_\perp=-\frac{U^{\alpha}\,p^{\beta}\,{\rm d}^2\!S_{\alpha\beta}}
{U^{\mu}\,p_\mu}\, .
\end{equation}
Here, ${\rm d}S_\perp$ is the projection of an element of area,
defined by ${\rm d}^2\!S_{\alpha\beta}$, on a spatial slice of an observer
with velocity $U^\alpha$ and perpendicular to light rays.
$U^\alpha$ is four-velocity of an observer measuring the
area ${\rm d}S_\perp$, and $p^\beta$ is four-momentum of the photon.
The\break proper area ${\rm d}S_\perp$ corresponding to the same flux tube
is identical for all observers\break (see \citealt{schneider1992}). This means that the
last equation holds
true for any four-velocity $U^\alpha$, and we can express it as
\begin{equation}
\label{dS3}
p^{\beta}\,{\rm d}^2\!S_{\alpha\beta}+p_\alpha\,{\rm d}S_\perp = 0\, ,
\quad\alpha = t,\,r,\,\theta,\,\varphi\, .
\end{equation}
For $\alpha=t$ (note that
${\rm d}^2\!S_{tr}={\rm d}^2\!S_{t\varphi}=0$) we get
\begin{equation}
\label{ratio_dS}
\frac{{\rm d}S_\perp}{{\rm d}S}=\left|\frac{1}{g^{\theta\theta}}
\frac{{\rm d}S_\perp}{{\rm d}^2\!S_{t\theta}}\right| =
\left|-\frac{p_\theta}{p_t}\right|=\frac{r\mu_{\rm e}}{g}\, .
\end{equation}
In the last equation we used the formula for the cosine of local emission
angle $\mu_{\rm e}$, see eq.~(\ref{cosine}), and the fact that we have chosen
such an affine parameter of the light geodesic that $p_t=-1$.
From eqs.~(\ref{N_o^S}), (\ref{dS}) and (\ref{ratio_dS}) we get for
the observed flux per unit solid angle
\begin{equation}
\label{emission1}
N^{\Omega}_{\rm o}(E)\equiv\frac{{\rm d}n(E)}{{\rm d}t\,{\rm d}\Omega_{\rm o}}=
{N_0}\int_{r_{\rm in}}^{r_{\rm out}}{\rm d}r\,\int_{\phi}^{\phi+{\Delta\phi}}
{\rm d}\varphi\,N_{\loc}(E/g)\,g\,l\,\mu_{\rm e}\,r,
\end{equation}
where $N_0$ is a normalization constant and
\begin{equation}
l=\frac{{\rm d}S_{\rm f}}{{\rm d}S_\perp}
\end{equation}
is the lensing factor in the limit $D\rightarrow\infty$
(keeping $D^2{\rm d}\Omega$ constant, see Fig.~\ref{denomination}c).

For the line emission, the normalization constant $N_0$ is chosen in
such a way that the total flux from the disc is unity. In the case of a
continuum model, the flux is normalized to unity  at a certain value of the
observed energy (typically at $E=1$~keV, as in other {\sc{}xspec}
models).

Finally, the integrated flux per energy bin, $\Delta E$, is
\begin{eqnarray}
\label{emission}
\nonumber
& & \hspace*{-3em} {\Delta}N^{\Omega}_{\rm o}(E,\Delta E,t) =
\int_{E}^{E+\Delta E}{\rm d}\bar{E}\,N^{\Omega}_{\rm o}(\bar{E},t)=\\
& & = N_0\int_{r_{\rm in}}^{r_{\rm out}}{\rm d}r\,
\int_{\phi}^{\phi+{\Delta\phi}}{\rm d}\varphi\,\int_{E/g}^{(E+\Delta E)/g}
{\rm d}E_{\loc}\,N_{\loc}(E_{\loc},r,\varphi,\mu_{\rm e},t-\Delta t)\,g^2\,l\,
\mu_{\rm e}\,r\, ,\hspace*{2em}
\end{eqnarray}
where $\Delta t$ is the relative time delay with which photons arrive to the
observer from different parts of the disc. The transfer functions
$g,\,l,\,\mu_{\rm e}$ and $\Delta t$ are read from the FITS file {\tt
KBHtablesNN.fits} described in Appendix~\ref{appendix3a}. This equation
is numerically integrated for a given local flux
$N_\loc(E_{\loc},r,\varphi,\mu_{\rm e},t-\Delta t)$ in all hereby
described new general relativistic {\it non-axisymmetric models}.

Let us assume that the local emission is stationary and the dependence on the
axial coordinate is only through the prescribed dependence on the local emission
angle $f(\mu_{\rm e})$ (limb darkening/brightening law) together with an
arbitrary radial dependence $R(r)$, i.e.
\begin{equation}
N_\loc(E_{\loc},r,\varphi,\mu_{\rm e},t-\Delta t)\equiv N_\loc(E_\loc)\,R(r)\,
f(\mu_{\rm e}).
\end{equation}
The observed flux $N_{\rm o}^{\Omega}(E)$
is in this case given by
\begin{equation}
N_{\rm o}^{\Omega}(E)=\int_{-\infty}^{\infty}{\rm
d}E_\loc\,N_\loc(E_\loc)\,G(E,E_\loc),
\end{equation}
where
\begin{equation}
G(E,E_\loc)=N_0\int_{r_{\rm in}}^{r_{\rm out}}{\rm
d}r\,R(r)\int_{0}^{2\pi}{\rm d}\varphi\,f(\mu_{\rm e})\,g^2\,l\,\mu_{\rm
e}\,r\,\delta(E-gE_\loc).
\end{equation}
In this case, the integrated flux can be expressed in the following way:
\begin{eqnarray}
\nonumber
{\Delta}N^{\Omega}_{\rm o}(E,\Delta E) & = &
\int_{E}^{E+\Delta E}{\rm d}\bar{E}\,N^{\Omega}_{\rm o}(\bar{E})\quad
=\hspace{19.5em}\\
\nonumber
& & \hspace*{-8.5em} = \int_{E}^{E+\Delta E}{\rm d}\bar{E}\,
{N_0}\int_{r_{\rm in}}^{r_{\rm out}}{\rm d}r\,R(r)\,\int_{0}^{2\pi}
{\rm d}\varphi\,f(\mu_{\rm e})\,N_{\loc}(\bar{E}/g)\,g\,l\,\mu_{\rm e}\,r\,
\int_{-\infty}^{\infty}{\rm d}E_\loc\,\delta(E_\loc-\bar{E}/g)=\\
\label{axisym_emission}
& & \hspace*{-8.5em} = N_0\int_{r_{\rm in}}^{r_{\rm out}}{\rm d}r\,R(r)\,
\int_{-\infty}^{\infty}{\rm d}E_{\loc}\,N_{\loc}(E_{\loc})
\int_{E/E_{\loc}}^{(E+\Delta E)/E_{\loc}}{\rm d}\bar{g}\,F(\bar{g})\, ,
\end{eqnarray}
where we substituted $\bar{g}=\bar{E}/E_\loc$ and
\begin{equation}
\label{conv_function}
F(\bar{g}) = \int_{0}^{2\pi}{\rm d}\varphi\,f(\mu_{\rm e})\,g^2\,l\,
\mu_{\rm e}\,r\,\delta(\bar{g}-g)\, .
\end{equation}
Eq.~(\ref{axisym_emission}) is numerically integrated in all
{\it axially symmetric models}. The function
${\rm d}F(\bar{g})\equiv {\rm d}\bar{g}\,F(\bar{g})$ has been
pre-calculated for several limb darkening/brightening laws
$f(\mu_{\rm e})$ and stored in separate files,
{\tt KBHlineNN.fits} (see Appendix~\ref{appendix3b}).

\section{Stokes parameters in a strong gravity regime}
\label{stokes_param}
For polarization studies, Stokes parameters are used. Let us define specific
Stokes parameters in the following way:
\begin{equation}
i_\nu\equiv \frac{I_{\nu}}{E}\, ,\quad q_\nu\equiv \frac{Q_{\nu}}{E}\, ,\quad
u_\nu\equiv \frac{U_{\nu}}{E}\, ,\quad v_\nu\equiv \frac{V_{\nu}}{E}\, ,
\end{equation}
where $I_{\nu}$, $Q_{\nu}$, $U_{\nu}$ and $V_{\nu}$ are Stokes
parameters for light with frequency $\nu$, $E$ is the energy of a photon at
this frequency. Further on, we drop the index $\nu$ but we will always
consider these quantities for light of a given frequency. We can calculate
the integrated specific Stokes parameters (per energy bin), i.e.\ $\Delta
i_{\rm o}$, $\Delta q_{\rm o}$, $\Delta u_{\rm o}$ and $\Delta v_{\rm
o}$. These are the quantities that the observer determines from the local
specific Stokes parameters $i_\loc$, $q_\loc$, $u_\loc$ and $v_\loc$ on the disc
in the following way:
\begin{eqnarray}
\label{S1}
{\Delta}i_{\rm o}(E,\Delta E) & = & N_0\int{\rm d}S\,\int{\rm d}E_{\loc}\,
i_{\loc}(E_{\loc})\,Fr\, ,\\
\label{S2}
{\Delta}q_{\rm o}(E,\Delta E) & = & N_0\int{\rm d}S\,\int{\rm d}E_{\loc}\,
[q_{\loc}(E_{\loc})\cos{2\Psi}-u_{\loc}(E_{\loc})\sin{2\Psi}]\,Fr\, ,\\
\label{S3}
{\Delta}u_{\rm o}(E,\Delta E) & = & N_0\int{\rm d}S\,\int{\rm d}E_{\loc}\,
[q_{\loc}(E_{\loc})\sin{2\Psi}+u_{\loc}(E_{\loc})\cos{2\Psi}]\,Fr\, ,\\
\label{S4}
{\Delta}v_{\rm o}(E,\Delta E) & = & N_0\int{\rm d}S\,\int{\rm d}E_{\loc}\,
v_{\loc}(E_{\loc})\,Fr\, .
\end{eqnarray}
Here, $F\equiv F(r,\varphi)=g^2\,l\,\mu_{\rm e}$ is a transfer
function, $\Psi$ is the angle by which a vector parallelly transported
along the light geodesic rotates. We refer to this angle also as a
change of the polarization angle, because the polarization vector is parallelly
transported along light geodesics. See Fig.~\ref{pol_angle} for an exact
definition of the angle $\Psi$. The integration boundaries are the
same as in eq.~(\ref{emission}). As can be seen from the
definition, the first specific Stokes parameter is equal to the photon
flux, therefore, eqs.~(\ref{emission}) and (\ref{S1}) are
identical. The local specific Stokes parameters may depend on $r$,
$\varphi$, $\mu_{\rm e}$ and $t-\Delta t$, which we did not state in the
eqs.~(\ref{S1})--(\ref{S4}) explicitly for simplicity.

The specific Stokes parameters that the observer measures may vary in time
in the case when the local parameters also depend on time. In eqs.\
(\ref{S1})--(\ref{S4}) we used a law of transformation of the Stokes
parameters by the rotation of axes (eqs.~(I.185) and (I.186) in
\citealt{chandrasekhar1960}).

An alternative way for expressing polarization of light is by using the
degree of polarization $P_{\rm o}$ and two polarization angles
$\chi_{\rm o}$ and $\xi_{\rm o}$, defined by
\begin{eqnarray}
P_{\rm o} & = & \sqrt{q_{\rm o}^2+u_{\rm o}^2+v_{\rm o}^2}/i_{\rm o}\, , \\
\tan{2\chi_{\rm o}} & = & u_{\rm o}/q_{\rm o}\, ,\\
\sin{2\xi_{\rm o}} & = & v_{\rm o}/\sqrt{q_{\rm o}^2+u_{\rm o}^2
+v_{\rm o}^2}\, .
\end{eqnarray}

\section{Local emission in lamp-post models}
\label{lamp-post}
The local emission from a disc is proportional to the incident illumination from
a power-law primary source placed on the axis at height $h$ above the black hole.
To calculate the incident illumination we need to integrate the geodesics from
the source to the disc.

The four-momentum of the incident photons which were emitted by a primary source
and which are striking the disc at radius $r$ is
(see eqs.~(\ref{carter1})--(\ref{carter4}) with $l=0$ and $\theta=\pi/2$
or also \citealt{carter1968} and \citealt{misner1973})
\begin{eqnarray}
p_{\rm i}^t & = & 1+2/r+4/\Delta\, ,\\
p_{\rm i}^r & = & {\rm R}^{\prime}_{\rm sgn}[(r^2+a^2)^2-\Delta
(a^2+q_{\rm L}^2)]^{1/2}/r^2\, ,\\
p_{\rm i}^\theta & = & q_{\rm L}/r^2\, ,\\
p_{\rm i}^\varphi & = & 2a/(r\Delta)\, ,
\end{eqnarray}
where
$q_{\rm L}^2 = \sin^2{\!\theta_{\rm L}}\,(h^2+a^2)^2/\Delta_{\rm L}-a^2$ is
Carter's constant of motion with $\Delta_{\rm L}=h^2-2h+a^2$, and with the angle
of emission $\theta_{\rm L}$ being the local angle
under which the photon is emitted from a primary source (it is measured in the rest
frame of the source). We define this angle by
$\tan{\theta_{\rm L}}=-{p_{\rm L}^{(\theta)}/p_{\rm L}^{(r)}}$, where
$p_{\rm L}^{(r)}=p_{\rm L}^\mu\,e_{{\rm L}\,\mu}^{(r)}$ and
$p_{\rm L}^{(\theta)}=p_{\rm L}^\mu\,e_{{\rm L}\,\mu}^{(\theta)}$ with
$p_{\rm L}^\mu$ and $e_{{\rm L}\,\mu}^{(a)}$ being the four-momentum of emitted
photons and the local tetrad connected with a primary source, respectively.
The angle is $0^\circ$ when the photon is emitted downwards and $180^\circ$ if
the photon is emitted upwards.

We denoted the sign of the radial component of the momentum by
${\rm R}^{\prime}_{\rm sgn}$. We have chosen such an affine parameter for the light
geodesic that the conserved energy of the light is
$-p_{{\rm i}\,t}=-p_{{\rm L}\,t}=1$. The conserved angular momentum of incident
photons is zero ($l_{\rm L}=0$).

The gravitational and Doppler shift of the photons striking the disc which were
emitted by a primary source is
\begin{equation}
\label{gfac_lamp}
g_{\rm L}= \frac{\nu_{\rm i}}{\nu_{\rm L}}=
\frac{p_{{\rm i}\,\mu} U^\mu}{p_{{\rm L}\,\alpha} U_{\rm L}^\alpha}=
-\frac{p_{{\rm i}\,\mu} U^\mu}{U_{\rm L}^t}\, .
\end{equation}
Here $\nu_{\rm i}$ and $\nu_{\rm L}$ denote the frequency of the incident and
emitted photons, respectively and $U_{\rm L}^\alpha$ is a four-velocity of the
primary source with the only
non-zero component \hbox{$U_{\rm L}^t=\sqrt{\frac{h^2+a^2}{\Delta_{\rm L}}}$}.

Cosine of the local incident angle is
\begin{equation}
\label{cosine_inc}
\mu_{\rm i}=|\cos{\delta_{\rm i}}\,|=
\displaystyle\frac{{p_{{\rm i}\,\alpha}\,n^{\alpha}}}
{{p_{{\rm i}\,\mu}\,U^{\mu}}}\, ,
\end{equation}
where $n^\alpha=-e_{(\theta)}^\alpha$ is normal to the disc with respect to the
observer co-moving with the matter in the disc.

We further define the azimuthal incident angle as the angle between the
projection of the three-momentum of the incident photon into the disc (in the
local rest frame co-moving with the disc) and the radial tetrad vector,
\begin{equation}
\label{azim_angle_inc}
\Phi_{\rm i}=-{\rm R}_{\rm sgn}^{\rm i}\arccos\left(
\frac{-1}{\sqrt{1-\mu_{\rm i}^2}}\frac{\,{p_{{\rm i}\,\alpha}}
\,e_{(\varphi)}^{\alpha}}
{p_{{\rm i}\,\mu}U^\mu}\right)+\frac{\pi}{2}\, ,
\end{equation}
where ${\rm R}_{\rm sgn}^{\rm i}$ is positive if the incident photon travels
outwards ($p_{\rm i}^{(r)}>0$) and negative if it travels inwards
($p_{\rm i}^{(r)}<0$) in the local rest frame of the disc.

In lamp-post models the emission of the disc will be proportional to the
incident radiation $N_{\rm i}^{S}(E_{\rm l})$ which comes from a primary source
\begin{equation}
\label{illumination}
N_{\rm i}^{S}(E_{\rm l})=N_{\rm L}^{\Omega}(E_{\rm L})\frac{{\rm d}
\Omega_{\rm L}}{{\rm d}S_{\loc}}\, .
\end{equation}
Here $N_{\rm L}^{\Omega}(E_{\rm L})=N_{0 {\rm L}}\,E_{\rm L}^{-\Gamma}$ is
an isotropic and stationary power-law emission from a primary source which is
emitted into
a solid angle ${\rm d}\Omega_{\rm L}$ and which illuminates local area
${\rm d}S_{\rm l}$ on the disc. The energy of the photon striking the disc
(measured in the local frame co-moving with the disc) will be redshifted
\begin{equation}
E_\loc=g_{\rm L}\,E_{\rm L}\, .
\end{equation}
The ratio ${\rm d}\Omega_{\rm L}/{\rm d}S_{\loc}$ is
\begin{equation}
\frac{{\rm d}\Omega_{\rm L}}{{\rm d}S_{\loc}}=\frac{{\rm d}\Omega_{\rm L}}
{{\rm d}S} \frac{{\rm d}S}{{\rm d}S_\loc}=\frac{\sin{\theta_{\rm L}{\rm d}
\theta_{\rm L}\,{\rm d}\varphi}}{{\rm d}r\,{\rm d}\varphi}\frac{{\rm d}S}
{{\rm d}S_\loc}\, ,
\end{equation}
where (see eqs.~(\ref{dS2}) and (\ref{dS3}))
\begin{equation}
{\rm d}S={\rm d}r\,{\rm d}\varphi=|{\rm d}^2{\!S_t}^{\theta}|=
-g^{\theta\theta}\,
\frac{p_{{\rm i} t}}{p_{\rm i}^\theta}\,{\rm d}S_\perp=
\,\frac{g^{\theta\theta}}{p_{\rm i}^\theta}\,{\rm d}S_\perp\, .
\end{equation}
Here we used
the same space-time slice as in the discussion above eq.~(\ref{dS}) and thus
the element
${\rm d}^2\!S_{\alpha\beta}$ is defined as before, see eq.~(\ref{dS2}). Note
that here the area ${\rm d}S_\perp$ is defined by the incident flux tube as
opposed to ${\rm d}S_\perp$ in eq.~(\ref{ratio_dS}) where it was defined by
the emitted flux tube.
The coordinate area ${\rm d}S$ corresponds to the proper area ${\rm d}S_\perp$
which is perpendicular to the incident light ray (in the local rest frame
co-moving with the disc). The corresponding proper area (measured in the same local
frame) lying in the equatorial plane will be
\begin{eqnarray}
\nonumber
{\rm d}S_{\loc} & = &
|{\rm d}^2{\!S_{(t)}}^{\!\!(\theta)}|=|e_{(t)}^\mu\,e^{{(\theta)}\,\nu}\,
{\rm d}^2\!S_{\mu\nu}|=|g_{\theta\theta}^{-1/2}\,U^{\mu}\,
{\rm d}^2\!S_{\mu\theta}|=\\
& = & -g_{\theta\theta}^{-1/2}\,
\frac{p_{{\rm i} \mu}\,U^\mu}{p_{\rm i}^\theta}\,{\rm d}S_\perp =
g_{\theta\theta}^{-1/2}\frac{U_{\rm L}^t}{p_{\rm i}^\theta}\,g_{\rm L}\,
{\rm d}S_\perp\, .
\label{dSl}
\end{eqnarray}
Here we have used eq.~(\ref{dS}) for the tetrad components of the element
${\rm d}^2\!S_{\alpha\beta}$, eqs.~(\ref{dSperp}) and (\ref{gfac_lamp}).

It follows from eqs.~(\ref{illumination})--(\ref{dSl}) that the incident
radiation will be again a power law with the same photon index
$\Gamma$ as in primary emission
\begin{equation}
N_{\rm i}^{S}(E_\loc)=N_{0 {\rm i}}\,E_\loc^{-\Gamma}\, ,
\end{equation}
with the normalization factor
\begin{equation}
N_{0 {\rm i}}=N_{0 {\rm L}}\,g_{\rm L}^{\Gamma-1}\,
\sqrt{1-\frac{2h}{h^2+a^2}}\,\frac{\sin{\theta_{\rm L}}\,
{\rm d}\theta_{\rm L}}{r\,{\rm d}r}\, .
\end{equation}
The emission of the disc due to illumination will be proportional to
this factor.

\chapter{New models for {\fontfamily{phv}\fontshape{sc}\selectfont xspec}}
\label{chapter3}
 \thispagestyle{empty}
We have developed several general relativistic models for line emission
and Compton reflection continuum. The line models are supposed to be
more accurate and versatile than the {\sc laor} model \citep{laor1991}, and
substantially faster than the {\sc kerrspec} model \citep{martocchia2000}.
Several models of intrinsic emissivity were employed, including the lamp-post
model \citep{matt1992}.  Among other features,
these models allow various parameters to be fitted such as the black-hole
angular momentum, observer inclination, accretion disc size and some of the
parameters characterizing disc emissivity and primary illumination
properties. They also
allow a change in the grid resolution and, hence, to control accuracy and
computational speed. Furthermore, we developed very general
convolution models. All these models are based on pre-calculated tables
described in Chapter~\ref{transfer_functions} and thus the geodesics
do not need to be calculated each time one integrates the disc emission.
These tables are calculated for the vacuum Kerr space-time and for a Keplerian
co-rotating disc plus matter that is freely falling below the marginally
stable orbit. The falling matter has the energy and angular momentum of
the matter at the marginally stable orbit. It is possible to use
different pre-calculated tables if they are stored in a specific FITS
file (see Appendix~\ref{appendix3a} for its detailed description).

There are two types of new models. The first type of model integrates
the local disc emission in both of the polar coordinates on the disc and thus
enables one to choose non-axisymmetric area of integration (emission
from spots or partially obscured discs). One can also choose the
resolution of integration and thus control the precision and speed of
the computation. The second type of model is axisymmetric -- the
axially dependent part of the emission from rings is pre-calculated and
stored in a FITS file (the function ${\rm d}F(\bar{g})={\rm
d}\bar{g}\,F(\bar{g})$ from eq.~(\ref{conv_function}) is integrated for
different radii with the angular grid having $20\,000$ points). These
models have less
parameters that can be fitted and thus are less flexible even though more
suited to the standard analysis approach. On the other
hand they are fast because the emission is integrated only in one
dimension (in the radial coordinate of the disc). It may be worth emphasizing
that the assumption about axial symmetry concerns only the form of intrinsic
emissivity of the disc, which cannot depend on the polar angle in this case, not
the shape of individual light rays, which is always
complicated near a rotating black hole.

\begin{table}[tbh]
\begin{center}
\dummycaption\label{common_par1}
\begin{tabular}[t]{l|c|c|c|c}
parameter & unit & default value & minimum value & maximum value \\ \hline
\hspace*{0.5em}{\tt a/M}       & $GM/c$   & 0.9982 &  0.      & 1.    \\
\hspace*{0.5em}{\tt theta\_o}  & deg      & 30.    &  0.      & 89.   \\
\hspace*{0.5em}{\tt rin-rh}    & $GM/c^2$ & 0.     &  0.      & 999.  \\
\hspace*{0.5em}{\tt ms}        & --       & 1.     &  0.      & 1.    \\
\hspace*{0.5em}{\tt rout-rh}   & $GM/c^2$ & 400.   &  0.      & 999.  \\
\hspace*{0.5em}{\tt zshift}    & --       & 0.     &  -0.999. & 10.   \\
\hspace*{0.5em}{\tt ntable}    & --       & 0.     &  0.      & 99.
\end{tabular}
\mycaption{Common parameters for all models.}
\end{center}
\end{table}

\begin{table}[tbh]
\begin{center}
\dummycaption\label{common_par2}
\begin{tabular}[h]{l|c|c|c|c}
parameter & unit & default value  & minimum value & maximum value \\ \hline
\hspace*{0.5em}{\tt phi}      & deg & 0.   &  -180. & 180.       \\
\hspace*{0.5em}{\tt dphi}     & deg & 360. &  0.    & 360.       \\
\hspace*{0.5em}{\tt nrad}     & --  & 200. &  1.    & 10000.     \\
\hspace*{0.5em}{\tt division} & --  & 1.   &  0.    & 1.         \\
\hspace*{0.5em}{\tt nphi}     & --  & 180. &  1.    & 20000.     \\
\hspace*{0.5em}{\tt smooth}   & --  & 1.   &  0.    & 1.         \\
\hspace*{0.5em}{\tt Stokes}   & --  & 0.   &  0.    & 6.
\end{tabular}
\mycaption{Common parameters for non-axisymmetric models.}
\end{center}
\end{table}

There are several parameters and switches that are common for all new models
(see Tab.~\ref{common_par1}):
\begin{description} \itemsep -2pt
 \item[{\tt a/M}] -- specific angular momentum of the Kerr black hole in units
 of $GM/c$ ($M$ is the mass of the central black hole),
 \item [{\tt theta\_o}] -- inclination of the observer in degrees,
 \item [{\tt rin-rh}] -- inner radius of the disc relative to the black-hole
 horizon in units of $GM/c^2$,
 \item [{\tt ms}] -- switch for the marginally stable orbit,
 \item [{\tt rout-rh}] -- outer radius of the disc relative to the black-hole
 horizon in units of $GM/c^2$,
 \item [{\tt zshift}] -- overall redshift of the object,
 \item [{\tt ntable}] -- number of the FITS file with pre-calculated tables to
 be used.
\end{description}
  The inner and outer radii are given relative to the black-hole horizon and,
therefore, their minimum value is zero. This becomes handy when one fits the
{\tt a/M} parameter, because the horizon of the black hole as well as the
marginally stable orbit change with {\tt a/M}, and so the lower limit for
inner and outer disc edges
cannot be set to constant values. The {\tt ms} switch determines whether
we also want to integrate emission below the marginally stable orbit.
If its value is set to zero and the inner radius of the disc is below this
orbit then the emission below the marginally stable orbit is taken
into account, otherwise it is not.

The {\tt ntable} switch determines which of the pre-calculated tables
should be used for intrinsic emissivity.
In particular, ${\tt ntable}=0$ for {\tt KBHtables00.fits}
({\tt KBHline00.fits}), ${\tt ntable}=1$ for {\tt KBHtables01.fits}
({\tt KBHline01.fits}), etc., corresponding to non-axisym\-met\-ric
(axisymmetric) models.

\begin{figure}[tbh]
\begin{center}
\dummycaption\label{sector}
\includegraphics[width=5cm]{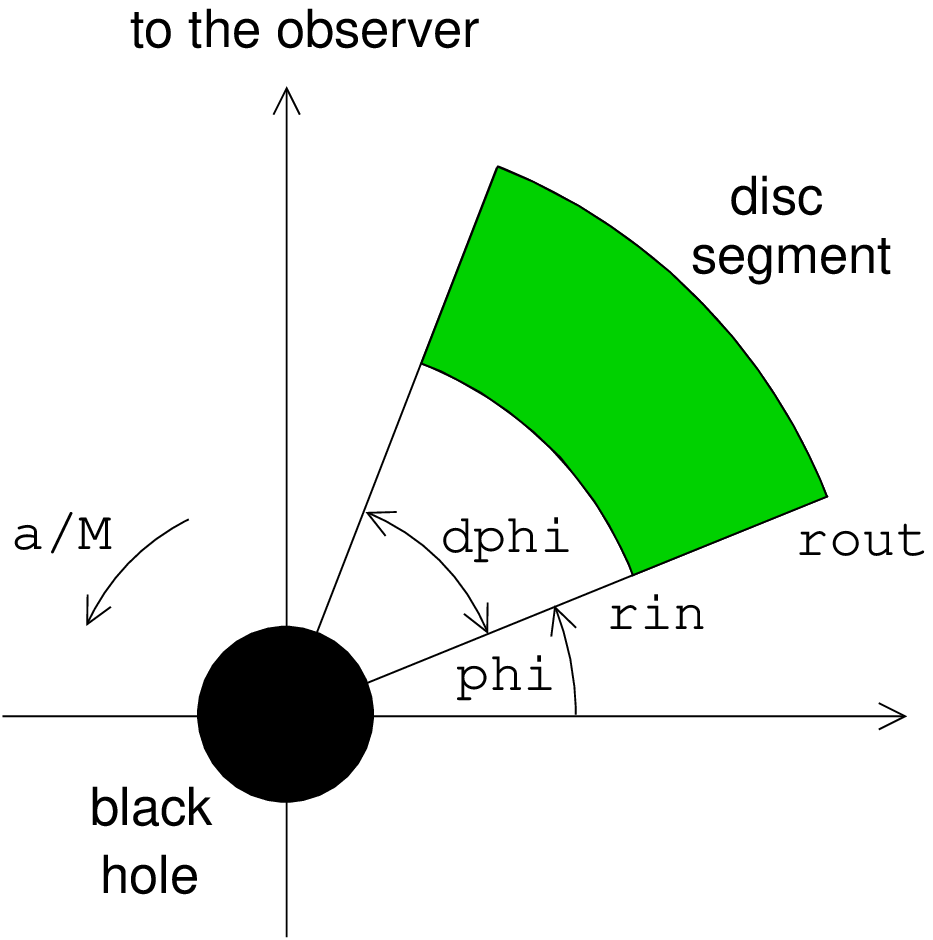}
\mycaption{Segment of a disc from which emission comes (view from above).}
\end{center}
\end{figure}

\begin{table}[tbh]
\begin{center}
\dummycaption\label{stokes}
\begin{tabular*}{11.5cm}{c|l}
value & photon flux array {\tt photar} contains$^{\dag}{}^{\ddag}$\\ \hline
0 & $i=I/E$, where $I$ is the first Stokes parameter (intensity) \\
1 & $q=Q/E$, where $Q$ is the second Stokes parameter \\
2 & $u=U/E$, where $U$ is the third Stokes parameter \\
3 & $v=V/E$, where $V$ is the fourth Stokes parameter \\
4 & degree of polarization, $P=\sqrt{q^2+u^2+v^2}/i$ \\
5 & angle $\chi$[deg] of polarization, $\tan{2\chi}=u/q$\\
6 & angle $\xi$[deg] of polarization, $\sin{2\xi}=v/\sqrt{q^2+u^2+v^2}$\\ \hline
\multicolumn{2}{l}{\parbox{11.cm}{\vspace*{1mm}\footnotesize
$^{\dag}$~the {\tt photar} array contains the values described in the table
and multiplied by the width of the corresponding energy bin\\
$^{\ddag}~E$ is the energy of the observed photons}}\\
\end{tabular*}
\mycaption{Definition of the {\tt Stokes} parameter.}
\end{center}
\end{table}

The following set of parameters is relevant only for non-axisymmetric models
(see Tab.~\ref{common_par2}):
\begin{description} \itemsep -2pt
 \item[{\tt phi}] -- position angle of the axial sector of the disc in degrees,
 \item[{\tt dphi}] -- inner angle of the axial sector of the disc in degrees,
 \item[{\tt nrad}] -- radial resolution of the grid,
 \item[{\tt division}] -- switch for spacing of radial grid
 ($0$ -- equidistant, $1$ -- exponential),
 \item[{\tt nphi}] -- axial resolution of the grid,
 \item[{\tt smooth}] -- switch for performing simple smoothing
 ($0$ -- no, $1$ -- yes),
 \item[{\tt Stokes}] -- switch for computing polarization
 (see Tab.~\ref{stokes}).
\end{description}
 The {\tt phi} and {\tt dphi} parameters determine the axial sector of the disc
from which emission comes (see Fig.~\ref{sector}). The {\tt nrad} and {\tt nphi}
parameters determine the grid for numerical integration.
If the {\tt division} switch is zero, the radial grid is equidistant;
if it is equal to unity then the radial grid is exponential
(i.e.\ more points closer to the black hole).

If the {\tt smooth} switch is set to unity then a simple smoothing
is applied to the final spectrum. Here $N_{\rm o}^{\Omega}(E_{\rm j})=
[N_{\rm o}^{\Omega}(E_{\rm j-1})+2N_{\rm o}^{\Omega}(E_{\rm j})+
N_{\rm o}^{\Omega}(E_{\rm j+1})]/4$.

If the {\tt Stokes} switch is different from zero, then the model also
calculates polarization. Its value determines which of the Stokes parameters
should be computed by {\sc{}xspec}, i.e.\ what will be stored in the output
array for the photon flux {\tt photar}; see Tab.~\ref{stokes}.
(If ${\tt Stokes}\neq0$ then a new {\sf ascii} data file {\tt stokes.dat} is
created in the current directory,
where values of energy $E$ together with all Stokes parameters
$i,\,q,\,u,\,v,\,P,\,\chi$[deg] and $\xi$[deg] are stored, each in one
column.)

A realistic model of polarization has been currently implemented only in the
{\sc kyl1cr} model (see Section~\ref{kyl1cr}
below). In other models, a simple assumption is made -- the local emission
is assumed to be linearly polarized in the direction perpendicular to
the disc (i.e.\ $q_{\loc}=i_{\loc}=N_\loc$ and $u_\loc=v_\loc=0$). In all
models (including {\sc kyl1cr}) there is always no final circular polarization
(i.e.\ $v=\xi=0$), which follows from the fact that the fourth local
Stokes parameter is zero in each model.

\section{Models for a relativistic spectral line}
Three general relativistic line models are included in the new set of
{\sc{}xspec} routines -- non-axisymmetric Gaussian line model
{\sc kyg1line}, axisymmetric Gaussian line model {\sc kygline} and
fluorescent lamp-post line model {\sc kyf1ll}.

\subsection{Non-axisymmetric Gaussian line model
{\fontfamily{phv}\fontshape{sc}\selectfont kyg1line}}
\label{section_kyg1line}
The {\sc kyg1line} model computes the integrated flux from the
disc according to eq.~(\ref{emission}). It assumes that the local
emission from the disc is
\begin{myeqnarray}
\label{line_emiss1a}
N_{\loc}(E_\loc) \hspace*{-0.7em} & = & \hspace*{-0.7em}
\frac{1}{r^{\tt alpha}}\,f(\mu_{\rm e})\,\exp{\left
[-\left (1000\,\frac{E_\loc-{\tt Erest}}{\sqrt{2}\,{\tt sigma}}\right )^2
\right ]} & {\rm for} & \hspace*{-0.5em} r\ge r_{\rm b}\, ,\\
\label{line_emiss1b}
N_{\loc}(E_\loc) \hspace*{-0.7em} & = & \hspace*{-0.7em}
{\tt jump}\ r_{\rm b}^{{\tt beta} - {\tt alpha}}\,
\frac{1}{r^{\tt beta}}\,f(\mu_{\rm e})\,\exp{\left [-\left (1000\,
\frac{E_\loc-{\tt Erest}}{\sqrt{2}\,{\tt sigma}}\right )^2\right ]} &
{\rm for} & \hspace*{-0.5em} r<r_{\rm b}\, .\hspace{12mm}
\end{myeqnarray}
The local emission is assumed to be a Gaussian line with its peak flux
depending on the radius
as a broken power law. The line is defined by nine points equally spaced with
the central point at its maximum.
The normalization constant $N_0$ in eq.~(\ref{emission}) is such that the total
integrated flux of the line is unity.
The parameters defining the Gaussian line are (see Tab.~\ref{kyg1line_par}):
\begin{description} \itemsep -2pt
 \item[{\tt Erest}] -- rest energy of the line in keV,
 \item[{\tt sigma}] -- width of the line in eV,
 \item[{\tt alpha}] -- radial power-law index for the outer region,
 \item[{\tt beta}] -- radial power-law index for the inner region,
 \item[{\tt rb}] -- parameter defining the border between regions with different
 power-law indices,
 \item[{\tt jump}] -- ratio between flux in the inner and outer regions at
 the border radius,
 \item[{\tt limb}] -- switch for different limb darkening/brightening laws.
\end{description}
There are two regions with different power-law dependences with indices
{\tt alpha} and {\tt beta}. The power law changes at the border radius
$r_{\rm b}$ where the local emissivity does not need to be continuous
(for ${\tt jump}\neq 1$). The {\tt rb} parameter defines this radius in the
following way:
\begin{myeqnarray}
\label{rb1}
r_{\rm b} &=& {\tt rb} \times r_{\rm ms} & & {\rm for}\quad {\tt rb}\ge
0\, ,\\
\label{rb2}
r_{\rm b} &=& -{\tt rb}+r_{\rm h} & & {\rm for}\quad {\tt rb}< 0\, ,
\end{myeqnarray}
where $r_{\rm ms}$ is the radius of the marginally stable orbit and $r_{\rm h}$
is the radius of the horizon of the black hole.

\begin{table}[tbh]
\begin{center}
\dummycaption\label{kyg1line_par}
\begin{tabular}[h]{r@{}l|l|c|c|c}
\multicolumn{2}{c|}{parameter} & unit & default value  & minimum value &
maximum value \\ \hline
&{\tt a/M}       & $GM/c$   & 0.9982  & 0.       & 1.        \\
&{\tt theta\_o}  & deg      & 30.     & 0.       & 89.       \\
&{\tt rin-rh}    & $GM/c^2$ & 0.      & 0.       & 999.      \\
&{\tt ms}        & --       & 1.      & 0.       & 1.        \\
&{\tt rout-rh}   & $GM/c^2$ & 400.    & 0.       & 999.      \\
&{\tt phi}       & deg      & 0.      & -180.    & 180.      \\
&{\tt dphi}      & deg      & 360.    & 0.       & 360.      \\
&{\tt nrad}      & --       & 200.    & 1.       & 10000.    \\
&{\tt division}  & --       & 1.      & 0.       & 1.        \\
&{\tt nphi}      & --       & 180.    & 1.       & 20000.    \\
&{\tt smooth}    & --       & 1.      & 0.       & 1.        \\
&{\tt zshift}    & --       & 0.      & -0.999   & 10.       \\
&{\tt ntable}    & --       & 0.      & 0.       & 99.       \\
{*}&{\tt Erest}  & keV      & 6.4     & 1.       & 99.       \\
{*}&{\tt sigma}  & eV       & 2.      & 0.01     & 1000.     \\
{*}&{\tt alpha}  & --       & 3.      & -20.     & 20.       \\
{*}&{\tt beta}   & --       & 4.      & -20.     & 20.       \\
{*}&{\tt rb}     & $r_{\rm ms}$ & 0.  & 0.       & 160.      \\
{*}&{\tt jump}   & --       & 1.      & 0.       & 1e6       \\
{*}&{\tt limb}   & --       & -1.     & -10.     & 10.       \\
&{\tt Stokes}    & --       & 0.      & 0.       & 6.        \\
\end{tabular}
\mycaption{Parameters of the non-axisymmetric Gaussian line model {\sc{}kyg1line}.
Model parameters that are not common for all non-axisymmetric models are
denoted by asterisk.}
\end{center}
\end{table}

The function $f(\mu_{\rm e})=f(\cos{\delta_{\rm e}})$ in eqs.~(\ref{line_emiss1a})
and (\ref{line_emiss1b}) describes the limb darkening/ brightening law, i.e.\
the dependence of the local emission on the local emission angle. Several limb
darkening/brightening laws are implemented:
\begin{myeqnarray}
\label{isotropic}
f(\mu_{\rm e}) & = & 1 & &  {\rm for} \quad {\tt limb}=0\, , \\
\label{laor}
f(\mu_{\rm e}) & = & 1 + 2.06 \mu_{\rm e} & & {\rm for}  \quad {\tt limb}=
-1\, ,\\
\label{haardt}
f(\mu_{\rm e}) & = & \ln{(1+\mu_{\rm e}^{-1})} & & {\rm for}  \quad {\tt limb}=
-2\, ,\\
\label{other_limb}
f(\mu_{\rm e}) & = & \mu_{\rm e}^{\tt limb} & & {\rm for} \quad {\tt limb}
\ne 0,-1,-2\, .
\end{myeqnarray}
Eq.~(\ref{isotropic}) corresponds to the isotropic local emission,
eq.~(\ref{laor}) corresponds to
limb\break darkening in an optically thick electron scattering atmosphere
(used by Laor, see\break \citealt{phillips1986,laor1990,laor1991}),
and eq.~(\ref{haardt}) corresponds
to limb brightening predicted by some models of a fluorescent line emitted
by an accretion disc due to X-ray illumination
\citep{haardt1993a,ghisellini1994}.

\begin{figure}
\dummycaption\label{fig:example0}
\includegraphics[width=0.5\textwidth]{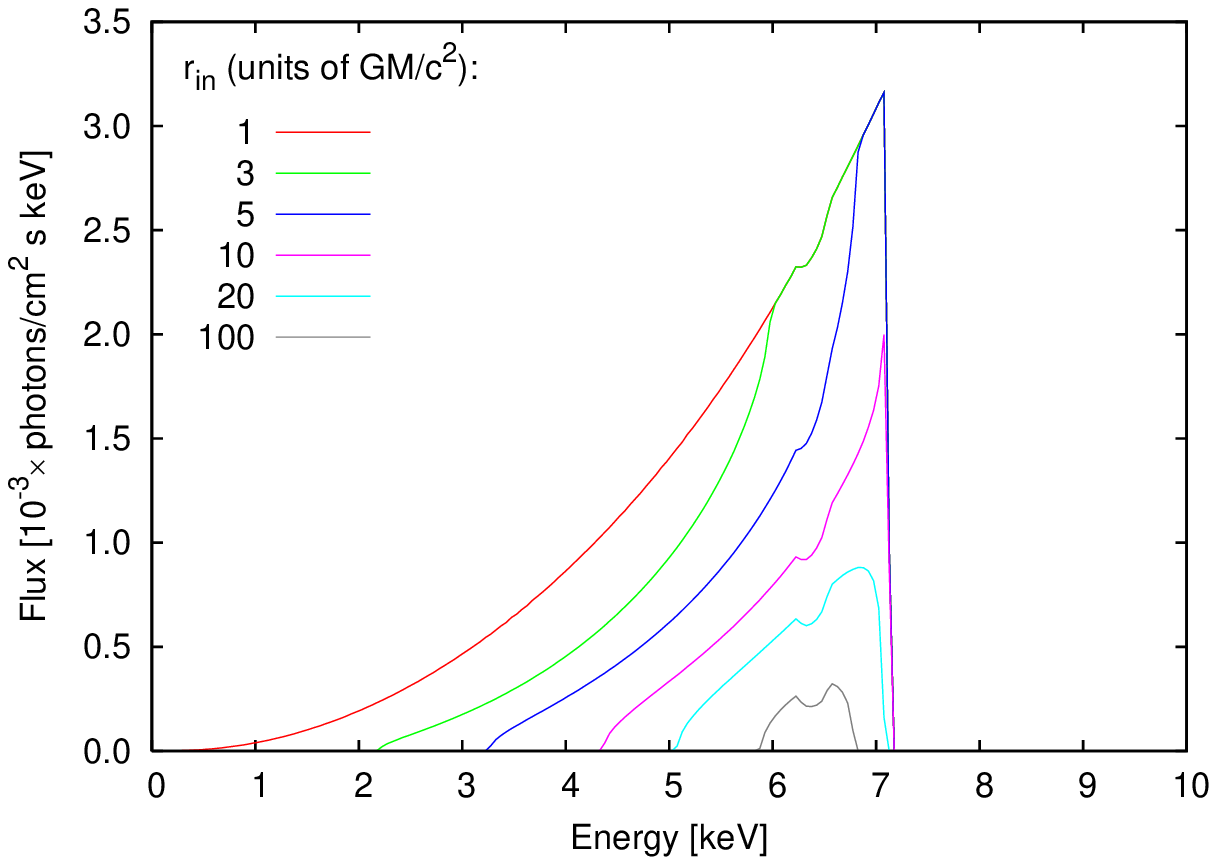}
\includegraphics[width=0.5\textwidth]{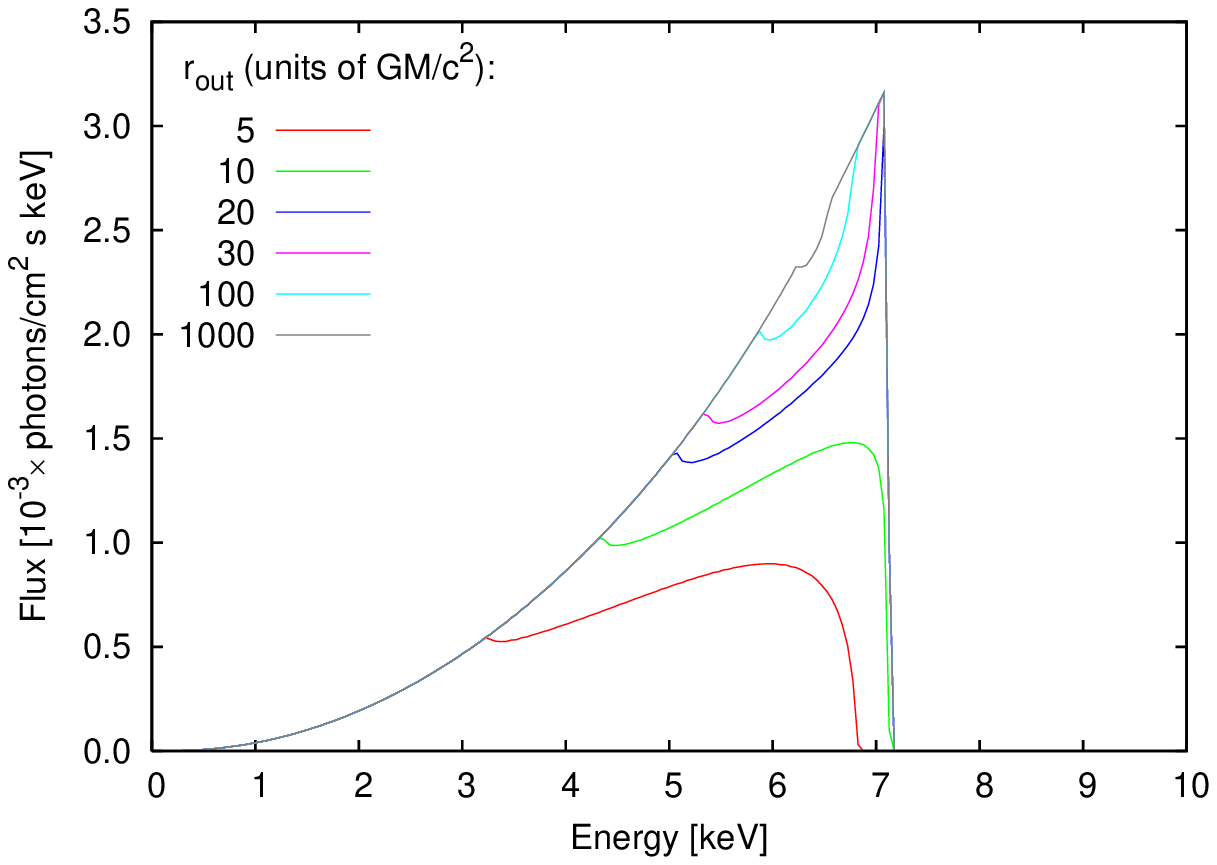}
\mycaption{Comparative examples of simple line profiles,
showing a theoretical line ($E_{\rm rest}=6.4\,$keV) with relativistic
effects originating from a black-hole accretion disc. Different sizes of
the annular region (axially symmetric) have been considered, assuming
that the intrinsic emissivity obeys a power law in the radial direction
($\alpha=3$). Resolution of the line-emitting region was
$n_r{\times}n_\varphi=3000\times1500$ with a non-equidistant layout
of the grid in Kerr ingoing coordinates, as described in the text. Left:
Dependence on the inner edge. Values of $r_{\rm{}in}$ are indicated in the
plot (the outer edge has been fixed at the maximum
radius covered by our tables, $r_{\rm{}out}=10^3$). Right: Dependence
on $r_{\rm{}out}$ (with the inner edge at horizon,
$r_{\rm{}in}=r_{\rm{}h}$). Other key parameters are:
$\theta_{\rm{}o}=45^{\circ}$, $a=1.0$. Locally isotropic emission was
assumed in the disc co-rotating frame.}
\vspace*{2.5\bigskipamount}
\dummycaption\label{fig:example1}
\includegraphics[width=0.5\textwidth]{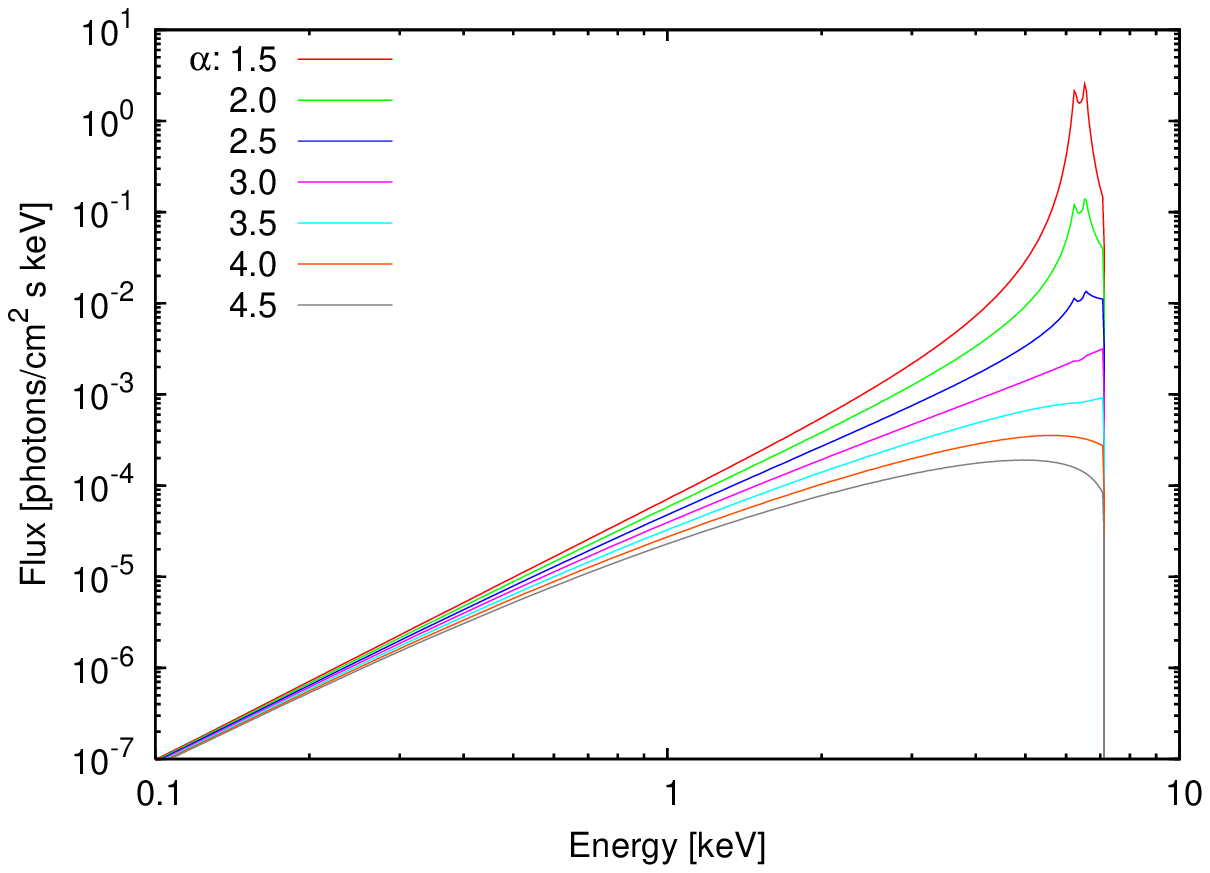}
\includegraphics[width=0.5\textwidth]{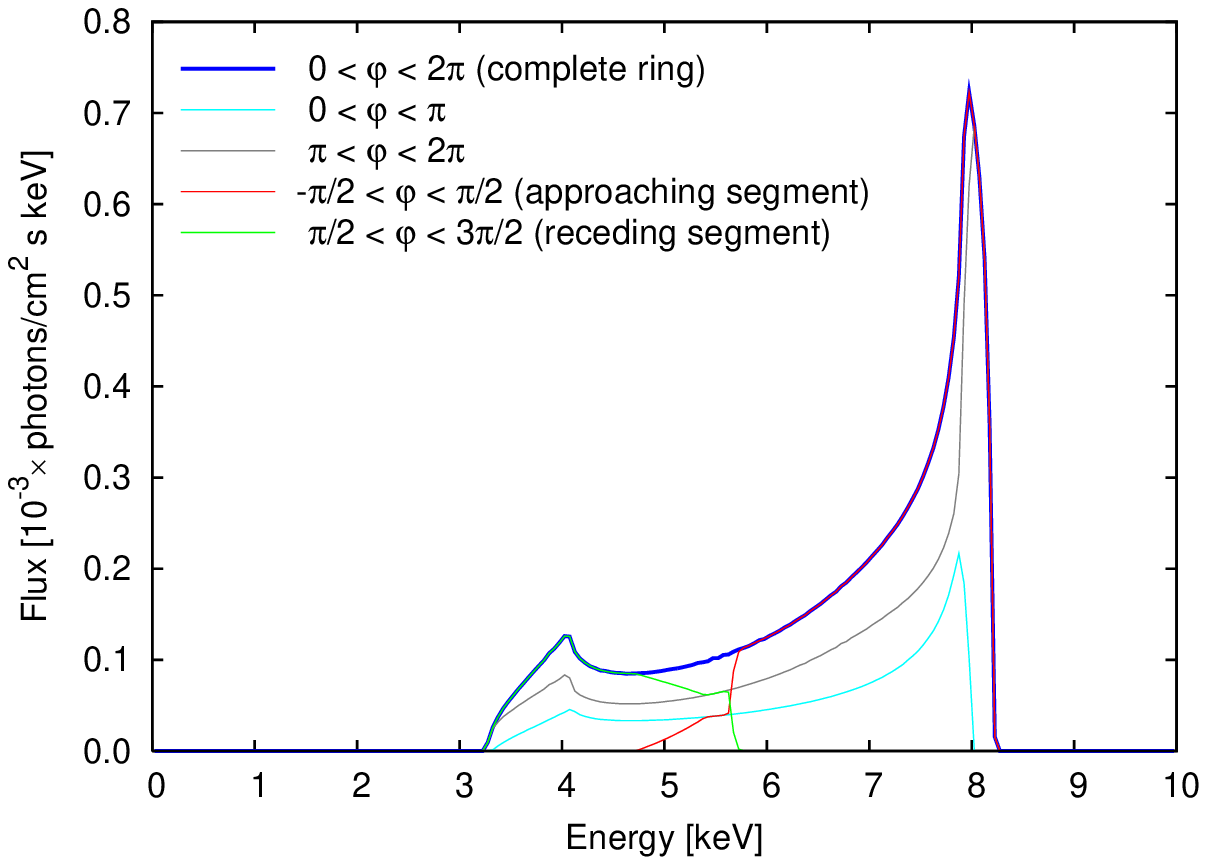}
\mycaption{
More calculated line profiles, as in the previous figure. Left: Line
profiles for different values of $\alpha$. Notice the enhanced red tail
of the line when the intrinsic emission is concentrated to the centre of
the disc. Right: Line emission originating from four different azimuthal
segments of the disc. This plot can serve as a toy model of
non-axisymmetric emissivity or to illustrate the expected  effects of
disc obscuration. Obviously, the receding segment of the disc
contributes mainly to the low-energy tail of the line while the
approaching segment constitutes the prominent high-energy peak. These
two plots illustrate a mutual interplay between the effect of  changing
$\alpha$ and the impact of obscuration, which complicates interpretation
of time-averaged spectra. The radial range is $r_{\rm{}h}<r<10^3$ in both
panels.}
\end{figure}

There is also a similar model {\sc kyg2line} present among the new {\sc{}xspec}
models, which is useful when fitting two general relativistic lines
simultaneously. The parameters are the same as in the {\sc kyg1line} model
except that there are two sets of those parameters
describing the local Gaussian line emission. There is one more parameter
present, {\tt ratio21}, which is the ratio of the maximum of the second local
line to the maximum of the first local line. Polarization computations are
not included in this model.

\begin{figure}[tb]
\dummycaption\label{fig:example3}
\includegraphics[width=0.5\textwidth]{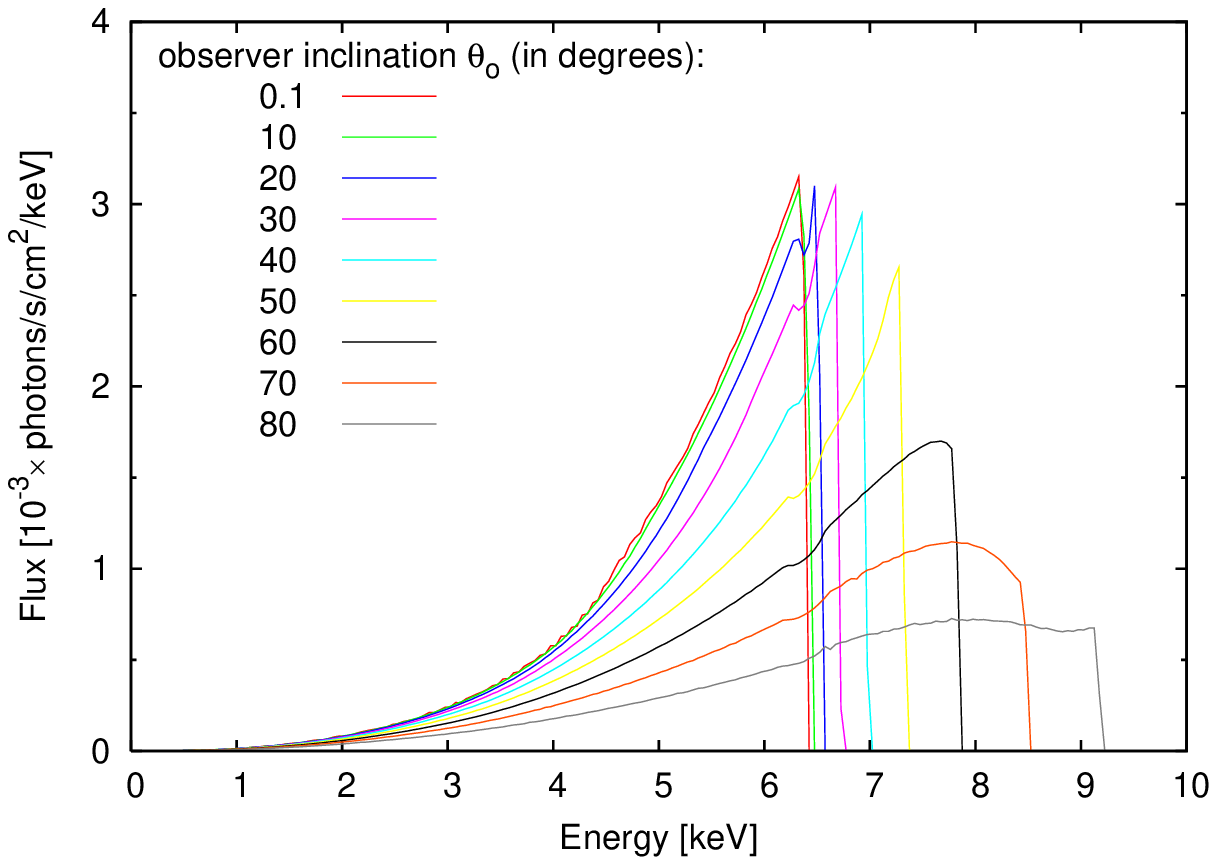}
\includegraphics[width=0.5\textwidth]{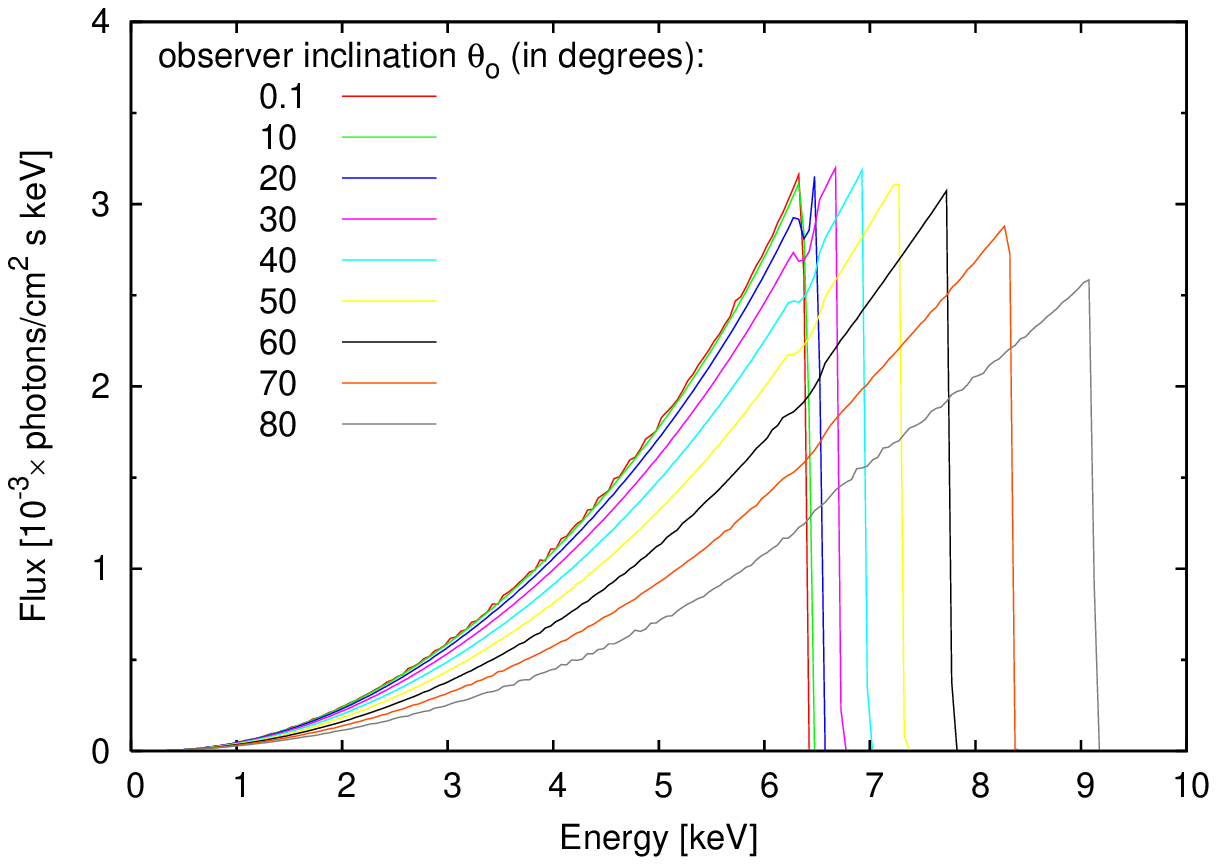}\\
\mycaption{Dependence on the observer inclination, $\theta_{\rm{}o}$, as given
in the plot. Left: Non-rotating black hole, $a=0$. Right: Maximally rotating
black hole, $a=1$. Other parameters as in Fig.~\ref{fig:example0} and
\ref{fig:example1}.}
\end{figure}

Results of an elementary code test are shown in
Figs.~\ref{fig:example0}--\ref{fig:example3}. The intrinsic emissivity
was assumed to be a narrow Gaussian line (width
$\sigma\dot{=}0.42\,\mbox{FWHM}=2$\,eV) with the amplitude decreasing
$\propto{}r^{-\alpha}$ in the local frame co-moving with the disc
medium. Typically, the slope of the low-energy wing is rather sensitive to
 the radial dependence of emissivity.
No background continuum is included here, so these lines can be
compared with similar pure disc-line profiles obtained in previous
papers \citep{kojima1991,laor1991} which also imposed the assumption of
axially symmetric and steady emission from an irradiated thin disc.
Again, the intrinsic width of the line is assumed to be much less than
the effects of broadening due to bulk Keplerian motion and the central
gravitational field.

\begin{figure}[tb]
\dummycaption\label{fig:example4}
\includegraphics[width=0.5\textwidth]{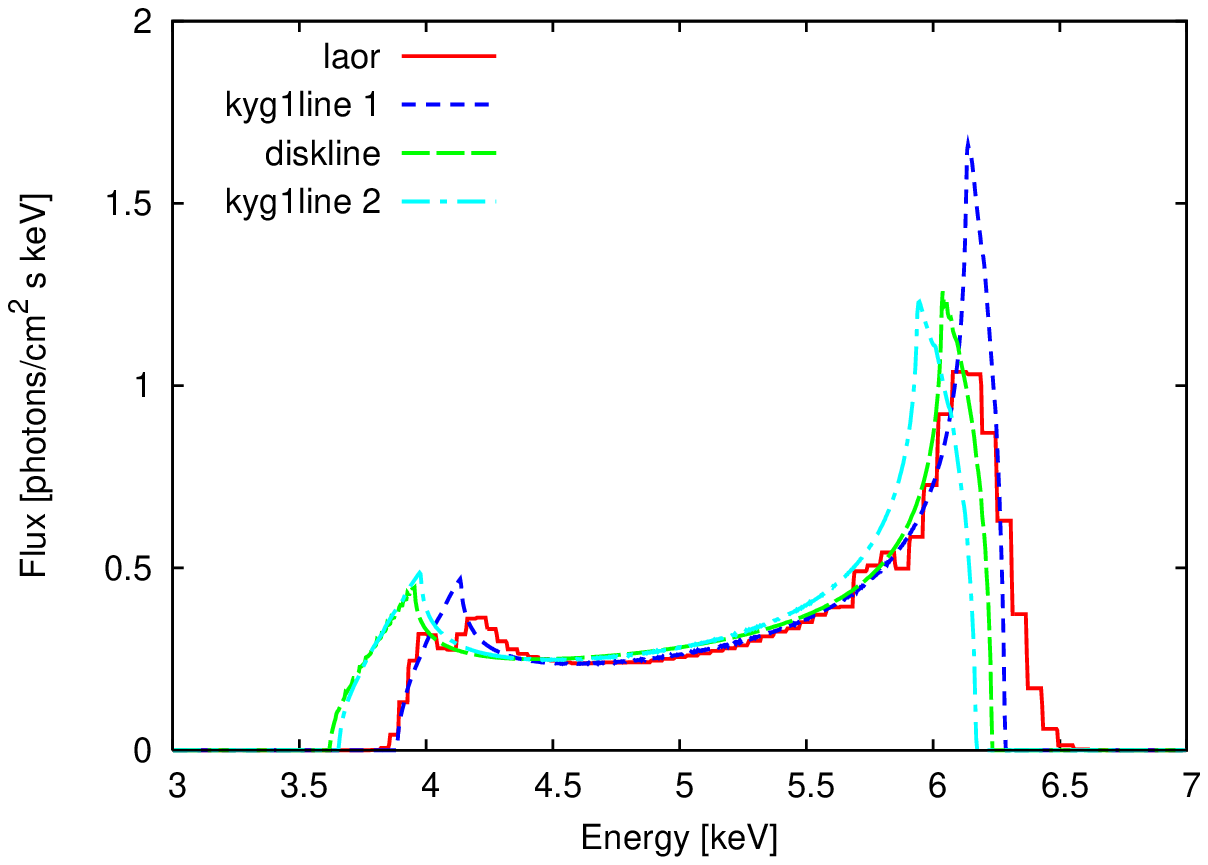}
\includegraphics[width=0.5\textwidth]{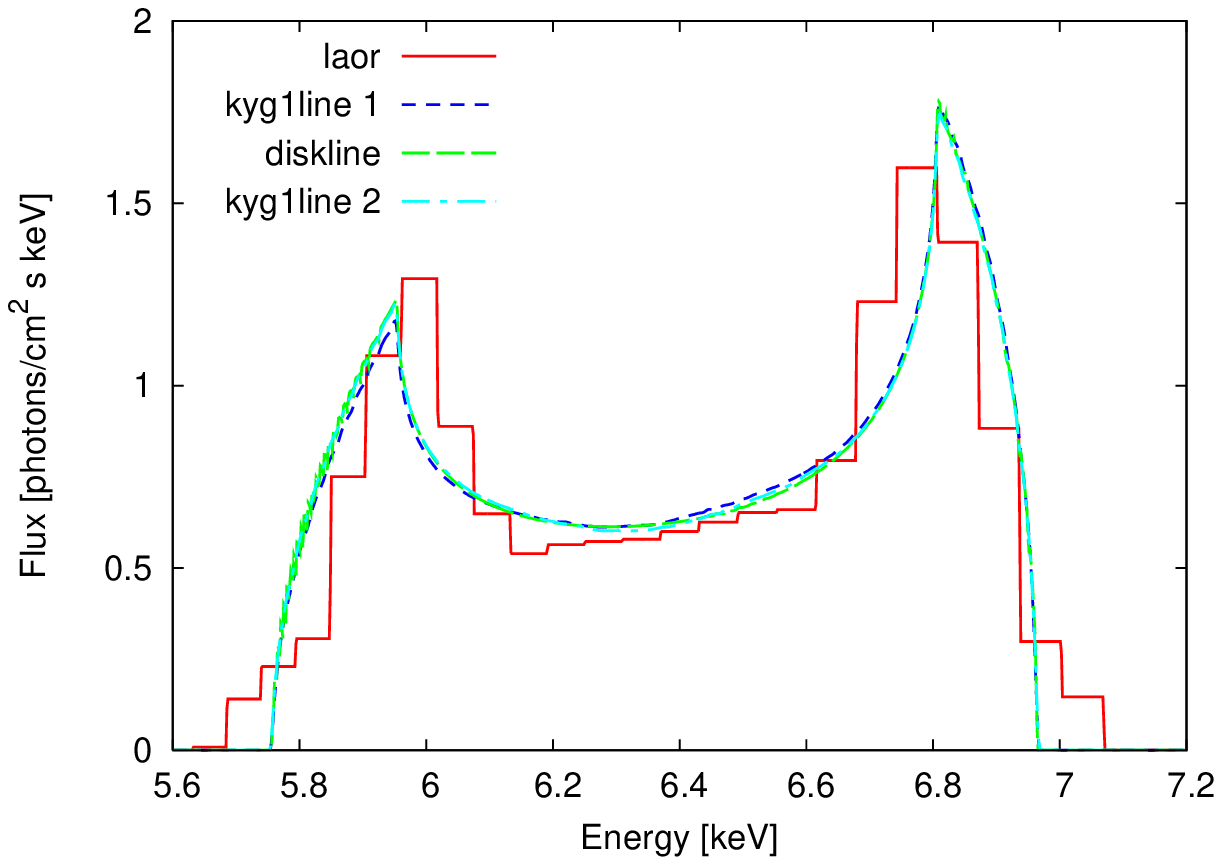}
\mycaption{Comparison of the output from {\sc{}xspec} models
for the disc-line problem: {\sc{}laor}, {\sc{}diskline}, and
{\sc{}kyg1line} (line 1 corresponds to the same limb darkening
law and $a=0.9982$ as in {\sc{}laor}; line 2 assumes locally
isotropic emission and $a=0$ as in {\sc{}diskline}). Left panel:
$\theta_{\rm{}o}=30^{\circ}$, $r_{\rm{}in}=6$, $r_{\rm{}out}=7$.
Right panel: $\theta_{\rm{}o}=70^{\circ}$, $r_{\rm{}in}=100$, $r_{\rm{}out}=200$.
Radial decay of intrinsic emissivity follows $\alpha=3$ power law.}
\vspace*{0.9\bigskipamount}
\dummycaption\label{fig:example5}
\includegraphics[width=0.5\textwidth]{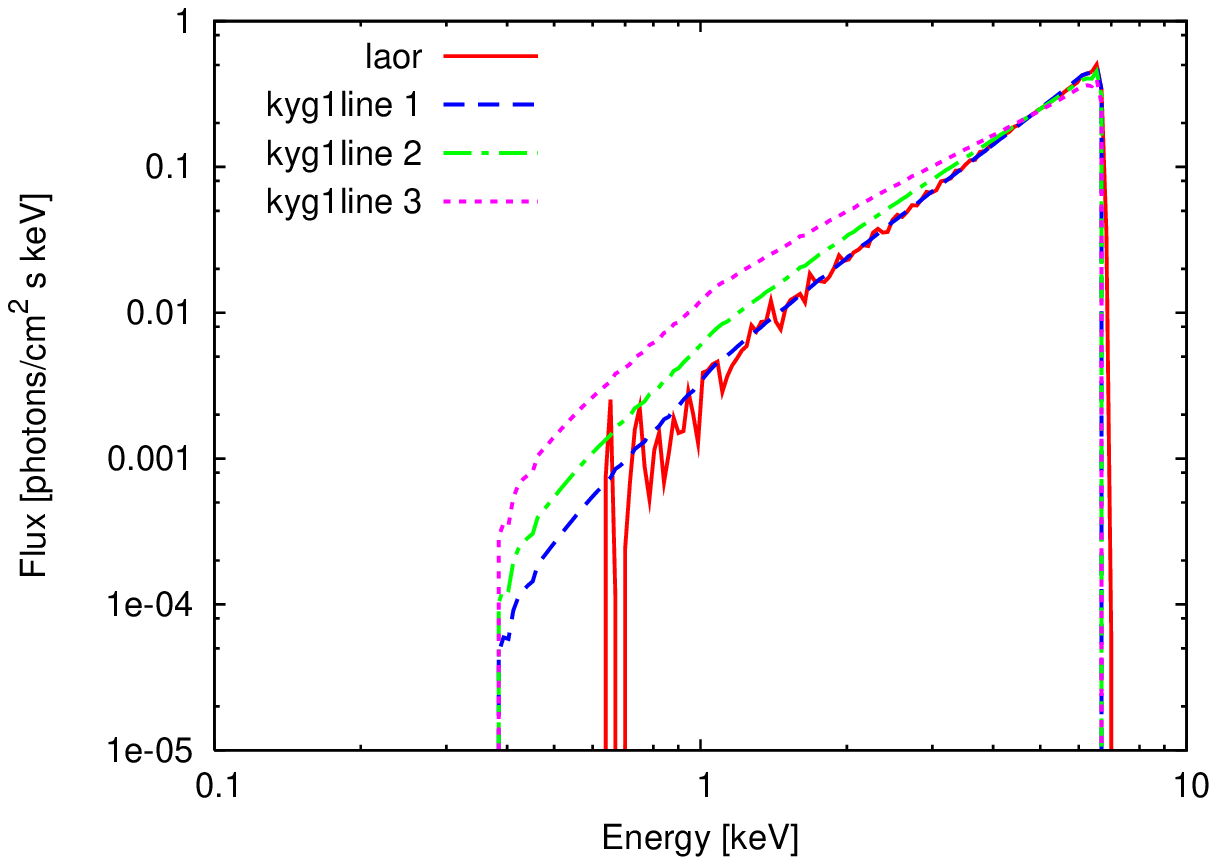}
\includegraphics[width=0.5\textwidth]{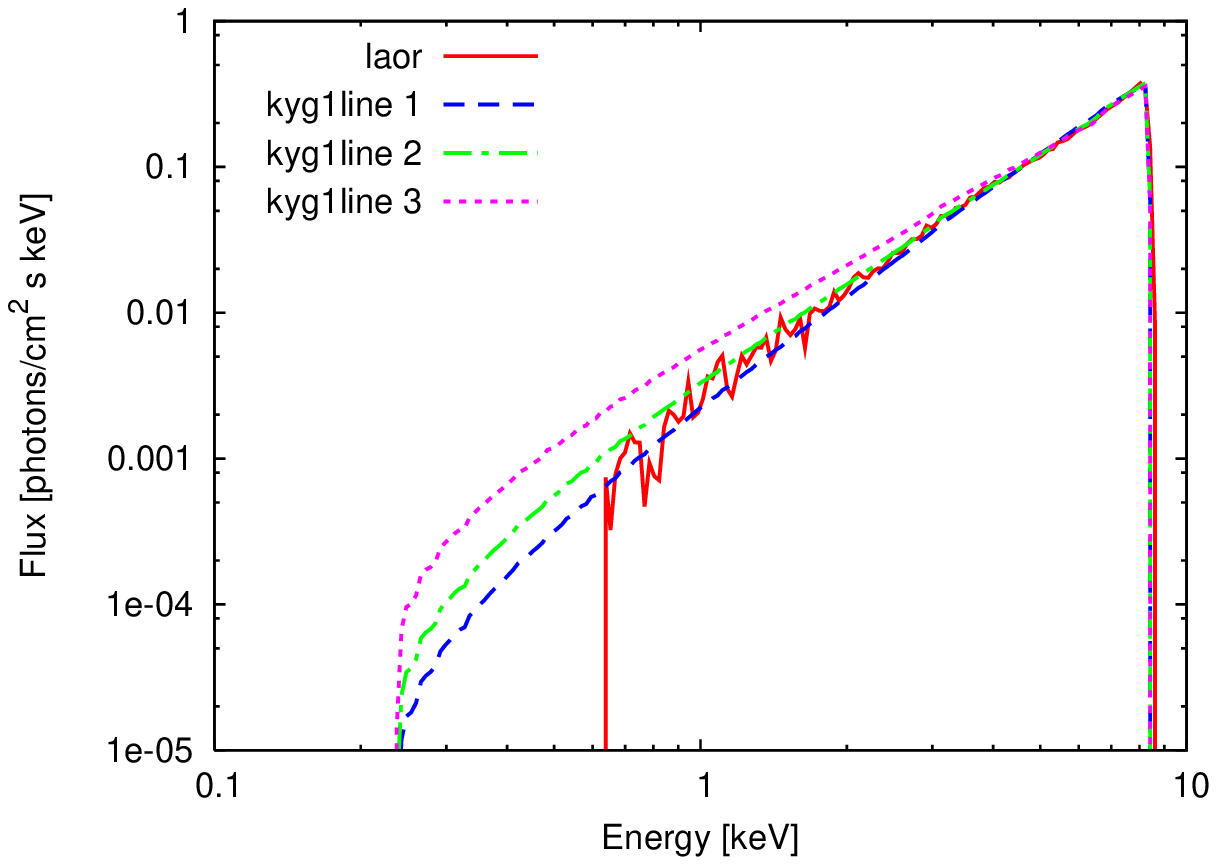}
\mycaption{Similar to previous figure but in logarithmic scale
and for three choices of the darkening law in {\sc{}kyg1line}
-- (1)~$f(\mu_{\rm e})=1+2.06\mu_{\rm e}$; (2)~$f(\mu_{\rm e})=1$;
(3) $f(\mu_{\rm e})=\log(1+1/\mu_{\rm e})$. Left panel: $\theta_{\rm{}o}=
30^{\circ}$; Right panel:  $\theta_{\rm{}o}=70^{\circ}$. In both panels,
$r_{\rm{}in}=r_{\rm{}ms}$, $r_{\rm{}out}=400$, $a=0.9982$.}
\end{figure}

Furthermore,
Figs.~\ref{fig:example4}--\ref{fig:example5} compare model spectra
of widely used {\sc{}xspec} models. In these examples one can see
that the {\sc{}laor} model gives zero contribution at energy below
$0.1E_{\rm{}rest}$ (in the disc local frame). This is because its grid has
only $35$ radial points distributed in the whole range $1.23{\leq}r{\leq}400$.
That is also why, in spite of a very efficient interpolation and smoothing of
the final spectrum, the {\sc{}laor} model does not accurately reproduce the
line originating from a narrow ring. Also, the dependence on the limb
darkening/brightening cannot be examined with this model, because the
form of directionality of the intrinsic emission is hard-wired in the
code, together with the position of the inner edge at
$r{\geq}r_{\rm{}ms}$. This affects especially the spectrum originating
near the black hole, where
radiation is expected to be very anisotropic and flow lines non-circular.
The {\sc{}diskline} model has also been frequently
used in the context
of spectral fitting, assuming a disc around a non-rotating black hole.
This model is analytical, and so it has clear advantages in
{\sc{}xspec}. Notice, however, that photons move along straight lines in this
model and that the lensing effect is neglected.

\subsection{Axisymmetric Gaussian line model
{\fontfamily{phv}\fontshape{sc}\selectfont kygline}}
This model uses eq.~(\ref{axisym_emission}) for computing the
disc emission with local flux being
\begin{eqnarray}
N_\loc(E_\loc) & = & \delta(E_\loc-{\tt Erest})\, ,\\
R(r) & = & r^{-{\tt alpha}}\, .
\end{eqnarray}
The function ${\rm d}F(\bar{g})\equiv {\rm d}\bar{g}\,F(\bar{g})$ in
eq.~(\ref{conv_function}) was pre-calculated for three different limb
darkening/brightening laws (see eqs.~(\ref{isotropic})--(\ref{haardt})) and stored in
corresponding FITS files {\tt KBHline00.fits} -- {\tt KBHline02.fits}. The local
emission is a delta function with its maximum depending on the radius as a power
law with index {\tt alpha} and also depending on the local emission angle. The
normalization constant $N_0$ in eq.~(\ref{axisym_emission}) is such that the
total integrated flux of the line is unity.

There are less parameters defining the line in this model than in
the previous one (see Tab.~\ref{kygline_par}):
\begin{description} \itemsep -2pt
 \item[{\tt Erest}] -- rest energy of the line in keV,
 \item[{\tt alpha}] -- radial power-law index.
\end{description}
Note that the limb darkening/brightening law can be chosen by means of
the {\tt ntable} switch.
\begin{table}[tbh]
\begin{center}
\dummycaption\label{kygline_par}
\begin{tabular}[h]{r@{}l|c|c|c|c}
\multicolumn{2}{c|}{parameter} & unit & default value  & minimum value &
maximum value \\ \hline
&{\tt a/M}      & $GM/c$   & 0.9982  & 0.     & 1.   \\
&{\tt theta\_o} & deg      & 30.     & 0.     & 89.  \\
&{\tt rin-rh}   & $GM/c^2$ & 0.      & 0.     & 999. \\
&{\tt ms}       & --       & 1.      & 0.     & 1.   \\
&{\tt rout-rh}  & $GM/c^2$ & 400.    & 0.     & 999. \\
&{\tt zshift}   & --       & 0.      & -0.999 & 10.  \\
&{\tt ntable}   & --       & 1.      & 0.     & 99.  \\
{*}&{\tt Erest} & keV      & 6.4     & 1.     & 99.  \\
{*}&{\tt alpha} & --       & 3.      & -20.   & 20.  \\
\end{tabular}
\mycaption{Parameters of the axisymmetric Gaussian line model {\sc{}kygline}.
Model parameters that are not common for all axisymmetric models are denoted
by asterisk.}
\end{center}
\end{table}

This model is much faster than the non-axisymmetric {\sc kyg1line} model.
Although it is not possible to change the resolution grid on the disc,
it is hardly needed because the resolution is set to be very large,
corresponding to ${\tt nrad}=500$,
${\tt division}=1$ and ${\tt nphi}=20\,000$ in the {\sc kyg1line} model,
which is more than sufficient in most cases.
(These values apply if the maximum range of radii is selected,
i.e.\ {\tt rin}=0, {\tt ms}=0 and {\tt rout}=999; in case of
a smaller range the number of points decreases accordingly.)
This means that the resolution of the {\sc kygline} model is much
higher than what can be achieved with the {\tt laor} model, and the
performance is still very good.

\subsection{Non-axisymmetric fluorescent lamp-post line model
{\fontfamily{phv}\fontshape{sc}\selectfont kyf1ll}}
\label{section_kyf1ll}
The line in this model is
induced by the illumination of the disc from the primary power-law source
located on the axis at {\tt height} above the black hole.
This model computes the final spectrum according to eq.~(\ref{emission})
with the local photon flux
\begin{eqnarray}
\nonumber
N_{\loc}(E_\loc) & = & g_{\rm L}^{{\tt PhoIndex}-1}\frac{\sin\theta_{\rm L}
{\rm d}\theta_{\rm L}}{r\,{\rm d}r}\,\sqrt{1-\frac{2\,{\tt height}}
{{\tt height}^2+{\tt (a/M)}^2}}\,f(\mu_{\rm i},\mu_{\rm e})\\
\label{fl_emission}
 & & \times\ \exp{\left [-\left
(1000\,\frac{E_\loc-{\tt Erest}}{\sqrt{2}\,{\tt sigma}}\right )^2\right ]}.
\end{eqnarray}
Here, $g_{\rm L}$ is ratio of the energy of a photon received by the
accretion disc to the energy of the same photon when emitted from a source
on the axis, $\theta_{\rm L}$ is an angle under which the photon is emitted
from the source (measured in the local frame of the source) and
$\mu_{\rm i}\equiv\cos{\,\delta_{\rm i}}$ is the cosine of the incident angle
(measured in the local frame of the disc) -- see
Fig.~\ref{compton_reflection}.
All of these functions depend on {\tt height} above the black hole at which
the source is located and on the rotational parameter {\tt a/M} of the black
hole. Values of $g_{\rm L}$, $\theta_{\rm L}$ and $\mu_{\rm i}$ for a given
height and rotation are read from the lamp-post tables {\tt lamp.fits}
(see Appendix~\ref{appendix3c}). At present, only tables for
${\tt a/M}=0.9987492$ (i.e.\ for the horizon of the black hole $r_{\rm h}=1.05$)
 and ${\tt height}=$ $2$,$\,3,\,4,\,5,\,6,\,8,\,10,\,12$,
$\,15,\,20,\,30,\,50,\,75$ and $100$ are included in {\tt lamp.fits},
therefore, the {\tt a/M} parameter is used only for the negative values
of {\tt height} (see below).

The factor in front of the function $f(\mu_{\rm i},\mu_{\rm e})$ gives
the radial dependence of the disc emissivity, which is different from the
assumed broken power law in the {\sc kyg1line} model.
For the derivation of this factor, which characterizes the illumination from
a primary source on the axis see Section~\ref{lamp-post}.

\begin{table}[tbh]
\begin{center}
\dummycaption\label{kyf1ll_par}
\begin{tabular}[h]{r@{}l|c|c|c|c}
\multicolumn{2}{c|}{parameter} & unit & default value  & minimum value &
maximum value \\ \hline
&{\tt a/M}         & $GM/c$   & 0.9982  & 0.       & 1.        \\
&{\tt theta\_o}    & deg      & 30.     & 0.       & 89.       \\
&{\tt rin-rh}      & $GM/c^2$ & 0.      & 0.       & 999.      \\
&{\tt ms}          & --       & 1.      & 0.       & 1.        \\
&{\tt rout-rh}     & $GM/c^2$ & 400.    & 0.       & 999.      \\
&{\tt phi}         & deg      & 0.      & -180.    & 180.      \\
&{\tt dphi}        & deg      & 360.    & 0.       & 360.      \\
&{\tt nrad}        & --       & 200.    & 1.       & 10000.    \\
&{\tt division}    & --       & 1.      & 0.       & 1.        \\
&{\tt nphi}        & --       & 180.    & 1.       & 20000.    \\
&{\tt smooth}      & --       & 1.      & 0.       & 1.        \\
&{\tt zshift}      & --       & 0.      & -0.999   & 10.       \\
&{\tt ntable}      & --       & 0.      & 0.       & 99.       \\
{*}&{\tt PhoIndex} & --       & 2.      & 1.5      & 3.        \\
{*}&{\tt height}   & $GM/c^2$ & 3.      & -20.     & 100.      \\
{*}&{\tt Erest}    & keV      & 6.4     & 1.       & 99.       \\
{*}&{\tt sigma}    & eV       & 2.      & 0.01     & 1000.     \\
&{\tt Stokes}      & --       & 0.      & 0.       & 6.        \\
\end{tabular}
\mycaption{Parameters of the fluorescent lamp-post line model {\sc{}kyf1ll}.
Model parameters that are not common for all non-axisymmetric models are
denoted by asterisk.}
\end{center}
\end{table}

The function $f(\mu_{\rm i},\mu_{\rm e})$ is a coefficient of
reflection. It depends on the incident and reflection angles.
Although the normalization of this function also depends on the photon
index of the power-law emission from a primary source,
we do not need to take this into account because the final spectrum
is always normalized to unity. Values of this function are read
from a pre-calculated table which is stored in
{\tt fluorescent\_line.fits} file (see \citealt{matt1991} and
Appendix~\ref{appendix3d}).

The local emission (\ref{fl_emission}) is defined in nine points
of local energy $E_\loc$ that are equally spaced with the
central point at its maximum. The normalization constant
$N_0$ in the formula (\ref{emission}) is such that the total
integrated flux of the line is unity. The parameters defining
local emission in this model are (see Tab.~\ref{kyf1ll_par}):
\begin{description} \itemsep -2pt
 \item[{\tt PhoIndex}] -- photon index of primary power-law illumination,
 \item[{\tt height}] -- height above the black hole where the primary source
 is located for ${\tt height}>0$, and radial power-law index for
 ${\tt height}\le0$,
 \item[{\tt Erest}] -- rest energy of the line in keV,
 \item[{\tt sigma}] -- width of the line in eV.
\end{description}
If positive, the {\tt height} parameter works as a switch -- the exact
value present in the tables {\tt lamp.fits} must be chosen.
If the {\tt height} parameter is negative, then this model assumes that
the local emission is the same as in the {\sc kyg1line} model with the
parameters ${\tt alpha}=-{\tt height}$, ${\tt rb=0}$ and ${\tt limb}=-2$
({\tt PhoIndex} parameter is unused in this case).

\begin{figure}[tb]
\begin{center}
\dummycaption\label{kyf}
  \begin{tabular}{cc}
   \hspace*{-2.5mm}\includegraphics[width=6.75cm]{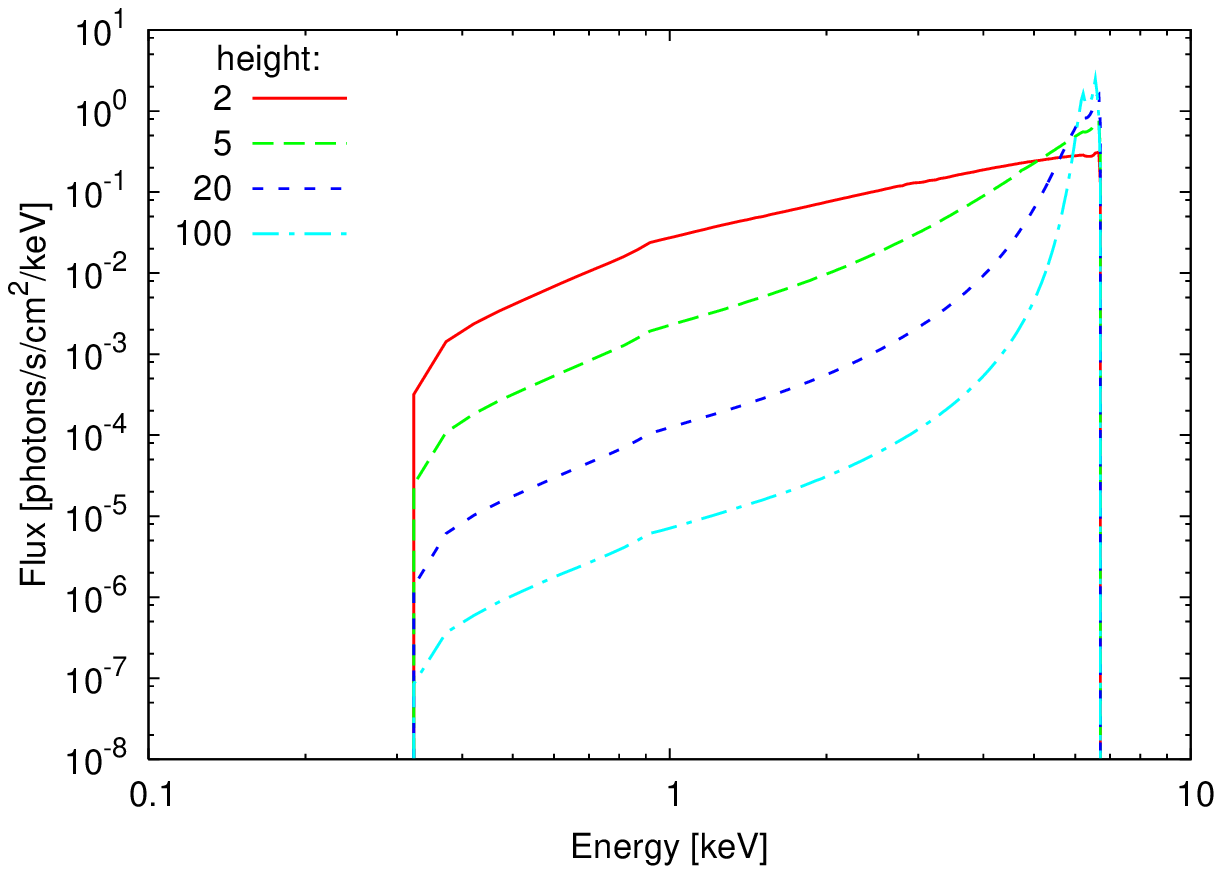} &
   \hspace*{-2.5mm}\includegraphics[width=6.75cm]{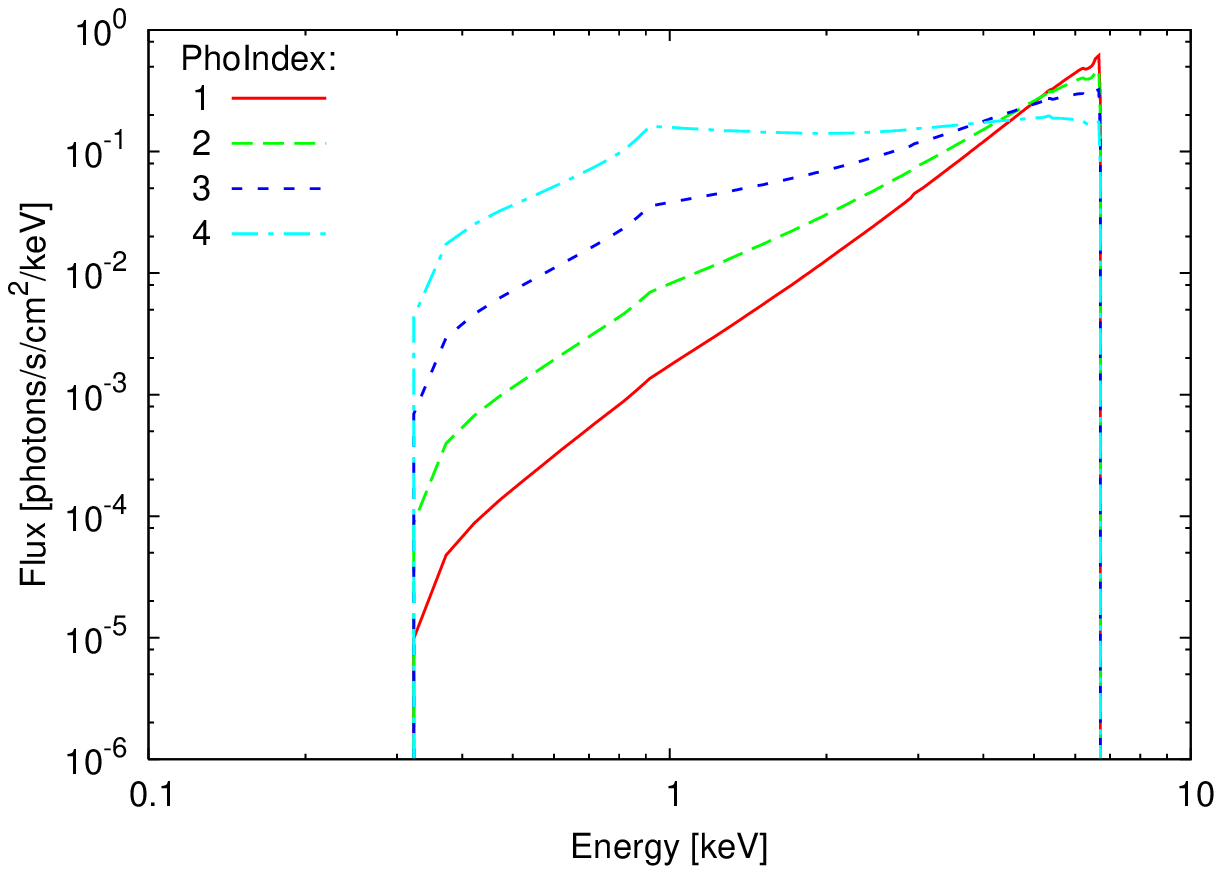}
  \end{tabular}
\mycaption{An example of a line profile originating from a disc in
equatorial plane of a Kerr black hole ($a=0.9987$, i.e.\ $r_{\rm h}=1.05$)
due to the illumination
from a primary source on the axis. The {\sc kyf1ll} model was used. Left:
Dependence of the line profile on the
height (in $GM/c^2$) of a primary source with photon index $\Gamma=2$.
Right: Dependence of the line profile on the photon index of the primary
emission with a source at height $3\,GM/c^2$ above the black hole.}
\end{center}
\end{figure}

In Fig.~\ref{kyf} we demonstrate that the broad iron emission lines due to
illumination from the source placed on the axis depend heavily on the height
where the ``lamp'' is located (left), as well as on the photon index of the
primary emission (right). These graphs correspond to the iron K$\alpha$ line
with the rest energy of $6.4$~keV.

\section{Compton reflection models}
We have developed two new relativistic continuum models -- the lamp-post Compton
reflection model {\sc kyl1cr} and the {\sc kyh1refl} model which is a
relativistically blurred {\sc hrefl} model that is already present in
{\sc{}xspec}. Both of these models are non-axisymmetric.

\subsection{Non-axisymmetric lamp-post Compton reflection model
{\fontfamily{phv}\fontshape{sc}\selectfont kyl1cr}}
\label{kyl1cr}
The emission in this model is induced by the illumination of the disc from the
primary power-law source located on the axis at {\tt height} above the
black hole. As in every non-axisymmetric model, the observed spectrum is
computed according to eq.~(\ref{emission}). The local emission is
\begin{myeqnarray}
\label{cr_emission}
N_{\loc}(E_\loc) \hspace*{-0.8em} & = & \hspace*{-0.8em}
g_{\rm L}^{{\tt PhoIndex}-1}\frac{\sin\theta_{\rm L}
{\rm d}\theta_{\rm L}}{r\,{\rm d}r}\,\sqrt{1-\frac{2\,{\tt height}}
{{\tt height}^2+{\tt (a/M)}^2}}\,f(E_\loc;\mu_{\rm i},\mu_{\rm e}) & &
\hspace*{-1.3em} {\rm for}\ {\tt height} > 0\, ,\hspace*{10mm}\\
\label{cr_emission_neg}
N_{\loc}(E_\loc) \hspace*{-0.8em} & = & \hspace*{-0.8em}
r^{\tt height}\,\bar{f}(E_\loc;\mu_{\rm e}) & &
\hspace*{-1.3em} {\rm for}\ {\tt height \le 0}\, .
\end{myeqnarray}
For the definition of $g_{\rm L}$, $\theta_{\rm L}$ and $\mu_{\rm i}$ see
Section~\ref{section_kyf1ll} and Appendix~\ref{appendix3c}, where pre-calculated
tables of these functions in {\tt lamp.fits} are described.

\begin{table}[tbh]
\begin{center}
\dummycaption\label{kyl1cr_par}
\begin{tabular}[h]{r@{}l|c|c|c|c}
\multicolumn{2}{c|}{parameter}  & unit & default value  & minimum value &
maximum value \\ \hline
&{\tt a/M}          & $GM/c$   & 0.9982   & 0.       & 1.        \\
&{\tt theta\_o}     & deg      & 30.      & 0.       & 89.       \\
&{\tt rin-rh}       & $GM/c^2$ & 0.       & 0.       & 999.      \\
&{\tt ms}           & --       & 1.       & 0.       & 1.        \\
&{\tt rout-rh}      & $GM/c^2$ & 400.     & 0.       & 999.      \\
&{\tt phi}          & deg      & 0.       & -180.    & 180.      \\
&{\tt dphi}         & deg      & 360.     & 0.       & 360.      \\
&{\tt nrad}         & --       & 200.     & 1.       & 10000.    \\
&{\tt division}     & --       & 1.       & 0.       & 1.        \\
&{\tt nphi}         & --       & 180.     & 1.       & 20000.    \\
&{\tt smooth}       & --       & 1.       & 0.       & 1.        \\
&{\tt zshift}       & --       & 0.       & -0.999   &  10.      \\
&{\tt ntable}       & --       & 0.       & 0.       & 99.       \\
{*}&{\tt PhoIndex}  & --       & 2.       & 1.5      & 3.        \\
{*}&{\tt height}    &$GM/c^2$  & 3.       & -20.     & 100.      \\
{*}&{\tt line}      & --       & 0.       & 0.       & 1.        \\
{*}&{\tt E\_cut}    & keV      & 300.     & 1.       & 1000.     \\
&{\tt Stokes}       & --       & 0.       & 0.       & 6.        \\
\end{tabular}
\mycaption{Parameters of the lamp-post Compton reflection model {\sc{}kyl1cr}.
Model parameters that are not common for all non-axisymmetric models are
denoted by asterisk.}
\end{center}
\end{table}

The function $f(E_\loc;\mu_{\rm i},\mu_{\rm e})$ gives the dependence of the
locally
emitted spectrum on the angle of incidence and the angle of emission, assuming a
power-law illumination. This function depends on the photon index {\tt PhoIndex}
of the power-law emission from a primary source.
Values of this function for various photon indices of primary emission
are read from the pre-calculated tables stored in {\tt refspectra.fits}
(see Appendix~\ref{appendix3e}). These tables were calculated by the Monte Carlo
simulations of Compton scattering in\break \cite{matt1991}. At present,
tables for ${\tt PhoIndex}=1.5,\ 1.6,\ \dots,\ 2.9,\ 3.0$  and for local
energies in the range from $2\,$keV to $300$~keV are available.
The normalization constant $N_0$ in eq.~(\ref{emission}) is such
that the final photon flux at an energy of $3$~keV is equal to unity,
which is different from what is usual for
continuum models in {\sc xspec} (where the photon flux is unity at $1\,$keV).
The choice adopted is due to the fact that current tables in
{\tt refspectra.fits} do not extend below $2\,$keV.

The function $\bar{f}(E_\loc;\mu_{\rm e})$, which is used for negative
{\tt height}, is an averaged function $f(E_\loc;\mu_{\rm i},\mu_{\rm e})$ over
$\mu_{\rm i}$
\begin{equation}
\bar{f}(E_\loc;\mu_{\rm e})\equiv\int_0^1 {\rm d}\mu_{\rm i}\,f(E_\loc;
\mu_{\rm i},\mu_{\rm e})\, .
\end{equation}
The local emission (\ref{cr_emission_neg}) can be interpreted as emission induced
by illumination from clouds localized near above the disc rather than from a
primary source on the axis (see Fig.~\ref{compton_reflection}). In this case
photons strike the disc from all directions.

For positive values of {\tt height} the {\sc kyl1cr} model includes a physical
model of polarization based on Rayleigh scattering in single scattering
approximation. The specific local Stokes parameters describing local
polarization of light are
\begin{eqnarray}
\label{polariz1}
i_\loc(E_\loc) & = & \frac{I_{\rm l}+I_{\rm r}}{\langle I_{\rm l}+
I_{\rm r}\rangle}\, N_\loc(E_\loc)\, ,\\[1mm]
q_\loc(E_\loc) & = & \frac{I_{\rm l}-I_{\rm r}}{\langle I_{\rm l}+
I_{\rm r}\rangle}\, N_\loc(E_\loc)\, ,\\[1mm]
u_\loc(E_\loc) & = & \frac{U}{\langle I_{\rm l}+
I_{\rm r}\rangle}\,N_\loc(E_\loc)\, ,\\[1mm]
\label{polariz4}
v_\loc(E_\loc) & = & 0\, ,
\end{eqnarray}
where the functions $I_{\rm l}$, $I_{\rm r}$ and $U$ determine the angular
dependence of the Stokes parameters in the following way
\begin{eqnarray}
\label{polariz2}
\nonumber
I_{\rm l} & = & \mu_{\rm e}^2(1+\mu_{\rm i}^2)+2(1-\mu_{\rm e}^2)
(1-\mu_{\rm i}^2)-4\mu_{\rm e}\mu_{\rm i}\sqrt{(1-\mu_{\rm e}^2)
(1-\mu_{\rm i}^2)}\,\cos{(\Phi_{\rm e}-\Phi_{\rm i})}\hspace*{3em} \\
 & & -\mu_{\rm e}^2(1-\mu_{\rm i}^2)\cos{[2(\Phi_{\rm e}-\Phi_{\rm i})]}\, ,\\
I_{\rm r} & = & 1+\mu_{\rm i}^2+(1-\mu_{\rm i}^2)\cos{[2(\Phi_{\rm e}-
\Phi_{\rm i})]}\, ,\\
U & = & -4\mu_{\rm i}\sqrt{(1-\mu_{\rm e}^2)(1-\mu_{\rm i}^2)}\,
\sin{(\Phi_{\rm e}-\Phi_{\rm i})}-2\mu_{\rm e}(1-\mu_{\rm i}^2)
\sin{[2(\Phi_{\rm e}-\Phi_{\rm i})]}\, .
\end{eqnarray}
Here $\Phi_{\rm e}$ and $\Phi_{\rm i}$ are the azimuthal emission
and the incident angles in the local rest frame co-moving with the accretion
disc (see Sections~\ref{azimuth_angle} and \ref{lamp-post} for their
definition). For the derivation of these formulae see the definitions (I.147)
and eqs.~(X.172) in \cite{chandrasekhar1960}.
We have omitted a common multiplication factor, which
would be cancelled anyway in eqs.~(\ref{polariz1})--(\ref{polariz4}).
The symbol $\langle\ \rangle$ in definitions of the local Stokes parameters
means value averaged over
the difference of the azimuthal angles $\Phi_{\rm e}-\Phi_{\rm i}$. We divide
the parameters by
$\langle I_{\rm l}+I_{\rm r}\rangle$ because the function
$f(E_\loc;\mu_{\rm i},\mu_{\rm e})$, and thus
also the local photon flux $N_\loc(E_\loc)$, is averaged over the difference of
the azimuthal angles.

The parameters defining local emission in this model are
(see Tab.~\ref{kyl1cr_par}):
\begin{description} \itemsep -2pt
 \item[{\tt PhoIndex}] -- photon index of primary power-law illumination,
 \item[{\tt height}] -- height above the black hole where the primary source
 is located for ${\tt height}>0$, and radial power-law index for
 ${\tt height}\le0$,
 \item[{\tt line}] -- switch whether to include the iron lines
 (0 -- no, 1 -- yes),
 \item[{\tt E\_cut}] -- exponential cut-off energy of the primary source in keV.
\end{description}
The tables {\tt refspectra.fits} for the function
$f(E_\loc;\mu_{\rm i},\mu{\rm e})$ also contain the emission in the iron lines
K$\alpha$ and K$\beta$. The two lines can be excluded from
computations if the {\tt line} switch is set to zero.
The {\tt E\_cut} parameter sets the upper boundary in energies
where the emission from a primary source ceases to follow a power-law
dependence. If the {\tt E\_cut} parameter is lower than both the maximum energy
of the considered dataset and the maximum energy in the tables
for $f(E_\loc;\mu_{\rm i},\mu{\rm e})$ in {\tt refspectra.fits} (300~keV),
then this model is not valid.

\begin{figure}[tb]
 \begin{center}
  \dummycaption\label{kyl_kyl}
  \begin{tabular}{cc}
   \includegraphics[width=6.5cm]{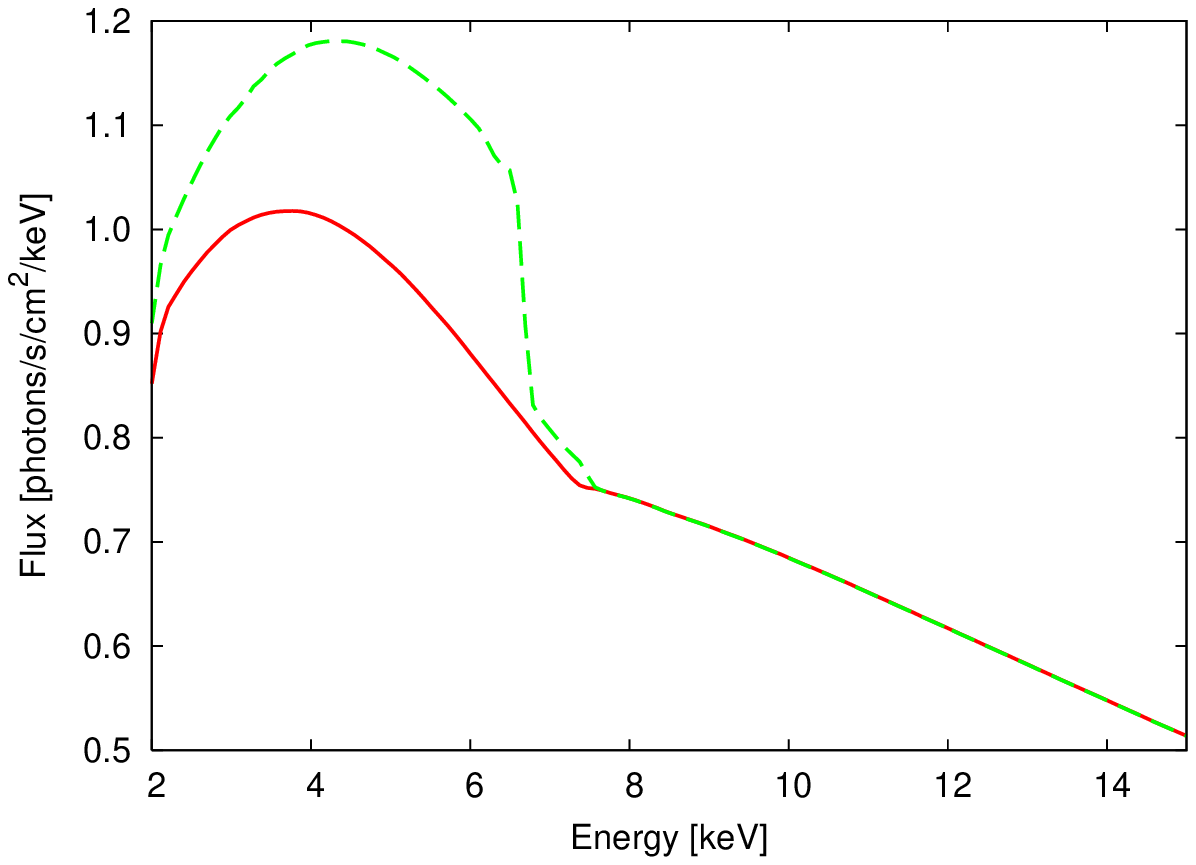} &
   \includegraphics[width=6.5cm]{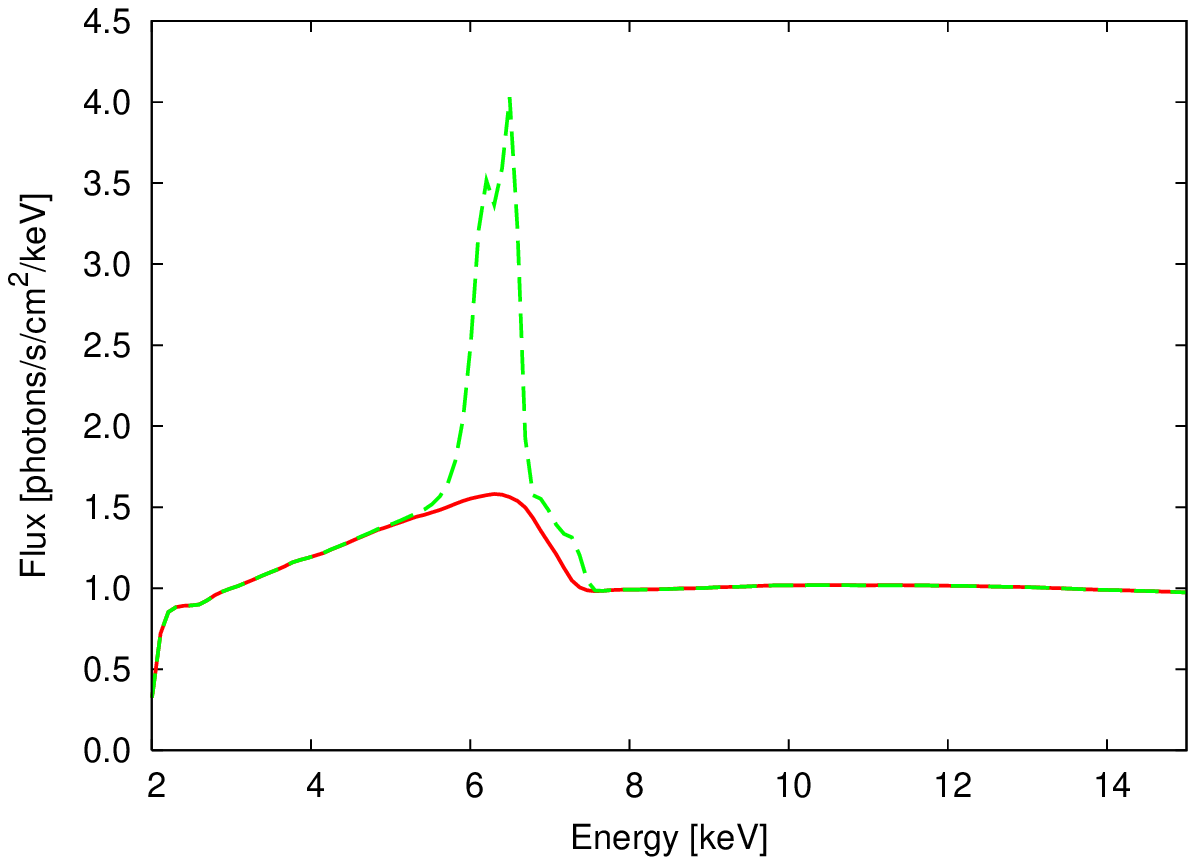}
  \end{tabular}
 \mycaption{General relativistic lamp-post Compton reflection model {\sc kyl1cr}
 with (dashed) and without (solid) iron lines K$\alpha$ and K$\beta$.
 The emission from the disc is induced by illumination from a primary source
 placed $2\,GM/c^2$ (left) and $100\,GM/c^2$ (right) above the black hole.}
 \end{center}
\end{figure}

\begin{table}[tbh]
\begin{center}
\dummycaption\label{kyh1refl_par}
\begin{tabular}[h]{r@{}l|c|c|c|c}
\multicolumn{2}{c|}{parameter} & unit & default value  & minimum value &
maximum value \\ \hline
&{\tt a/M}         & $GM/c$   & 0.9982 & 0.      & 1.      \\
&{\tt theta\_o}    & deg      & 30.    & 0.      & 89.     \\
&{\tt rin-rh}      & $GM/c^2$ & 0.     & 0.      & 999.    \\
&{\tt ms}          & --       & 1.     & 0.      & 1.      \\
&{\tt rout-rh}     & $GM/c^2$ & 400.   & 0.      & 999.    \\
&{\tt phi}         & deg      & 0.     & -180.   & 180.    \\
&{\tt dphi}        & deg      & 360.   & 0.      & 360.    \\
&{\tt nrad}        & --       & 200.   & 1.      & 10000.  \\
&{\tt division}    & --       & 1.     & 0.      & 1.      \\
&{\tt nphi}        & --       & 180.   & 1.      & 20000.  \\
&{\tt smooth}      & --       & 1.     & 0.      & 1.      \\
&{\tt zshift}      & --       & 0.     & -0.999  & 10.     \\
&{\tt ntable}      & --       & 0.     & 0.      & 99.     \\
{*}&{\tt PhoIndex} & --       & 1.     & 0.      & 10.     \\
{*}&{\tt alpha}    & --       & 3.     & -20.    & 20.     \\
{*}&{\tt beta}     & --       & 4.     & -20.    & 20.     \\
{*}&{\tt rb}       & $r_{\rm ms}$ & 0. & 0.      & 160.    \\
{*}&{\tt jump}     & --       & 1.     & 0.      & 1e6     \\
{*}&{\tt Feabun}   & --       & 1.     & 0.      & 200.    \\
{*}&{\tt FeKedge}  & keV      & 7.11   & 7.0     & 10.     \\
{*}&{\tt Escfrac}  & --       & 1.     & 0.      & 1000.   \\
{*}&{\tt covfac}   & --       & 1.     & 0.      & 1000.   \\
&{\tt Stokes}      & --       & 0.     & 0.      & 6.      \\
\end{tabular}
\mycaption{Parameters of the reflection {\sc{}kyh1refl} model. Model parameters
that are not common for all non-axisymmetric models are denoted by asterisk.}
\end{center}
\end{table}

Examples of the Compton reflection emission component of the spectra with and
without the fluorescent K$\alpha$ and K$\beta$ lines are shown in
Fig.~\ref{kyl_kyl}. It can be seen that originally narrow lines can
contribute substantially to the continuum component.

\subsection{Non-axisymmetric Compton reflection model
{\fontfamily{phv}\fontshape{sc}\selectfont kyh1refl}}
This model is based on an existing multiplicative {\sc hrefl} model in
combination with the {\sc powerlaw} model, both of which are present in
{\sc xspec}. Local emission in eq.~(\ref{emission}) is the same as the spectrum
given by the model {\sc hrefl*powerlaw} with the parameters ${\tt thetamin}=0$ and
${\tt thetamax}=90$ with a broken power-law radial dependence added:
\begin{myeqnarray}
N_\loc(E_\loc) & = & r^{-{\tt alpha}}\,\textsc{hrefl*powerlaw} & & {\rm for}
\quad r\ge r_{\rm b}\, ,\hspace*{1em}\\
N_\loc(E_\loc) & = & {\tt jump}\ r_{\rm b}^{{\tt beta}-{\tt alpha}}\,
r^{-{\tt beta}}\,\textsc{hrefl*powerlaw} & & {\rm for} \quad r<r_{\rm b}\, .
\end{myeqnarray}
\begin{figure}[tb]
 \begin{center}
 \dummycaption\label{kyh_href}
  \begin{tabular}{cc}
   \includegraphics[width=6.5cm]{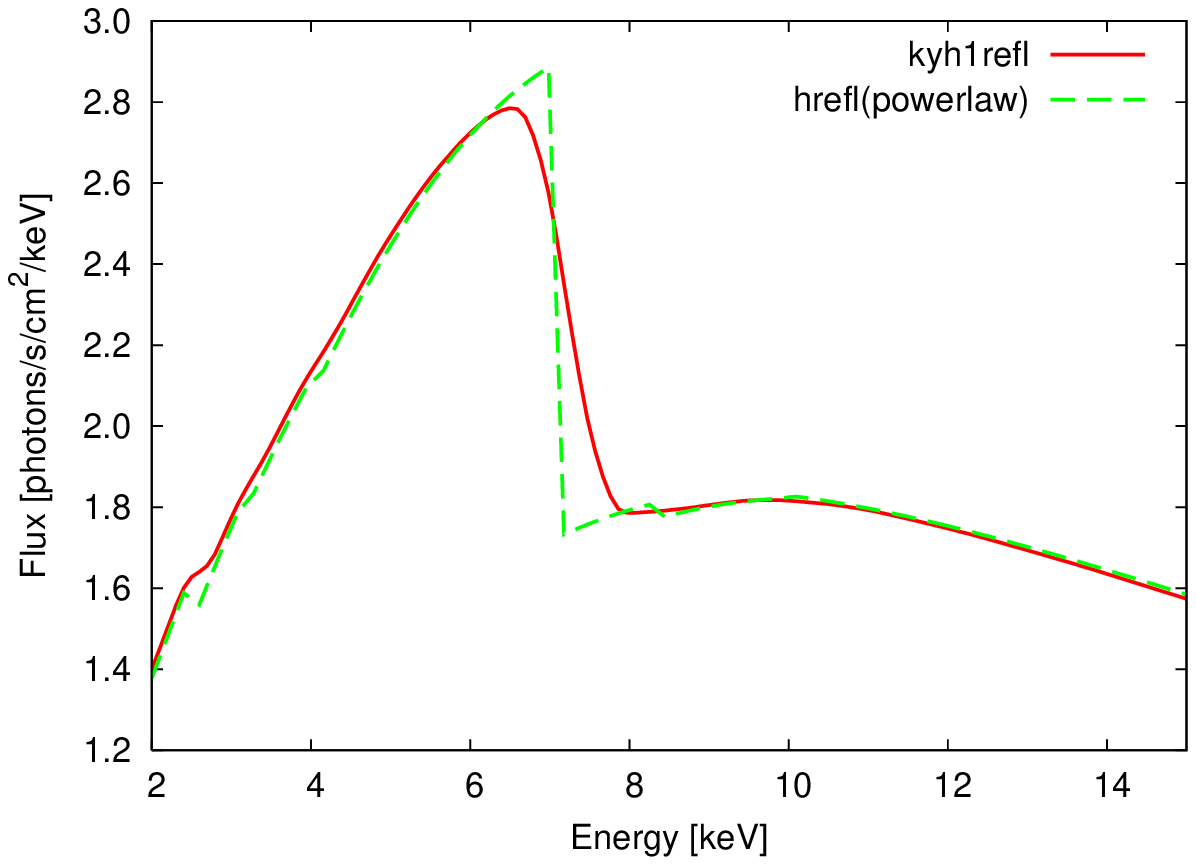} &
   \includegraphics[width=6.5cm]{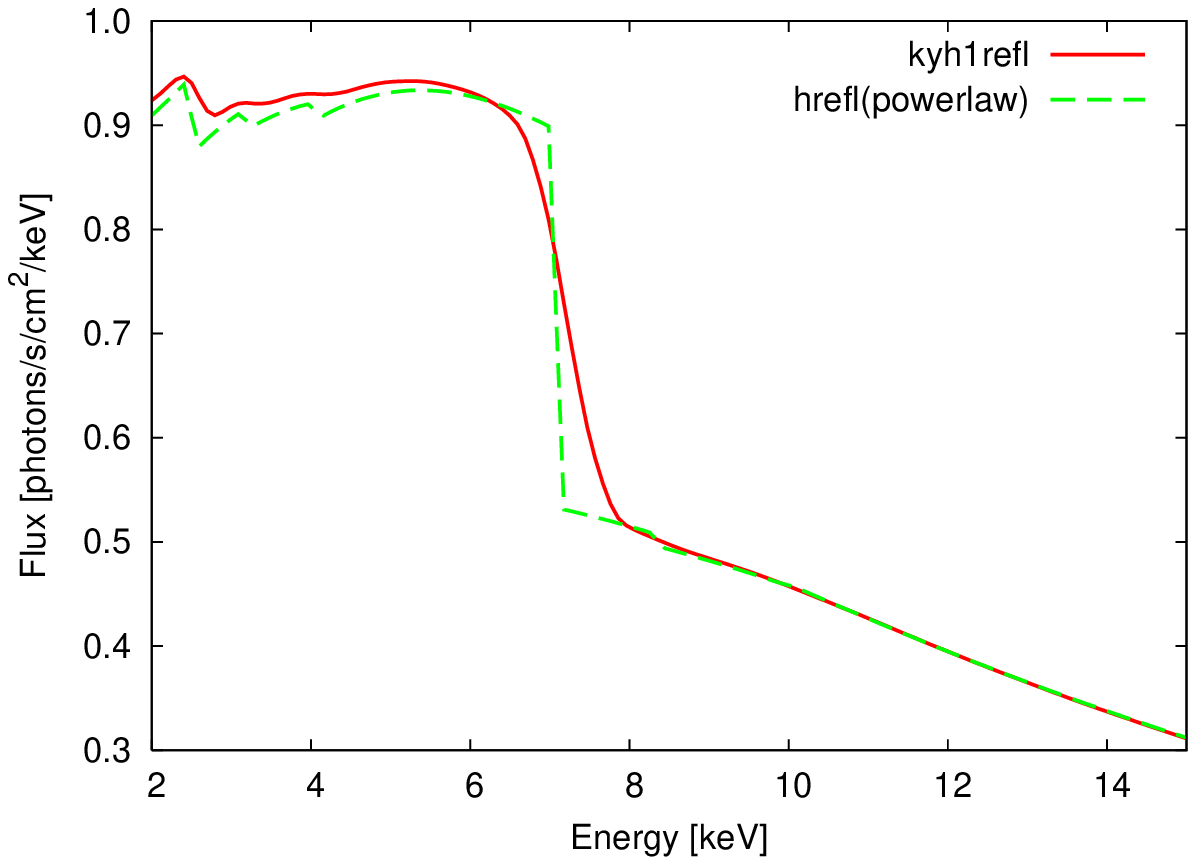}
  \end{tabular}
 \mycaption{Comparison of the general relativistic {\sc kyh1refl} model with the
 non-relativistic {\sc hrefl(powerlaw)}. The relativistic blurring of the iron
 edge is clearly visible. The power-law index of the primary source is
 {\tt PhoIndex}=2 (left) and {\tt PhoIndex}=2.6 (right).}
 \end{center}
\bigskip
 \begin{center}
 \dummycaption\label{kyl_kyh}
  \begin{tabular}{cc}
   \includegraphics[width=6.5cm]{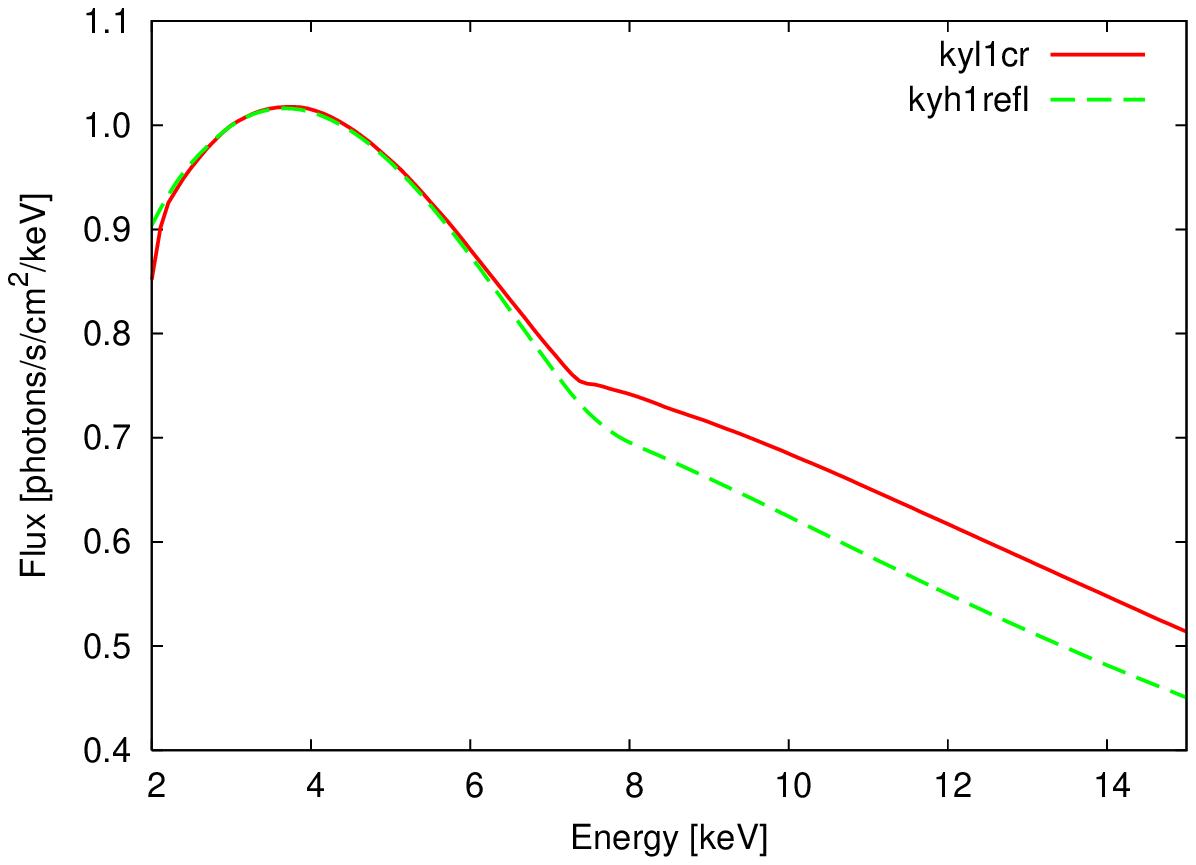} &
   \includegraphics[width=6.5cm]{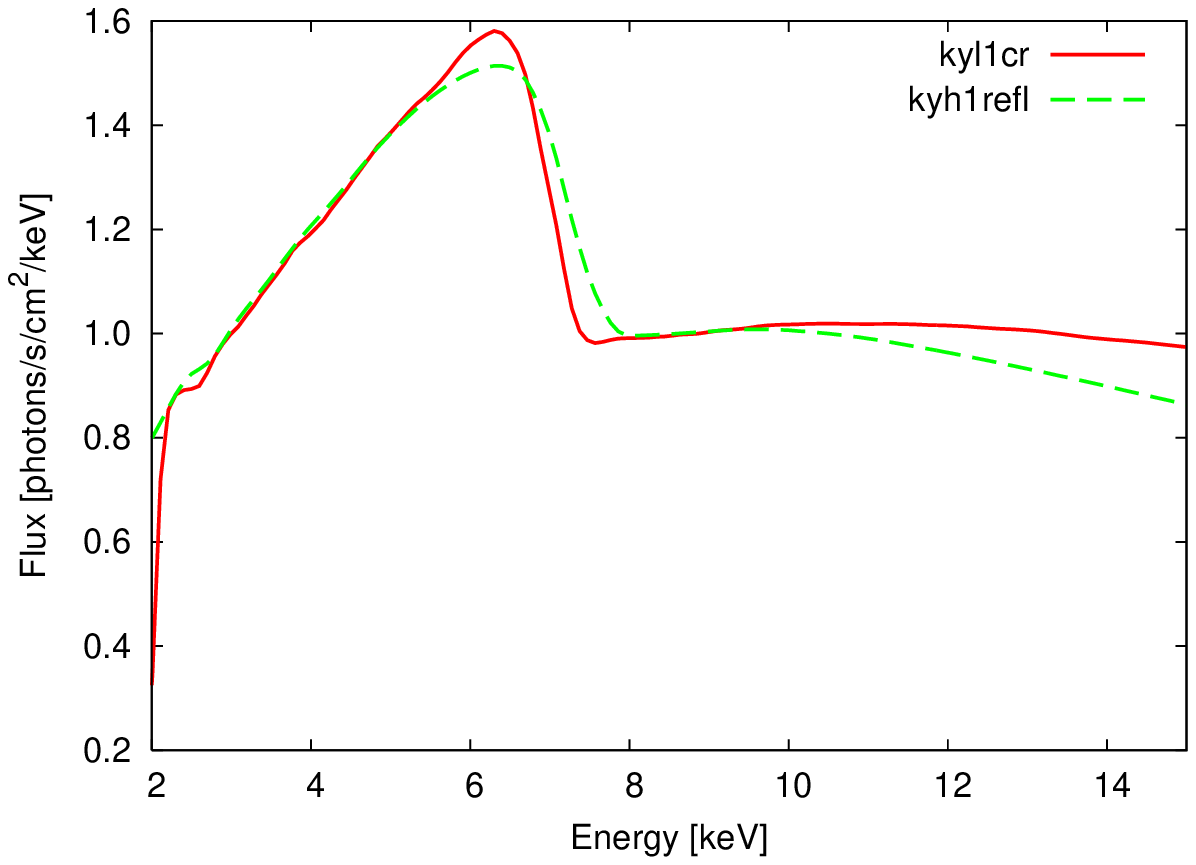}
  \end{tabular}
 \mycaption{Comparison of the two new general relativistic Compton reflection
 models {\sc kyl1cr} and
 {\sc kyh1refl}. The lamp-post {\sc kyl1cr} model is characterized by the height
 $h$ above the disc where a primary source of emission is placed, the
 reflection {\sc kyh1refl} model is characterized by the radial power-law index
 $\alpha$. Left: $h=2\,GM/c^2$, $\alpha=3.4\,$. Right:
 $h=100\,GM/c^2$, $\alpha=1.5\,$.}
 \end{center}
\end{figure}
For a definition of the boundary radius $r_{\rm b}$ by the {\tt rb} parameter
see eqs.~(\ref{rb1})--(\ref{rb2}), and
for a detailed description of the {\sc hrefl} model see \cite{dovciak2004}
and the {\sc xspec} manual. The {\sc{}kyh1refl} model can be interpreted
as a Compton-reflection model for which the source of primary
irradiation is near above the disc, in contrast to the lamp-post
scheme with the source on the axis (see Fig.~\ref{compton_reflection}).
The approximations for Compton reflection used in {\sc{}hrefl}
(and therefore also in {\sc{}kyh1refl})
are valid below $\sim15$~keV in the disc rest-frame.
The normalization of the final spectrum in this model is the same as
in other continuum models in {\sc xspec}, i.e.\ photon flux is unity at
the energy of $1$~keV.

The parameters defining the local emission in {\sc kyh1refl}
(see Tab.~\ref{kyh1refl_par}) are
\begin{description} \itemsep -2pt
 \item[{\tt PhoIndex}] -- photon index of the primary power-law illumination,
 \item[{\tt alpha}] -- radial power-law index for the outer region,
 \item[{\tt beta}] -- radial power-law index for the inner region,
 \item[{\tt rb}] -- parameter defining the border between regions with different
 power-law indices,
 \item[{\tt jump}] -- ratio between flux in the inner and outer regions at
 the border radius,
 \item[{\tt Feabun}] -- iron abundance relative to solar,
 \item[{\tt FeKedge}] -- iron K-edge energy,
 \item[{\tt Escfrac}] -- fraction of the direct flux from the power-law primary
 source seen by the observer,
 \item[{\tt covfac}] -- normalization of the reflected continuum.
\end{description}

  Smearing of sharp features in continuum (iron edge) by relativistic effects
is demonstrated in Fig.~\ref{kyh_href}, where
non-relativistic reflection model {\sc hrefl(powerlaw)} is compared with
our relativistic {\sc kyh1refl} model. Here, we set the radial power-law index
$\alpha=1$ in {\sc kyh1refl}. Other parameters defining these models were set to
their default values.

We compare the two new relativistic reflection models {\sc kyl1cr} and
{\sc kyh1refl} in Fig.~\ref{kyl_kyh}. Note that the {\sc kyl1cr} model is
valid only above approximately $2\,{\rm keV}$ and the {\sc kyh1refl} model only
below approximately $15\,{\rm keV}$.

\section{General relativistic convolution models}
We have also produced two convolution-type
models, {\sc ky1conv} and {\sc kyconv}, which can be applied to any existing
{\sc{}xspec} model for the intrinsic X-ray emission from a disc around a Kerr
black hole. We must stress
that these models are substantially more powerful than the usual
convolution models in {\sc{}xspec} (these are commonly
defined in terms of one-dimensional integration over energy bins).
Despite the fact that our convolution models still use the standard
{\sc{}xspec} syntax in evaluating the observed spectrum
(e.g.\ {\sc kyconv(powerlaw)}), our code
accomplishes a more complex operation. It still performs ray-tracing
across the disc surface so that the intrinsic model contributions are
integrated from different radii and azimuths on the disc.

There are several restrictions that arise from the fact that we use existing
{\sc xspec} models:
\begin{itemize} \itemsep -2pt
 \item[--] by local {\sc xspec} models only the energy dependence of the photon
 flux can be defined,
 \item[--] only a certain type of radial dependence of the local photon flux can
 be imposed -- we have chosen to use a broken power-law radial dependence,
 \item[--] there is no azimuthal dependence of the local photon flux, except
 through limb darkening law,
 \item[--] local flux depends on the binning of the data because it is defined
 in the centre of each bin, a large number of bins is needed for highly varying
 local flux.
\end{itemize}

For emissivities that cannot be defined by existing {\sc xspec} models,
or where the limitations mentioned above are too restrictive, one has
to add a new user-defined model to {\sc{}xspec} (by adding a new
subroutine to {\sc xspec}). This method is more flexible and faster than
convolution models (especially when compared with non-axisymmetric
one), and hence it is recommended even for cases when these
prefabricated models could be used. In any new model for {\sc
xspec} one can use the common ray-tracing driver for relativistic smearing
of the local emission: {\tt ide} for non-axisymmetric
models and {\tt idre} for axisymmetric ones. For a detailed description
see Appendixes~\ref{appendix4a} and \ref{appendix4b}.

\subsection{Non-axisymmetric convolution model
{\fontfamily{phv}\fontshape{sc}\selectfont kyc1onv}}
The local emission in this model is computed according to the
eq.~(\ref{emission}) with the local emissivity equal to
\begin{table}[tbh]
\begin{center}
\dummycaption\label{kyc1onv_par}
\begin{tabular}[h]{r@{}l|c|c|c|c}
\multicolumn{2}{c|}{parameter} & unit & default value  & minimum value &
maximum value \\ \hline
&{\tt a/M}         & $GM/c$   & 0.9982   & 0.       & 1.        \\
&{\tt theta\_o}    & deg      & 30.      & 0.       & 89.       \\
&{\tt rin-rh}      & $GM/c^2$ & 0.       & 0.       & 999.      \\
&{\tt ms}          & --       & 1.       & 0.       & 1.        \\
&{\tt rout-rh}     & $GM/c^2$ & 400.     & 0.       & 999.      \\
&{\tt phi}         & deg      & 0.       & -180.    & 180.      \\
&{\tt dphi}        & deg      & 360.     & 0.       & 360.      \\
&{\tt nrad}        & --       & 200.     & 1.       & 10000.    \\
&{\tt division}    & --       & 1.       & 0.       & 1.        \\
&{\tt nphi}        & --       & 180.     & 1.       & 20000.    \\
&{\tt smooth}      & --       & 1.       & 0.       & 1.        \\
{*}&{\tt normal}   & --       & 1.       & -1.      & 100.      \\
&{\tt zshift}      & --       & 0.       & -0.999   & 10.       \\
&{\tt ntable}      & --       & 0.       & 0.       & 99.       \\
{*}&{\tt ne\_loc}  & --       & 100.     & 3.       & 5000.     \\
{*}&{\tt alpha}    & --       & 3.       & -20.     & 20.       \\
{*}&{\tt beta}     & --       & 4.       & -20.     & 20.       \\
{*}&{\tt rb}       & $r_{\rm ms}$ & 0.   & 0.       & 160.      \\
{*}&{\tt jump}     & --       & 1.       & 0.       & 1e6       \\
{*}&{\tt limb}     & --       & 0.       & -10.     & 10.       \\
&{\tt Stokes}      & --       & 0.       & 0.       & 6.        \\
\end{tabular}
\mycaption{Parameters of the non-axisymmetric convolution model {\sc{}kyc1onv}.
Model parameters that are not common for all non-axisymmetric models are
denoted by asterisk.}
\end{center}
\end{table}
\begin{myeqnarray}
N_\loc(E_\loc) & = & r^{-{\tt alpha}}\,f(\mu_{\rm e})\,\textsc{model} & &
{\rm for}\quad r > r_{\rm b} \, ,\\
N_\loc(E_\loc) & = & {\tt jump}\ r_{\rm b}^{{\tt beta}-{\tt alpha}}\,
r^{-{\tt beta}}\,f(\mu_{\rm e})\,\textsc{model} & & {\rm for}\quad r
\le r_{\rm b} \, .\
\end{myeqnarray}
For a definition of the boundary radius $r_{\rm b}$ by the {\tt rb} parameter
see eqs.~(\ref{rb1})--(\ref{rb2}) and for the definition of different limb
darkening laws $f(\mu_{\rm e})$ see eqs.~(\ref{isotropic})--(\ref{other_limb}).
The local emission is given by the
{\sc model} in the centre of energy bins used in {\sc xspec} with the
broken power-law radial dependence and limb darkening law added. Apart from the
parameters of the {\sc model}, the local emission is defined also by the
following parameters (see Tab.~\ref{kyc1onv_par}):
\begin{description} \itemsep -2pt
 \item[{\tt normal}] -- switch for the normalization of the final spectrum,\\
   $=$ 0 -- total flux is unity (usually used for the line),\\
   $>$ 0 -- flux is unity at the energy = {\tt normal} keV (usually used for
   the continuum),\\
   $<$ 0 -- flux is not normalized,
 \item[{\tt ne\_loc}] -- number of points in the energy grid where the local
 photon flux is defined,
 \item[{\tt alpha}] --  radial power-law index for the outer region,
 \item[{\tt beta}] -- radial power-law index for the inner region,
 \item[{\tt rb}] -- parameter defining the border between regions with different
 power-law indices,
 \item[{\tt jump}] -- ratio between the flux in the inner and outer regions at
 the border radius,
 \item[{\tt limb}] -- switch for different limb darkening/brightening laws.
\end{description}
\begin{table}[tbh]
\begin{center}
\dummycaption\label{kyconv_par}
\begin{tabular}[h]{r@{}l|c|c|c|c}
\multicolumn{2}{c|}{parameter} & unit & default value  & minimum value &
maximum value \\ \hline
&{\tt a/M}         & $GM/c$   & 0.9982   & 0.       & 1.        \\
&{\tt theta\_o}    & deg      & 30.      & 0.       & 89.       \\
&{\tt rin-rh}      & $GM/c^2$ & 0.       & 0.       & 999.      \\
&{\tt ms}          & --       & 1.       & 0.       & 1.        \\
&{\tt rout-rh}     & $GM/c^2$ & 400.     & 0.       & 999.      \\
&{\tt zshift}      & --       & 0.       & -0.999   & 10.       \\
&{\tt ntable}      & --       & 0.       & 0.       & 99.       \\
{*}&{\tt alpha}    & --       & 3.       & -20.     & 20.       \\
{*}&{\tt ne\_loc}  & --       & 100.     & 3.       & 5000.     \\
{*}&{\tt normal}   & --       & 1.       & -1.      & 100.      \\
\end{tabular}
\mycaption{Parameters of the axisymmetric convolution model {\sc{}kyconv}.
Model parameters that are not common for all axisymmetric models are denoted
by asterisk.}
\end{center}
\end{table}
The local emission in each {\sc ky} model has to by defined either on
equidistant or exponential (i.e.\ equidistant in logarithmic scale)
energy grid. Because the energy grid used in the convolution model depends on
the binning of the data, which may be arbitrary, the flux has to be
rebinned. It is always rebinned into an exponentially spaced
energy grid in {\sc ky} convolution models.
The {\tt ne\_loc} parameter defines the number of points in which the rebinned
flux will be defined.

\subsection{Axisymmetric convolution model
{\fontfamily{phv}\fontshape{sc}\selectfont kyconv}}
The local emission in this model is computed according to
eq.~(\ref{axisym_emission}) with the local emissivity equal to
\begin{eqnarray}
N_\loc(E_\loc) & = & \textsc{model}\, ,\\
R(r) & = & r^{-{\tt alpha}}\, .
\end{eqnarray}
Except for the parameters of the {\sc model}, the local emission is defined also
by the following parameters (see Tab.~\ref{kyconv_par}):
\begin{description} \itemsep -2pt
 \item[{\tt alpha}] --  radial power-law index,
 \item[{\tt ne\_loc}] -- number of points in energy grid where local photon
 flux is defined,
 \item[{\tt normal}] -- switch for the normalization of the final spectrum,\\
   $=$ 0 -- total flux is unity (usually used for the line),\\
   $>$ 0 -- flux is unity at the energy = {\tt normal} keV (usually used for
   the continuum),\\
   $<$ 0 -- flux is not normalized.
\end{description}
Note that the limb darkening/brightening law can be chosen through the
{\tt ntable} switch. This model is much faster than the non-axisymmetric
convolution model {\sc kyc1onv}.

\section{Non-stationary model
{\fontfamily{phv}\fontshape{sc}\selectfont kyspot}}

Emission in this model originates from a localized spot on the disc. The spot
may either move along a stable circular orbit or fall from the vicinity of the
marginally stable orbit down to the horizon.
In the former case the spot is orbiting with a Keplerian velocity, in the latter
it falls with energy and angular momentum of the matter on the marginally
stable orbit.
Because the emission changes in time this model cannot be included into
{\sc{}xspec} which
cannot handle non-stationary problems. In spite of this we may include
integrated emission received by the observer in a certain time span
(corresponding to a certain part of the orbit) in future.

In this model it is assumed that the corona above the disc is heated by a flare.
Thus a hot cloud is formed, which illuminates the disc below it by X-rays.
The disc reflects this radiation by Compton scattering and by fluorescence.
In the current model we consider only the fluorescent part of the reflection.

The local emission from the disc is
\begin{eqnarray}
\label{spot1a}
N_{\loc}(E_\loc)\hspace*{-0.5em} & = & \hspace*{-0.5em}
f(\mu_{\rm e})\,\exp{\left
[-\left (\frac{E_\loc-{\tt Erest}}{\sqrt{2}\,\sigma}\right )^2
\right ]}\,\exp{\left[-{\tt beta}\,(\Delta r)^2\right]}\ \
{\rm for}\ \ {\tt beta}\,(\Delta r)^2 < 4\, ,\hspace*{10mm}\\
\label{spot1b}
N_{\loc}(E_\loc)\hspace*{-0.5em} & = & \hspace*{-0.5em}
0\quad {\rm for}\quad {\tt beta}\,(\Delta r)^2 \ge 4\, .
\end{eqnarray}
Here the function $f(\mu_{\rm e})$ describes the limb darkening/brightening
law (see Section~\ref{section_kyg1line} for more details), {\tt Erest} is
the local energy of the
fluorescent line, $\sigma=2\,$eV is its width, {\tt beta} determines the size of
the spot and $(\Delta r)^2=r^2+r_{\rm spot}^2-2\,r\,r_{\rm spot}\cos{(\varphi-
\varphi_{\rm spot})}$ with $r_{\rm spot}$ and $\varphi_{\rm spot}$ being polar
coordinates of the centre of the spot. The emission is largest at the centre
of the spot and decreases towards its edge as is obvious from eq.~(\ref{spot1a}).
The Gaussian line in energies is defined by nine points equally spaced with
the central point at its maximum.

\begin{figure}[tb]
\dummycaption\label{fig:spot1}
\includegraphics[width=0.5\textwidth]{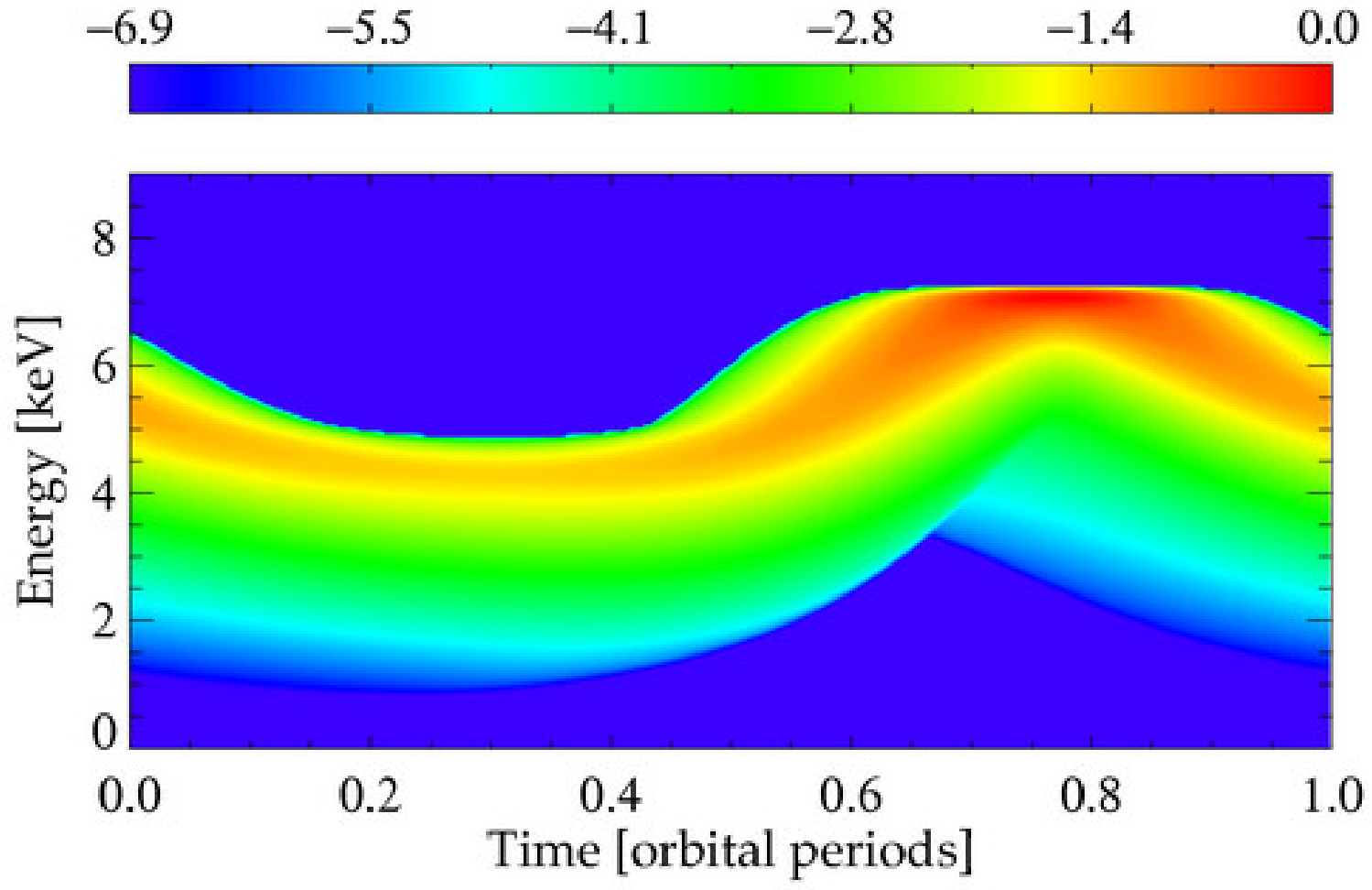}
\includegraphics[width=0.5\textwidth]{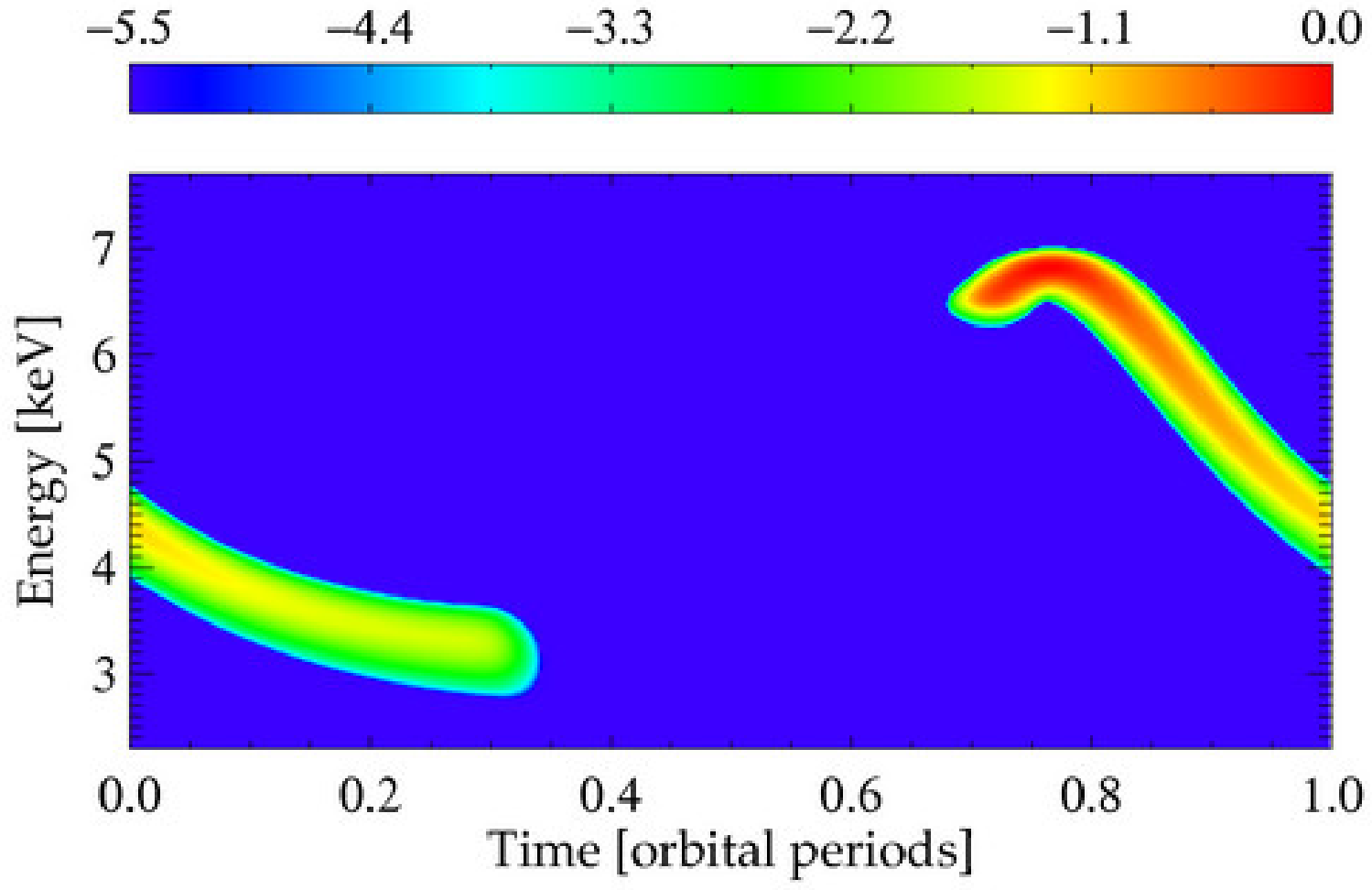}
\mycaption{Dynamical profile of an iron line ($E_{\rm rest}=6.4\,$keV) produced by
an orbiting spot. Energy is on the ordinate, time on
the abscissa. The horizontal range spans the interval of one
orbital period. Time zero corresponds to the time when the observer receives
photons from the spot at the closest approach to the observer. Left: Large spot
($\beta=0.1$) at the radius $r_{\rm spot}=8GM/c^2$.
Right: Obscured small spot ($\beta=4$) at the radius $r_{\rm spot}=5GM/c^2$.
Obscuration occurs between $315^\circ-540^\circ$ measured from the closest
approach. In both cases the dimensionless angular momentum parameter
of the black hole is $a=1$ and observer inclination is
$\theta_{\rm{}o}=45^{\circ}$. The observed photon flux
is colour-coded (logarithmic scale with arbitrary units).}
\end{figure}

\begin{table}[tbh]
\begin{center}
\dummycaption\label{kyspot_par}
\begin{tabular}[h]{r@{}l|c|c|c|c}
\multicolumn{2}{c|}{parameter} & unit & default value  & minimum value &
maximum value \\ \hline
&{\tt a/M}         & $GM/c$   & 0.9982   & 0.       & 1.        \\
&{\tt theta\_o}    & deg      & 30.      & 0.       & 89.       \\
&{\tt rin-rh}      & $GM/c^2$ & 0.       & 0.       & 999.      \\
&{\tt rout-rh}     & $GM/c^2$ & 20.      & 0.       & 999.      \\
&{\tt phi}         & deg      & 0.       & -180.    & 180.      \\
&{\tt dphi}        & deg      & 360.     & 0.       & 360.      \\
&{\tt nrad}        & --       & 300.     & 1.       & 10000.    \\
&{\tt division}    & --       & 0.       & 0.       & 1.        \\
&{\tt nphi}        & --       & 750.     & 1.       & 20000.    \\
&{\tt smooth}      & --       & 1.       & 0.       & 1.        \\
&{\tt zshift}      & --       & 0.       & -0.999   & 10.       \\
&{\tt ntable}      & --       & 0.       & 0.       & 99.       \\
{*}&{\tt Erest}    & keV      & 6.4      & 1.       & 99.       \\
{*}&{\tt sw}       & --       & 1.       & 1.       & 2.        \\
{*}&{\tt beta}     & --       & 0.1      & 0.001    & 1000.     \\
{*}&{\tt rsp}      & $GM/c^2$ & 8.       & 0.       & 1000.     \\
{*}&{\tt psp}      & deg      & 90.      & -360.    & 360.     \\
{*}&{\tt Norbits}  & --       & 1.       & 0.       & 10        \\
{*}&{\tt nt}       & --       & 500.     & 1.       & 1e6       \\
{*}&{\tt limb}     & --       & 0.       & -10.     & 10.       \\
{*}&{\tt polar}    & --       & 0.       & 0.       & 1.        \\
\end{tabular}
\mycaption{Parameters of the non-stationary model {\sc{}kyspot}.
Model parameters that are not common for all non-axisymmetric models are denoted
by asterisk.}
\end{center}
\end{table}

The local emission is defined by the following parameters
(see Tab.~\ref{kyspot_par}):
\begin{description} \itemsep -2pt
 \item[{\tt Erest}] -- rest energy of the line in keV,
 \item[{\tt sw}] -- switch for choosing the type of spot (1 -- orbiting, 2 --
 falling),
 \item[{\tt beta}] -- parameter defining the size of the spot,
 \item[{\tt rsp}] -- radius (in $GM/c^2$) at which the spot is orbiting
 or the initial radius in units of the marginally stable orbit
 $r_{\rm ms}$ if spot is in-falling,
 \item[{\tt psp}] -- azimuthal angle in degrees where the spot is initially
 located ($90^\circ$ for the closest approach),
 \item[{\tt Norbits}] -- number of orbits for an in-falling spot (not used for
 an orbiting spot),
 \item[{\tt nt}] -- time resolution of the grid,
 \item[{\tt limb}] -- switch for different limb darkening/brightening laws,
 \item[{\tt polar}] -- switch for polarization calculations (0 -- not performed,
 1 -- performed).
\end{description}

The {\sc kyspot} model creates an {\sf ascii} file {\tt kyspot.dat}, where the
dependence of the observed spectrum (columns) on time (rows) are stored. If the
{\tt polar} switch is set to unity then two other {\sf ascii} files are created
-- {\tt kyspot\_poldeg.dat} and {\tt kyspot\_psi.dat}. Here, the dependence of
the degree of polarization and the angle of polarization on time are stored.
The polarization calculations in this model are based on the same assumptions
as in the {\sc kyg1line} model.

\begin{figure}[tb]
\dummycaption\label{fig:example2}
\hspace*{0.05\textwidth}
\includegraphics[width=0.33\textwidth]{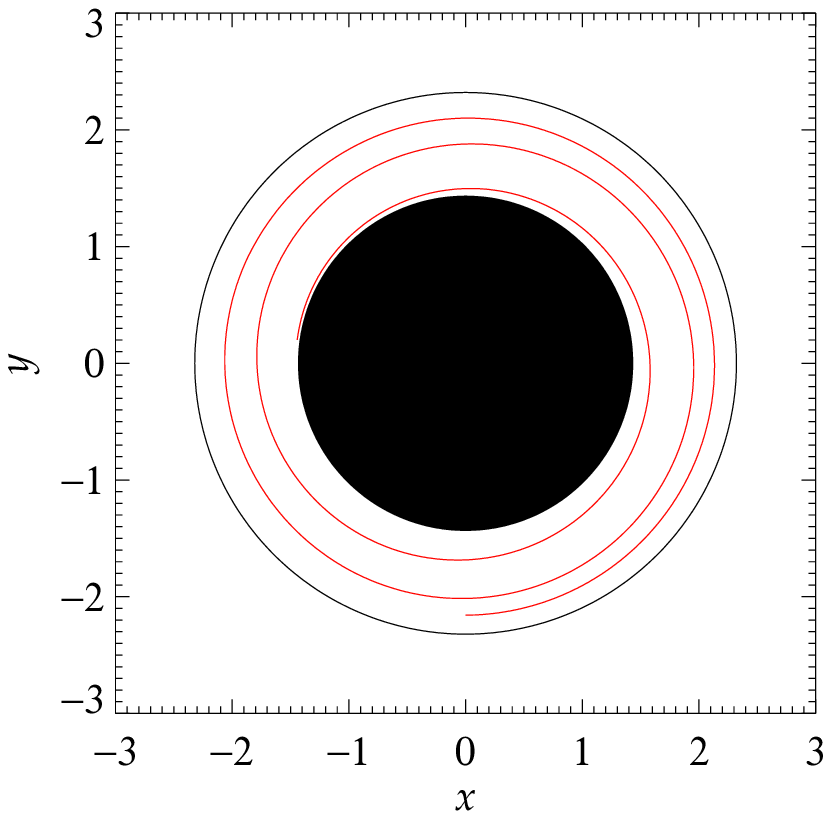}
\hspace*{0.05\textwidth}
\includegraphics[width=0.5\textwidth]{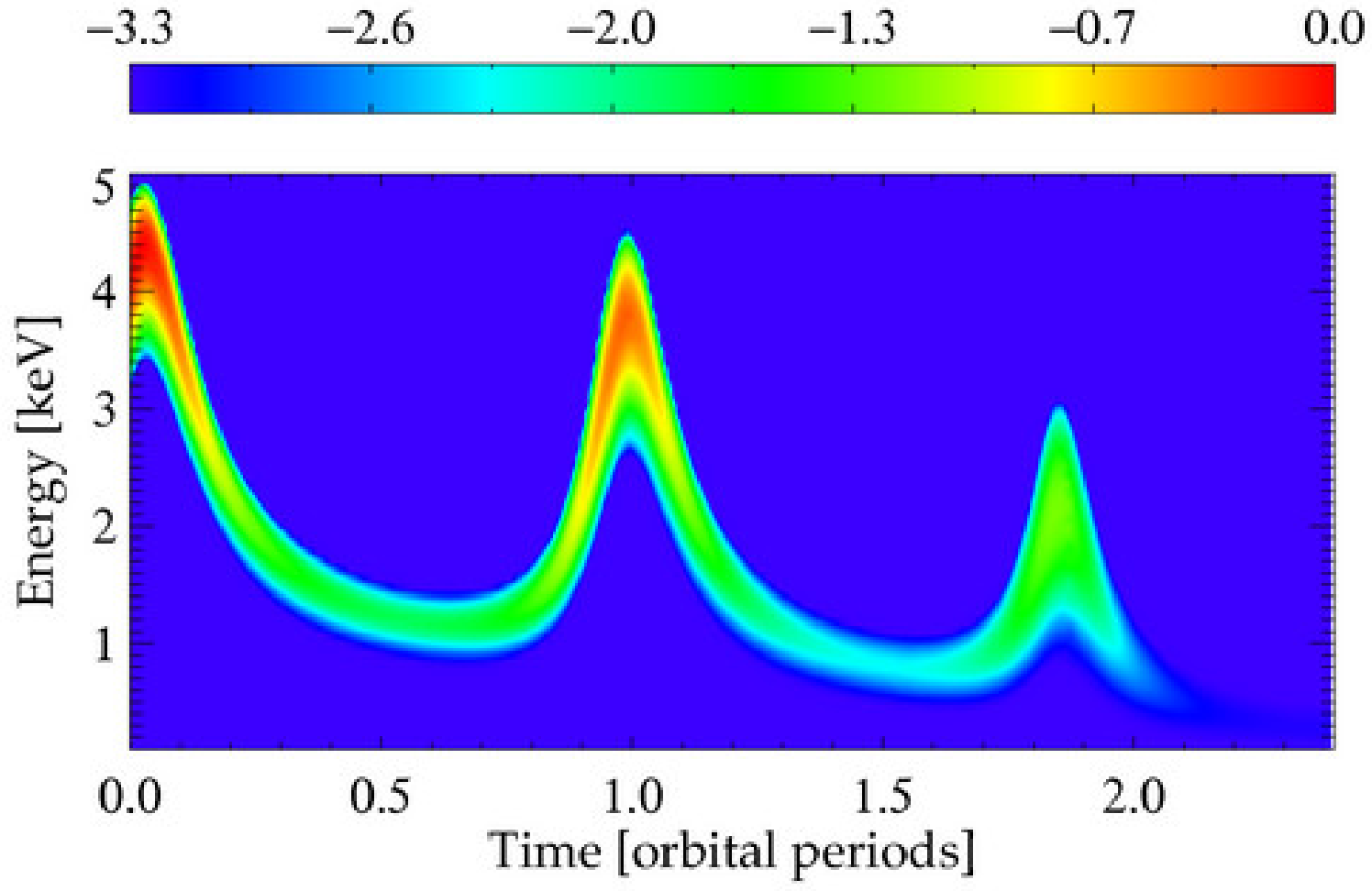}
\mycaption{Dynamical profile of an iron line produced by
an in-spiralling small spot ($\beta=100$) in free fall. Left: Trajectory
(in the equatorial plane) of the spot,
defined by a constant energy and angular momentum during the in-fall.
The observer is located to the top of the figure.
Right: Spectrum of the spot. Energy is on the ordinate, time on
the abscissa. The horizontal range spans the interval of $2.4$
orbital periods at the corresponding initial radius,
$r_{\rm spot}=0.93\,r_{\rm{}ms}(a)$. Here, the dimensionless angular
momentum parameter
of the black hole is $a=0.9$, observer inclination is
$\theta_{\rm{}o}=45^{\circ}$. The observed photon flux
is colour-coded (logarithmic scale
with arbitrary units). Gradual decay of the signal and an
increasing centroid redshift can be observed as the spot
completes over two full revolutions and eventually plunges
into the black hole.}
\end{figure}
Examples of dynamical spectra of orbiting and in-falling
spots can be seen in Figs.~\ref{fig:spot1}--\ref{fig:example2}.
The first figure represents an evolving spectral line
in a dynamical diagram for an orbiting spot of two different sizes
one of which is being obscured on part of its orbit around the black hole.
The characteristic radius of the larger spot is $\sim6.3GM/c^2$ and
the radius of the smaller spot is $\sim1GM/c^2$. The second figure represents
the dynamical spectra of the spot falling from the marginally stable orbit down
towards the black-hole horizon. The
characteristic radius of the spot is $\sim0.2GM/c^2$. This figure reflects the
effect of changing the energy shift along the spot trajectory.
The in-spiral motion starts just below the
marginally stable orbit (${\tt rsp}=0.93$) and it proceeds down
to the horizon, maintaining the specific energy and
angular momentum of its initial circular orbit at $r_{\rm ms}(a)$.
One can easily recognize that variations in the redshift could
hardly be recovered from data if only time-averaged spectrum were
available. In particular, the motion above $r_{\rm{}ms}(a)$ near a
rapidly spinning black hole is difficult to distinguish from a
descent below $r_{\rm{}ms}(0)$ in the non-rotating case.

\begin{table}[tbh]
\begin{center}
\dummycaption\label{tab:versions}
\begin{tabular*}{\textwidth}{p{0.1\textwidth}p{0.13\textwidth}p{0.20\textwidth}
                             p{0.46\textwidth}}
\hline\hline
Name  \rule[-1.5ex]{0mm}{4.5ex}& Type$^{\dag}$~ & Symmetry & Usage \\
\hline\hline
{\sf{}KYGline} \break {\sf{}KYG1line} & additive & axisymmetric \break
non-axisymmetric & spectral line from a black-hole disc \\
\hline
{\sf{}KYF1ll} & additive & non-axisymmetric & lamp-post fluorescent line\\
\hline
{\sf{}KYL1cr} & additive & non-axisymmetric & lamp-post Compton reflection
                model, physical model of polarization is included \\
\hline
{\sf{}KYH1refl} & additive & non-axisymmetric & Compton reflection with an
incident power-law (or a broken power-law) continuum \\
\hline
{\sf{}KYConv} \break {\sf{}KYC1onv} & convolution & axisymmetric \break
non-axisymmetric & convolution of the relativistic kernel with intrinsic
emissivity across the disc \\
\hline
{\sf{}KYSpot} & \multicolumn{2}{p{0.33\textwidth}}{to be used as an independent
code outside {\sc{}xspec}} & time-dependent spectrum of a pattern co-orbiting
with the disc \\
\hline\hline
\end{tabular*}
\parbox{\textwidth}{\vspace*{1mm}\footnotesize{}$^{\dag}$~Different
model types correspond to the {\sc{}xspec} syntax and are
defined by the way they act in the overall model and form the
final spectrum. According to the usual convention in {\sc{}xspec},
additive models represent individual {\it emission}\/ spectral components,
which may, for instance, originate in different regions of the source.
Additive models are simply superposed in the total signal.
Multiplicative components (e.g.\ {\sc{}hrefl}) multiply the current model by an
energy-dependent
factor. Convolution models modify the model in a more non-trivial
manner. See \cite{arnaud1996} for details.}
\mycaption{Different versions of the {\sc{}ky} model.}
\end{center}
\end{table}

\section{Summary of the new model}
The new {\sc{}ky} model is suited for use with the {\sc{}xspec}
package \citep{arnaud1996}. Several  mutations of {\sc{}ky} were developed,
with the aim to specifically provide different applications, and linked
with a common ray-tracing subroutine, which therefore does not have to
be modified when the intrinsic emissivity  function in the model is
changed.

When the relativistic line distortions are computed, the new model is
more accurate than the {\sc{}laor} model \citep{laor1991} and faster than
{\sc{}kerrspec} \citep{martocchia2000}. These are the other
two {\sc{}xspec} models with a similar usage (see also Pariev \& Bromley 
\citeyear{pariev1998};
\citealt{gierlinski2001,beckwith2004,schnittman2004}). It is also important
to compare the results from fully independent relativistic codes since
the calculations are sufficiently complex for significant differences
to arise. Our code is more general than the currently available
alternatives in {\sc xspec}.

In our {\sc{}ky} model it is possible to choose from various limb
darkening/brightening laws and thus change the angular distribution
of the local emission. The problem of directional distribution of the reflected
radiation is quite complicated and the matter has not been completely
settled yet (e.g.\ \citealt{george1991,zycki1994,magdziarz1995}, and
references  cited therein). A specific angular
dependence is often assumed in  models, such as limb darkening
$f(\mu_{\rm e})=1+2.06\mu_{\rm e}$ in the {\sc{}laor} model.  However,
it has been argued that limb brightening may  actually occur in the case
of strong primary irradiation of the disc. This is relevant for
accretion discs near black holes, where the effects of emission
anisotropy are crucial \citep{beckwith2004,czerny2004}. We
find that the spectrum of the inner disc turns out to be very sensitive
to the adopted angular dependence of the emission, and so the
possibility to modify this profile and examine the results using
{\sc{}ky} appears to be rather useful.

Among its useful features, the {\sc{}ky} model allows one to fit various
parameters such as the black-hole angular momentum ($a$), the observer
inclination angle relative to the disc axis ($\theta_{\rm{}o}$),
and the size and shape of the emission area on the disc,
which can be non-axisymmetric. A straightforward modification of a single
subroutine suffices to alter the prescription for the disc emissivity,
which is specified either by an analytical formula or in a tabular form.
Our code allows one to change the mesh spacing and resolution for
the (two-dimensional) polar grid that covers the disc plane, as well
as the energy vector (the output resolution is eventually determined
by the detector in use when the model is folded through the instrument
response). Hence, there is sufficient control of the (improved) accuracy and
computational speed.

Furthermore, {\sc{}ky} can be run as a stand-alone program
(detached from {\sc{}xspec}). In this mode there is an option
for time-variable sources such as orbiting spots,
spiral waves or evolving flares (e.g.\ \citealt{czerny2004},
who applied a similar approach to compute the predicted {\sf{}rms}
variability in a specific flare/spot model).
The improved accuracy of the new model has been achieved in several
ways: (i)~photon rays are integrated in Kerr ingoing coordinates
which follow principal photons, (ii)~simultaneous integration of
the geodesic deviation equations ensures accurate evaluation of the
lensing effect, and (iii)~non-uniform and rather fine grids
have been carefully selected.

Several versions of the routine have been prefabricated for different
types of sources (Tab.~\ref{tab:versions}): (i)~an intrinsically narrow
line produced by a disc, including the lamp-post fluorescent line model
\citep{martocchia2000}, (ii)~the Compton reflection model in two
variants -- a relativistically blurred Compton-reflection continuum
including a primary power-law component and a lamp-post model
\citep{martocchia2000}, (iii) general convolution models,
and (iv) the time-dependent spectrum of an orbiting or a free-falling spot.
Default parameter values for the line model correspond to those in  the
{\sc{}laor} model, but numerous options have been added. For example, in
the new model one is able to set the emission  inner radius below the
marginally stable orbit, $r_{\rm{}in}<r_{\rm{}ms}(a)$. One can also
allow $a$ to vary independently, in which case the horizon  radius,
$r_{\rm{}h}(a)\equiv1+\sqrt{1-a^2}$, has to be, and indeed is,  updated
at each step of the fit procedure. We thus define emission radii in
terms of their offset from the horizon. Several arguments have been
advocated in favour of having $r_{\rm{}in}{\neq}r_{\rm{}ms}$ for the disc
emission, but this possibility has never been tested rigorously against
observational data. The set of {\sc{}ky-}routines introduced above
provide the tools to explore black-hole disc models and to actually fit
for their key parameters, namely, $a$, $\theta_{\rm{}o}$, and
$r_{\rm{}in}$.

We have also produced a convolution-type
model, {\sc{}kyconv}, which can be applied to any existing
{\sc{}xspec} model of intrinsic X-ray emission (naturally,
a meaningful combination of the models is the responsibility of
the user). We remind the reader that {\sc{}kyconv} is substantially more
powerful than the usual convolution models in {\sc{}xspec}, which are
defined in terms of a one-dimensional integration over energy bins.
Despite the fact that {\sc{}kyconv}
still uses the standard {\sc{}xspec} syntax in evaluating the observed
spectrum (e.g.\ \hbox{\sc{}kyconv(gaussian)}), our code
performs a more complex operation. It still performs
ray-tracing across the disc surface so that the intrinsic model
contributions are integrated from different radii. Thus the
{\sc{}kyconv(gaussian)} model gives the same results as the {\sc{}kygline}
model if corresponding parameters are set to the same values. The price that
one has to pay for the enhanced functionality is a higher
demand on computational power.

Other user-defined emissivities can be easily adopted. This can be
achieved either by using the above mentioned convolution model
or by adding a new user-defined model to {\sc{}xspec}. The latter method is
more flexible and faster, and hence recommended. In both approaches, the
ray-tracing routine is linked and used for relativistic blurring.


\chapter{Applications}
\label{chapter4}
 \thispagestyle{empty}
  \section{Seyfert galaxy MCG--6-30-15}
The Seyfert~1 galaxy MCG--6-30-15 is a unique source in which the evidence
of a broad and skewed \fekalfa line has led to a wide acceptance of
models with an accreting black hole in the nucleus
\citep{tanaka1995,iwasawa1996,nandra1997,guainazzi1999,fabian2003}.
Being a nearby AGN (the galaxy redshift is $z=0.0078$), this source offers
an unprecedented opportunity to explore directly the pattern of
the accretion flow onto the central hole.
The Fe K line shape and photon redshifts indicate that a large
fraction of the emission originates from $r\lesssim10$ ($GM/c^2$).
The mean line profile derived from {\it{}XMM-Newton}\/ observations
is similar to the one observed previously using {\it ASCA}.
The X-ray continuum shape in the hard spectral band was well
determined from {\it BeppoSAX}\/ data \citep{guainazzi1999}.

\begin{table}[tbh]
\begin{center}
\newlength{\mylen}
\settowidth{\mylen}{{\mbox{$2.1\pm0.2^{\,\dag}$}~}}
\dummycaption\label{tab:best-fit-models}
\begin{tabular}{ccccc@{}c@{}c}
\hline
\multicolumn{1}{c}{{\normalsize{}\#}\rule[-1.5ex]{0mm}{4.5ex}} &
\multicolumn{1}{c}{{\normalsize$a$}} &
\multicolumn{1}{c}{{\normalsize$\theta_{\rm{}o}$}} &
\multicolumn{2}{c}{{\normalsize{}Continuum}} &
\multicolumn{1}{c}{} &
\multicolumn{1}{c}{{$\!$\normalsize$\chi^2$}} \\
\cline{4-5}\cline{7-7}
 & & & $\Gamma_{\rm{}c}$ & $\alpha_{\rm{}c}$ & &
 ~({\sf{}dof})\rule[-1ex]{0mm}{4ex}\\
\hline
1  & $0.35^{+0.57}_{-0.30}$ & $31.8\pm0.3$ & $2.01\pm0.02$ & $1.0^{+9}_{-1}$ &
   & {\small$\frac{368.8}{(329)}$}\rule[0ex]{0mm}{4ex}\\
2a & $0.99\pm0.01$ & $40.4\pm0.6$ & $2.03\pm0.02$ & $5.5^{+6}_{-2}$ & &
     {\small$\frac{308.6}{(330)}$}\rule[0ex]{0mm}{4ex}\\
2b & $0.72^{+0.12}_{-0.30}$ & $28.5\pm0.5$ & $2.01\pm0.01$ & $0.1^{+2}_{-0.1}$ &
   & {\small$\frac{313.5}{(330)}$}\rule[0ex]{0mm}{4ex}\\
2c & $0.25\pm0.03$ & $27.6\pm0.6$ & $1.97\pm0.02$ & $3.1^{+0.3}_{-0.1}$ & &
     {\small$\frac{313.9}{(330)}$}\rule[0ex]{0mm}{4ex}\rule[-3ex]{0mm}{7ex}\\
\hline
\end{tabular}\\[\bigskipamount]
\begin{tabular}{ccp{\mylen}cccc}
\hline
\multicolumn{1}{c}{{\normalsize{}\#}\rule[-1.5ex]{0mm}{4.5ex}} &
\multicolumn{6}{c}{{\normalsize{}Broad \fekalfa line}} \\
\cline{2-7}
\rule[-0.5em]{0em}{1.6em}    & $r_{\rm{}in}\!-r_{\rm{}h}$ & $r_{\rm{}b}\!-r_{\rm{}h}$  &
	 $r_{\rm{}out}\!-r_{\rm{}h}$ & $\alpha_{\rm{}in}$ & $\alpha_{\rm{}out}$ & EW\\
\hline
\rule[-0.8em]{0em}{2.4em}  1 & $5.1\pm0.2$ & -- & $11.4\pm0.8$ & -- & $3.9\pm0.6$ &
 $258^{+26}_{-13}$\\
\rule[-0.9em]{0em}{2.4em} 2a & $0.67\pm0.04$ & $3.35\pm0.05$ & $40^{+960}_{-33}$ &
 $6.9^{+0.5}_{-0.4}$ & $9.7^{+0.3}_{-0.8}$ & $268\pm13$ \\
\rule[-1.1em]{0em}{2.4em} 2b & $0.65\pm0.35$ &
 \parbox{\mylen}{\mbox{$2.1\pm0.2^{\,\dag}$}\break\mbox{$7.2\pm0.2$}} & $48^{+200}_{-25}$ &
 $8.1^{+1.4}_{-0.9}$ & $4.9^{+0.4}_{-0.3}$ & $241^{+13}_{-10}$ \\
\rule[-1.2em]{0em}{2.4em} 2c & $1.23\pm0.06$ & $4\pm0.02$ & $109^{+20}_{-10}$ &
 $9.2\pm0.2$ & $3.1\pm0.1$ & $267\pm10$ \\
\hline
\end{tabular}
\vspace*{2mm}\par{}
{\parbox{0.95\textwidth}{\footnotesize{}
Best-fitting values of the important parameters and their
statistical errors for models \#1--2, described in the text.
The models include broad \fekalfa and \fekbeta emission
lines, a narrow Gaussian line at $\sim 6.9$~keV, and
a Compton-reflection continuum from a relativistic disc.
These models illustrate different assumptions
about intrinsic emissivity of the disc (the radial
emissivity law does not need to be a simple power law, but axial
symmetry has been still imposed here).
The inclination angle $\theta_{\rm{}o}$
is in degrees, relative to the rotation axis; radii are expressed
as an offset from the horizon (in $GM/c^2$);
the equivalent width, EW, is in electron volts.\\
$^{\dag}$~Two values of the transition radius define the interval
$\langle{}r_{\rm{}b-},r_{\rm{}b+}\rangle$ where reflection is
diminished.}}
\mycaption{Spectral fitting results for MCG--6-30-15 using the {\sc ky} model.}
\end{center}
\end{table}

To illustrate the new {\sc ky} model capabilities,
we used our code to analyze the mean EPIC PN spectrum which we
compiled from the long {\it{}XMM-Newton}\/ 2001
campaign (e.g., as described in \citealt{fabian2002}).
The data were cleaned and reduced using standard data reduction
routines, employing
SAS version 5.4.1.\footnote{See {\tt{}\href{http://xmm.vilspa.esa.es/sas/}
{http://xmm.vilspa.esa.es/sas/}.}}
We summed the EPIC PN data from five observations
made in the interval 11 July 2001 to 04 August 2001,
obtaining a total good exposure time of $\sim290$~ks (see
\citealt{fabian2002} for details of the observations).
The energy range was restricted to $3-10$~keV with $339$ energy
bins, unless otherwise stated, and models were fitted by minimizing
the $\chi^{2}$ statistic. Statistical errors quoted correspond to
$90$\% confidence for one interesting parameter
(i.e.\ $\Delta\chi^{2}=2.706$), unless otherwise stated.

No absorber was taken into account; the assumption
here is that any curvature in the spectrum above 3~keV is
not due to absorption, and only due to the Fe K line.
We remind the reader that the Fe~K emission line dominates
around the energy $6-7$~keV, but it has been supposed to stretch
down to $\sim3$~keV or even further.
We emphasize that our aim here is to test the hypothesis that
{\it{}if all of the curvature is entirely due to the broad
Fe~K feature, is it possible to constrain $a$
of the black hole?} Obviously, if the answer to this is `no', it
will also be negative if some of the spectral curvature
between $\sim3-5$~keV is due to processes other than
the Fe~K line emission.

We considered \fekalfa and \fekbeta
iron lines (with their rest-frame energy fixed at $E_{\rm{}rest}=6.400$
and $7.056$~keV, respectively), a narrow-line feature
(observed at $E_{\rm{}obs}\sim6.9$~keV) modelled as a Gaussian with the central
energy, width, and intensity as free parameters, plus a power-law
continuum. We used a superposition of two {\sc kygline} models
to account for the broad relativistic K$\alpha$ and K$\beta$ lines,
{\sc zgauss} for
the narrow, high-energy line (which we find to be centred at
$6.86$~keV in the source frame, slightly redshifted relative
to $6.966$~keV, the rest-energy of Fe~Ly$\alpha$),
and {\sc kyh1refl} for the Comptonized continuum convolved
with the relativistic kernel. We assumed that the high-energy narrow
line is non-relativistic. Note that an alternative interpretation of the
data (as pointed out by \citealt{fabian2002}) is that there
is not an emission line at $\sim 6.9$~keV, but He-like resonance
absorption at $\sim 6.7$~keV. This would not affect our conclusions.
The ratio of the intensities of the two relativistic lines was fixed at
$I_{\rm{K\beta}}/I_{\rm{K\alpha}}=0.1133$ ($=17/150$), and the iron
abundance for the Compton-reflection continuum was assumed to be three times
the solar value \citep{fabian2002}. We used the angles
$\theta_{\rm{}min}=0^{\circ}$ and $\theta_{\rm{}max}=90^{\circ}$ for
local illumination in {\sc kyh1refl} (these values are equivalent to a central
illumination of an infinite disc).
The normalization of the Compton-reflection continuum relative
to the direct continuum is controlled by the
effective reflection `covering factor', $R_{\rm{}c}$, which is the
ratio of the actual reflection normalization to that expected
from the illumination of an infinite disc.
Also $r_{\rm{}out}$ (outer radius of the disc), $\alpha_{\rm{}c}$
(slope of power-law continuum radial emissivity in the {\sc kyh1refl}
model component), as well as $R_{\rm{}c}$, were included among the
free parameters, but we found them to be only very poorly constrained.
Weak constraints on $\alpha_{\rm{}c}$ and
$R_{\rm{}c}$ are actually expected in this model for two reasons.
Firstly, in these data the continuum is indeed rather featureless. Secondly,
the model continuum was blurred with the relativistic kernel of {\sc kyh1refl},
and so the dependence of the final spectrum on the exact form of
the emissivity distribution over the disc must be quite weak.

\begin{figure}[tb]
\dummycaption\label{fig:confcont1}
\includegraphics[width=0.5\textwidth]{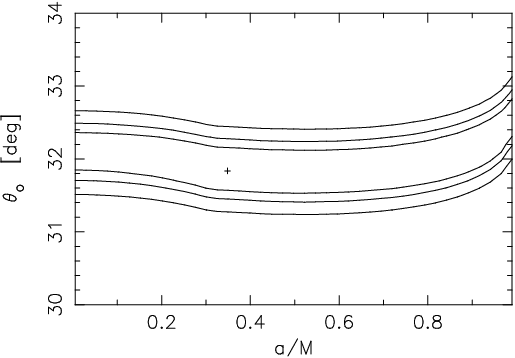}
\includegraphics[width=0.5\textwidth]{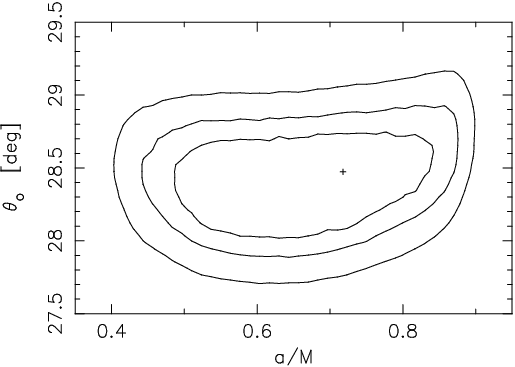}
\mycaption{Confidence contours around the best-fitting parameter
values (indicated by a cross). Left: the case of a single
line-emitting region (model~\#1 with zero emissivity for
$r<r_{\rm{}ms}$). Right:
the case of a non-monotonic radial emissivity, model~\#2b.
The joint two-parameter contour levels
for $a$ versus $\theta_{\rm{}o}$ correspond to
68\%, 95\% and 99\% confidence.}
\end{figure}

\textit{Model \#1.}
Using a model with a plain power-law radial emissivity
on the disc, we obtained the best-fitting values for the
following set of parameters: $a$ of the black hole, disc
inclination angle $\theta_{\rm{}o}$, radii of the line-emitting region,
$r_{\rm{}in}<r<r_{\rm{}out}$, the corresponding radial emissivity
power-law index $\alpha$ of the line emission, as well as the radial
extent of the continuum-emitting region, its photon index $\Gamma_{\rm{}c}$,
and the corresponding $\alpha_{\rm{}c}$ for the continuum.
Notice that $\Gamma_{\rm{}c}$ and $\alpha_{\rm{}c}$ refer
to the continuum component {\it{}before\/} the relativistic
kernel was applied to deduce the observed spectrum.
As a result of the integration across the disc, the model
weakly constrains these parameters.

There are two ways to interpret this. A model which is
over-parameterized is undesirable from the point of view of deriving
unique model parameters from modelling the data. However, another
interpretation is that a model with a greater number of parameters may
more faithfully  reflect the real physics and it is the actual physical
situation which leads to degeneracy in the model parameters. The latter
implies that some model parameters can never be constrained uniquely,
regardless of the quality of the data. In practice one must apply both
interpretations and assess the approach case by case, taking into
account the quality of the data, and which parameters can be constrained
by the data and which cannot. If preliminary fitting shows that large
changes in a parameter do not affect the fit, then that parameter can be
fixed at some value obtained by invoking physically reasonable arguments
pertaining to the situation.

We performed various fits with the inner edge tied
to the marginally stable orbit and also fits where $r_{\rm{}in}$
was allowed to vary independently. Free-fall motion with constant
angular momentum was assumed below $r_{\rm{}ms}$, if the emitting region
extended that far. Tab.~\ref{tab:best-fit-models}
gives best-fitting values of the
key relativistic line and continuum model parameters
for the case in which $r_{\rm{}in}$ and $a$ were independent.

Next, we froze some of the parameters at their best-fit
values and examined the $\chi^2$ space by varying the remaining
free parameters. That way we constructed joint confidence contours
in the plane $a$ versus $\theta_{\rm{}o}$
(see left panel of Fig.~\ref{fig:confcont1}).
These representative plots demonstrate that $\theta_{\rm{}o}$
appears to be tightly constrained, while $a$ is allowed to vary
over a large interval around the best fit, extending down to $a=0$.

\textit{Model \#2.}
We explored the possibility that the broad-line
emission does not conform to a unique power-law radial
emissivity but that, instead,
the line is produced in two concentric rings (a `dual-ring' model).
This case can be considered as a toy model for a more complex
(non-power law) radial dependence of the line emission than the
standard monotonic decline, which we represent here by
allowing for different values of $\alpha$ in the inner and outer
rings: $\alpha_{\rm{}in}$ and $\alpha_{\rm{}out}$.
Effectively, large values of the power-law index
represent two separate rings.
The two regions are matched at the transition radius, $r=r_{\rm{}b}$, and
so this is essentially a broken power law. We explored both cases of
continuous and discontinuous line emissivity at $r=r_{\rm{}b}$. Notice that
the double power-law emissivity arises naturally in the lamp-post model
\citep{martocchia2000} in which the disc irradiation and
the resulting \fekalfa reflection are substantially anisotropic
due to fast orbital motion in the inner ring. Although the lamp-post
model is very simplified in several respects, namely the way in which the
primary source is set up on the rotation axis, one can expect fairly
similar irradiation to arise from more sophisticated schemes of
coronal flares distributed above the disc plane.
Also, in order to provide a physical picture of the steep emissivity
found in {\it{}XMM-Newton}\/ data of MCG--6-30-15,
\cite{wilms2001} invoked strong magnetic stresses acting
in the innermost part of the system, assuming that they are able to
dissipate a considerable amount of energy in the disc
at very small radii. Intense self-irradiation
of the inner disc may further contribute to the effect.

This more complex model is consistent
with the findings of \cite{fabian2002}. Indeed, the
fit is improved relative to
models with a simple emissivity law because the
enormous red wing and relatively sharp core
of the line are better reproduced thanks to
the contribution from a highly redshifted inner disc
(see Tab.~\ref{tab:best-fit-models}, model~\#2a).
For the same reason that the more
complex model reproduces the line core along
with the red wing well, the model prefers higher values of
$a$ and $\theta_{\rm{}o}$ than what we found for
case \#1. Notice that $a\rightarrow1$
implies that all radiation is produced above $r_{\rm{}ms}$.
Maximum rotation is favoured with
both $a$ and $\theta_{\rm{}o}$
appearing to be tightly constrained near their
best-fit values. Likewise for the continuum radial emissivity
indices. There is a certain freedom
in the parameter values that can be accommodated by this model.
By scanning the remaining parameters,
we checked that the reduced $\chi^2\sim1$ can be achieved
also for $a$ going down to $\sim0.9$ and
$\theta_{\rm{}o}\sim37^{\circ}$. This conclusion is also consistent with
the case for large $a$ in \cite{dabrowski1997}; however, we actually
do not support the claim that the current data {\it{}require}\/
a large value of $a$. As shown below, reasonable
assumptions about the intrinsic emissivity can fit the data with small $a$
equally well.

In model \#2a, small residuals remain near $E\sim4.8$~keV
(at about the $\sim1$\% level),
the origin of which cannot be easily clarified with
the time-averaged data that we employ now. The
excess is reminiscent of a Doppler horn typical of relativistic line
emission from a disc, so it may
also be due to \fekalfa emission which is locally enhanced
on some part of the disc. We were able to reproduce the peak
by modifying the emissivity at the transition radius,
where the broken power-law emissivity changes its slope (model~\#2b).
We can even allow non-zero emissivity below $r_{\rm{}ms}$ (the inner ring)
with a gap of zero emissivity between the outer edge of the inner
ring and the inner edge of the outer one. The inner ring,
$r_{\rm{}in}{\leq}r{\leq}r_{\rm{}b-}$, contributes to the red tail
of the line while the outer ring, $r_{\rm{}b+}{\leq}r{\leq}r_{\rm{}out}$,
forms the main body of the broad line. The resulting plot of
joint confidence contours of $\theta_{\rm{}o}$ versus $a$ is shown
in Fig.~\ref{fig:confcont1} (right panel).
In order to construct the confidence contours we scanned a broad
interval of parameters, $0<a<1$ and $0<\theta_{\rm{}o}<45^{\circ}$;
here, detail is shown only around the minimum $\chi^2$ region.

\begin{figure}[tb]
\dummycaption\label{fig:dualline3-10kev_v3}
\includegraphics[width=0.5\textwidth]{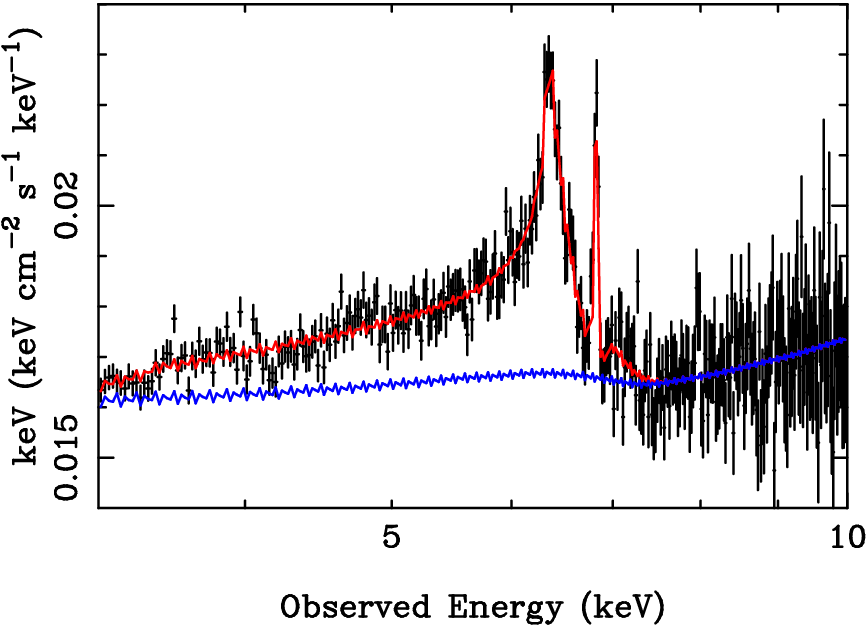}
\includegraphics[width=0.5\textwidth]{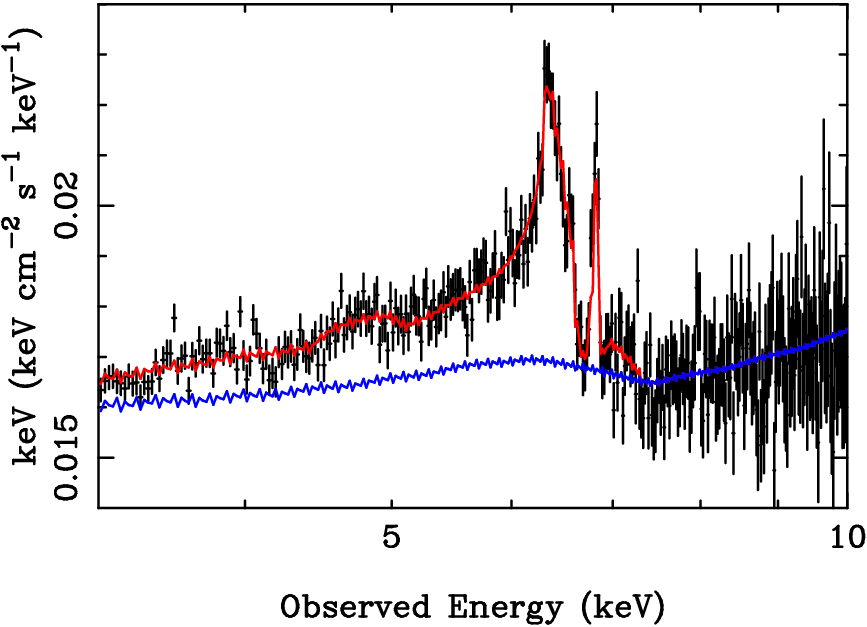}
\mycaption{Spectrum and best-fitting model for the {\it XMM-Newton}\/
data for MCG--6-30-15 in which the Fe~K line originates in a dual-ring.
The models 2a (left) and 2b (right) are shown in comparison.
See Tab.~\ref{tab:best-fit-models} for the model parameters.
The continuum component ({\sc kyh1refl}) is also plotted.
The data points are not unfolded: the spectrum in these units was
made by multiplying the ratio of measured counts to the counts
predicted by the best-fitting model and then this ratio was multiplied by
the best-fitting model and then by $E^{2}$.}
\end{figure}

Two examples of the spectral profiles are shown in
Fig.~\ref{fig:dualline3-10kev_v3}. It can be seen that the overall
shapes are very similar and the changes concern mainly the red wing of
the profile. For completeness we also fitted several modifications of
the model \#2 and found that with this type of disc emissivity (i.e.\
one which does not decrease monotonically with radius) we can still
achieve comparably good fits which have small values of $a$. The case
\#2c in Tab.~\ref{tab:best-fit-models} gives another example.
This shows that one cannot draw firm conclusions about $a$ (and some of the
other model parameters) based on the redshifted part of the line
using current data. The data used here represent the highest signal-to-noise
relativistically broadened Fe K line profile yet available for any AGN.

The differences in $\chi^2$ between model \#1 and models
\#2a, 2b, and 2c  are very large ($\sim60$) for only one
additional free parameter, so all variations of model 2 shown
are better fits than the case \#1. We note that formally, model \#2a
and \#2c appear to constrain $a$ much more tightly than model \#1.
The issue here is that the values of $a$ can be completely different,
with the statistical errors on $a$ not overlapping (compare, for example,
model \#2a and \#2c in Tab.~\ref{tab:best-fit-models}).
This demonstrates that, at least for the parameter $a$,
it is not simply only the statistical error which determines
whether a small $a$ value is or is not allowed by the data.
We confirmed this conclusion with more complicated models obtained
in {\sc ky} by relaxing the assumption of axial symmetry;
additional degrees of freedom do not change the previous results.

One should bear in mind a well-known technical difficulty
which is frequently encountered while scanning
the parameter space of complex models
and producing confidence contour plots similar to
Fig.~\ref{fig:confcont1}. That is, in a rich parameter space
the procedure may be caught in a local minimum which produces
an acceptable statistical measure of the goodness of fit
and appears to tightly constrain parameters near
the best-fitting values. However, manual searching revealed
equally acceptable results in rather remote
parts of the parameter space. Indeed, as the
results above show, we were able to find
acceptable fits with the central
black hole rotating either slowly or rapidly (in terms of the $a$
parameter). This fact is not in contradiction with previous
results (e.g.\ \citealt{fabian2002}) because we assumed a
different radial profile of intrinsic emissivity, but it indicates
intricacy of unambiguous determination of
model parameters. We therefore need more observational
constraints on realistic physical mechanisms
to be able to fit complicated models to actual data with sufficient confidence
\citep{ballantyne2001,nayakshin2002,rozanska2002}.

We have seen that the models described above are able
to constrain parameters with rather different
degrees of uncertainty.
It turns out that the more complex type \#2 models (i.e.\ those
that have radial emissivity profiles which are
non-monotonic or even have an appreciable contribution from
$r<r_{\rm{}ms}$) provide better fits to the
data but a physical interpretation is not obvious.
\cite{ballantyne2003} also deduce a
dual-reflector model from the same data and propose that the
outer reflection is due to the disc being warped or flared
with increasing radius.
\clearpage
  \section[Relativistic spectral features from X-ray illuminated spots]
{Relativistic spectral features from X-ray illuminated \break spots}


Relativistic iron line profiles may provide a powerful tool for
measuring the mass of the black hole in active galactic nuclei
and Galactic black-hole candidates. For this aim,
\cite{stella1990} proposed to use temporal changes in the line
profile following variations of the illuminating primary source
(which at that time was assumed to be located on the disc axis
for simplicity). Along the same line of thought,
\cite{matt1992a} proposed to employ, instead, variations of the
integrated line properties such as equivalent width, centroid energy and
line width. These methods are very similar conceptually to the
classical reverberation mapping method, widely and successfully applied
to optical broad lines in AGNs. Sufficiently long
monitoring of the continuum and of the line emission is required,
as well as large enough signal-to-noise ratio. However, the
above-mentioned methods have not provided many results yet. Even in the
best studied case of the Seyfert galaxy MCG--6-30-15, the mass estimate is
hard to obtain due to the apparent lack of correlation between
the line and continuum emission \citep{fabian2002}. It was also suggested
that these complications are possibly caused by an interplay of
complex general relativistic effects \citep{miniutti2003}.
X-ray spectra from high throughput and high energy resolution
detectors should resolve the problem of interpretation of
observed spectral features. However, before such high quality data are
available it is desirable to examine existing spectra and attempt to
constrain physical parameters of the models.

A simple, direct and potentially robust way to measure the
black-hole mass would be available if the line emission originates
at a given radius and azimuth, as expected if the disc
illumination is provided by a localized flare just above the
disc (possibly due to magnetic reconnection), rather than a central
illuminator or an extended corona. If a resulting `hot spot' co-rotates with the
disc and lives for at least a significant part of an orbit, by fitting
the light curve and centroid energy of the line flux, the inclination
angle $\theta_{\rm{}o}$ and the orbit radius could be derived (radius in units of
the gravitational radius $r_{\rm{}g}$). Further, assuming Keplerian rotation,
the orbital period is linked with radius in a well-known manner. The equation
for the orbital period then contains the black-hole mass $\mbh$
explicitely, and so this parameter can be determined, as
discussed later.

Hot spots in AGN accretion discs were popular for a while, following the
finding of apparent periodicity in the X-ray emission of the Seyfert~1
galaxy NGC~6814. They, however, were largely abandoned when this periodicity
was demonstrated to be associated with an AM~Herculis system in the field of view
rather than the AGN itself\break \citep{madejski1993}. Periodicities in AGNs
were subsequently reported in a few sources \citep{iwasawa1998,lee2000,
boller2001}. The fact that they were not confirmed in different
observations of the same sources is not
surprising -- quite on the contrary, it would be hard to imagine a hot spot
surviving for several years.

Recently, the discovery of narrow emission features in the X-ray spectra
of several AGNs \citep{turner2002,guainazzi2003,yaqoob2003,turner2004}
has renewed interest in hot spots. There is a tentative
explanation (even if not the only one) for these features,
typically observed in the $5-6$~keV energy range, in terms of iron
emission produced in a small range of radii and distorted by
joint action of Doppler and gravitational shift of photon energy.
Iron lines would be produced by localized flares
which illuminate the underlying disc surface, producing the line by
fluorescence. Indeed, the formation of
magnetic flares on the disc surface is one of the
most promising scenarios for the X-ray emission of AGNs. A particularly
strong flare, or one with a very large anisotropic emission towards the
disc, could give rise to the observed features. Small width of the
observed spectral
features implies that the emitting region must be small, and that
it is seen for only a fraction of the entire orbit
(either because the flare dies out, or
because emission goes below detectability, see next section).
If the flares co-rotate with the disc and if they last
for a significant part of the orbit, it may be possible
by observing their flux and energy
variations with phase to determine the orbital parameters, and thence
$\mbh$.


The basic properties of line emission from the innermost regions of an
accretion disc around a black hole are well-known \citep[see e.g.]
[for recent reviews]{fabian2000,reynolds2003}. Let us here briefly summarize
several formulae most relevant to our purposes.

If $r$ is the orbital radius and $a$ is the dimensionless black-hole
angular momentum, the orbital period of matter co-rotating along a
circular trajectory $r=\mbox{const}$ around the black hole
is given by \citep{bardeen1972}
\begin{equation}
T_{\rm{}orb} \doteq 310~\left(r^\frac{3}{2}+a\right)
\frac{\mbh}{10^7M_{\odot}}\quad\mbox{[sec]},
\label{torb}
\end{equation}
as measured by a distant observer. We express lengths
in units of the gravitational radius
$r_{\rm{}g}{\equiv}G{\mbh}/c^2{\doteq}1.48\times10^{12}M_7$~cm, where
$M_7$ is the mass of the black hole in units of $10^7$ solar masses.
Angular momentum $a$ (per unit mass) is in geometrized units
($0\leq{a}\leq1$). See e.g.\ \cite{misner1973} for useful conversion
formulae between geometrized and physical units.

The innermost stable orbit,
$r_{\rm ms}$, occurs for an equatorial disc at radius
\begin{equation}
r_{\rm ms} = 3+Z_2-\big[\left(3-Z_1)(3+Z_1+2Z_2\right)\big]^{1 \over 2},
\label{velocity1}
\end{equation}
where
$Z_1 = 1+(1-a^2)^{1 \over 3}[(1+a)^{1 \over 3}+(1-a)^{1 \over 3}]$
and $Z_2 = (3a^2+Z_1^2)^{1 \over 2}$;
$r_{\rm ms}$ spans the range of radii from $r=1$ ($a=1$, i.e.\ the case
of a maximally rotating black hole) to $6$
($a=0$, a static black hole). Rotation of a black hole
is believed to be limited by an equilibrium value
$a\dot{=}0.998$ because of the capture of photons from the disc
\citep{thorne1974}.
This would imply $r_{\rm{}ms}\dot{=}1.23$. Different specific models
of accretion can result in somewhat different limiting values
of $a$ and the corresponding $r_{\rm{}ms}$. Notice that in the static case,
the radial dependence $T_{\rm{}orb}(r)_{{\mid}a=0}$ is identical to that
in purely Newtonian gravity.

\begin{figure}
\dummycaption\label{orbits_1}
\includegraphics[width=0.325\textwidth,height=6cm]{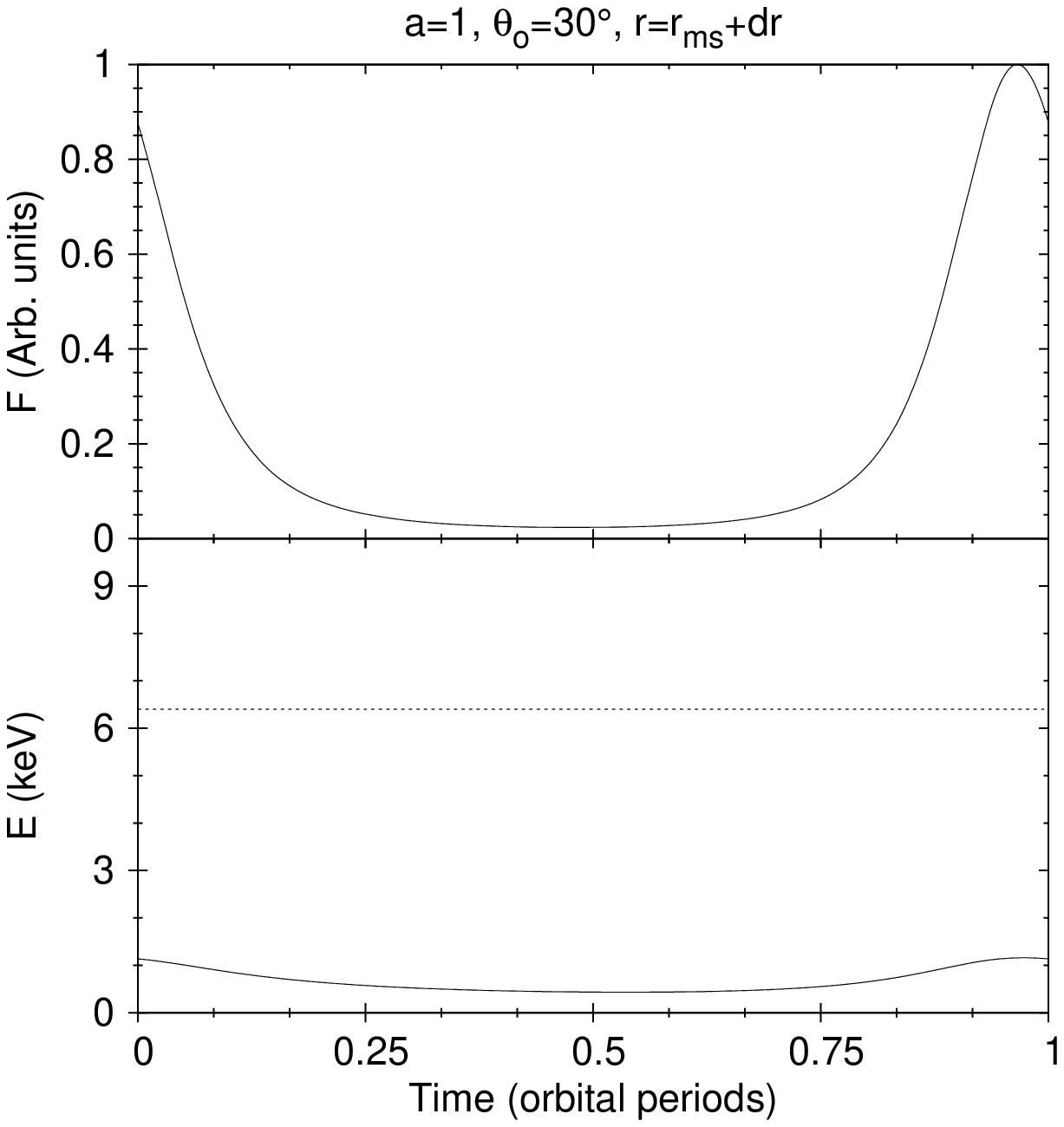}
\hfill
\includegraphics[width=0.325\textwidth,height=6cm]{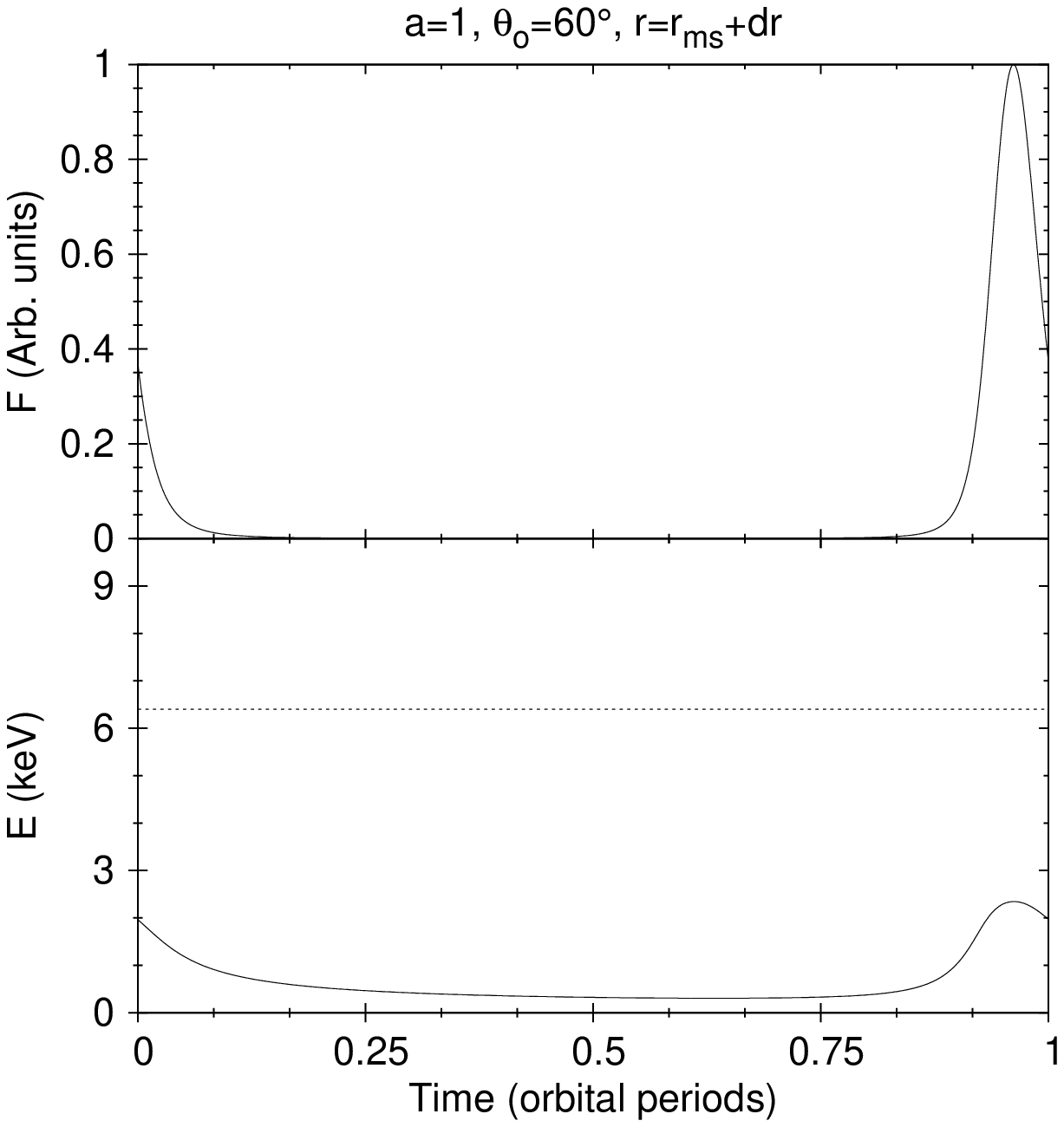}
\hfill
\includegraphics[width=0.325\textwidth,height=6cm]{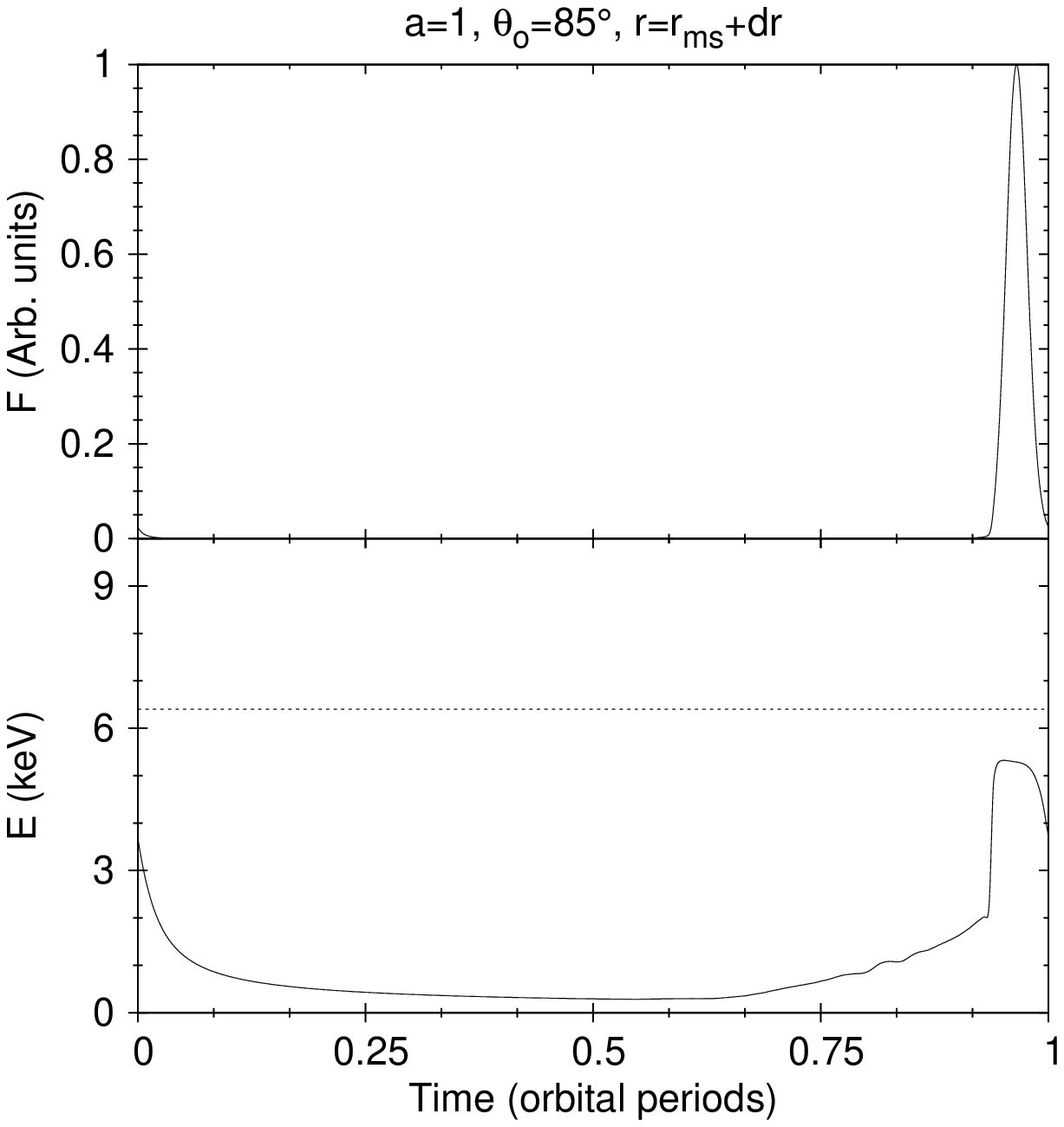}
\vspace*{3mm}\\
\includegraphics[width=0.325\textwidth,height=6cm]{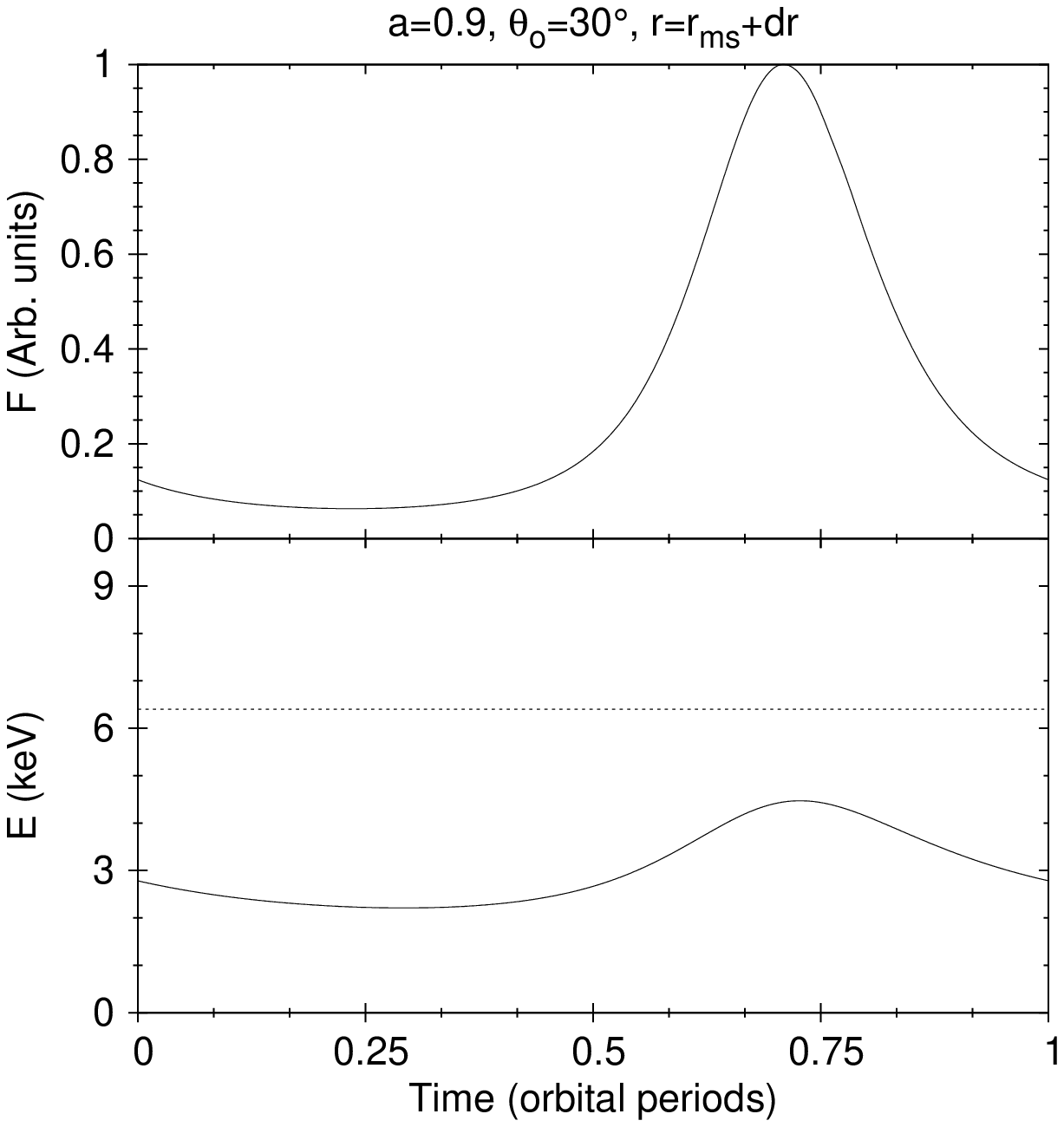}
\hfill
\includegraphics[width=0.325\textwidth,height=6cm]{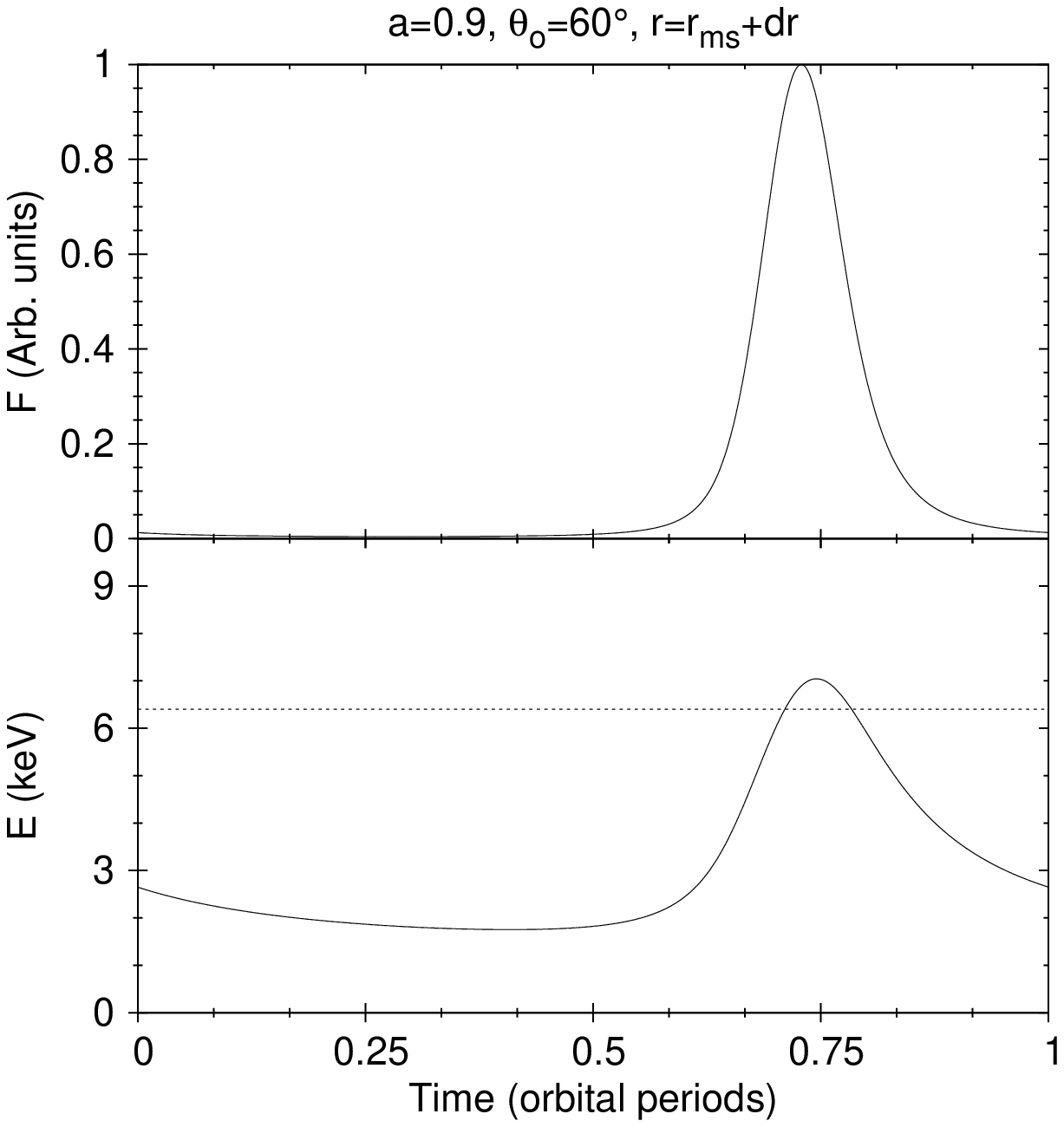}
\hfill
\includegraphics[width=0.325\textwidth,height=6cm]{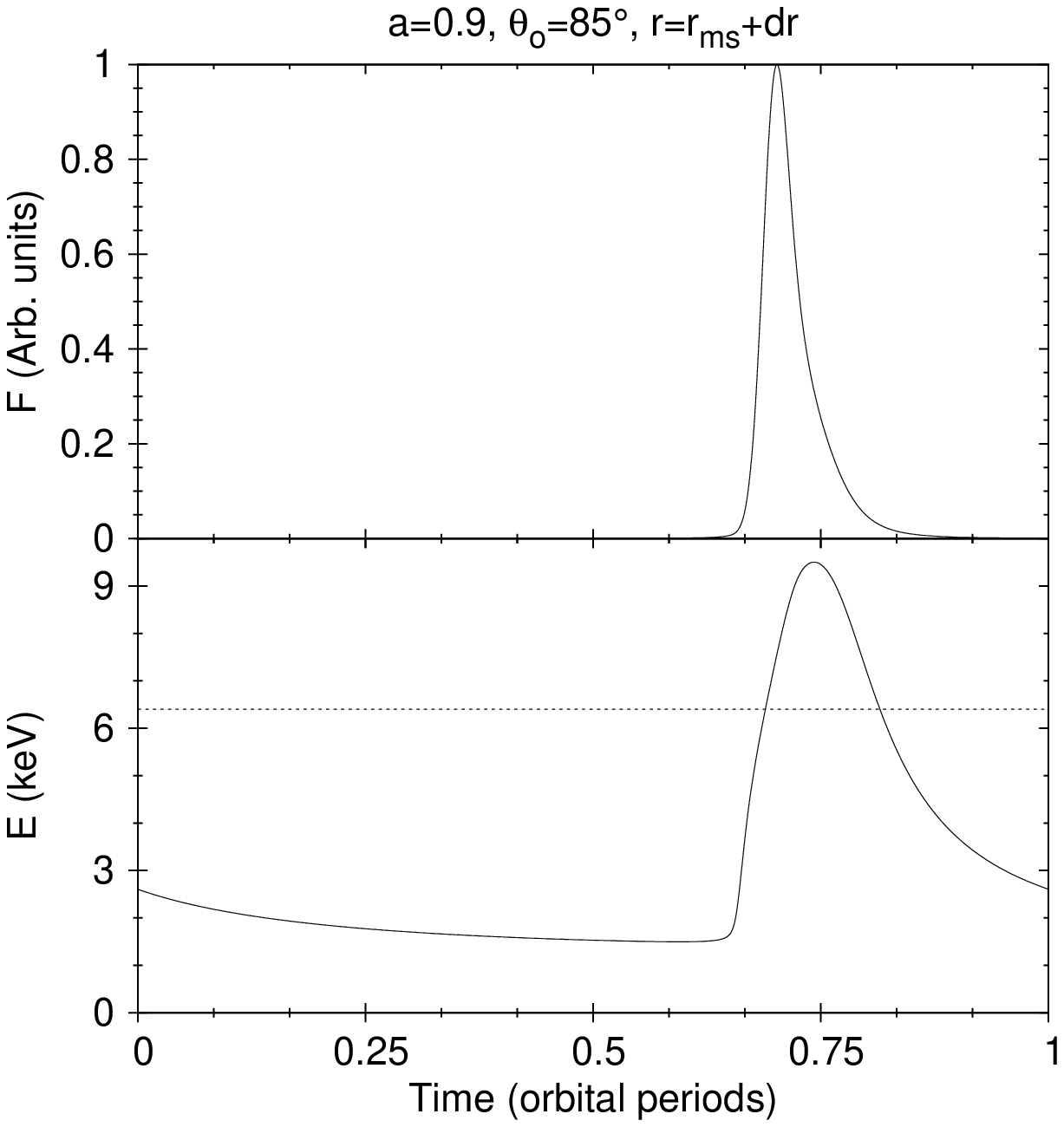}
\vspace*{3mm}\\
\includegraphics[width=0.325\textwidth,height=6cm]{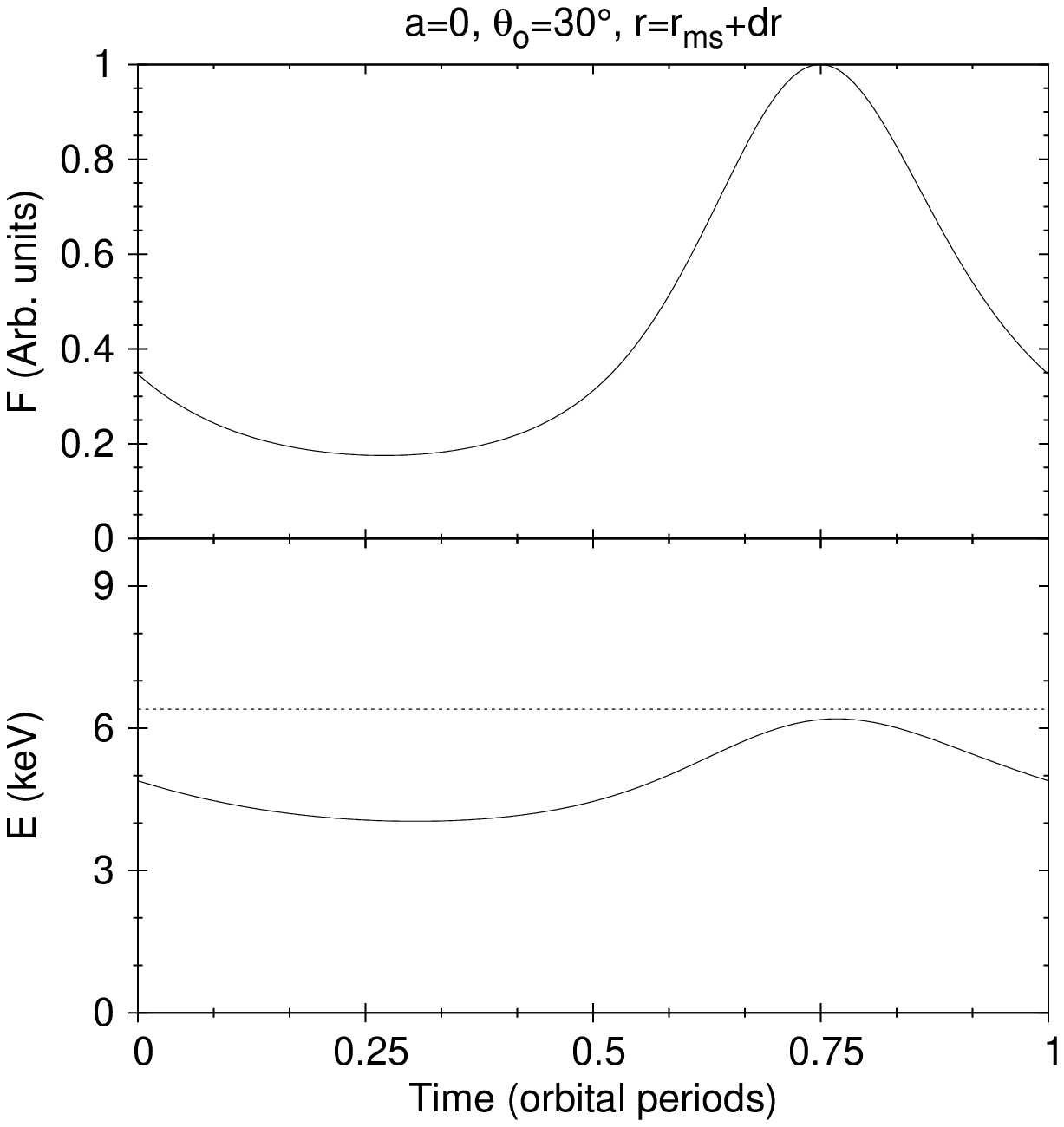}
\hfill
\includegraphics[width=0.325\textwidth,height=6cm]{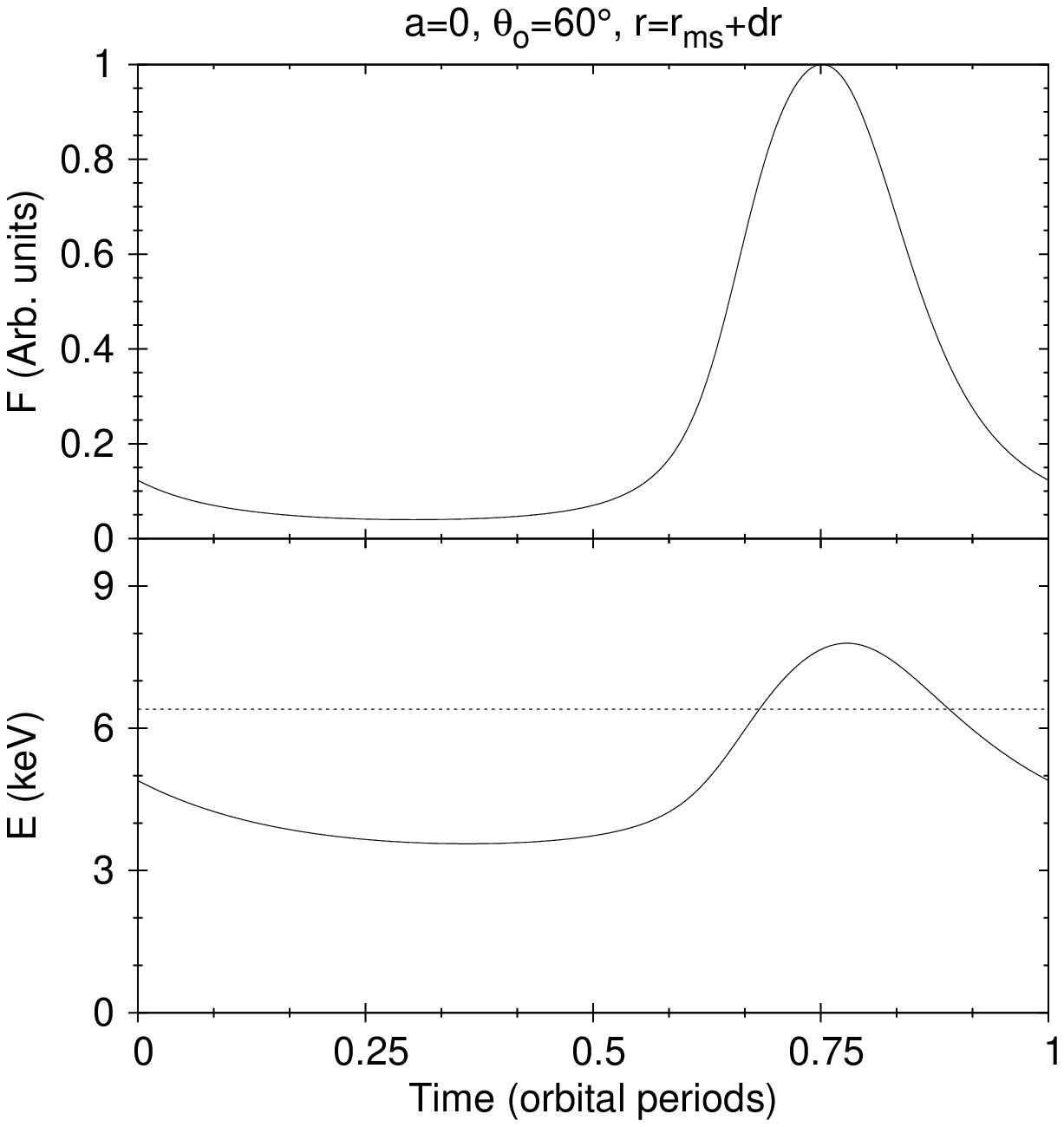}
\hfill
\includegraphics[width=0.325\textwidth,height=6cm]{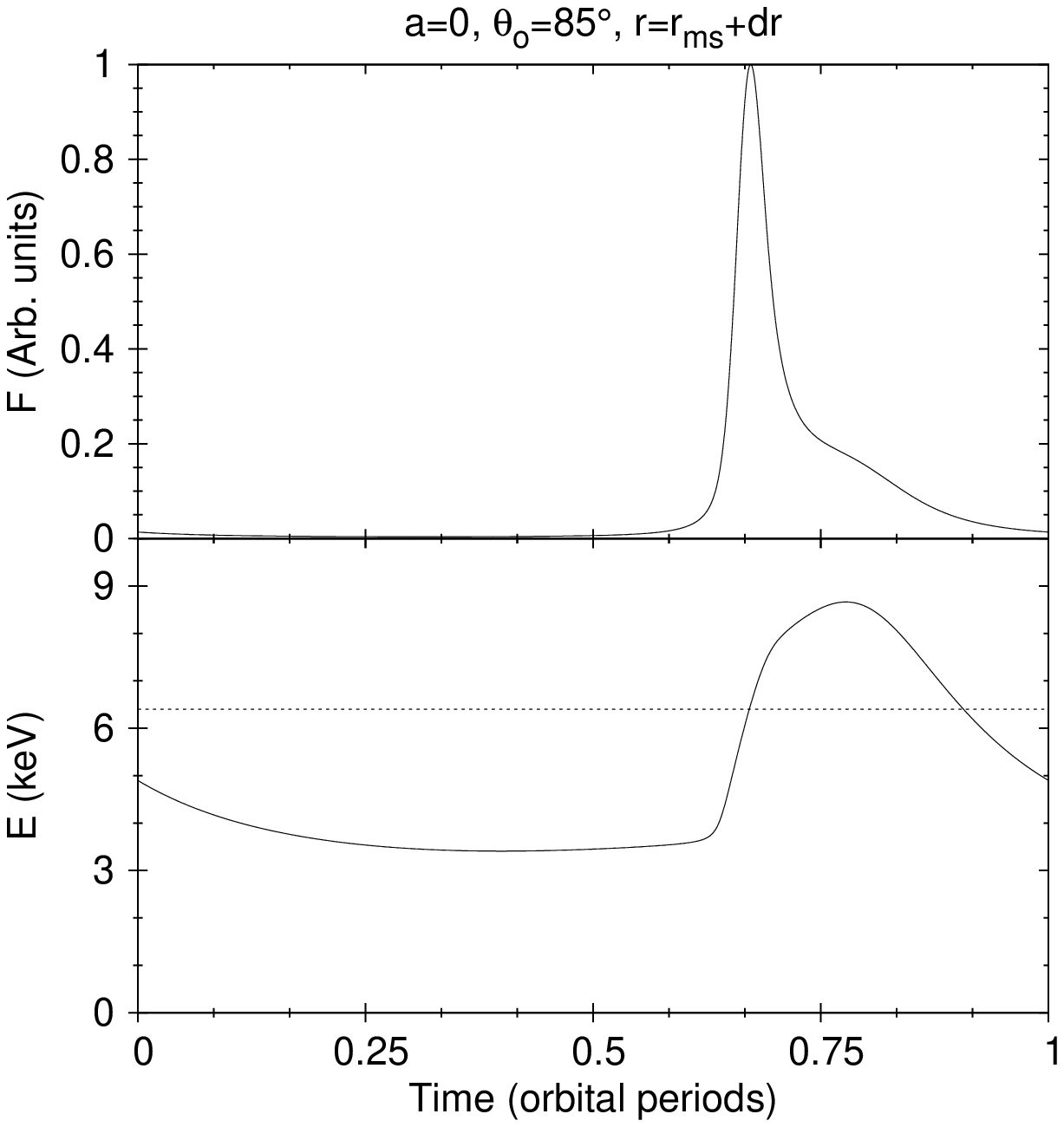}
\vspace*{0mm}
\mycaption{Line flux and centroid energy as functions of the
orbital phase of a spot, for three values of angular momentum
($a=1$, $0.9$, and $0$) and three inclination angles
($\theta_{\rm{}o}=30^{\circ}$, $60^{\circ}$, $85^{\circ}$).
The centre of the spot is located at radial distance $r$, which corresponds
to the last stable orbit $r_{\rm{}ms}(a)$ for that angular momentum
plus a small displacement given by the spot radius, ${\rm{}d}r$.
The intrinsic energy of the line emission is assumed to be at
$6.4$~keV (indicated by a dotted line). Prograde rotation is assumed.
Time is expressed in orbital periods.}
\end{figure}

\begin{figure}
\dummycaption\label{orbits_2}
\includegraphics[width=0.325\textwidth,height=6cm]{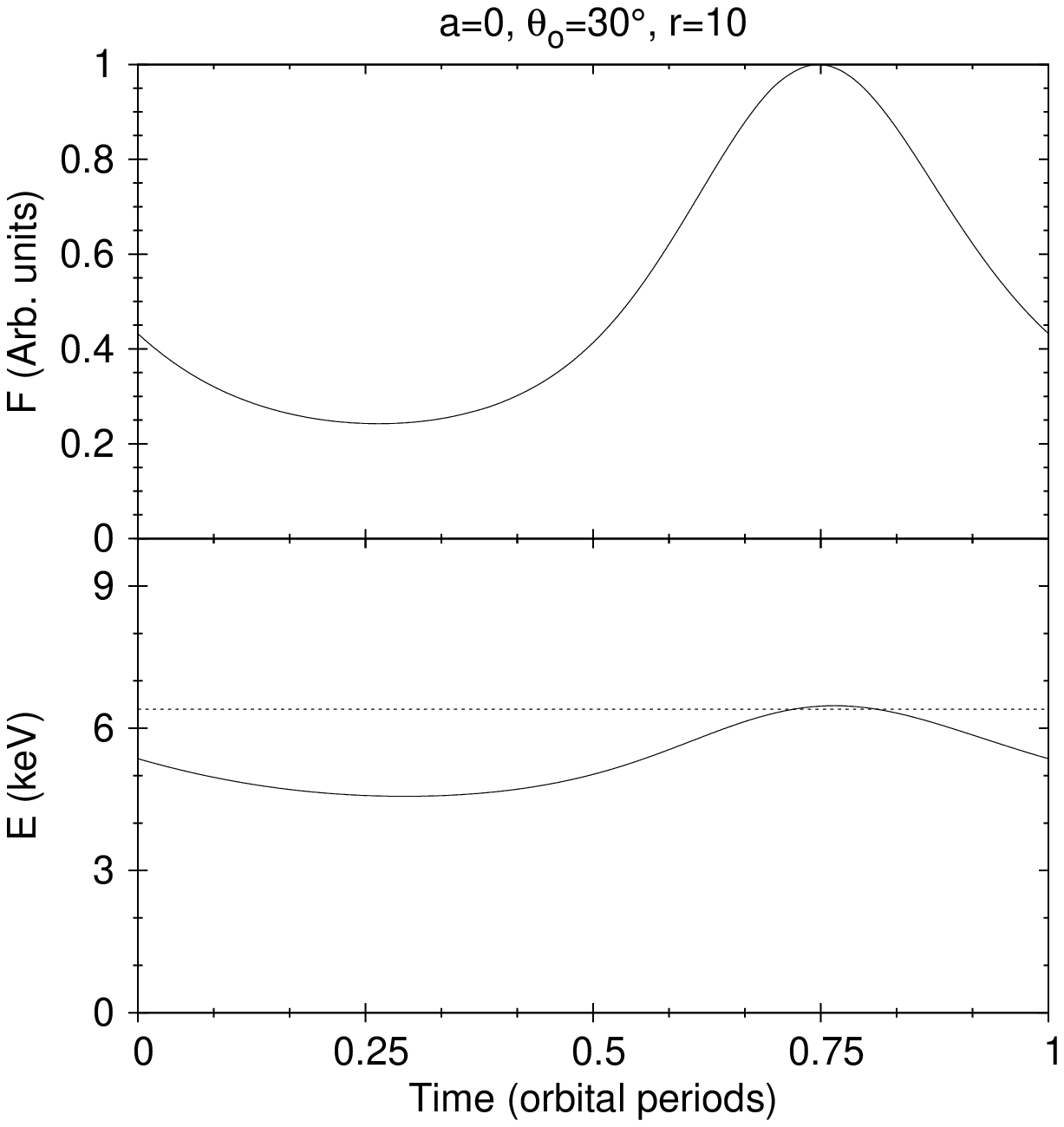}
\hfill
\includegraphics[width=0.325\textwidth,height=6cm]{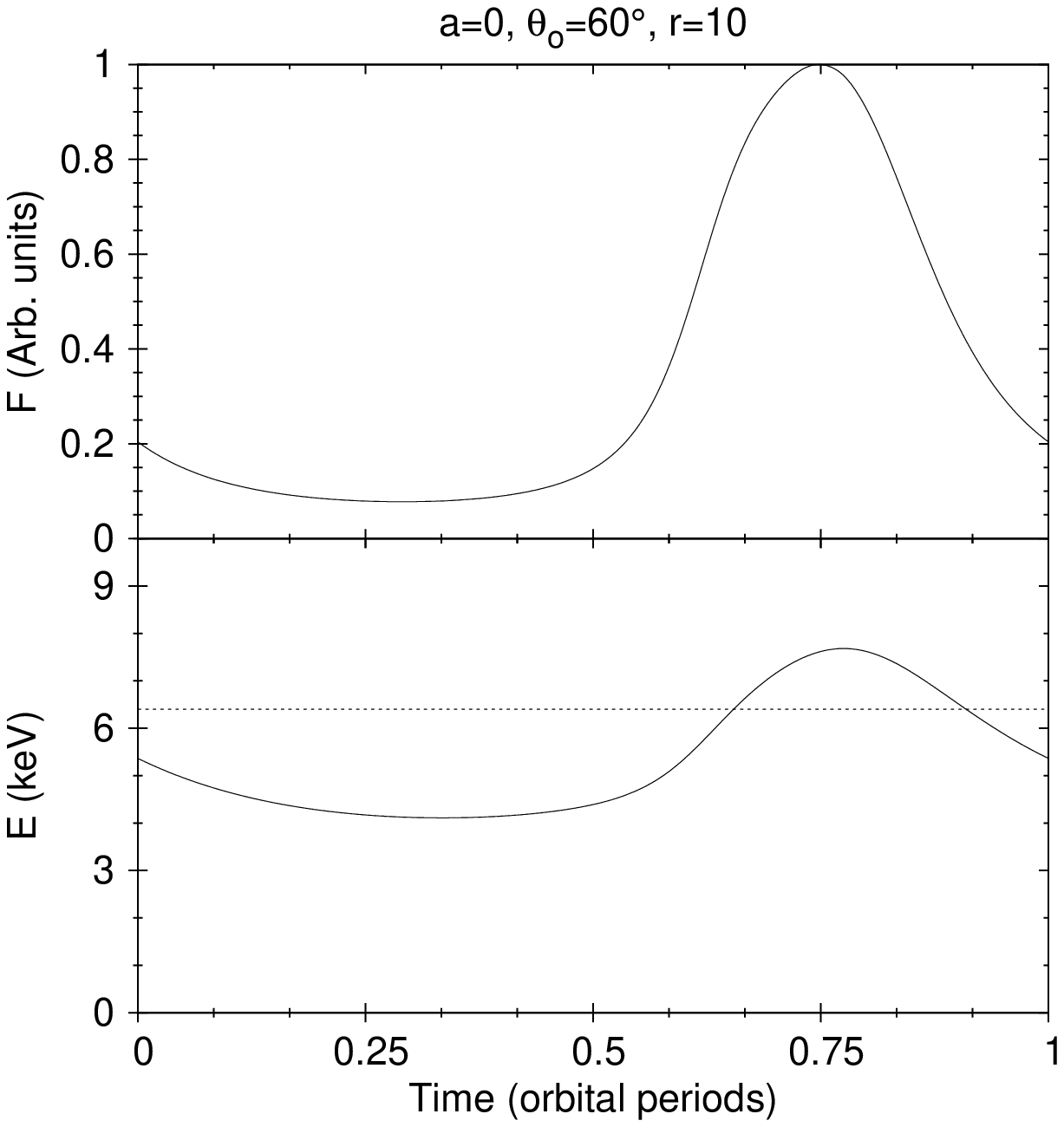}
\hfill
\includegraphics[width=0.325\textwidth,height=6cm]{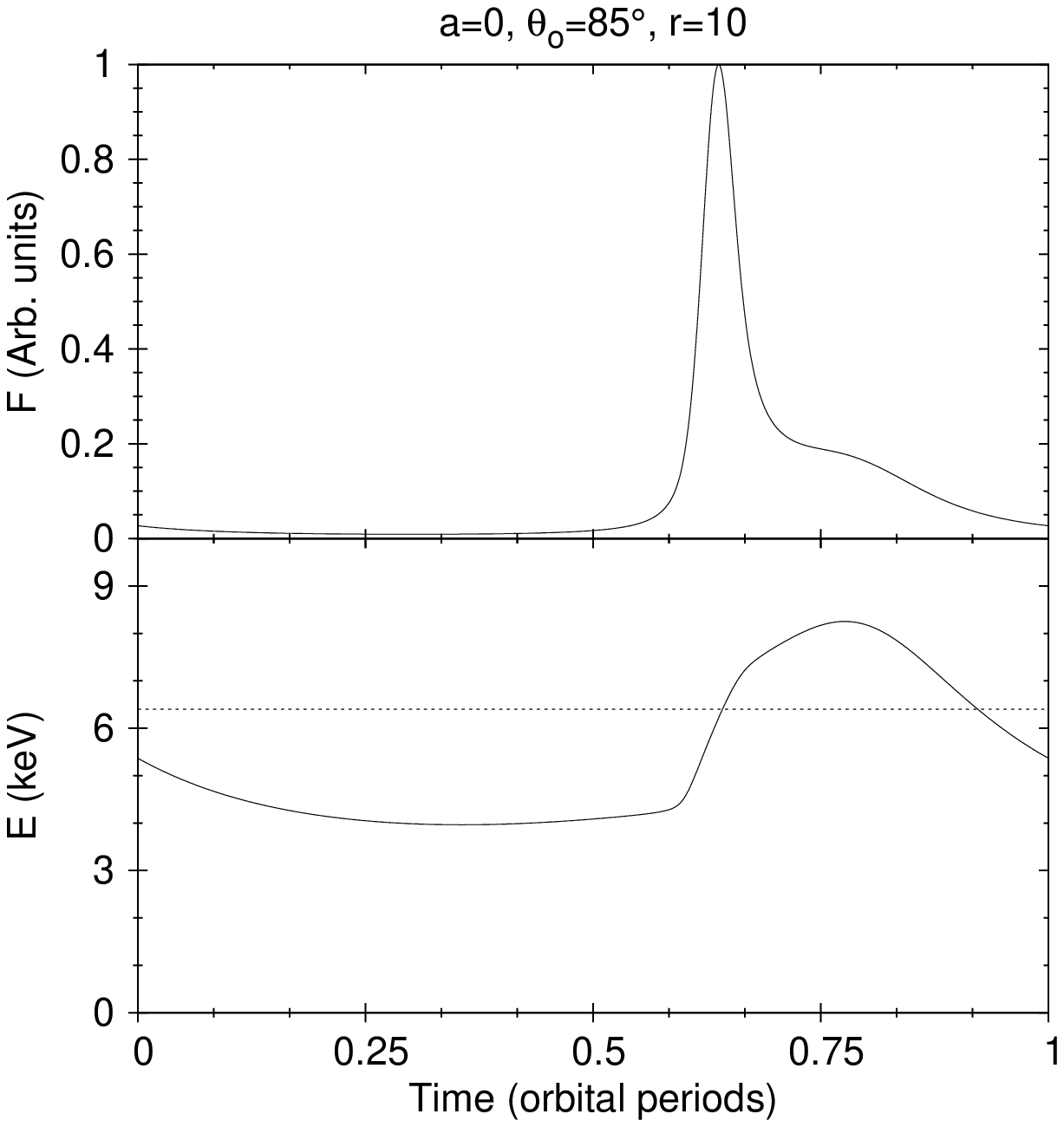}
\vspace*{3mm}\\
\includegraphics[width=0.325\textwidth,height=6cm]{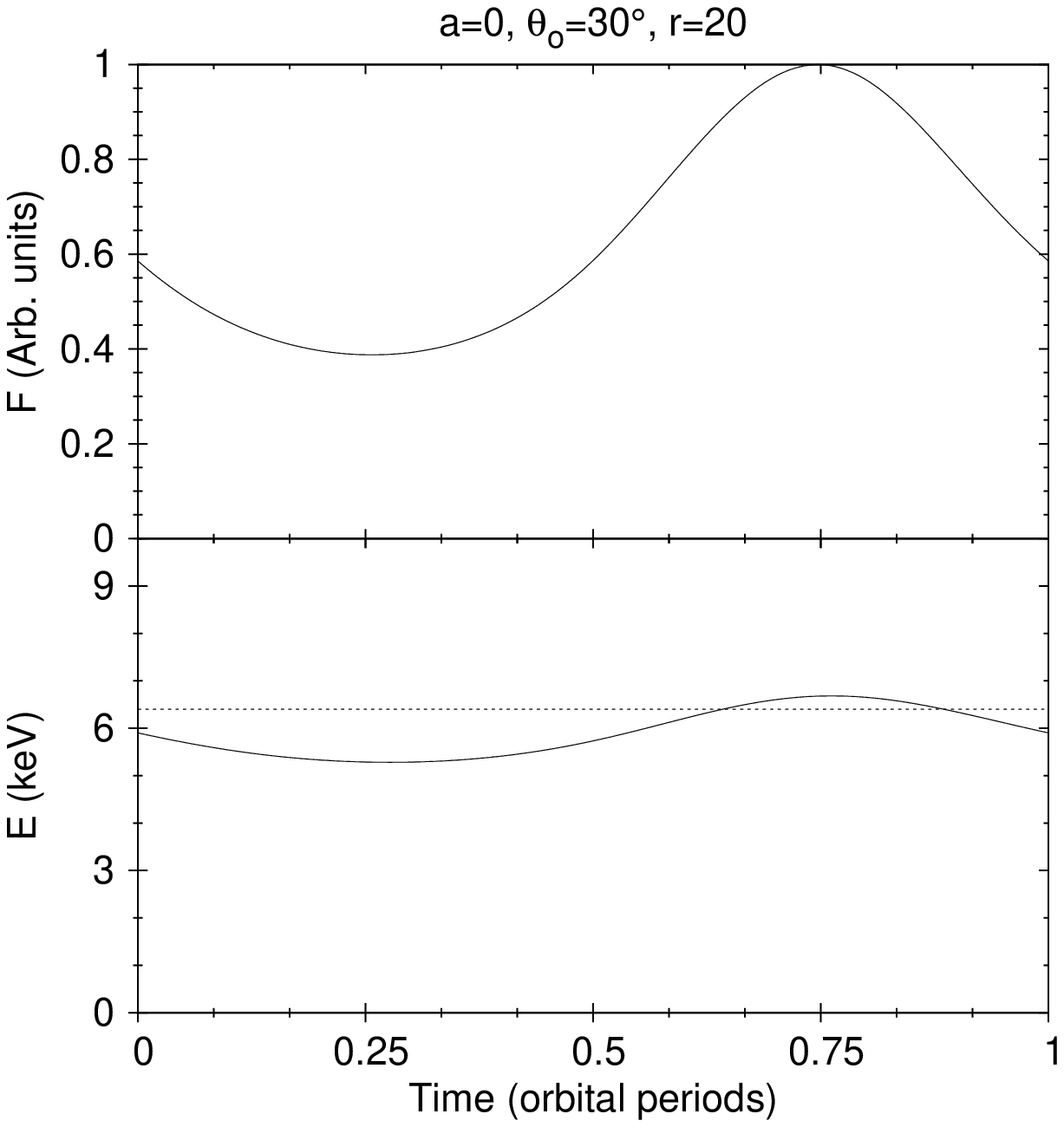}
\hfill
\includegraphics[width=0.325\textwidth,height=6cm]{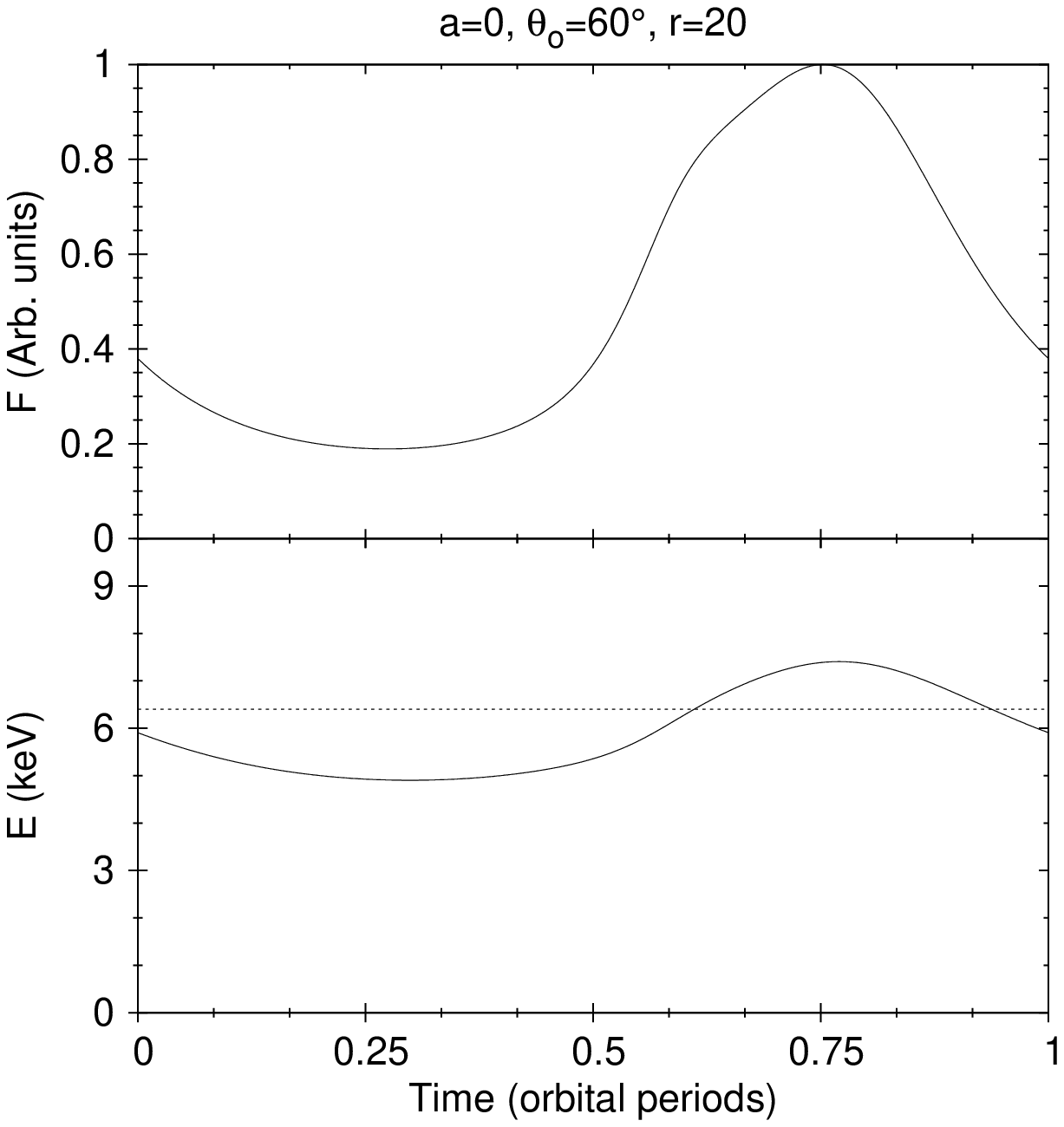}
\hfill
\includegraphics[width=0.325\textwidth,height=6cm]{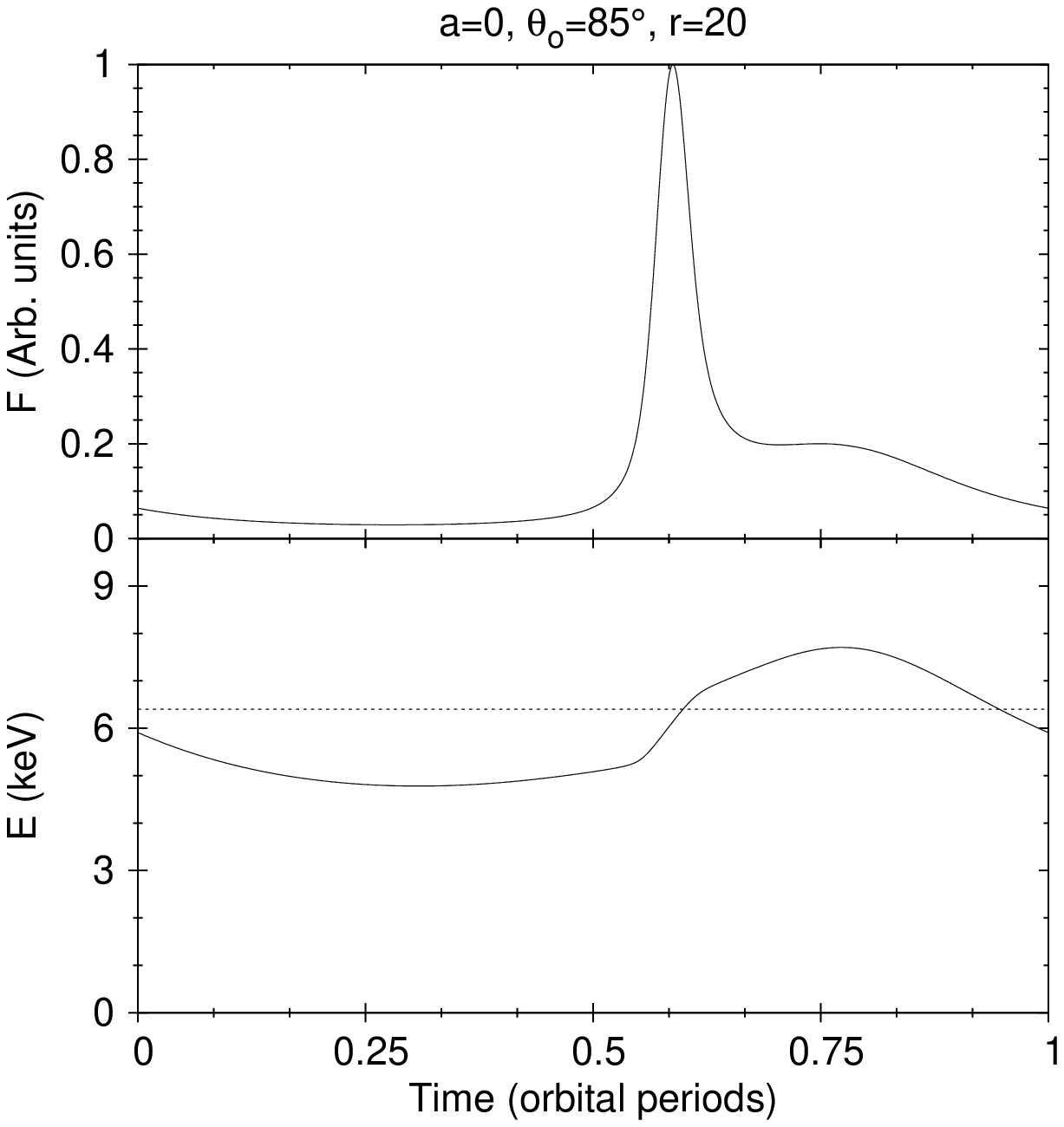}
\vspace*{3mm}\\
\includegraphics[width=0.325\textwidth,height=6cm]{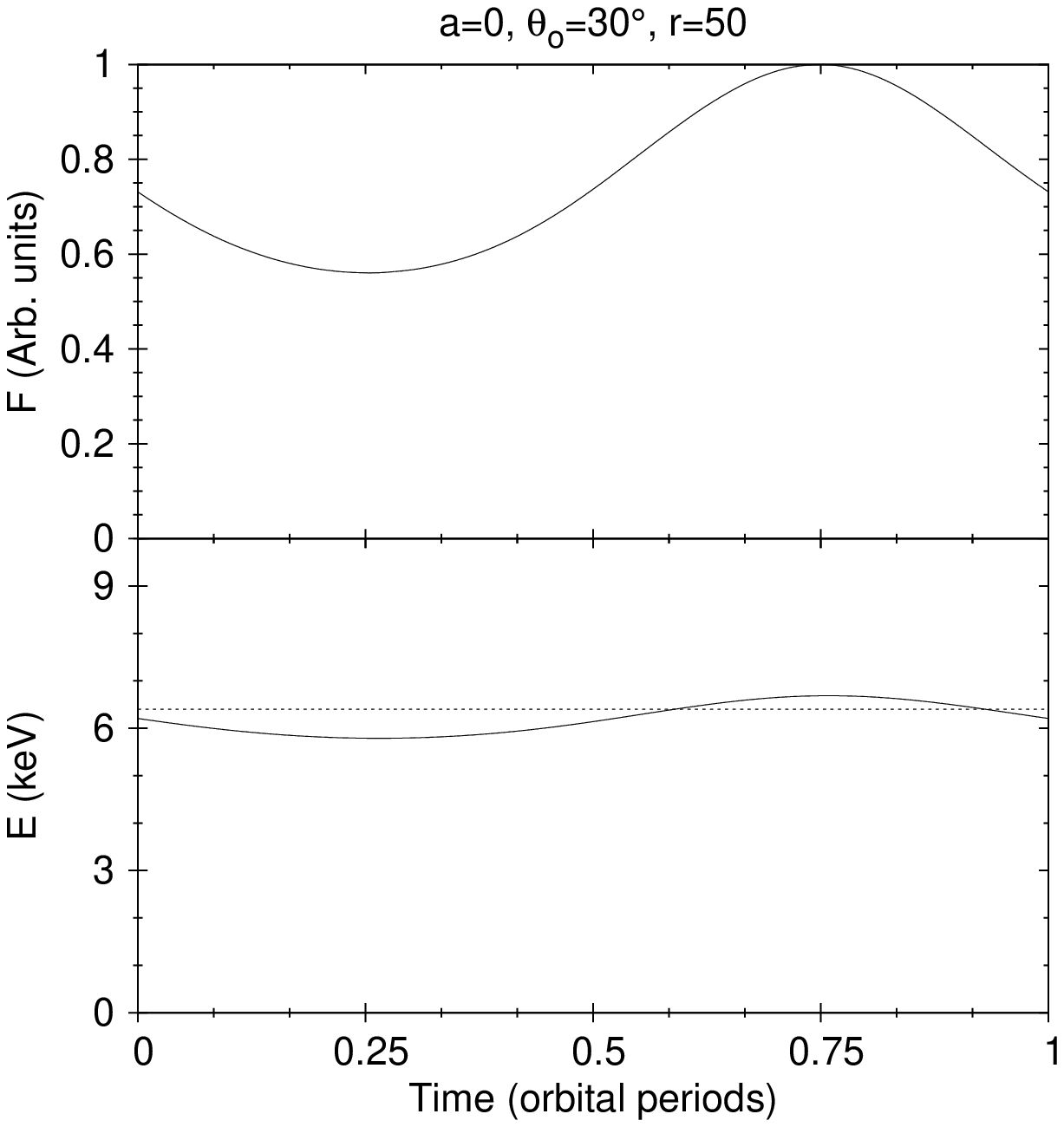}
\hfill
\includegraphics[width=0.325\textwidth,height=6cm]{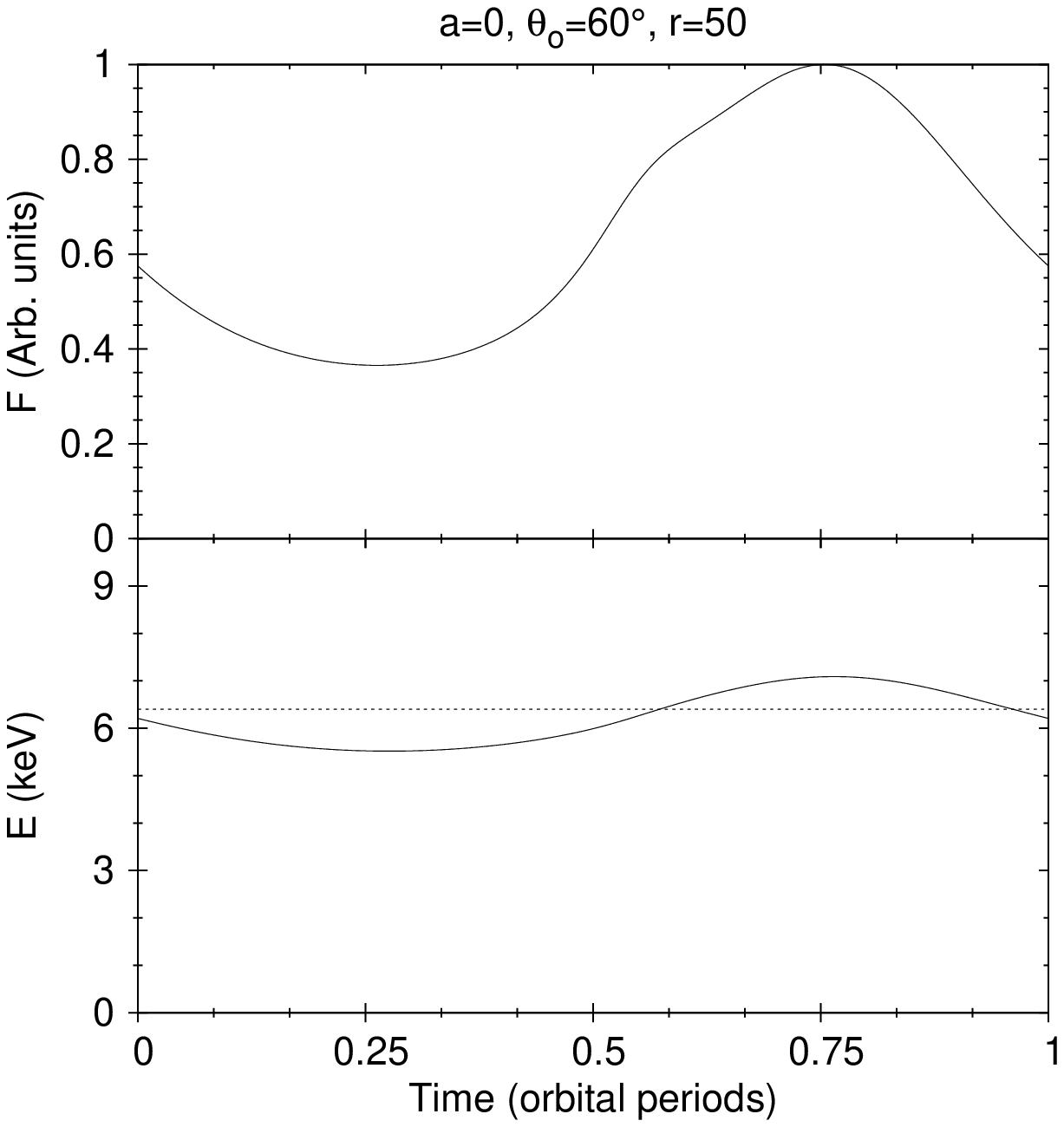}
\hfill
\includegraphics[width=0.325\textwidth,height=6cm]{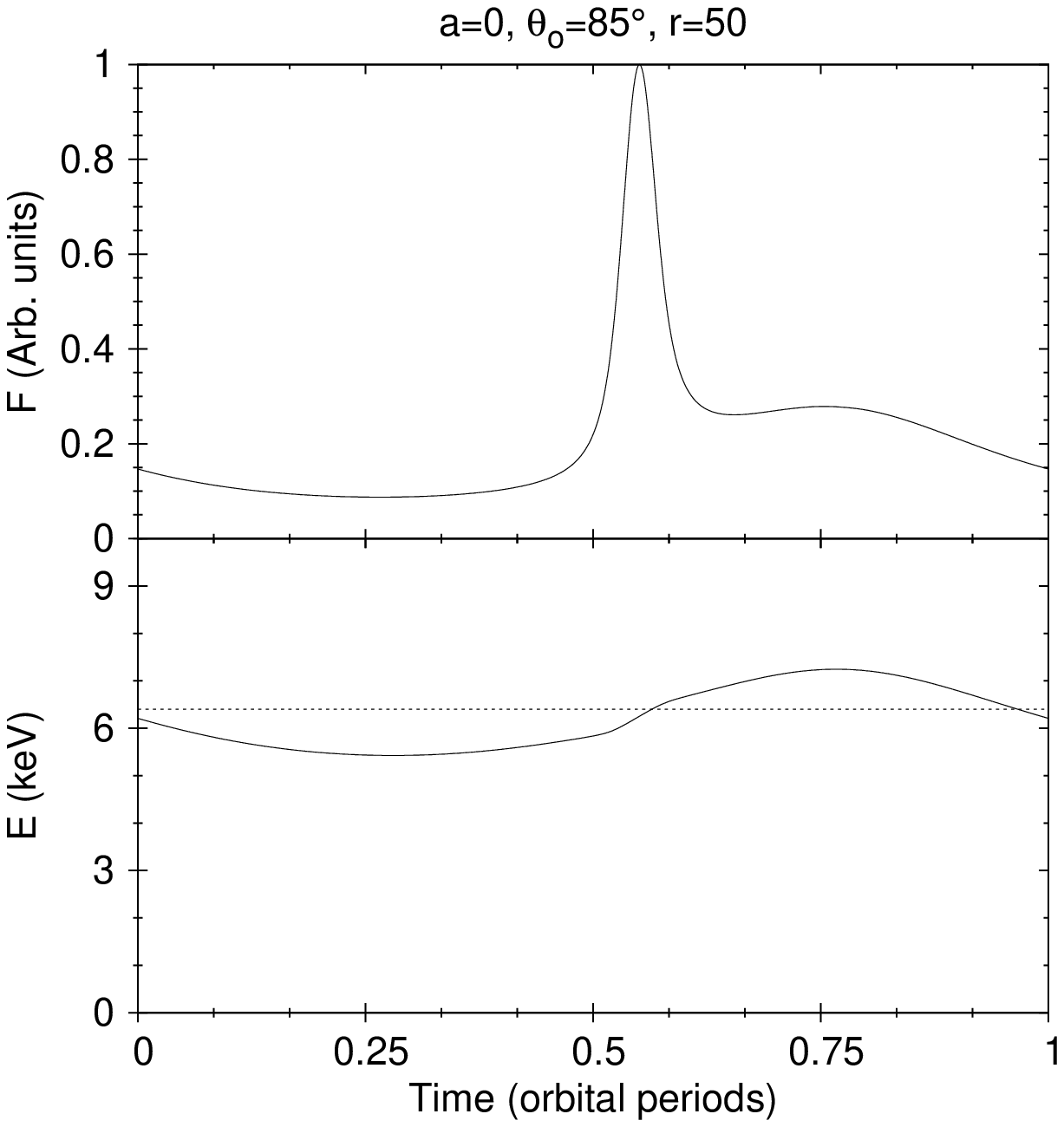}
\vspace*{0mm}
\mycaption{The same as the previous figure, but with $a=0$ and
$r=10$ (top), $20$ (middle), and $50$ (bottom).}
\end{figure}

\begin{figure}
\dummycaption\label{prof_phi_1}
\includegraphics[width=0.325\textwidth,height=6cm]{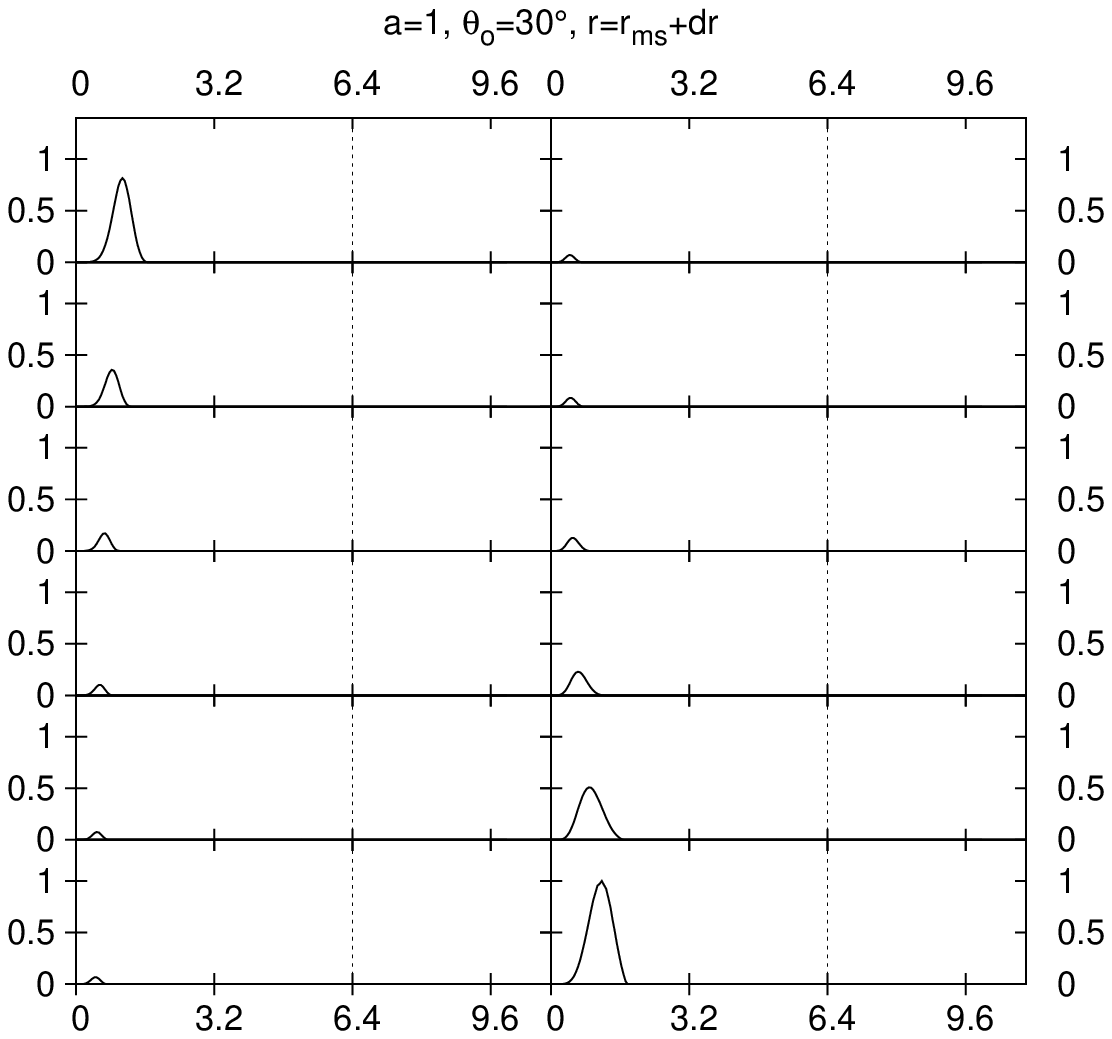}
\hfill
\includegraphics[width=0.325\textwidth,height=6cm]{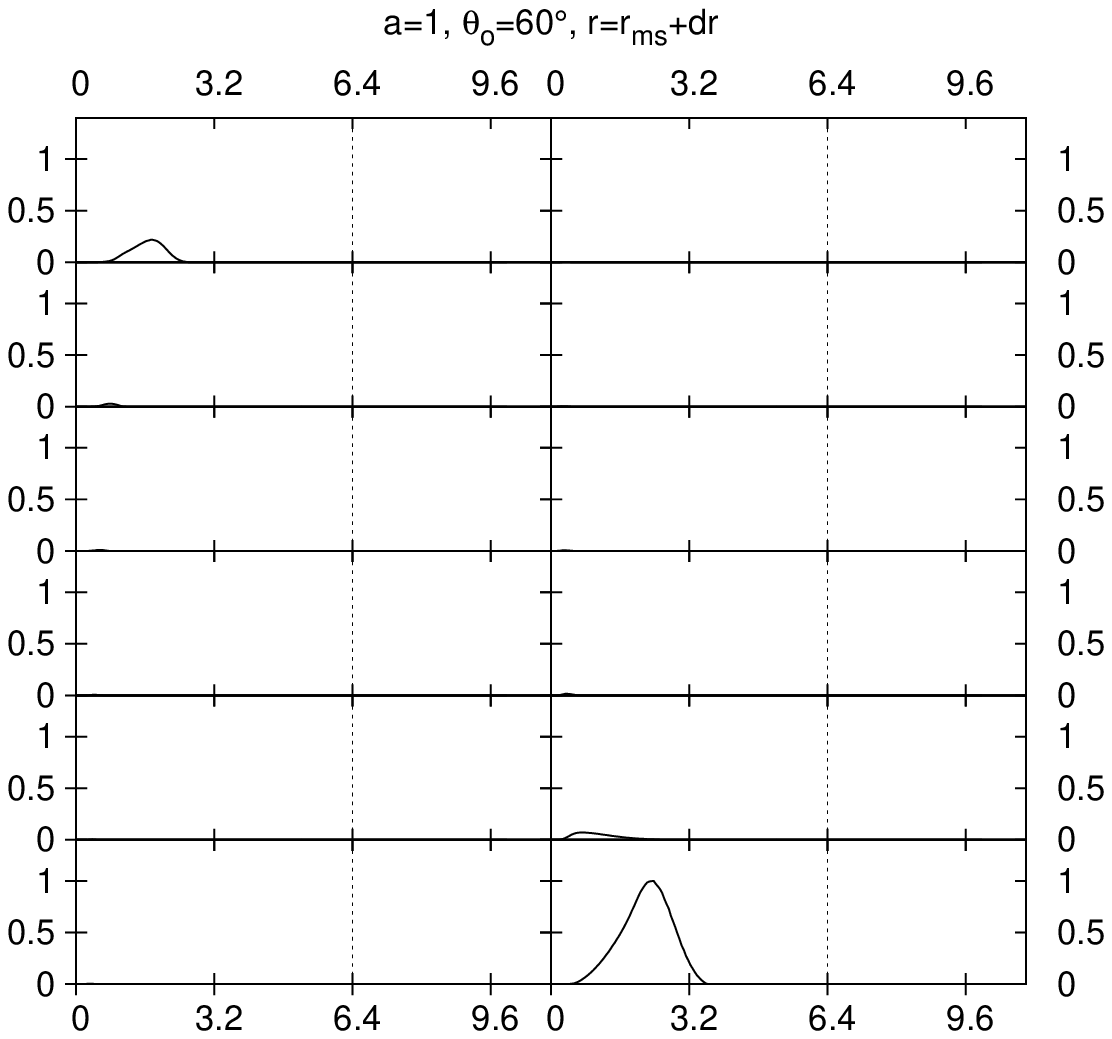}
\hfill
\includegraphics[width=0.325\textwidth,height=6cm]{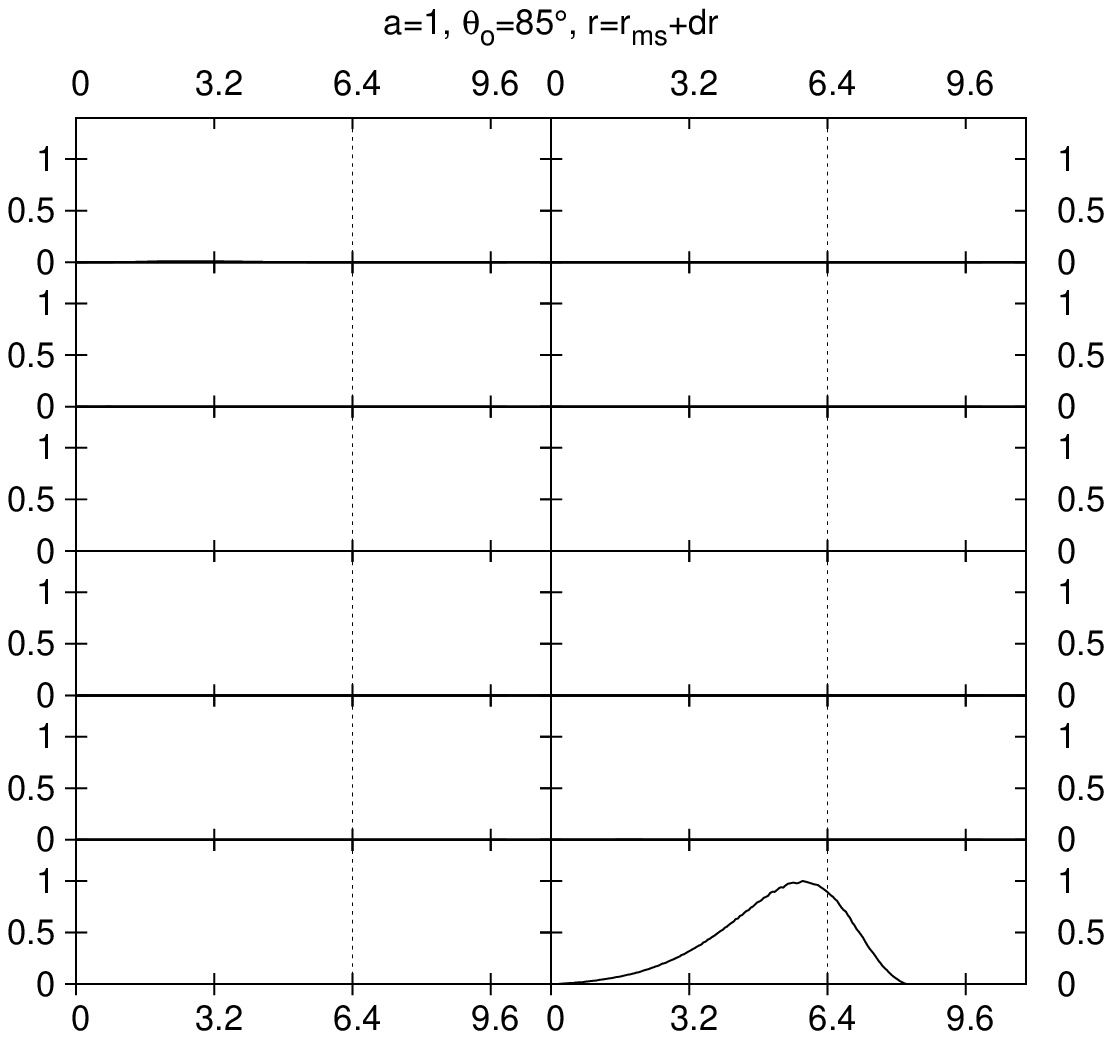}
\vspace*{5mm}\\
\includegraphics[width=0.325\textwidth,height=6cm]{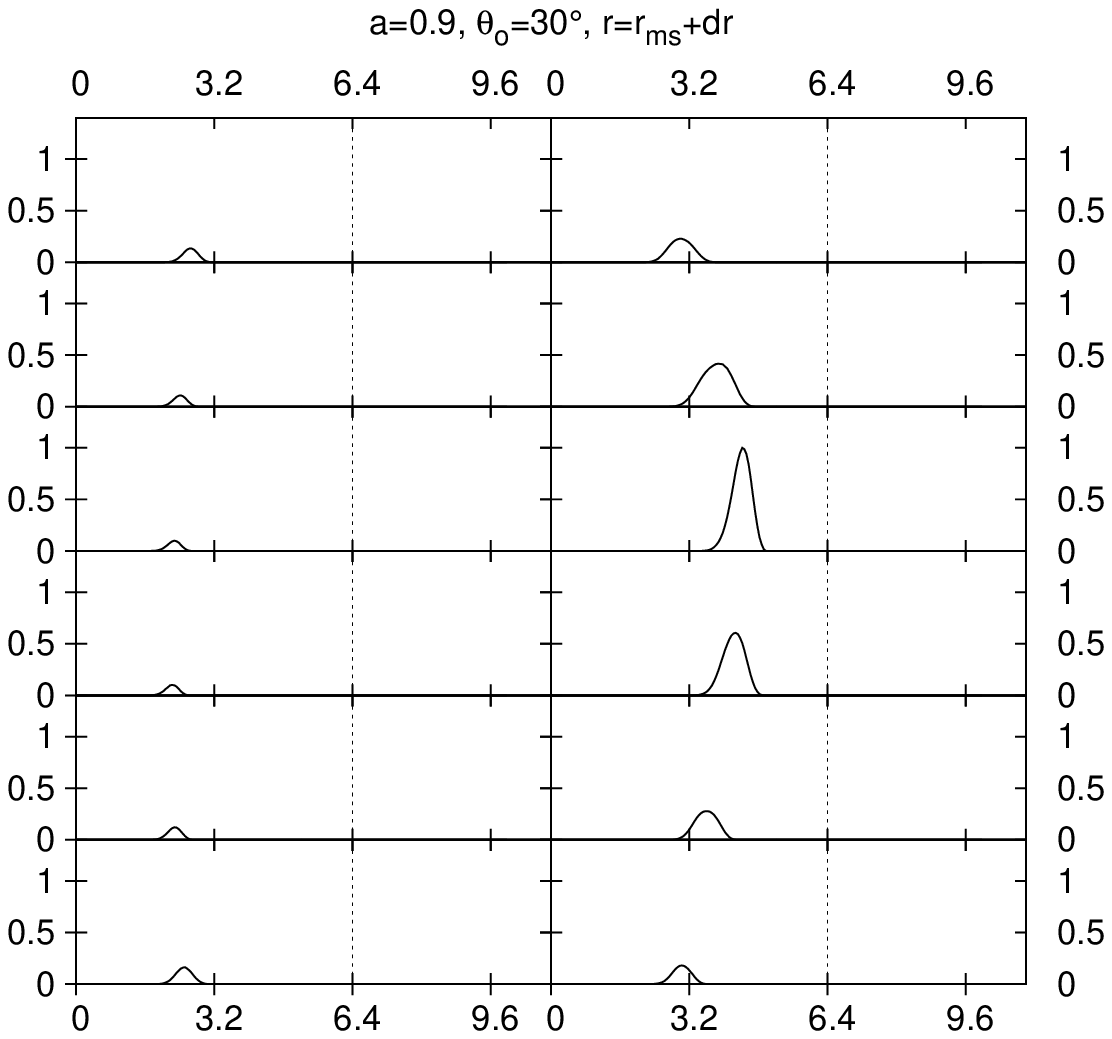}
\hfill
\includegraphics[width=0.325\textwidth,height=6cm]{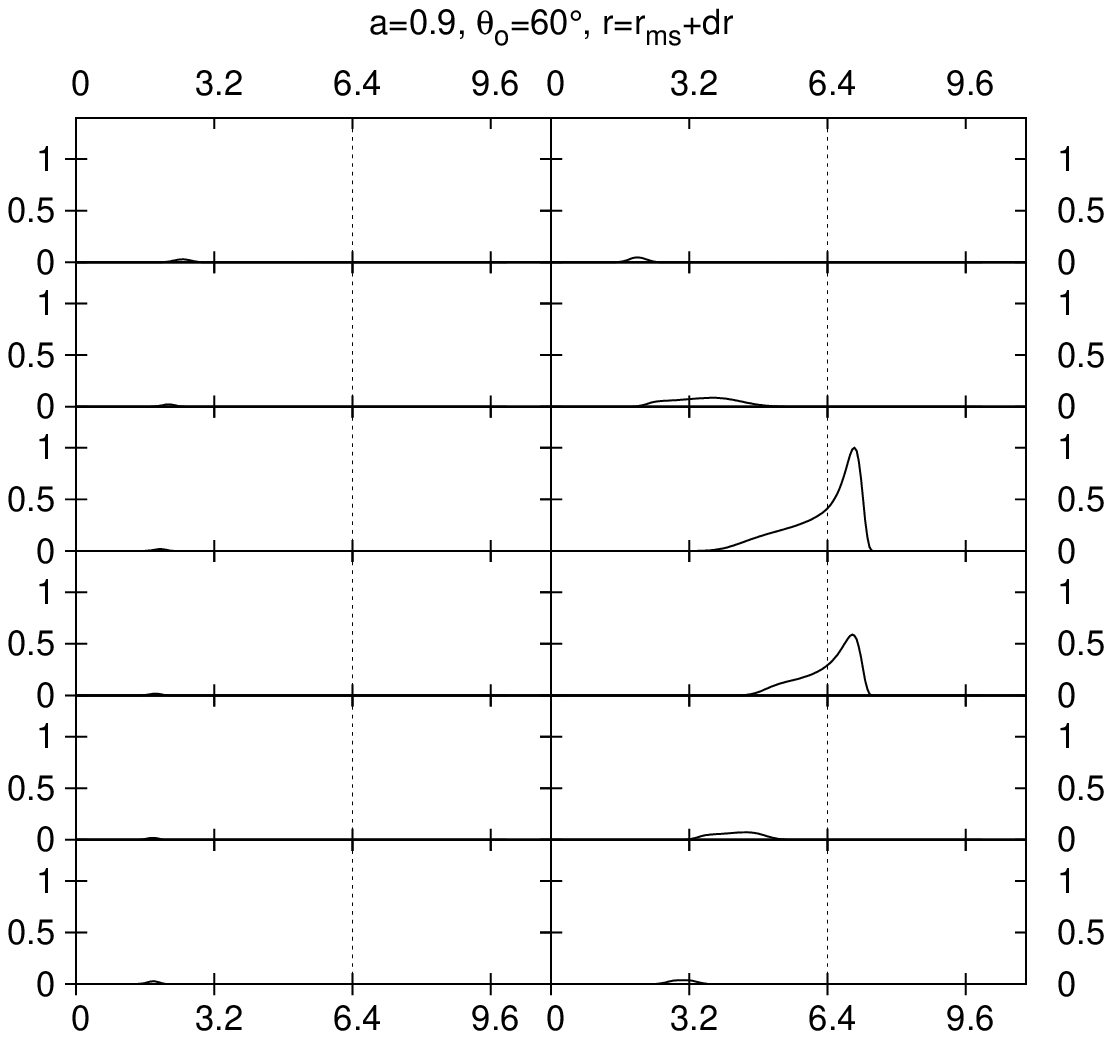}
\hfill
\includegraphics[width=0.325\textwidth,height=6cm]{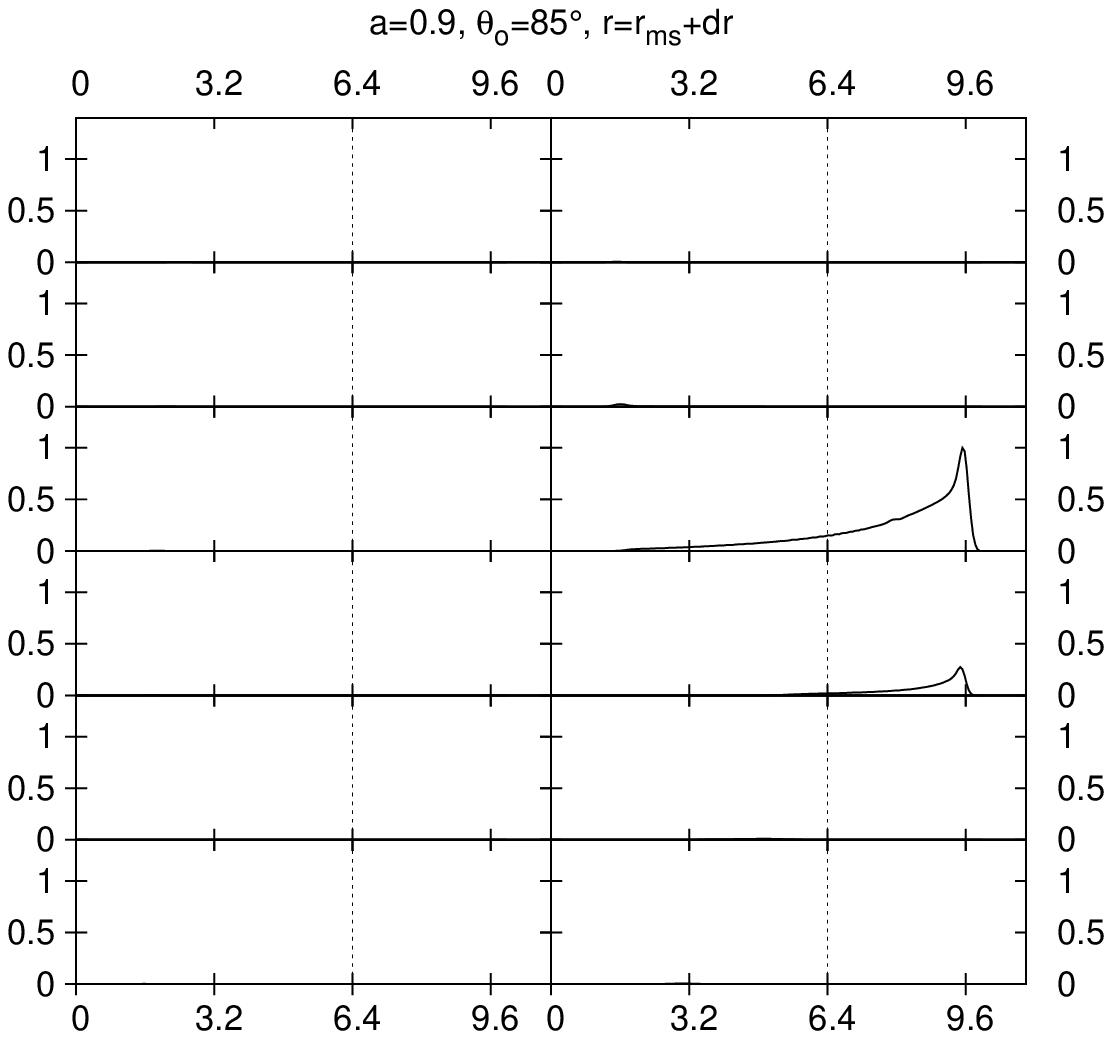}
\vspace*{5mm}\\
\includegraphics[width=0.325\textwidth,height=6cm]{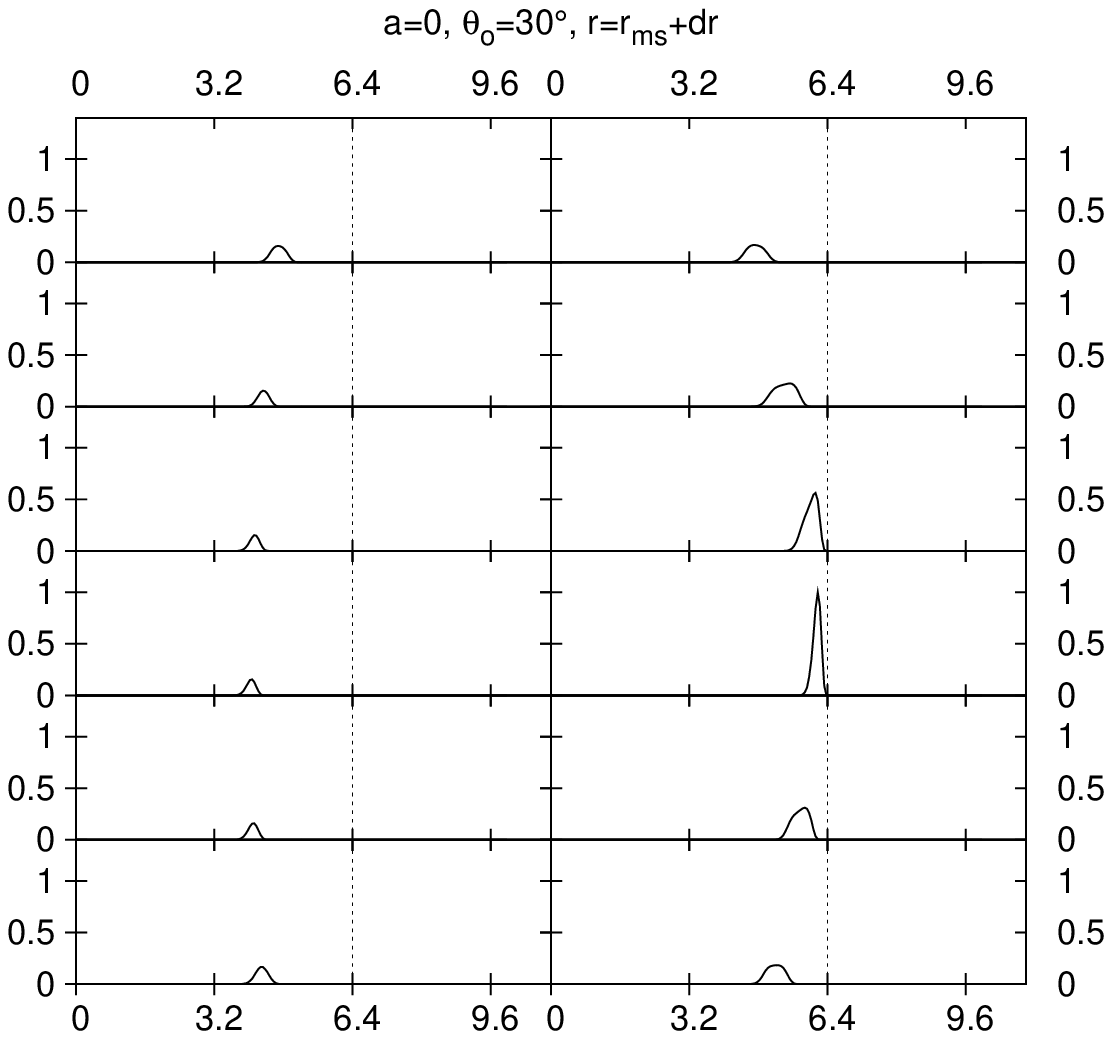}
\hfill
\includegraphics[width=0.325\textwidth,height=6cm]{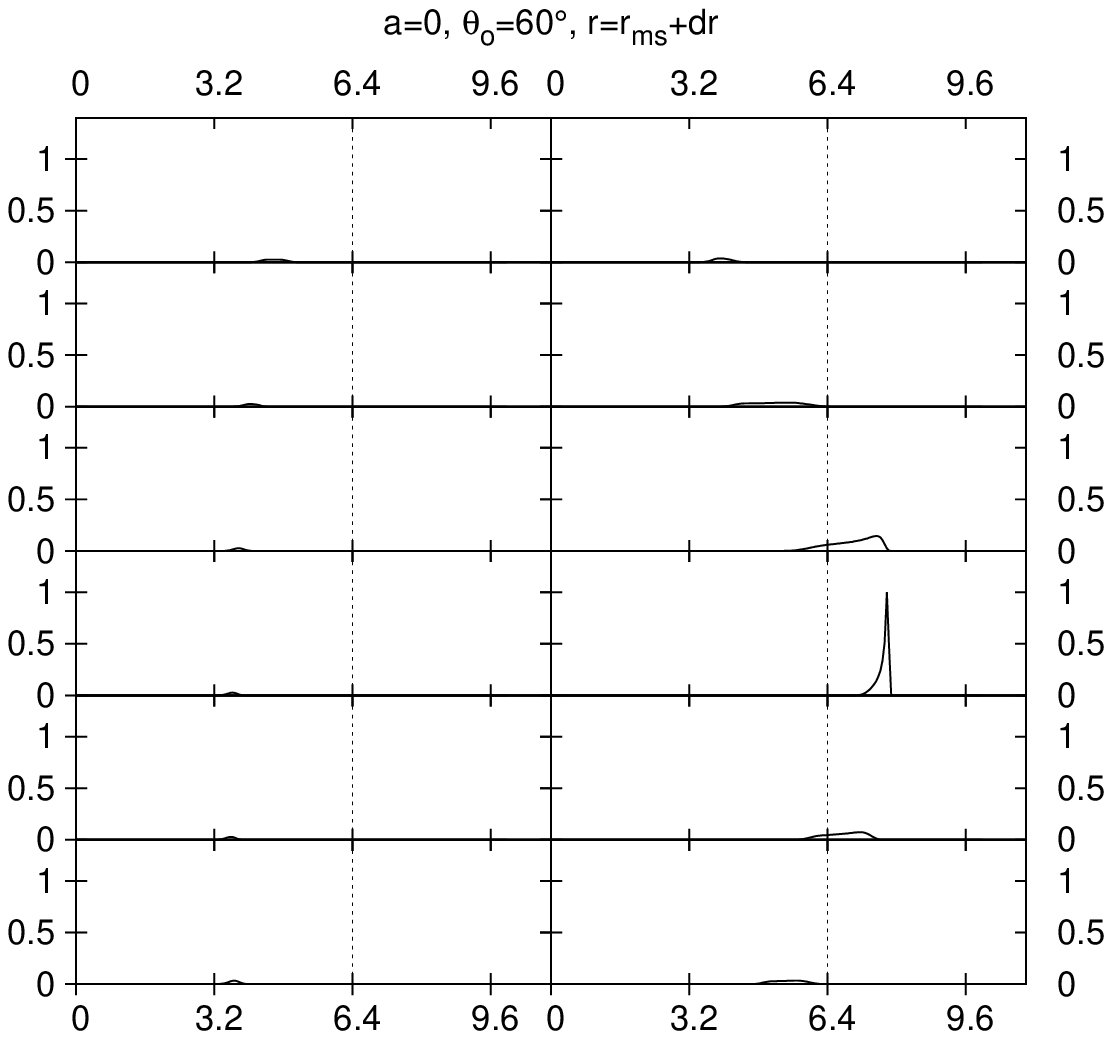}
\hfill
\includegraphics[width=0.325\textwidth,height=6cm]{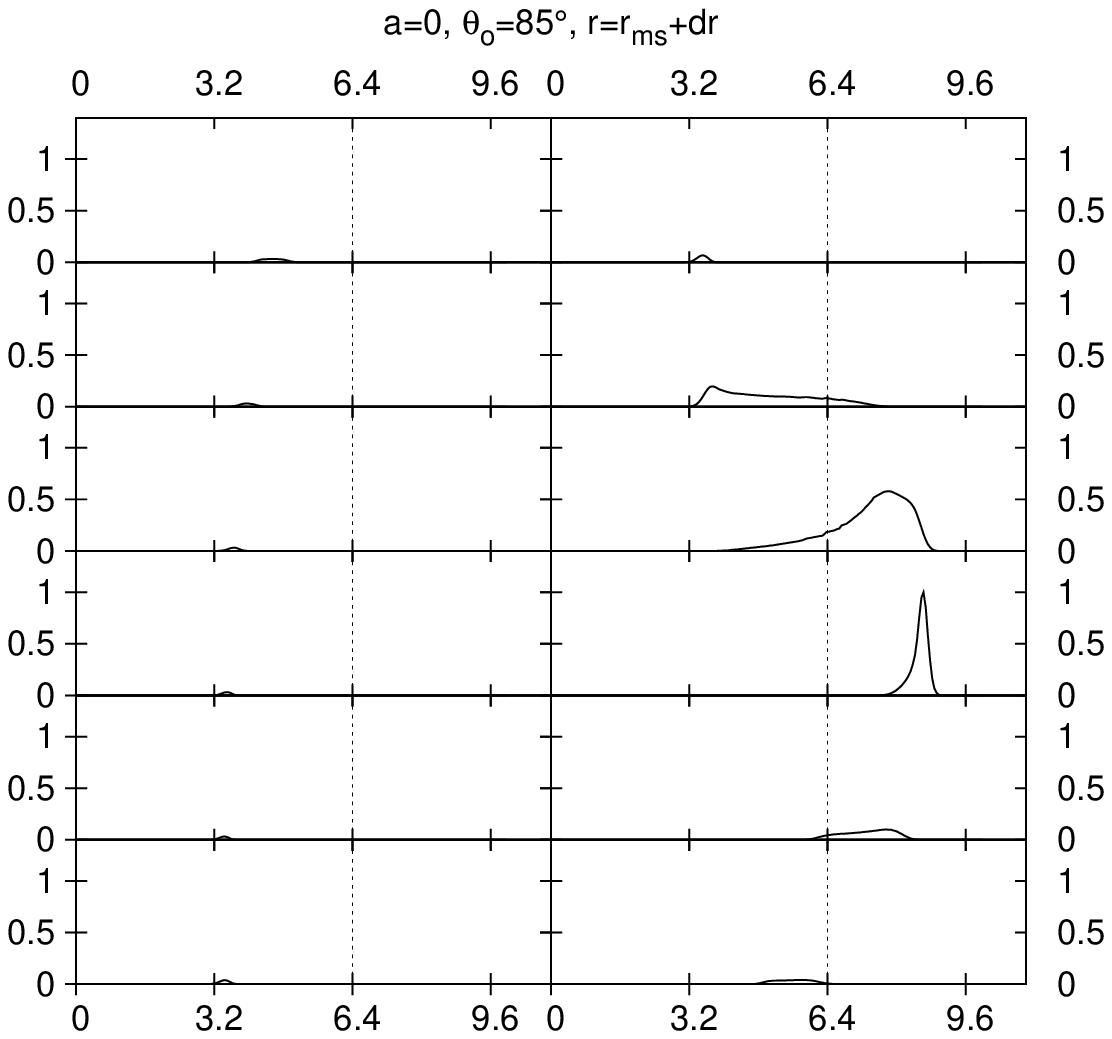}
\vspace*{0mm}
\mycaption{Line profiles integrated over twelve consecutive
temporal intervals of equal duration. Each interval covers $1/12$
of the orbital period at corresponding radius. As explained in the text,
top-left frame of each panel corresponds to the spot being observed
at the moment of passing through lower conjunction. Energy is on abscissa
(in keV). Observed photon flux is on ordinate (arbitrary units, scaled
to the maximum flux which is reached during the complete revolution
of the spot). Notice the occurrences of narrow and prominent peaks which
appear for relatively brief fraction of the total period.}
\end{figure}

\begin{figure}
\dummycaption\label{prof_phi_2}
\includegraphics[width=0.325\textwidth,height=6cm]{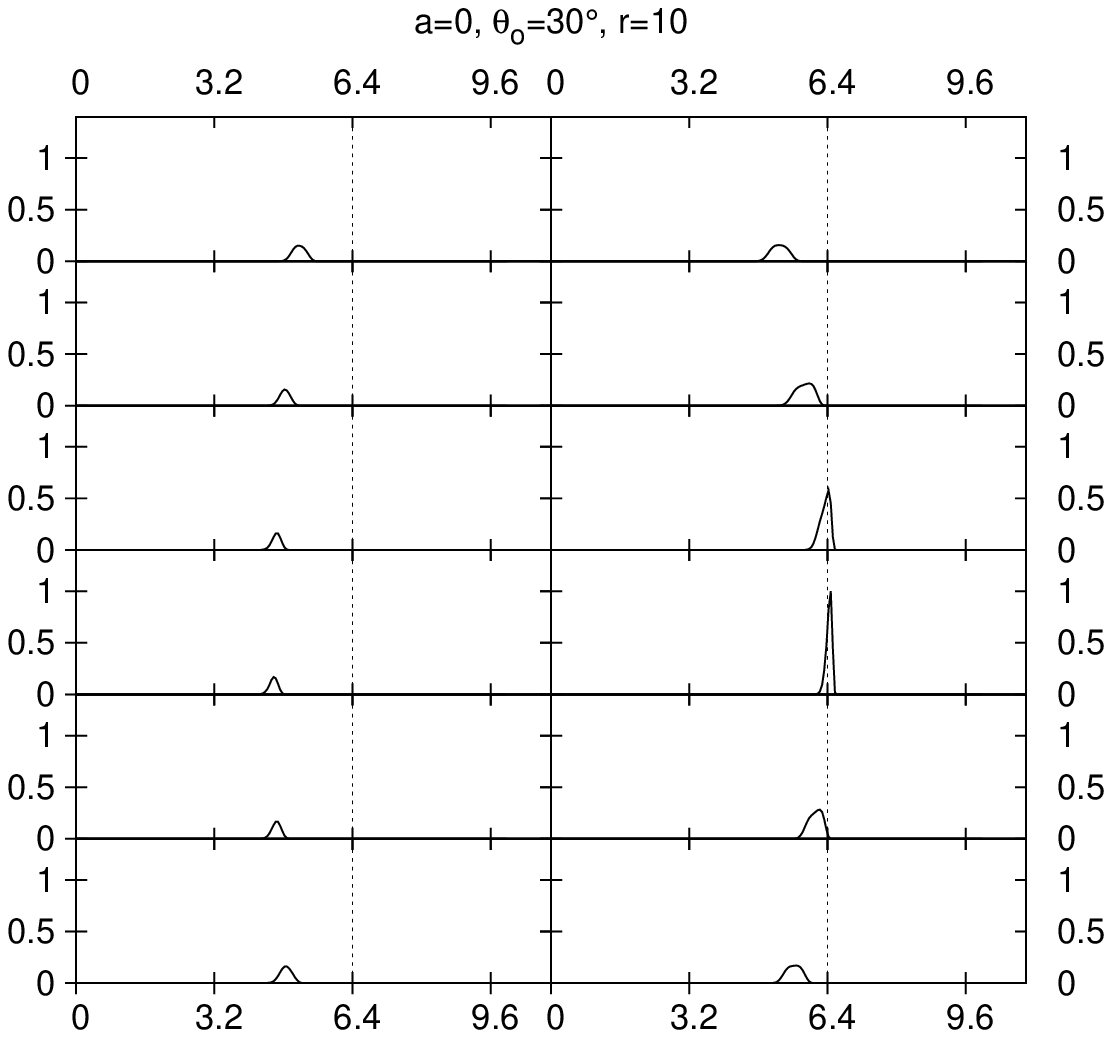}
\hfill
\includegraphics[width=0.325\textwidth,height=6cm]{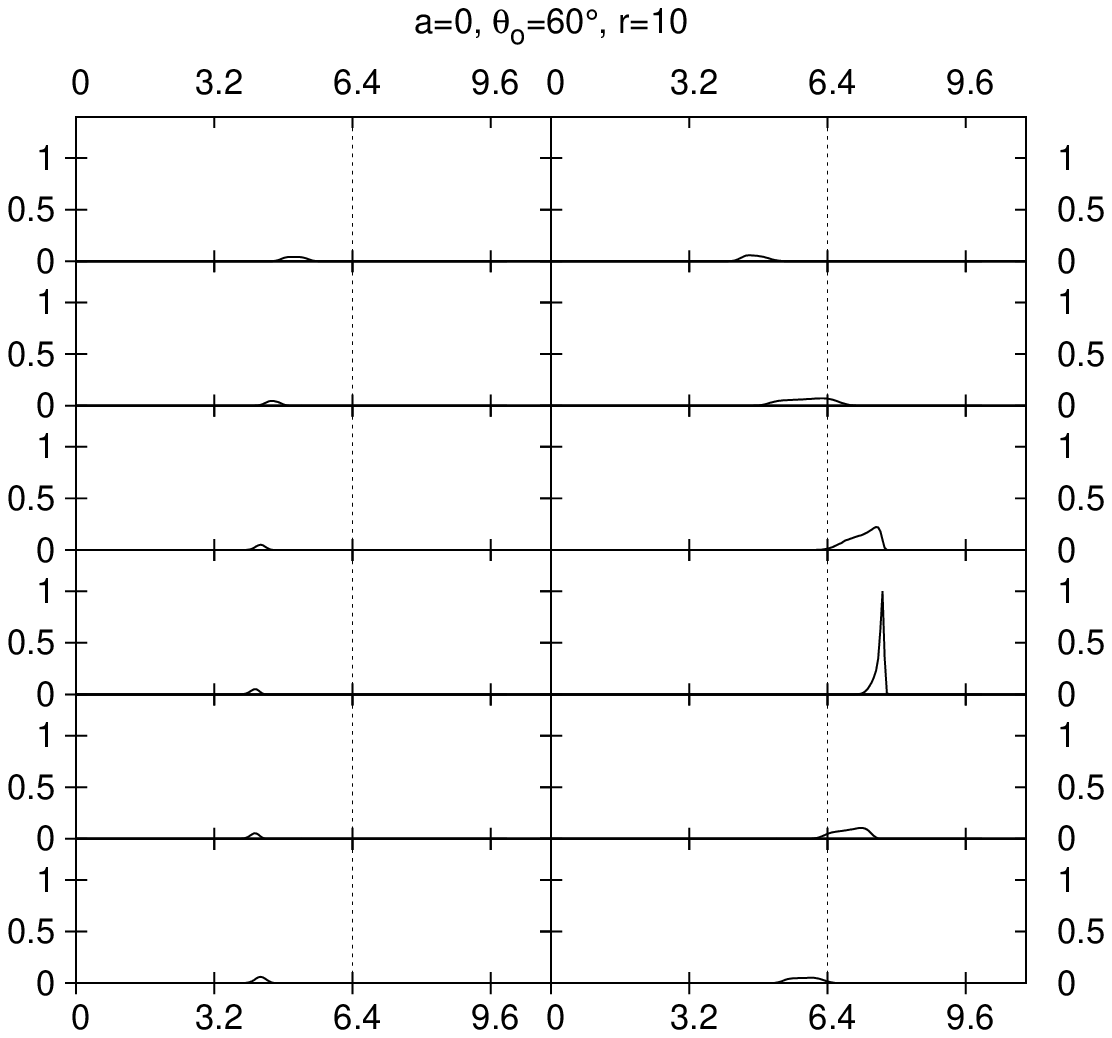}
\hfill
\includegraphics[width=0.325\textwidth,height=6cm]{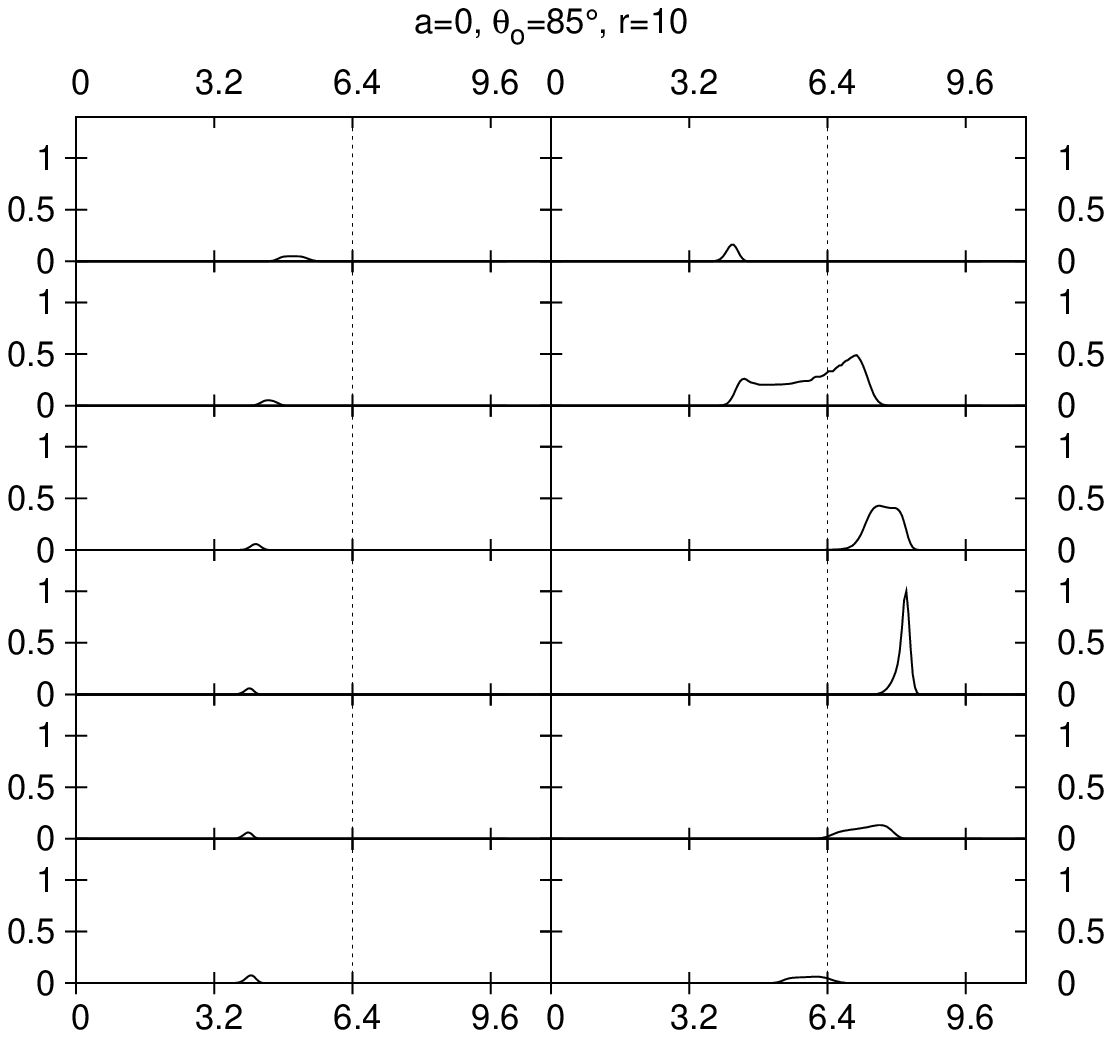}
\vspace*{5mm}\\
\includegraphics[width=0.325\textwidth,height=6cm]{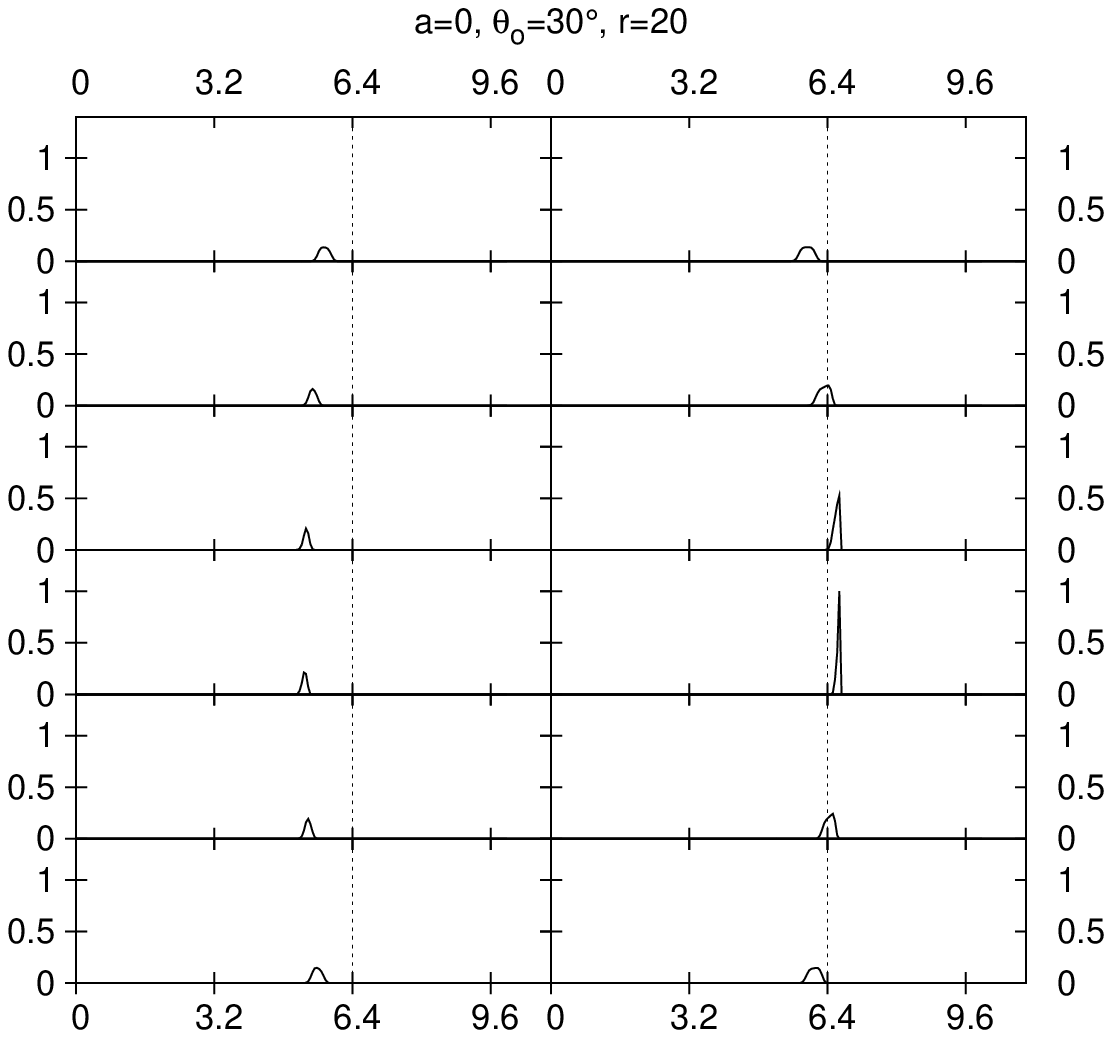}
\hfill
\includegraphics[width=0.325\textwidth,height=6cm]{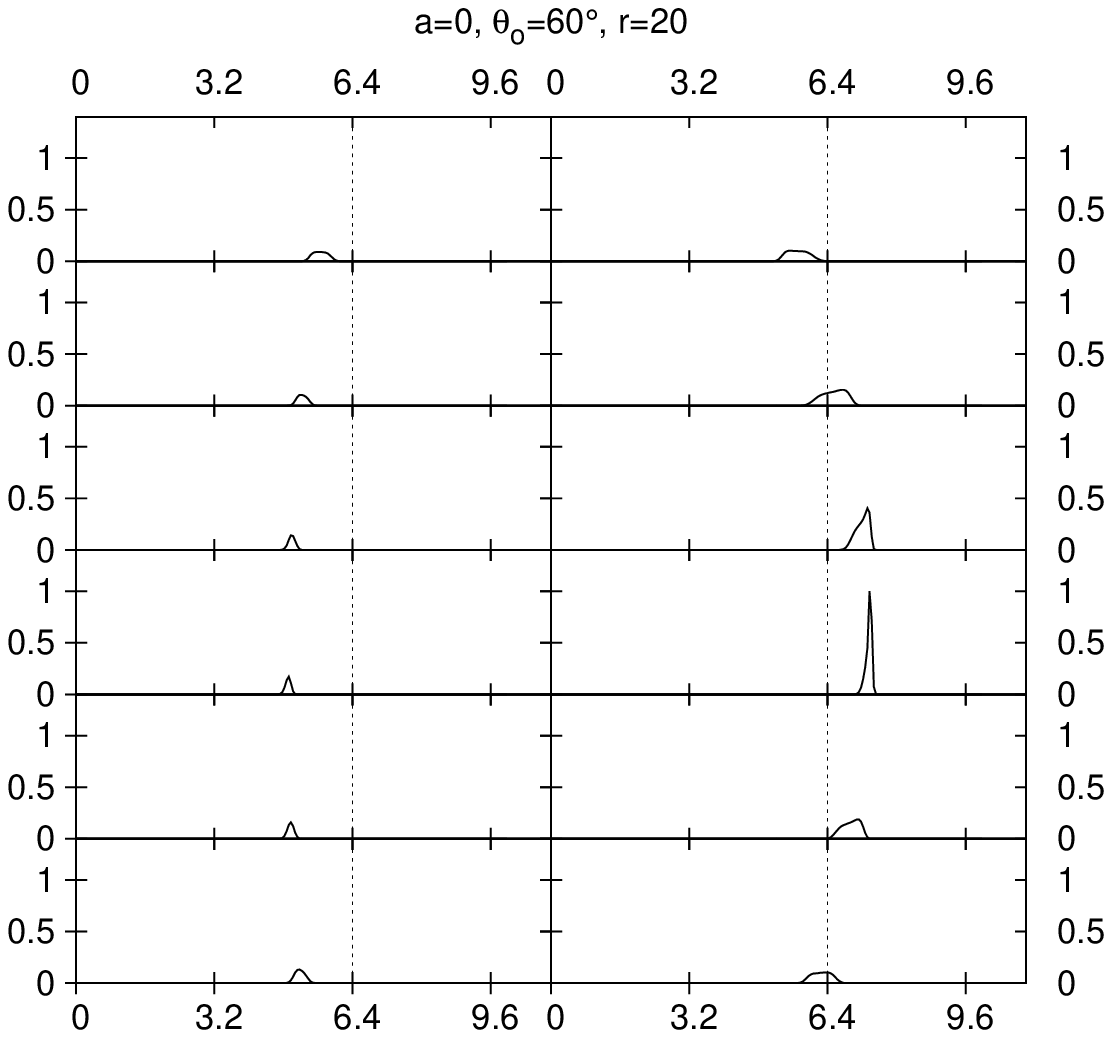}
\hfill
\includegraphics[width=0.325\textwidth,height=6cm]{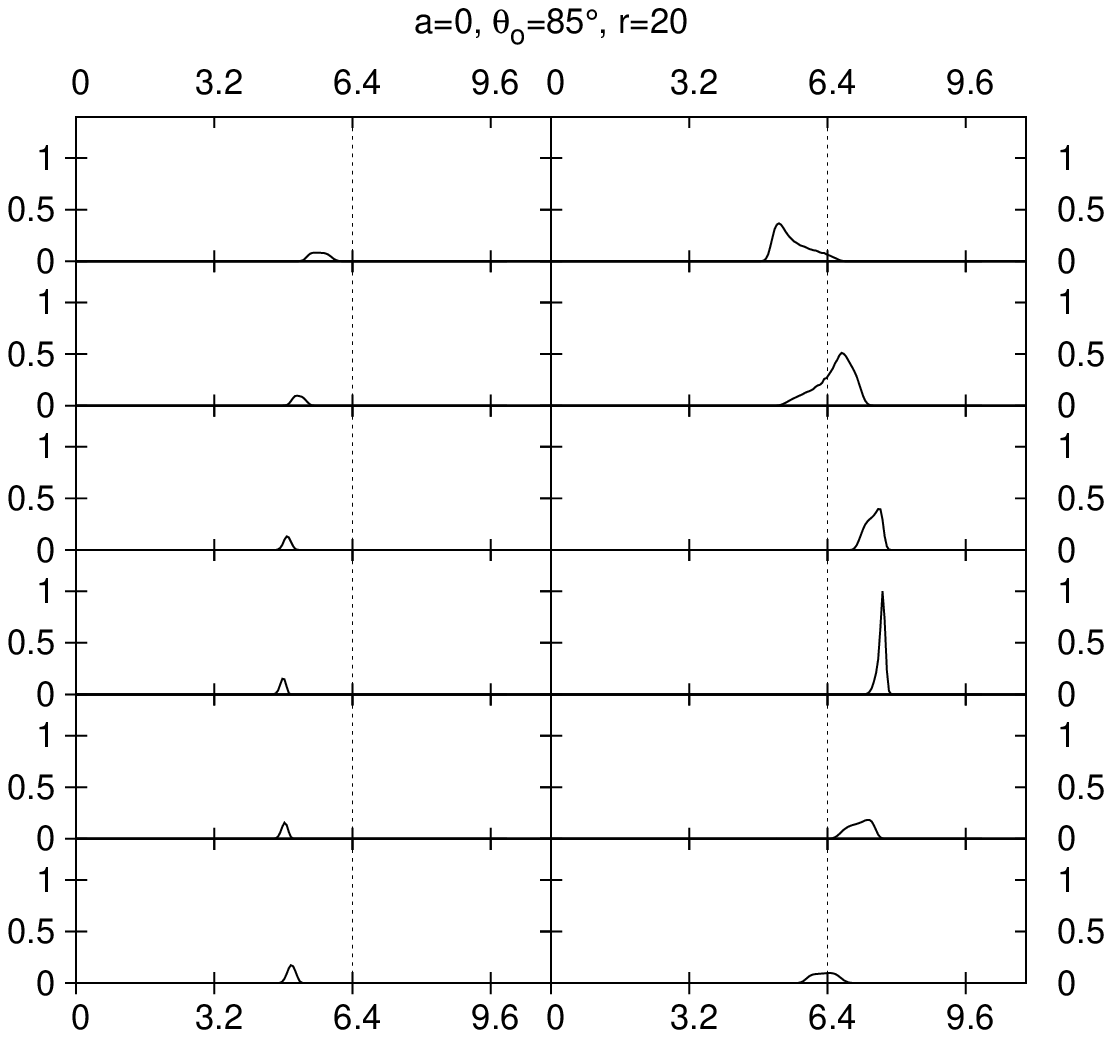}
\vspace*{5mm}\\
\includegraphics[width=0.325\textwidth,height=6cm]{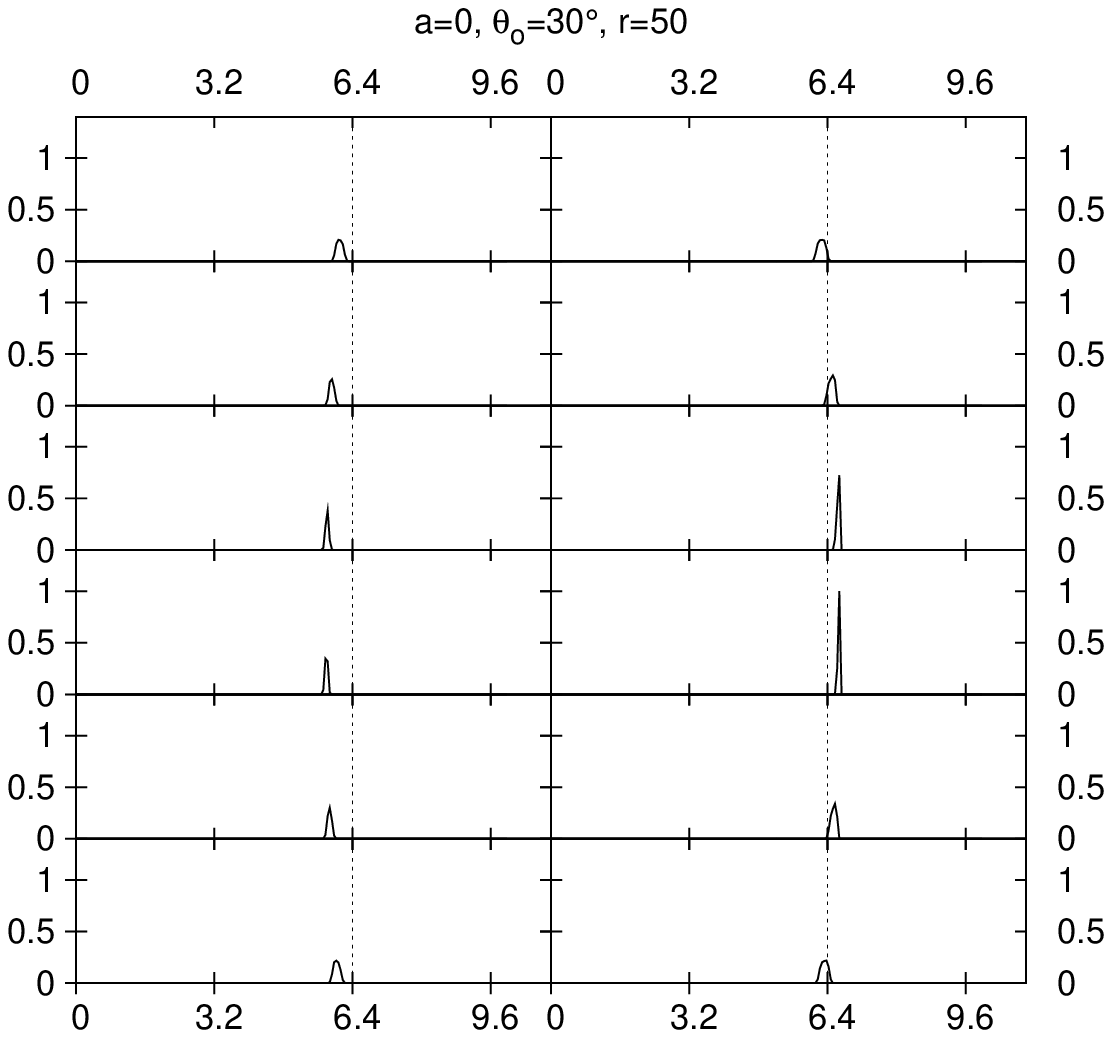}
\hfill
\includegraphics[width=0.325\textwidth,height=6cm]{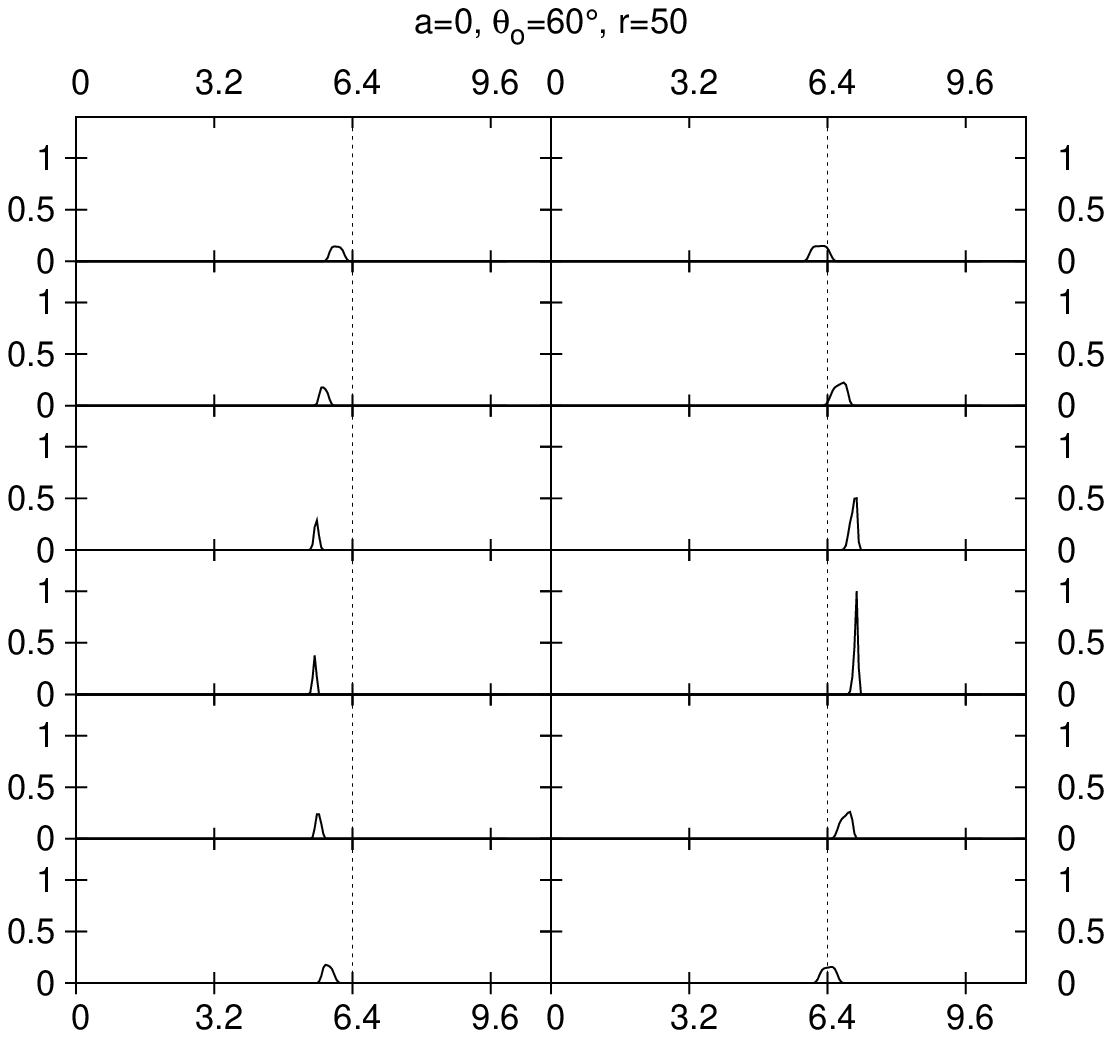}
\hfill
\includegraphics[width=0.325\textwidth,height=6cm]{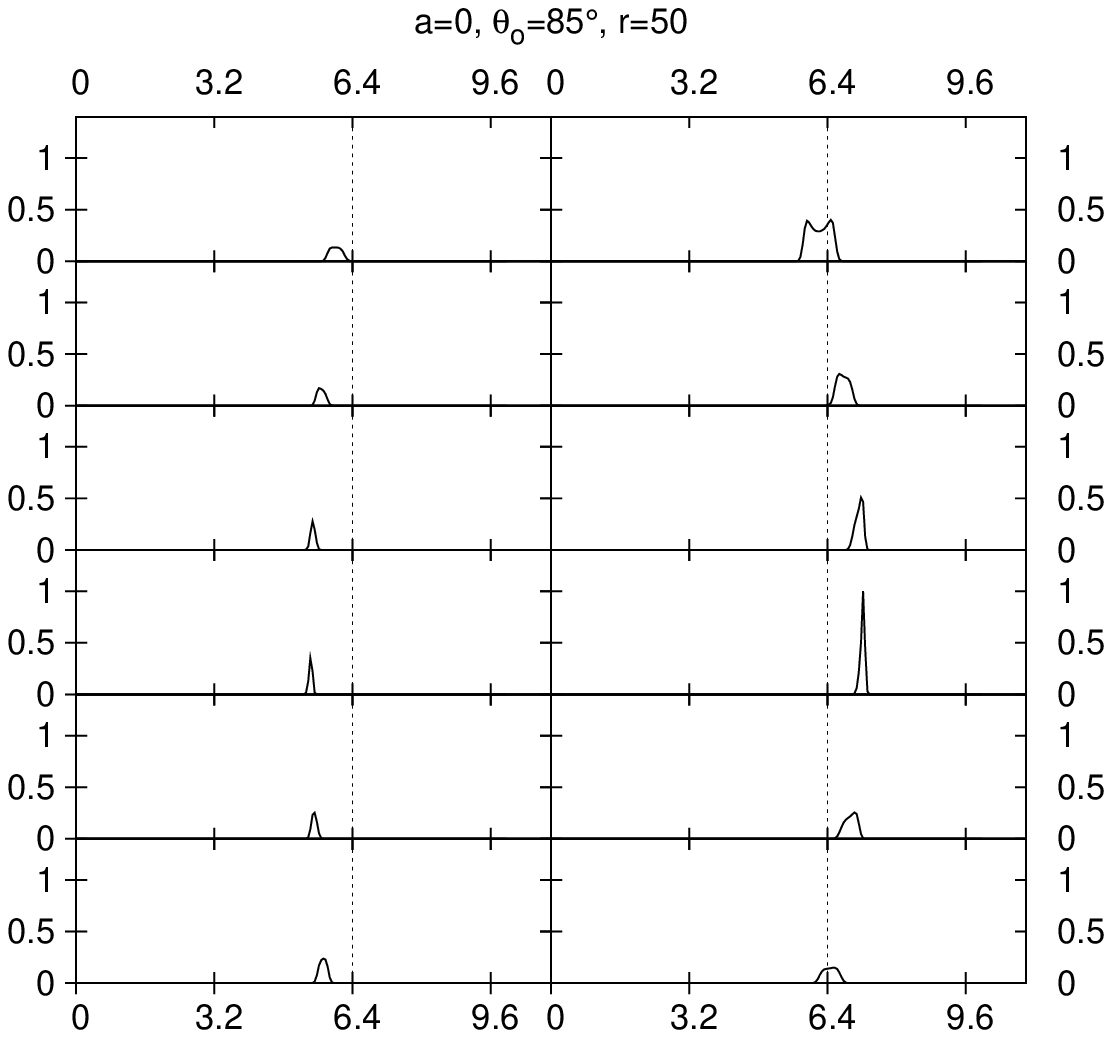}
\vspace*{0mm}
\mycaption{The same as in previous figure, but with $r=10$ (top), $20$
(middle), and $50$ (bottom). The black hole was assumed non-rotating,
$a=0$ in this figure.}
\end{figure}

In order to compute a synthetic profile of an observed spectral line
one has to link the points of emission in the disc with corresponding
pixels in the detector plane at spatial infinity. This can be achieved by
solving the ray-tracing problem in curved space-time of the black hole.
Appropriate methods were discussed by several authors; see\break 
\cite{reynolds2003} for a recent review and for further references.
This way one finds the redshift factor, which determines the
energy shift of photons, the lensing effect (i.e.\ the change of solid angle
due to strong gravity), and the effect of aberration (which influences
the emission direction of photons from the disc; this must be taken
into account if the intrinsic emissivity is non-isotropic). We consider these
effects in our computations, assuming a rotating (Kerr)
black-hole space-time \citep{misner1973}. We also consider the time of arrival
of photons originating at different regions of the disc plane.
Variable travel time results in mutual time delay between different
photons, which can be ignored when analyzing time-averaged data
but it may be important for time-resolved data.

Assuming purely azimuthal Keplerian motion of a spot, one obtains
for its orbital velocity (with respect to a locally non-rotating
observer at corresponding radius $r$):
\begin{equation}
 v^{(\phi)} = \frac{r^2-2a\sqrt{r}+{a}^2}{\sqrt{\Delta}\left(r^{3/2}
 +{a}\right)}.
\end{equation}
In order to derive time and frequency as measured by a distant observer,
one needs to take into account the Lorentz factor associated with
this orbital motion,
\begin{equation}
 \Gamma = \frac{\left(r^{3/2}+{a}\right)\sqrt{\Delta}}{r^{1/4}\;
  \sqrt{r^{3/2}-3r^{1/2}+2{a}}\;\sqrt{r^3+{a}^2r+2{a}^2}}\,.
\end{equation}
The corresponding angular velocity of orbital motion is
$\Omega=(r^{3/2}+{a})^{-1}$, which also
determines the orbital period in eq.~(\ref{torb}).
The redshift factor $g$ and the emission angle $\vartheta$ (with respect to
the normal direction to the disc) are then given by
\begin{equation}
 g  =  \frac{{\cal{C}}}{{\cal{B}}-r^{-3/2}\xi},
 \quad
 \vartheta  =  \arccos\frac{g\sqrt{\eta}}{r},
\end{equation}
where ${\cal{B}}=1+{a}r^{-3/2}$,
${\cal{C}}=1-3r^{-1}+2{a}r^{-3/2}$; $\xi$ and $\eta$ are constants
of motion connected with the photon ray in an axially symmetric
and stationary space-time.

For practical purposes formula (\ref{torb}) with $a=0$ is also accurate
enough in the case of a spinning black hole,
provided that $r$ is not
very small. For instance, even for $r=6$ (the last stable orbit in
Schwarzschild metric), $T_{\rm orb}(r_{\rm ms})$ calculated for a
static and for a maximally rotating ($a=1$) black hole differ by
about $6.8$\%. The relative difference decreases, roughly linearly, down to
$1.1$\% at $r=20$.
This implies that eq.~(\ref{torb}) can be used
in most cases to estimate the black-hole mass even if the angular momentum
is not known (deviations are relevant only for $r<6$, when the radius itself can be
used to constrain the allowed range of $a$).

Various
pseudo-Newtonian formulae have been devised for accreting black holes
to model their observational properties, which are connected with
the orbital motion of surrounding matter \citep[e.g.][]{abramowicz1996,
artemova1996,semerak1999}.
Although this approach
is often used and found to be practical, we do not employ it here because
error estimates are not possible within the pseudo-Newtonian scheme.

Due to Doppler and gravitational energy shift the line shape changes along
the orbit. Centroid energy is redshifted with respect to the rest
energy of the line emission for most of the orbit.
Furthermore, light aberration and bending cause the flux to
be strongly phase-dependent. These effects are shown in
Figs.~\ref{orbits_1}--\ref{orbits_2}. In these plots,
the arrival time of photons is defined in orbital periods,
i.e.\ scaled with $T_{\rm{}orb}(r;a)$. The
orbital phase of the spot is of course linked with the
azimuthal angle in the disc, but the relation is made complex
by time delays which cannot be neglected, given the large velocities of
the orbiting matter and frame-dragging effects near the black hole.
Here, zero time corresponds to the moment
when the centre of the spot was at the nearest point on its orbit with
respect to the observer (a lower conjunction).
The plots in Fig.~\ref{orbits_1} refer to the case of a spot
circulating at the innermost stable
orbit $r_{\rm{}ms}(a)$ for $a=0$, $0.9$ and $1$.
The effect of black-hole rotation becomes prominent for almost
extreme values of $a$; one can check, for example, that the
difference between cases $a=0$ and $a=0.5$ is very small.

Worth remarking is a large difference in the orbital phase of
maximum emission between the extreme case, $a\rightarrow1$, in
contrast to the non-rotating case, $a\rightarrow0$. The reason is
that for large $a$ the time delay and the effect of frame-dragging
on photons emitted behind the black hole are very substantial. It
is also interesting to note that, for very high inclination angles, most
of the flux comes from the far side of the disc, due to very strong
light bending, as pointed out by \cite{matt1992,matt1993a} and
examined further by many authors who performed detailed ray-tracing,
necessary to determine the expected variations of the line flux and
shape. A relatively simple fitting formula has also been derived
\citep{karas1996} and can be useful for practical computations.

Three more orbits (centred at $r=10$, $20$ and $50$) are shown for
$a=0$ (Fig.~\ref{orbits_2}). As said above, at
these radii differences between spinning and static black holes are
small. Indeed, it can be verified that the dependence on $a$ is only
marginal if $r\gtrsim20$, and so it can be largely neglected for
present-day measurements.

\begin{figure}[tb]
\dummycaption\label{profiles}
\includegraphics[width=0.325\textwidth]{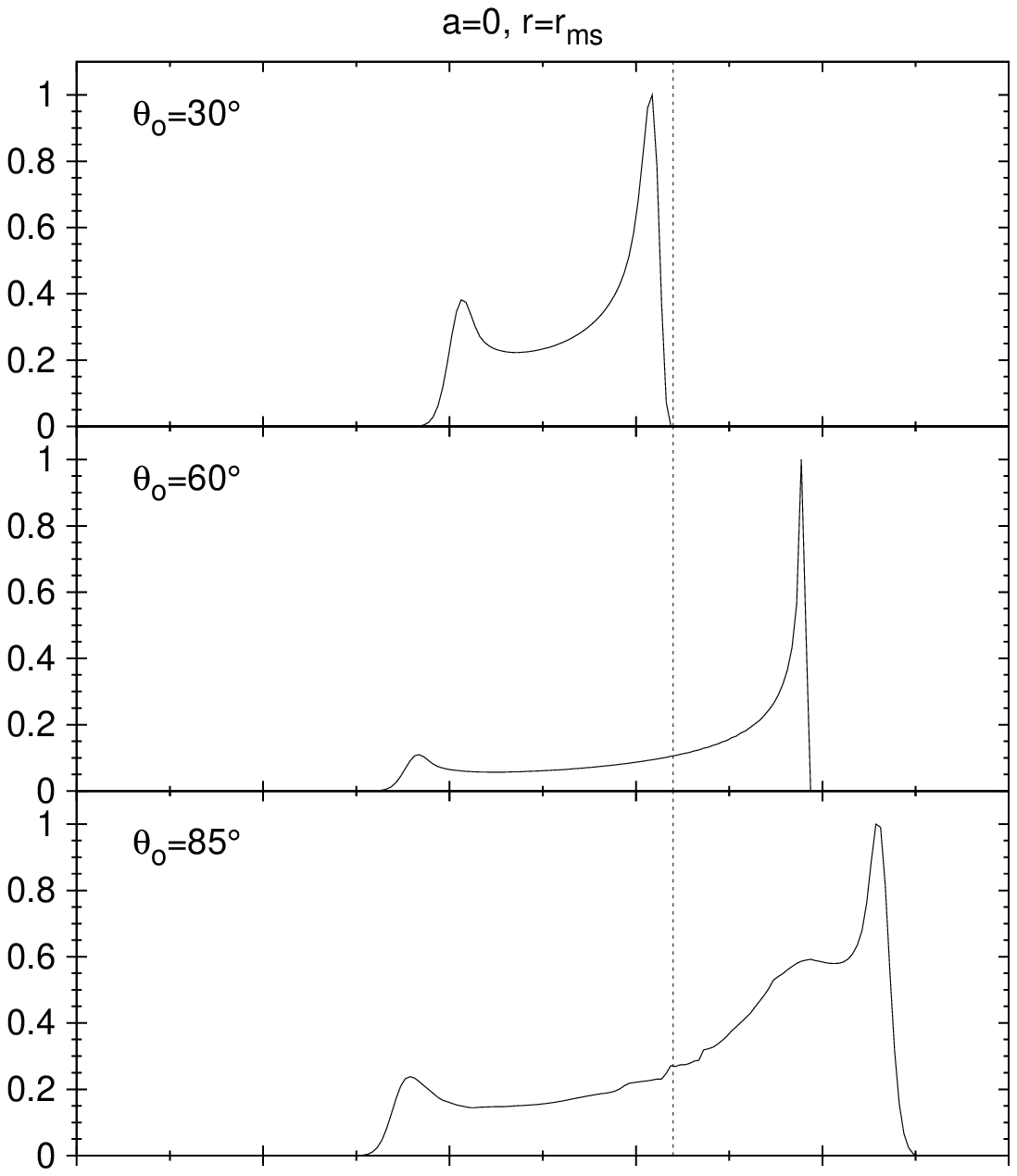}
\hfill
\includegraphics[width=0.325\textwidth]{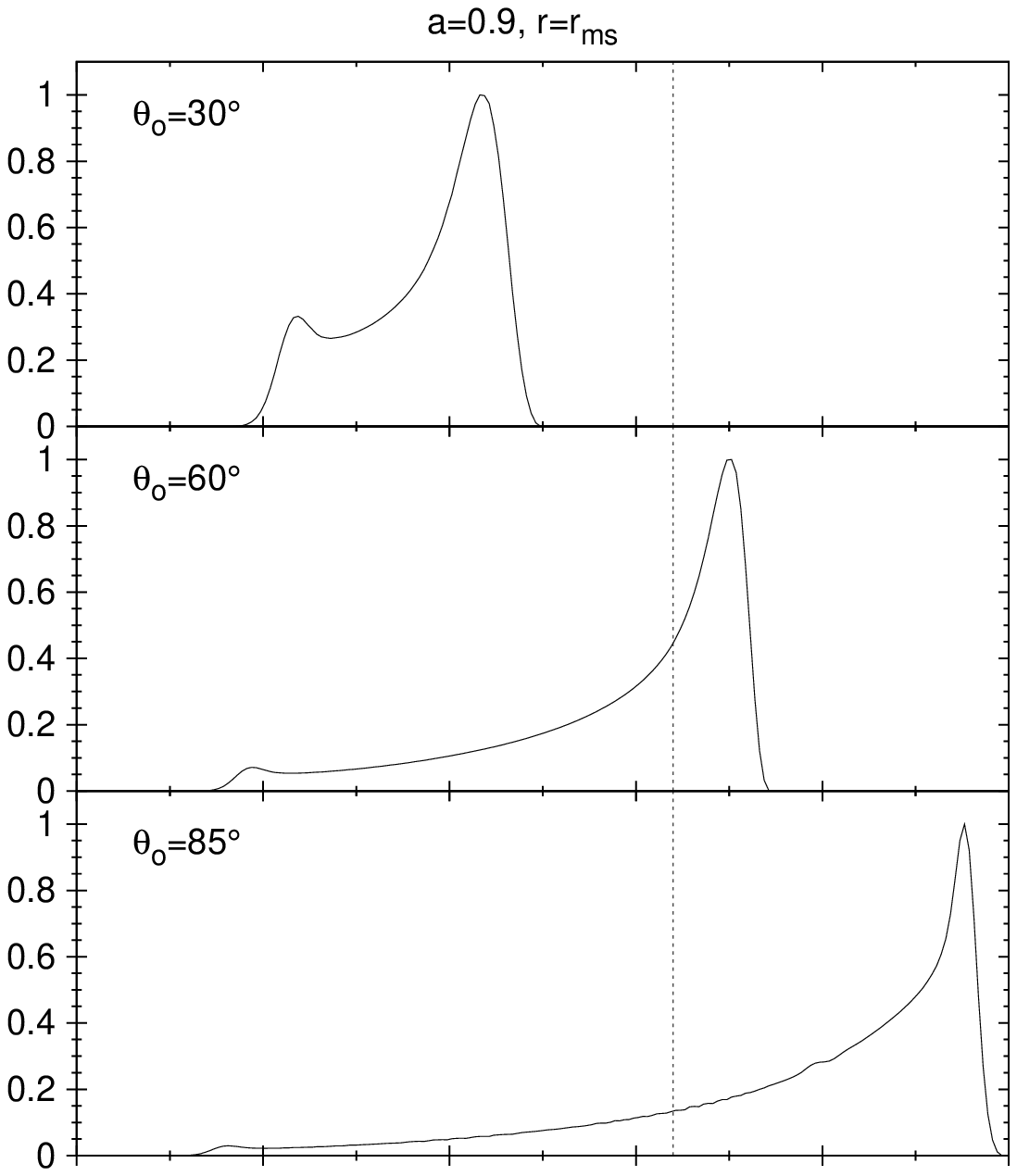}
\hfill
\includegraphics[width=0.325\textwidth]{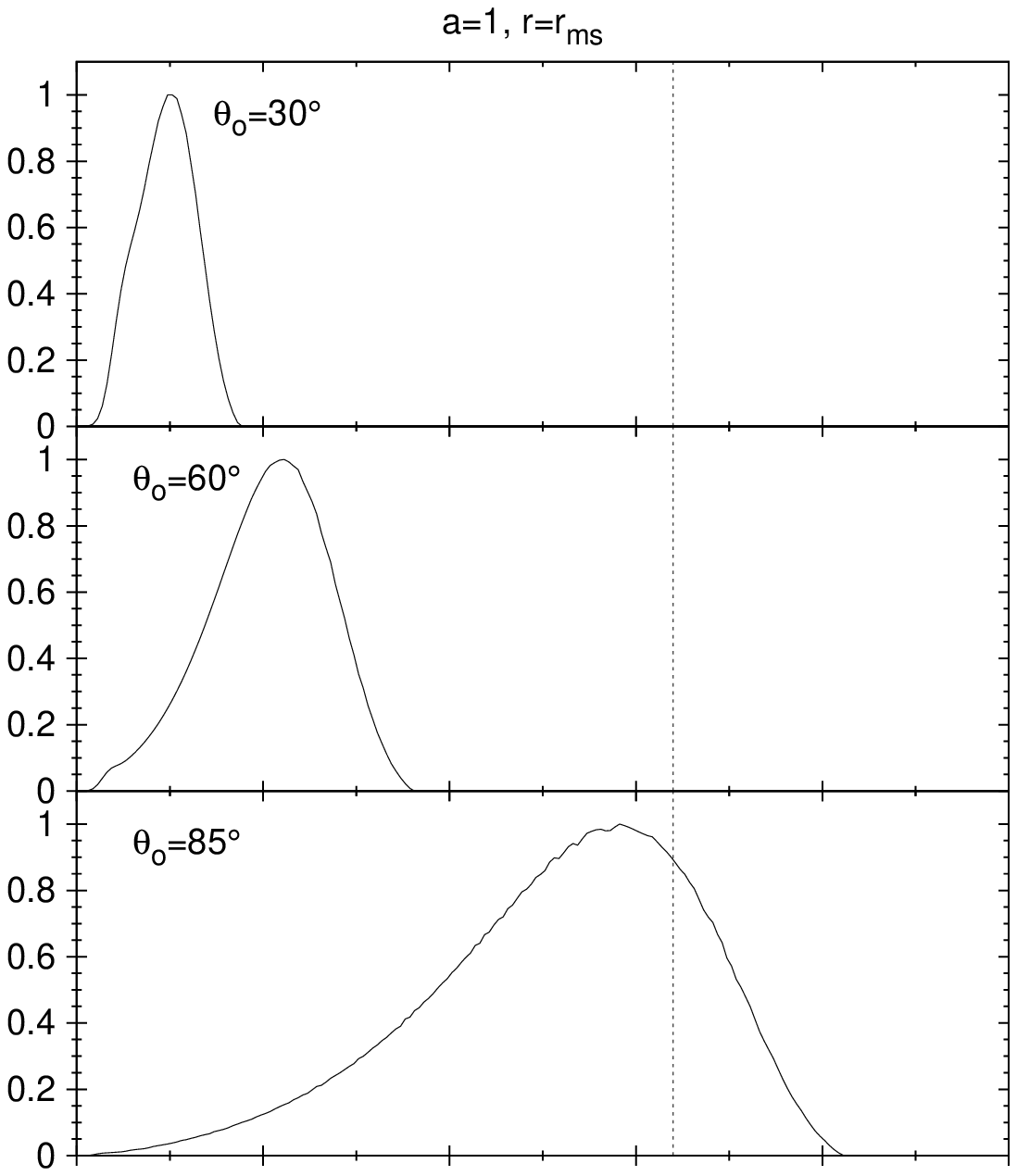}
\vspace*{3mm}\\
\includegraphics[width=0.325\textwidth]{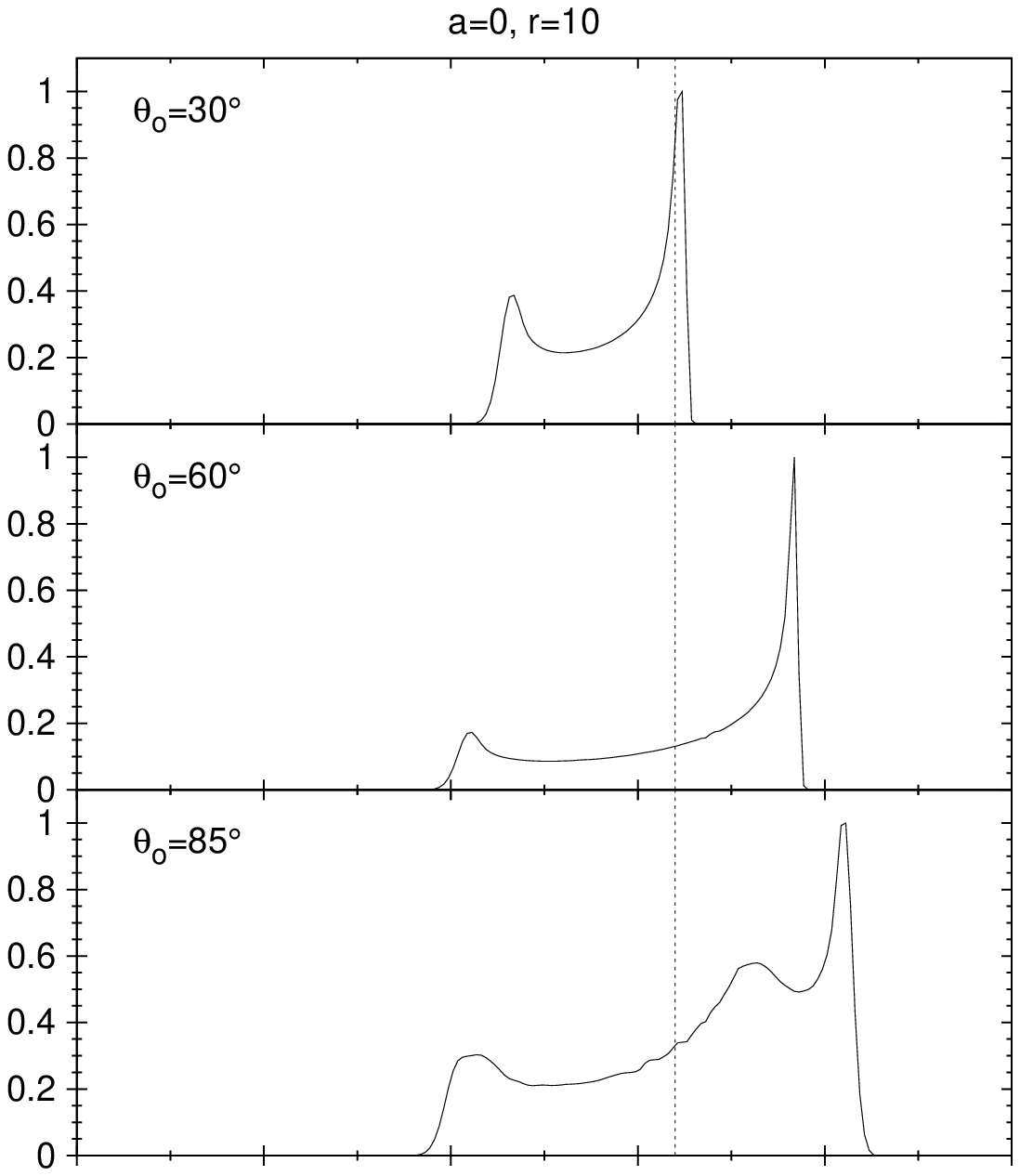}
\hfill
\includegraphics[width=0.325\textwidth]{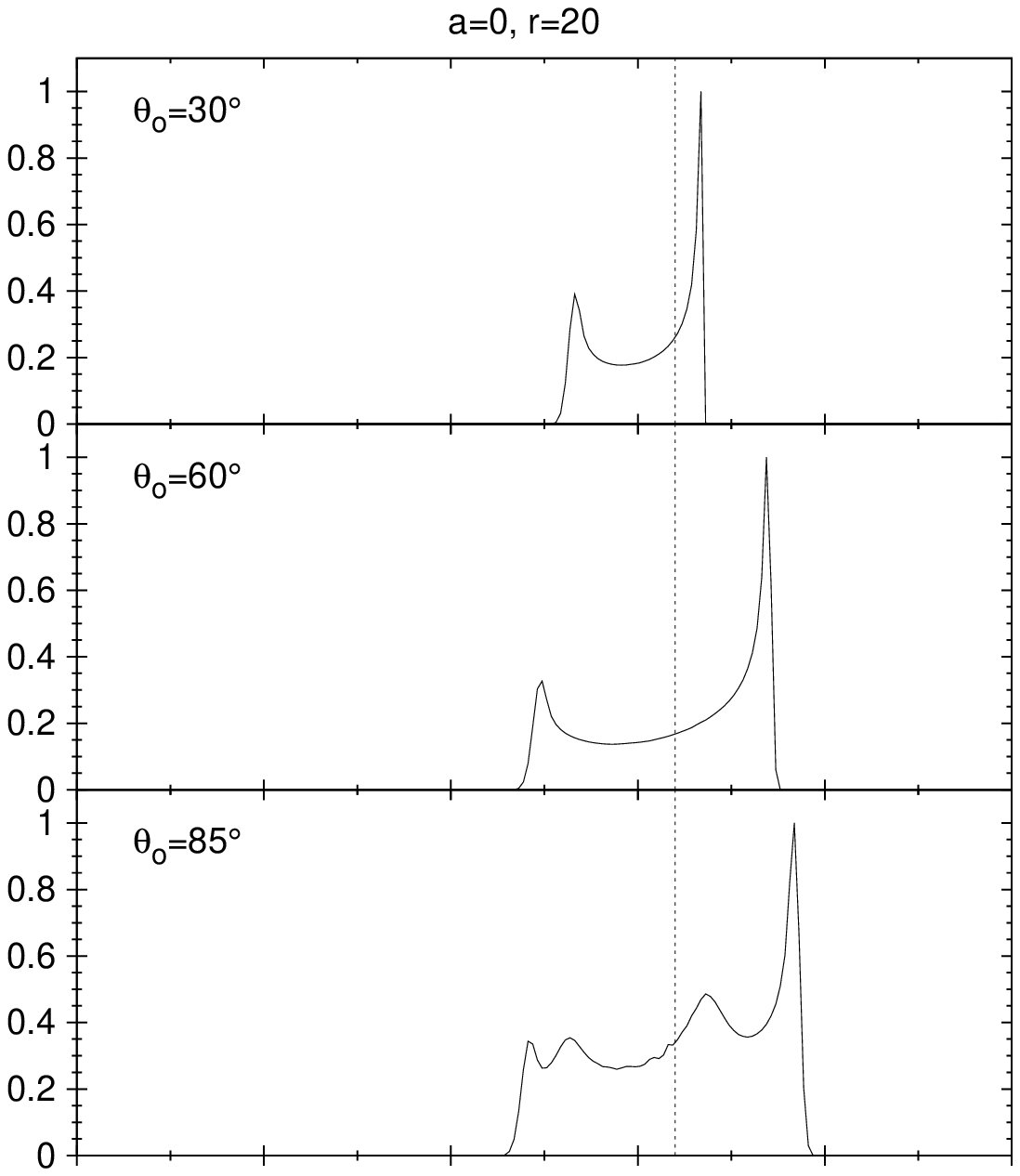}
\hfill
\includegraphics[width=0.325\textwidth]{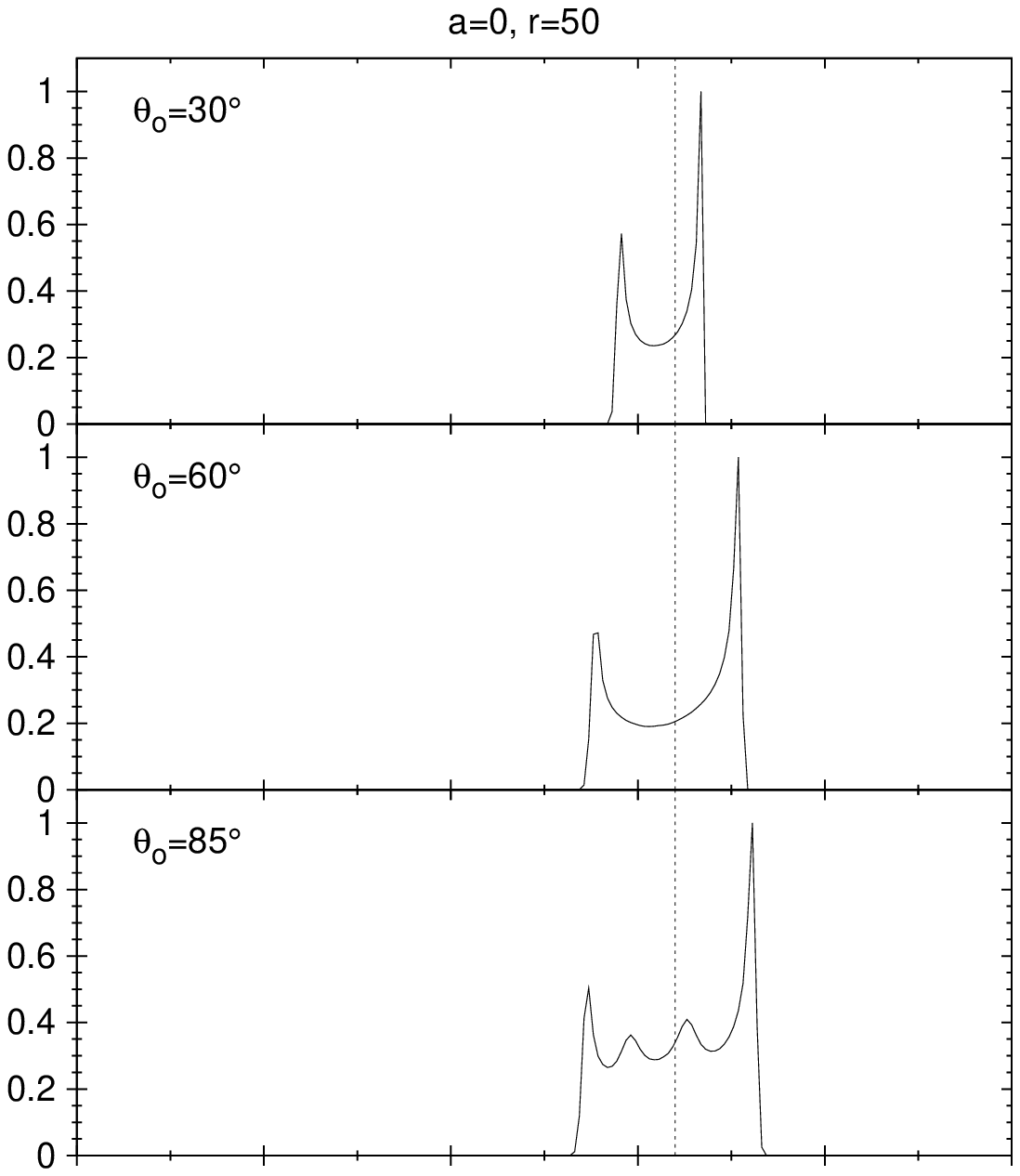}
\vspace*{0mm}
\mycaption{Time-averaged synthetic spectra in terms of photon
flux (in arbitrary units) versus energy (in keV). These
profiles represent the mean, background-subtracted spectra
of the Fe K$\alpha$ iron-line originating from spots at
different radii. Top panels correspond to $r=r_{\rm{}ms}$ and
$a=0$ (left), $a=0.9$ (middle), and $a=1$ (right). In bottom panels
we fix $a=0$ and choose $r=10$, $20$, and $50$, respectively
(other values of $a$ give very similar profiles).
Three consecutively increasing values of observer
inclination $\theta_{\rm{}o}$ are shown, as indicated inside
the frames.}
\end{figure}

In Figs.~\ref{prof_phi_1}--\ref{prof_phi_2} we show
the actual form of the line profiles for the same sets of parameters as
those explored in Figs.~\ref{orbits_1}--\ref{orbits_2}. The entire
revolution was split into
twelve different phase intervals. The intrinsic flux $I$ is
assumed to decrease exponentially with the distance ${\rm{}d}r$ from the
centre of the spot
(i.e.\ ${\log}I\propto-[\kappa{\rm{}d}r/r]^2$, where $r$ is the
location of the spot centre and $\kappa\sim10$ is a constant).
The illumination is supposed to
cease at ${\rm{}d}r=0.2r$, which also defines the illuminated area in the
disc. Let us remark that we concentrate on a spectral line
which is intrinsically narrow and unresolved in the rest frame of
the emitting medium. Such a line can be produced by a spot which
originates due to sharply
localized illumination by flares, as proposed and discussed
by various authors \citep[e.g.][]{haardt1994,poutanen1999,merloni2001}.
Very recently, \cite{czerny2004} have examined
the induced {\sf{}rms} variability in the flare/spot
model with relativistic effects.
In many cases, and especially for small radii and intermediate to large
inclination angles, the line emission comes from a relatively minor fraction
of the orbit. This implies in practice that for observations with
{\it{}moderate signal-to-noise ratios, only a narrow blue horn can be visible, and
only for a small part of the orbit.}
These large and rapid changes of the line shape get averaged
when integrating over the entire orbit, and so an important
piece of information is missing in the mean spectra.
The line profiles integrated over the whole revolution
are shown in Fig.~\ref{profiles}. Effectively, the mean
profile of a spot is identical
to the profile of an annulus whose radius is equal to the
distance of the spot centre and the width is equal to the spot
size.

We discussed the possibility that the narrow features in the
$5-6$~keV range, recently discovered in a few AGNs and usually interpreted as
redshifted iron lines, could be due to illumination by localized
orbiting spots just above the accretion disc. If this is indeed the
case, these features may provide a powerful and direct way to measure
the black-hole mass in active galactic nuclei. To achieve this aim, it is
necessary to follow the line emission along the orbit.
The orbital radius (in units of $r_{\rm{}g}$) and the disc inclination
can be inferred from the variations of the line flux and centroid energy.
Furthermore, $\mbh$ can be estimated by comparing the measured orbital
period with the value expected for the derived radius.
We must point out, however, that present-day X-ray
instruments do not have enough collecting area to perform this task
accurately. This capability should be achieved by the planned
high-performance X-ray missions such as
{\it{}Constellation-X}\/ and {\it{}Xeus}.
\clearpage
  \section{Polarization signatures of strong gravity in AGN accretion discs}

Accretion discs in central regions of active galactic nuclei are
subject to strong external illumination originating from some kind
of corona and giving rise to specific spectral features in the X-ray
band. In particular, the K-shell lines of iron are found to be prominent
around $6-7$~keV. It has been shown that the shape of the intrinsic
spectra must be further modified by the strong gravitational field of
the central mass, and so X-ray spectroscopy could allow us to explore
the innermost regions of accretion flows near supermassive black holes
\citep[for recent reviews see][]{fabian2000,reynolds2003}.
Similar mechanisms operate also in some Galactic black-hole candidates.

A rather surprising result from recent {\it XMM-Newton}\/ observations is
that relativistic iron lines are not as common as previously believed,
see \cite{bianchi2004} and references therein and also \cite{yaqoob2003}.
This does not necessarily mean that the iron line is not produced in
the innermost regions of accretion discs. The situation is likely to be more
complex than in simple, steady scenarios, and indeed some evidence for
line emission arising from orbiting spots is present in the
time-resolved spectra of a few AGNs \citep{dovciak2004a}. Even
when clearly observed, relativistic lines behave differently than
expected. The best example is the puzzling lack of correlation  between
line and continuum emission in MCG--6-30-15 \citep{fabian2002},
unexpected because the very broad line profile clearly indicates that
the line originates in the innermost regions of the accretion disc,
hence very close to the illuminating source.  \cite{miniutti2003}
have proposed a solution to this problem in terms of an illuminating
source moving along the black-hole rotation axis or very close to it.

In this section we show that polarimetric studies could provide additional
information about accretion discs in a strong gravity regime, which may be
essential to discriminate between different possible geometries of the
source. The idea of using polarimetry to gain additional information
about accreting compact objects is not a new one. In this context it was
proposed by \cite{rees1975} that polarized X-rays are of high 
relevance.\break
\cite{pozdnyakov1979} studied spectral profiles of iron
X-ray lines that result from multiple Compton scattering.  Later on,
various influences affecting polarization (due to magnetic fields,
absorption as well as strong gravity) were examined for black-hole
accretion discs \citep{agol1997}.  Temporal variations of polarization were
also discussed, in particular the case of orbiting spots near a black
hole (\citealt{connors1980};\break \citealt{bao1996}). With
the promise of new polarimetric detectors \citep{costa2001},
quantitative examination of specific models becomes timely.

\begin{figure}[tbh!]
\vspace*{1.em}
\dummycaption\label{pol}
\includegraphics[width=0.328\textwidth]{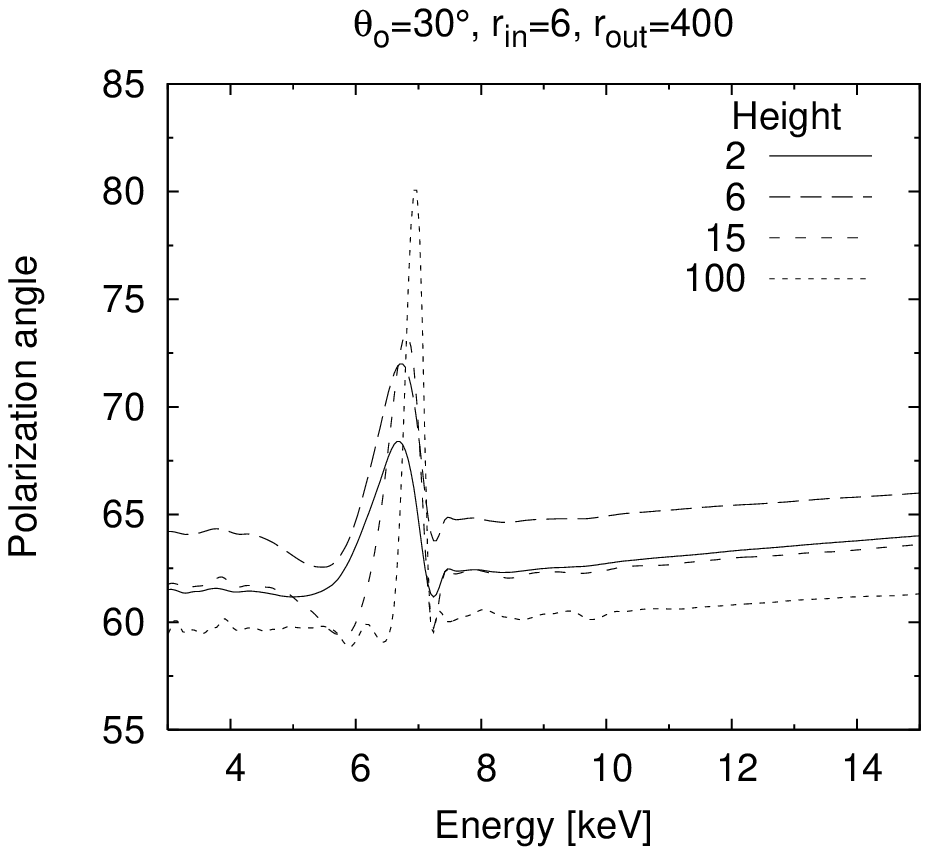}
\hfill
\includegraphics[width=0.328\textwidth]{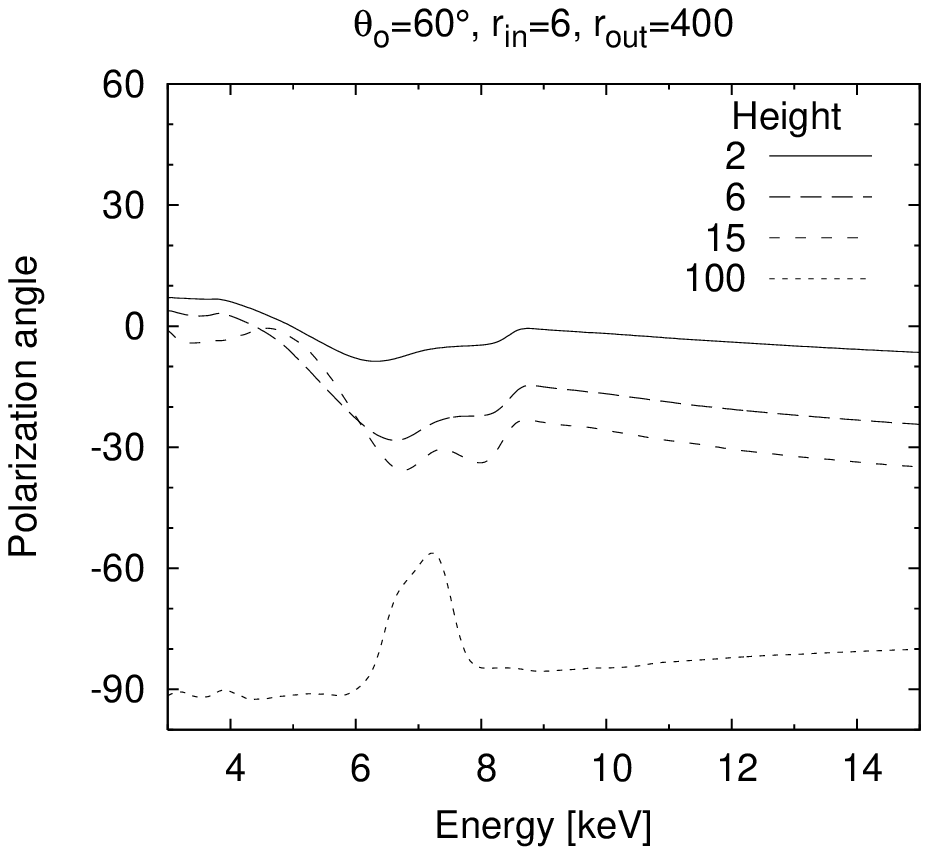}
\hfill
\includegraphics[width=0.328\textwidth]{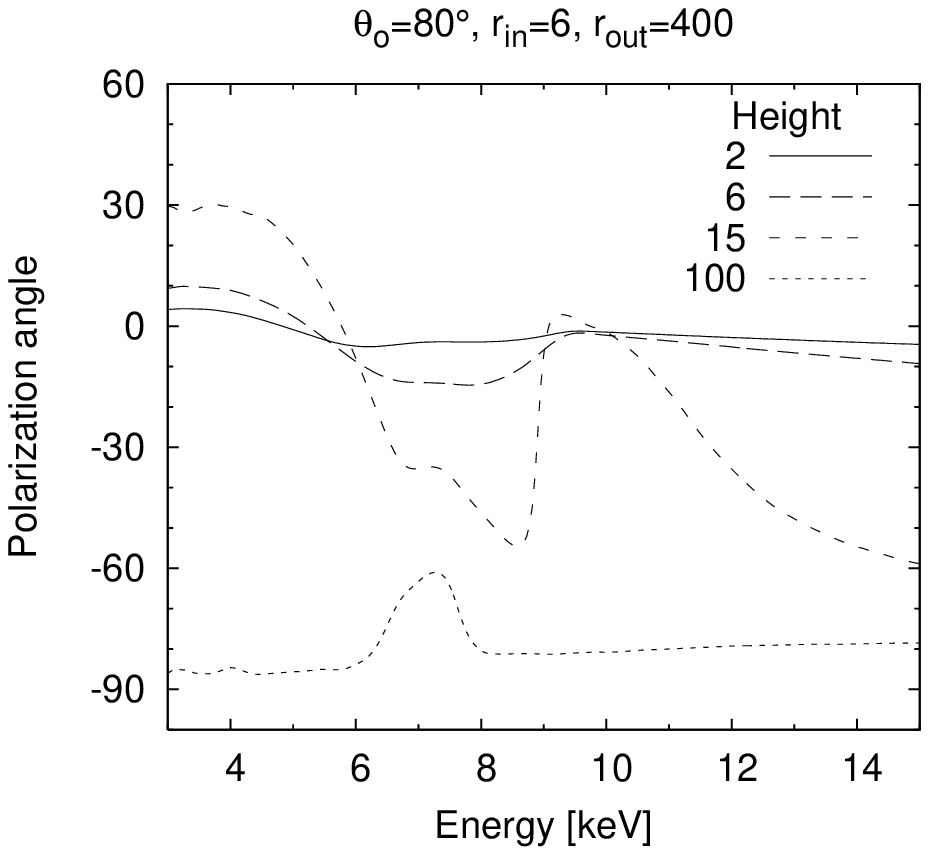}
\includegraphics[width=0.328\textwidth]{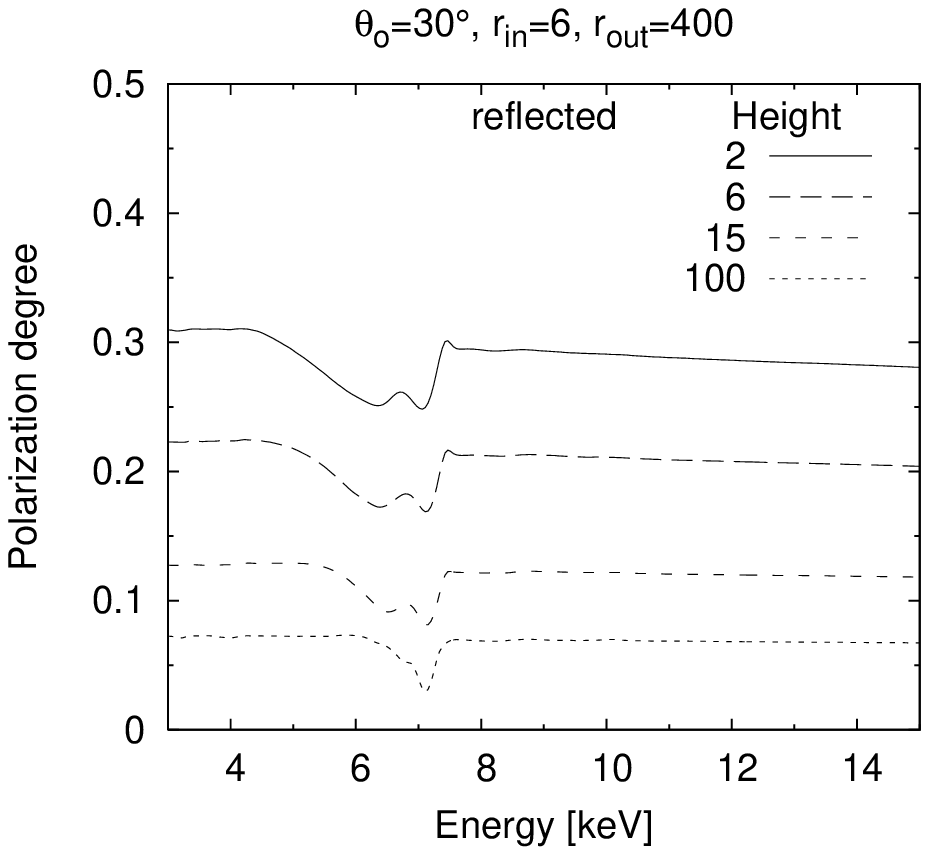}
\hfill
\includegraphics[width=0.328\textwidth]{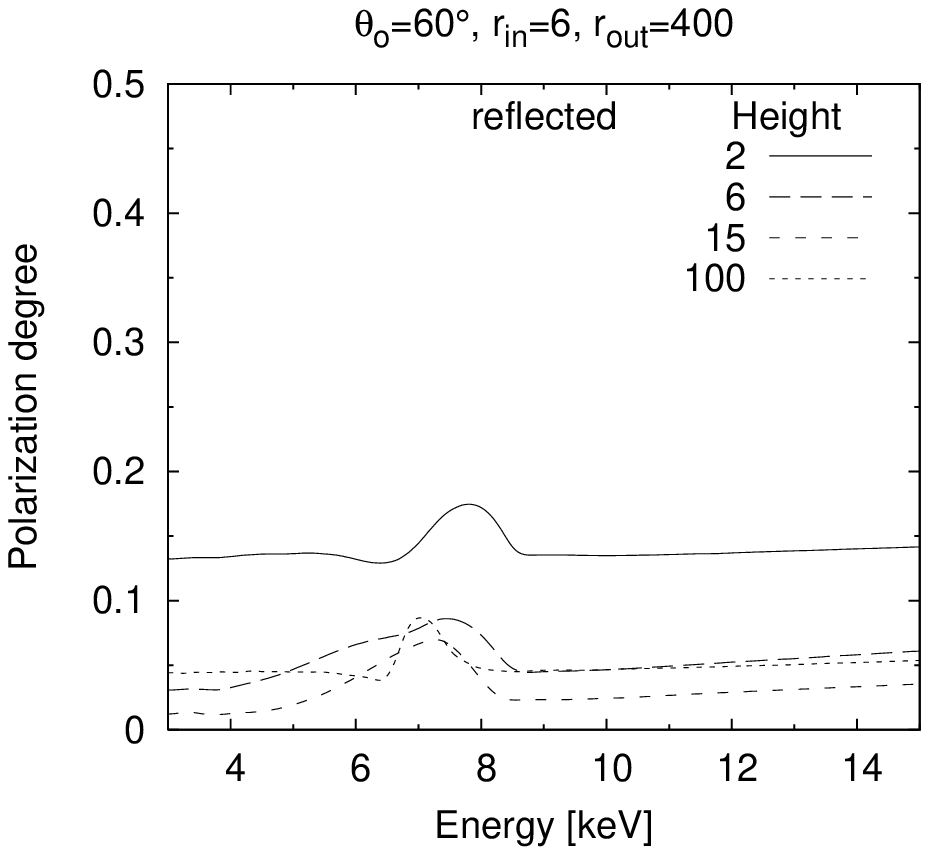}
\hfill
\includegraphics[width=0.328\textwidth]{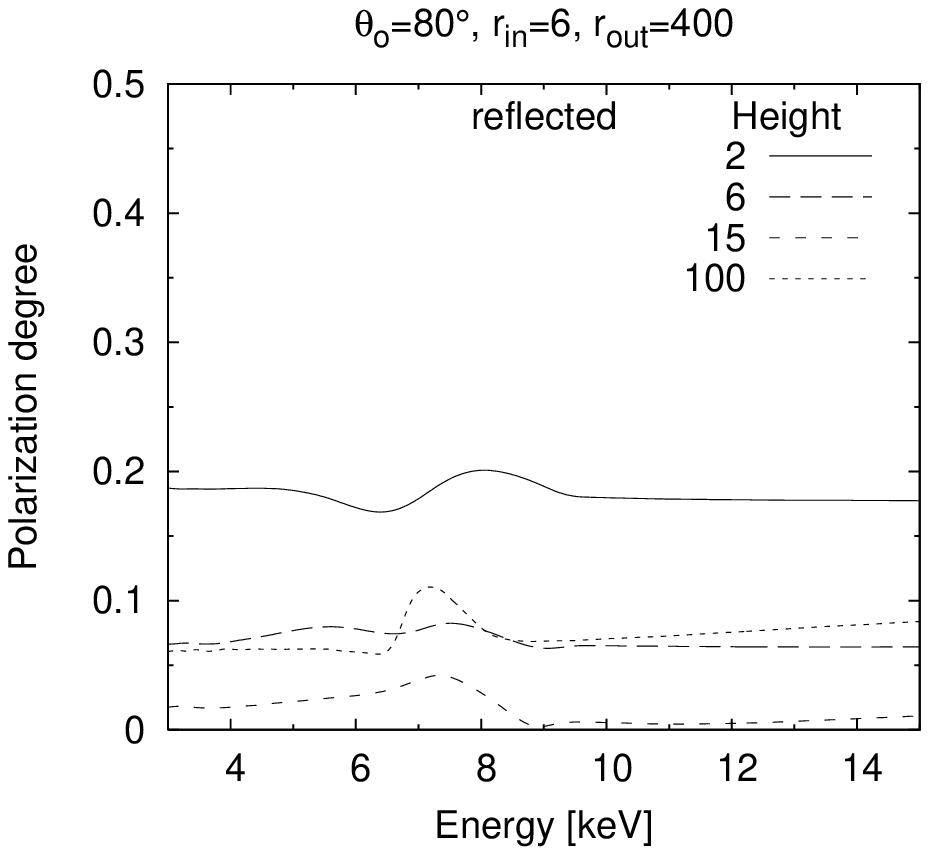}
\includegraphics[width=0.328\textwidth]{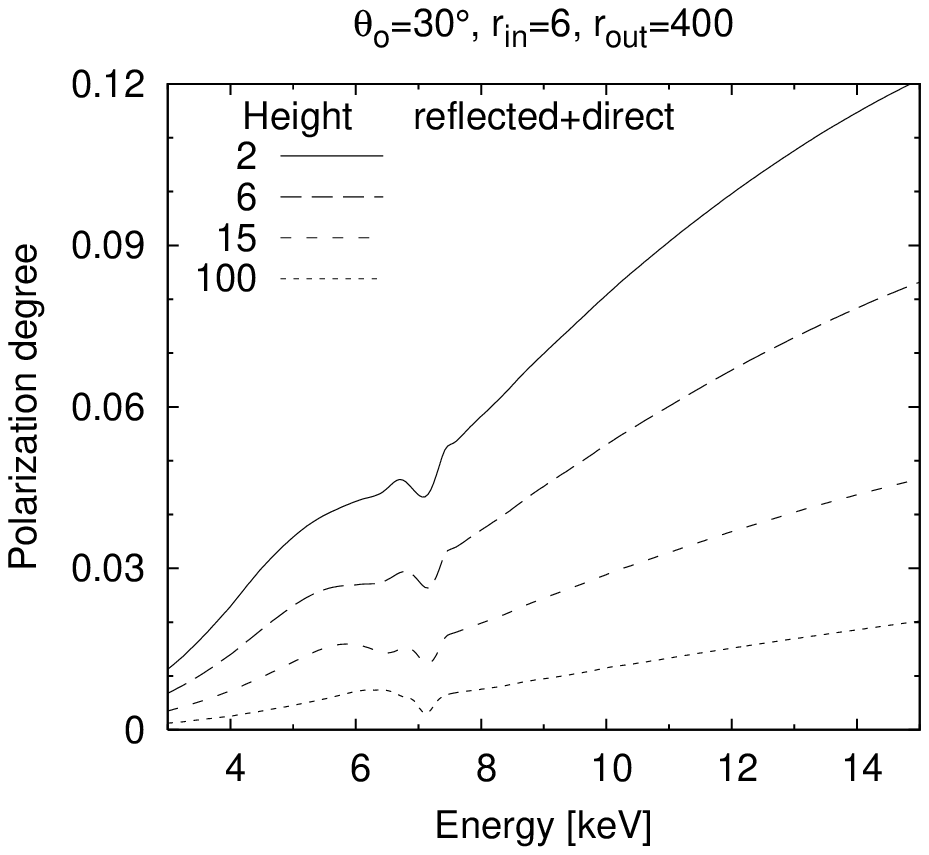}
\hfill
\includegraphics[width=0.328\textwidth]{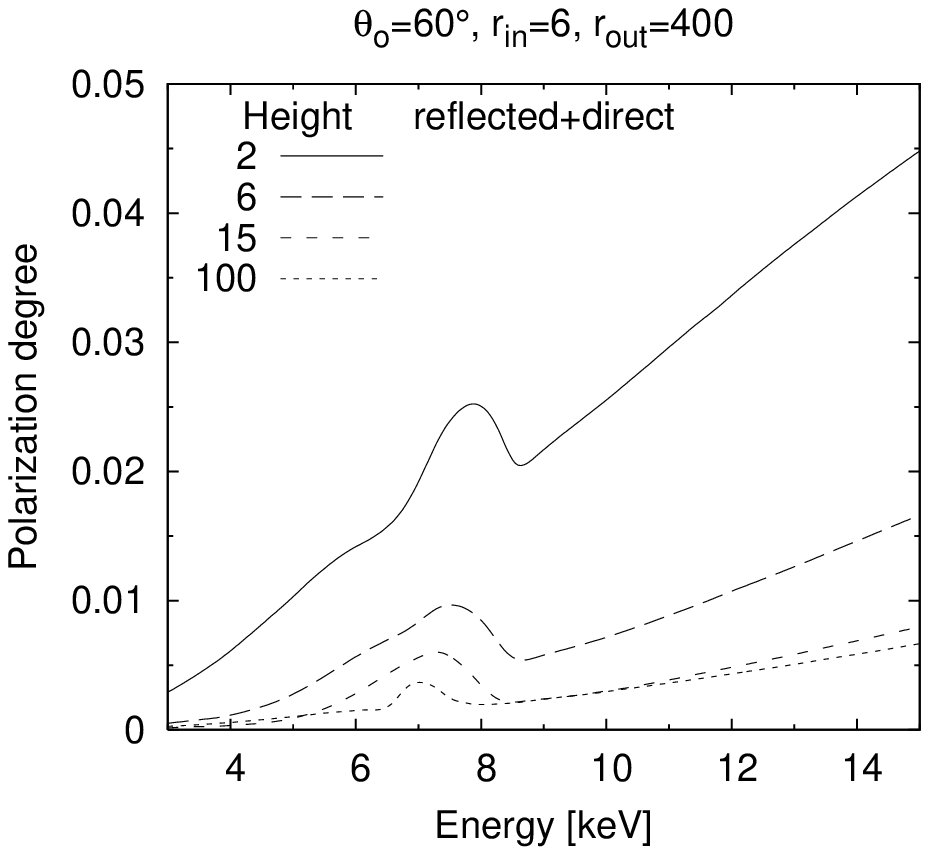}
\hfill
\includegraphics[width=0.328\textwidth]{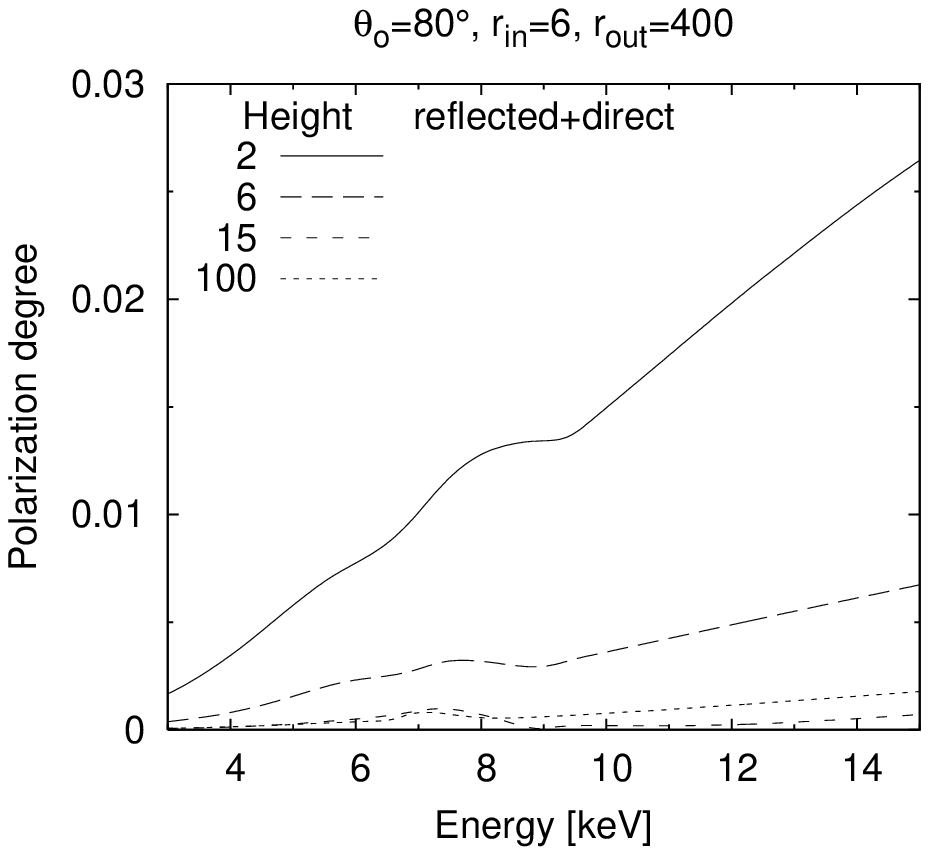}
\mycaption{Energy dependence of polarization angle (top panels) and polarization
degree (middle panels) due to reflected radiation for different observer's
inclination angles ($\theta_{\rm o}=30^\circ,\,60^\circ$ and $80^\circ$) and for
different heights of the primary source ($h=2,\,6,\,15$ and $100$).
Polarization degree for reflected plus direct radiation is also plotted (bottom
panels). The emission comes from a disc within $r_{\rm{}in}=6$ and
$r_{\rm out}=400$. Isotropic primary radiation with photon index $\Gamma=2$ and
angular momentum of the central black hole $a=0.9987$ were assumed.}
\end{figure}

Since the reflecting medium has a disc-like geometry, a substantial
amount of linear polarization is expected in the resulting spectrum
because of Compton scattering. Polarization properties of the disc
emission are modified by the photon propagation in a gravitational field,
providing additional information on its structure. Here we
calculate the observed polarization of the reflected radiation assuming
the lamp-post model for the stationary power-law illuminating source
\citep{martocchia1996, petrucci1997}.  We assume a rotating (Kerr) black hole
as the only source of the  gravitational field, having a common symmetry
axis with an accretion disc. The disc is
also assumed to be stationary and we restrict ourselves to the
time-averaged analysis. In other words, we examine processes that vary
at a much slower pace than the light-crossing time at the corresponding
radius. Intrinsic polarization of the emerging
light can be computed locally, assuming a plane-parallel scattering
layer which is illuminated by light radiated from the primary source.
This problem was studied extensively in various
approximations \citep[e.g.][]{chandrasekhar1960,sunyaev1985}.
Here we employ the Monte Carlo computations \citep{matt1991,
matt1993b} and thus we find the intrinsic emissivity of an
illuminated disc.
Then we integrate contributions to the total signal
across the disc emitting region using a general relativistic
ray-tracing technique described in previous chapters and we compute the
polarization angle and degree as measured by a distant observer (see
Section~\ref{stokes_param} and equations therein). We show the
polarization properties of scattered light as a function of model
parameters, namely, the height $z=h$ of the primary source on the symmetry
axis, the dimensionless angular momentum $a$ of the black hole, and the
viewing angle $\theta_{\rm{}o}$ of the observer.

\begin{figure}[tb]
\vspace*{1em}
\dummycaption\label{pol1}
\includegraphics[width=0.328\textwidth]{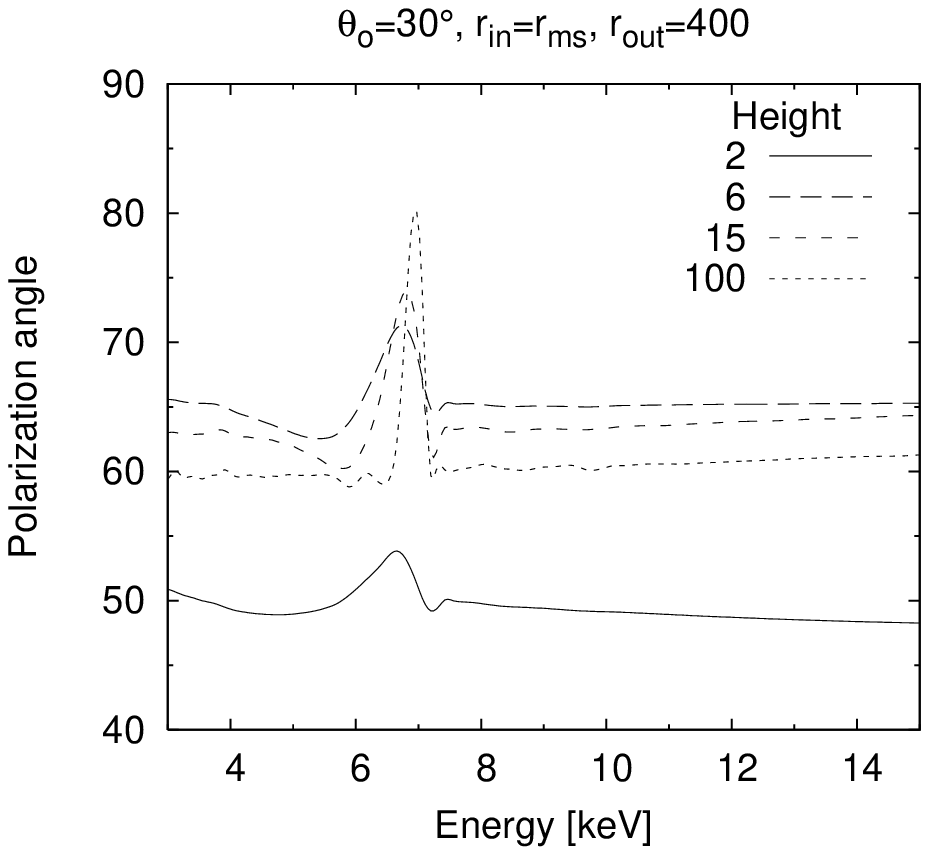}
\hfill
\includegraphics[width=0.328\textwidth]{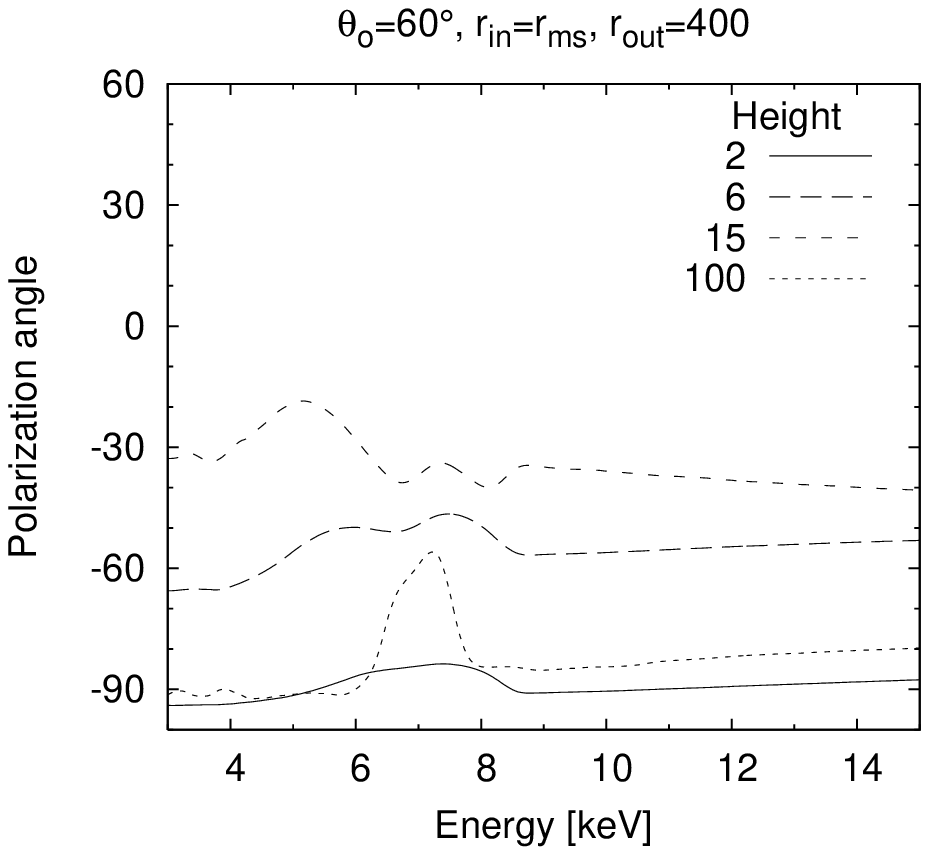}
\hfill
\includegraphics[width=0.328\textwidth]{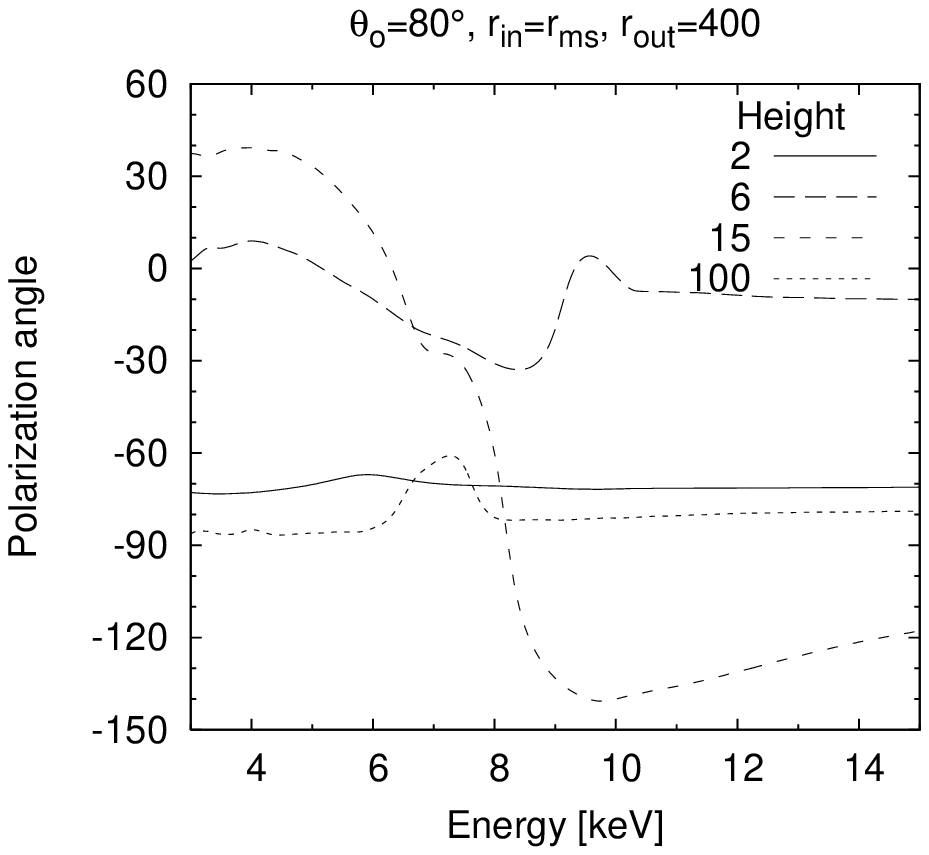}
\includegraphics[width=0.328\textwidth]{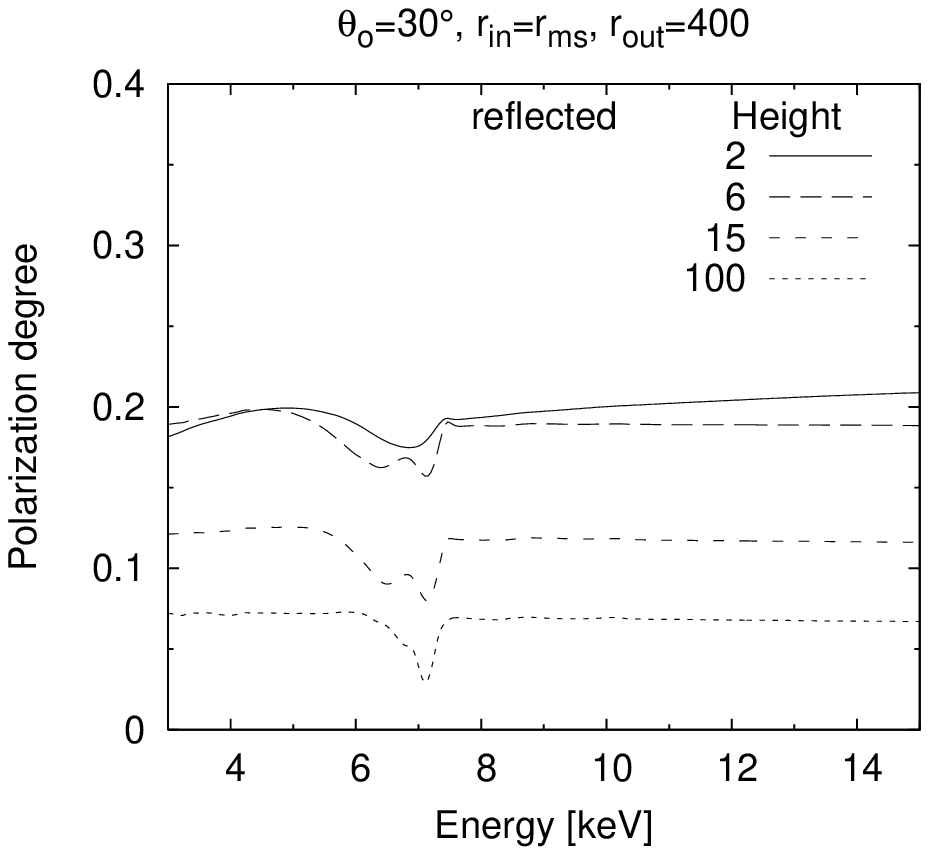}
\hfill
\includegraphics[width=0.328\textwidth]{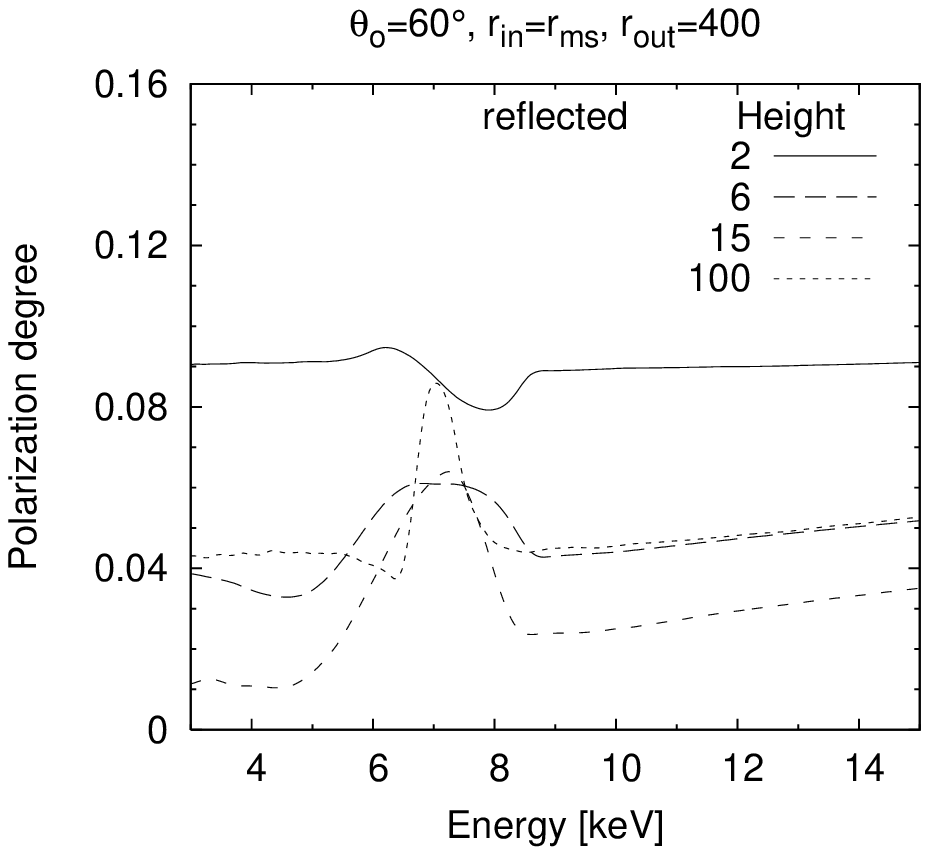}
\hfill
\includegraphics[width=0.328\textwidth]{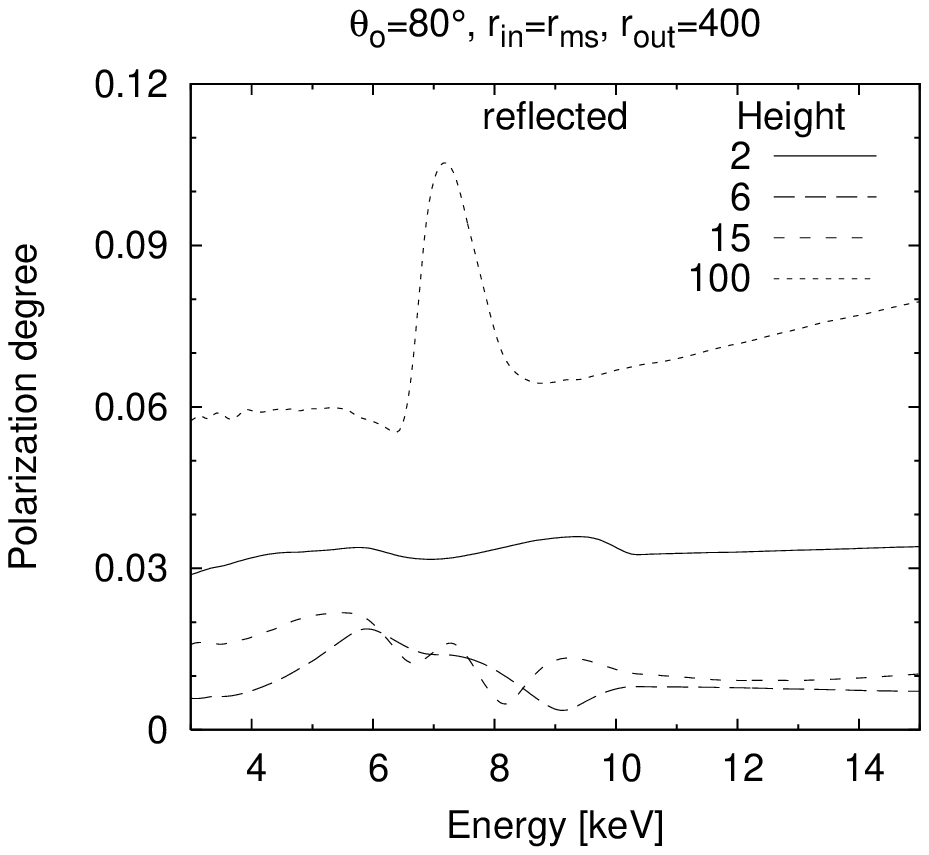}
\includegraphics[width=0.328\textwidth]{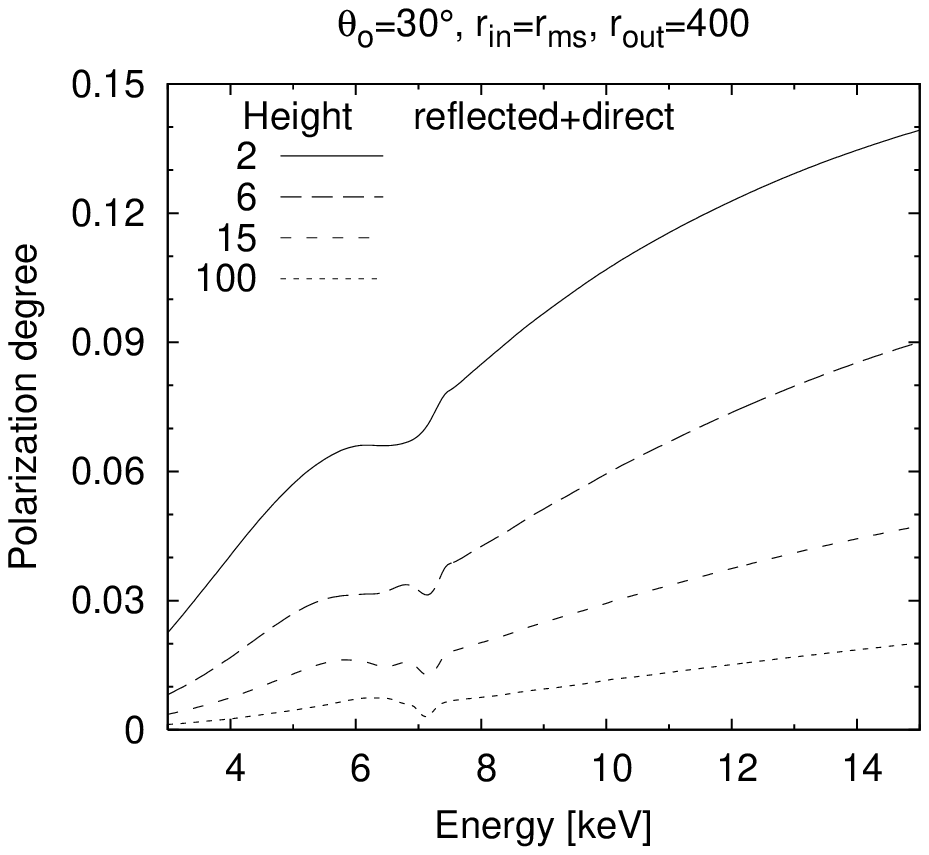}
\hfill
\includegraphics[width=0.328\textwidth]{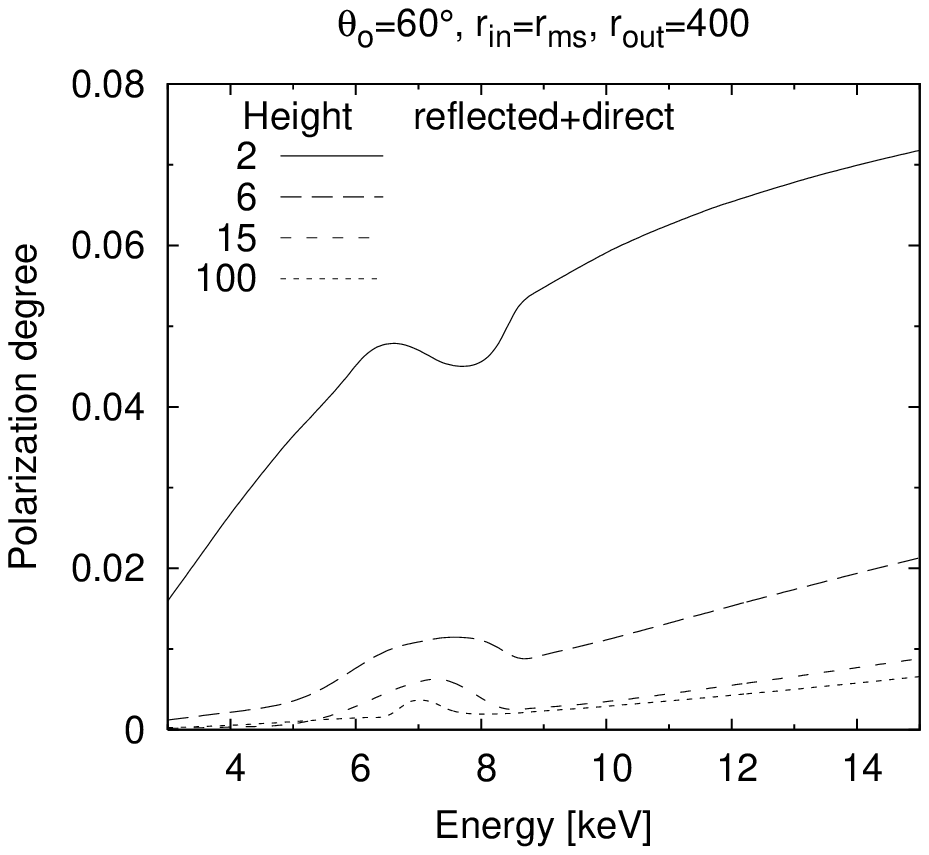}
\hfill
\includegraphics[width=0.328\textwidth]{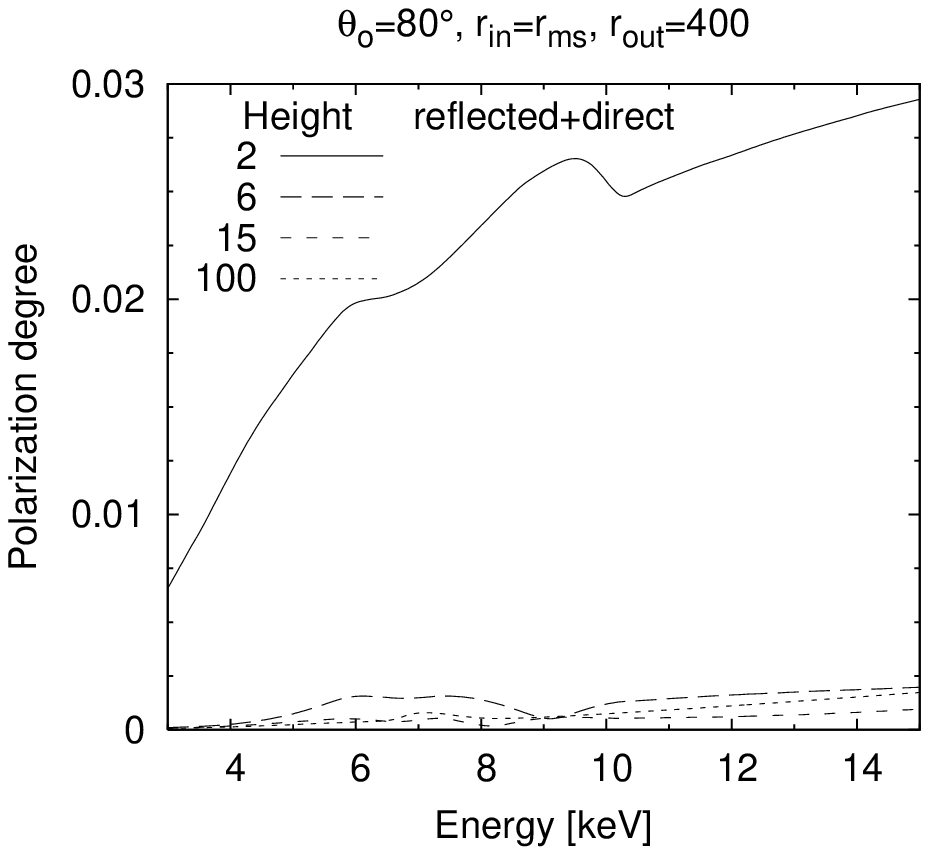}
\mycaption{Same as in the previous figure but for disc starting at
$r_{\rm in}=1.20\,$.}
\vspace*{1em}
\end{figure}

In the first set of figures (Figs.~\ref{pol}--\ref{pol1}) we show the energy
dependence of polarization angle and degree due to reflected and reflected
plus direct radiation for different inclination angles and different heights
of the primary source. One can see that the polarization of reflected radiation
can be as high as thirty percent or even more for small inclinations and
small heights. Polarization of the reflected radiation does not depend on energy
very much except for the region close to the iron edge at approximately
$7.2\,$keV, where it either decreases for small inclinations or increases
for large ones.

\begin{figure}
\dummycaption\label{poldeg}
\includegraphics[width=0.49\textwidth]{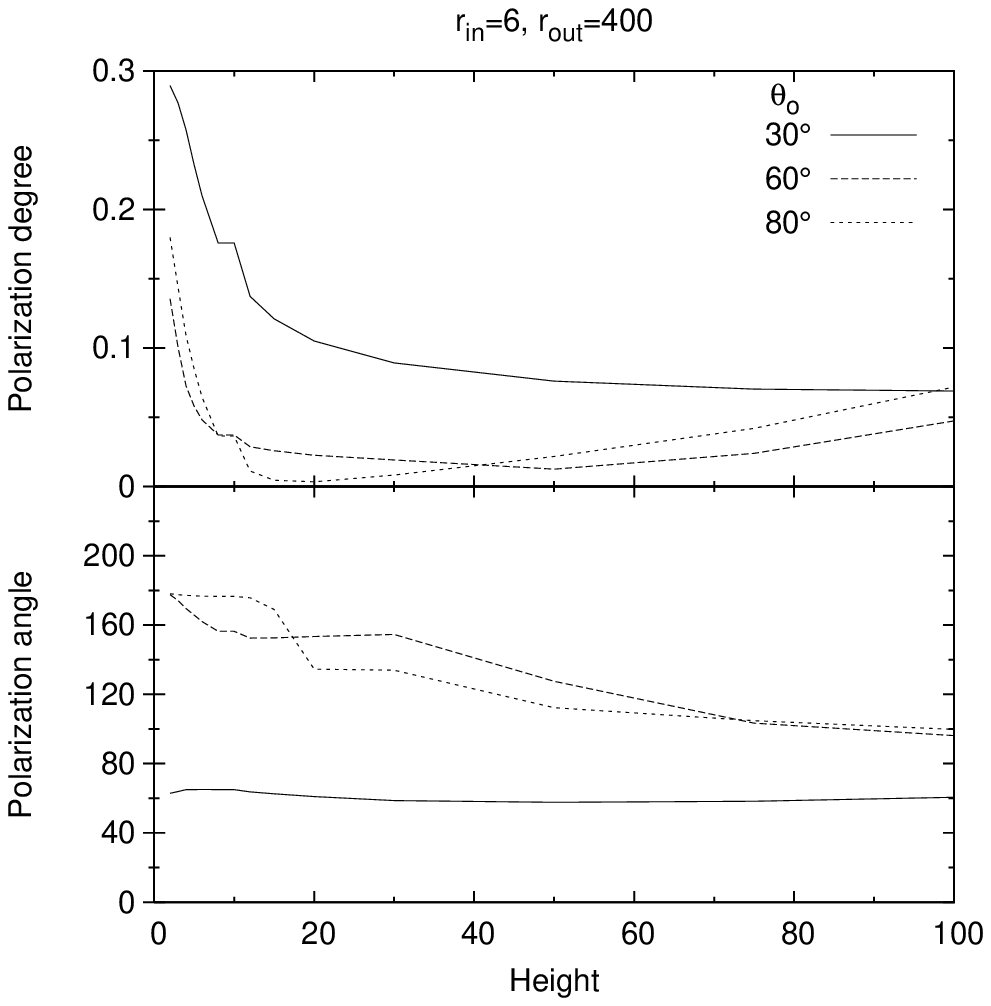}
\hfill
\includegraphics[width=0.49\textwidth]{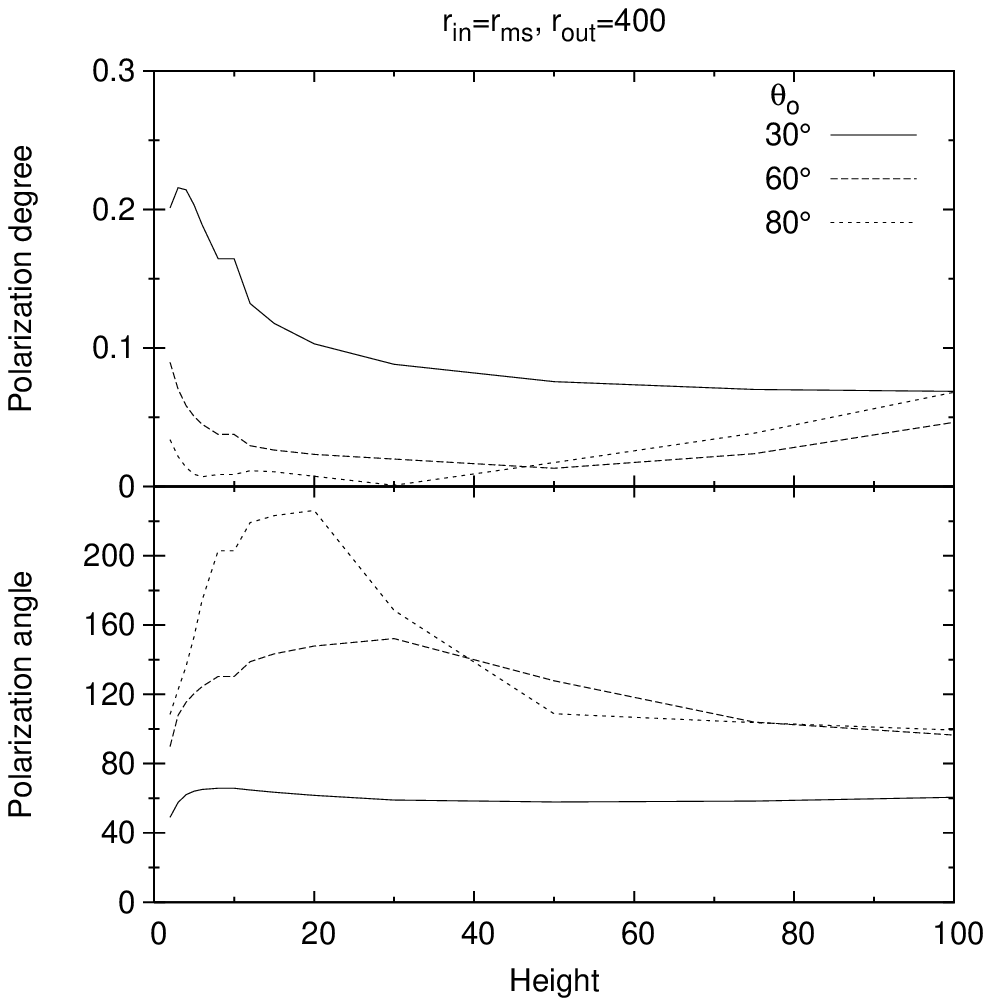}
\mycaption{Polarization degree and angle due to reflected radiation integrated
over the whole surface of the disc and propagated to
the point of observation. Dependence on height $h$ is plotted.
Left panel: $r_{\rm{}in}=6$; right panel: $r_{\rm{}in}=1.20$. In both
the panels the energy range was assumed $9-12$~keV, the photon index of
incident radiation $\Gamma=2$, the angular momentum $a=0.9987$.}
\end{figure}

\begin{figure}
\dummycaption\label{polangle1}
\includegraphics[width=0.48\textwidth]{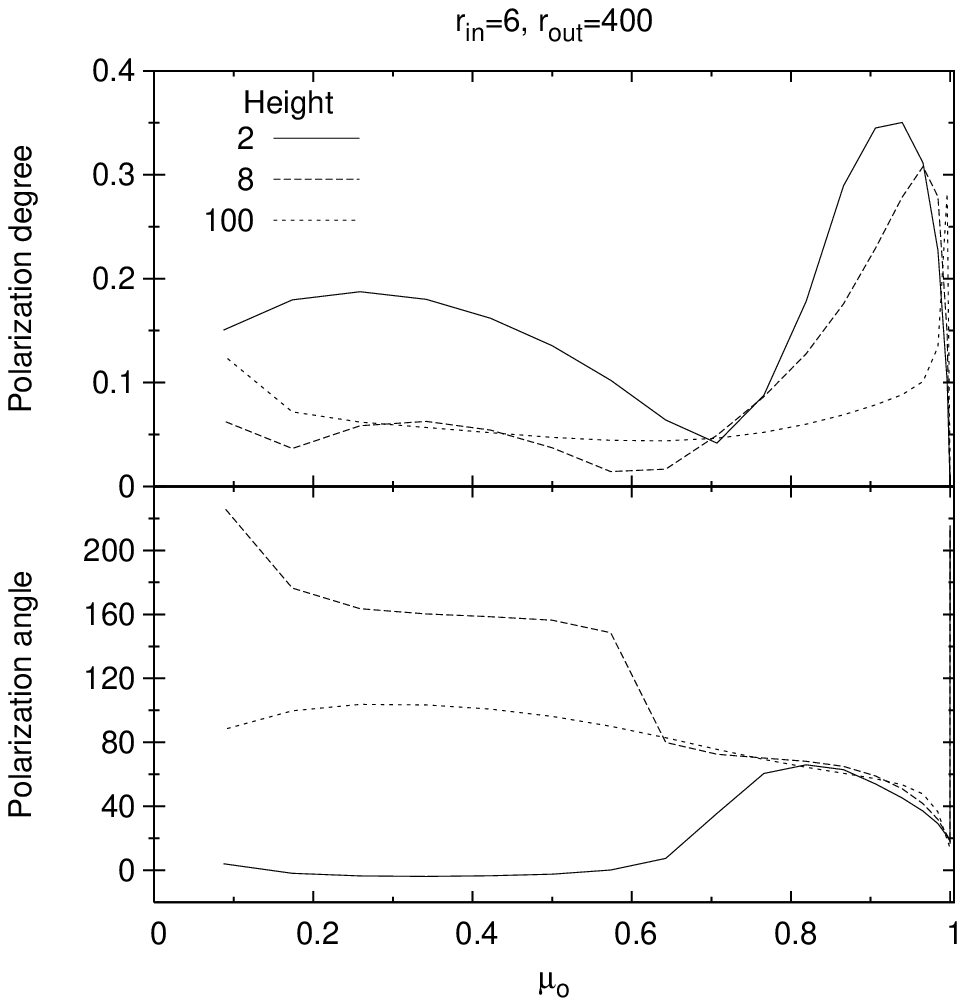}
\hfill
\includegraphics[width=0.48\textwidth]{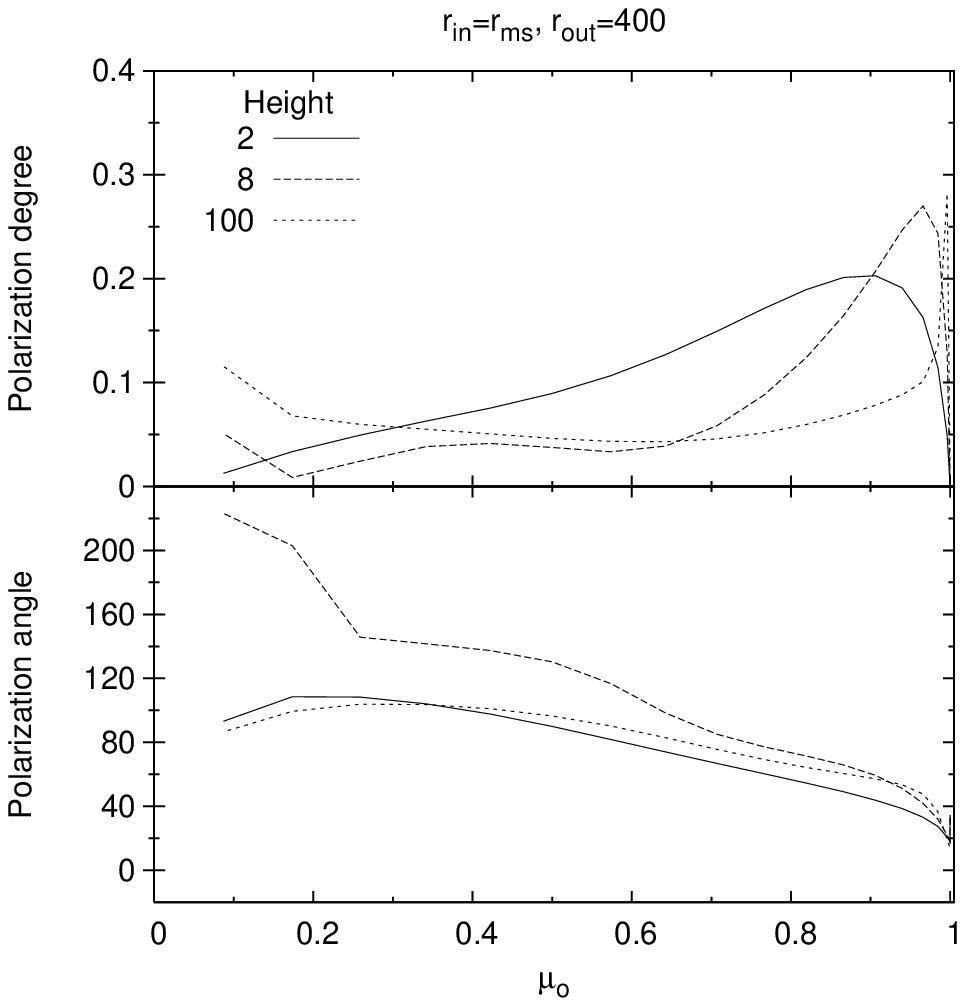}
\mycaption{Polarization degree and angle as functions of
$\mu_{\rm{}o}$ (cosine of observer inclination, $\mu_{\rm{}o}=0$
corresponds to the edge-on view of the disc). The same model
is shown as in the previous figure.}
\end{figure}

\begin{figure}
\dummycaption\label{poldeg1}
\includegraphics[width=0.48\textwidth]{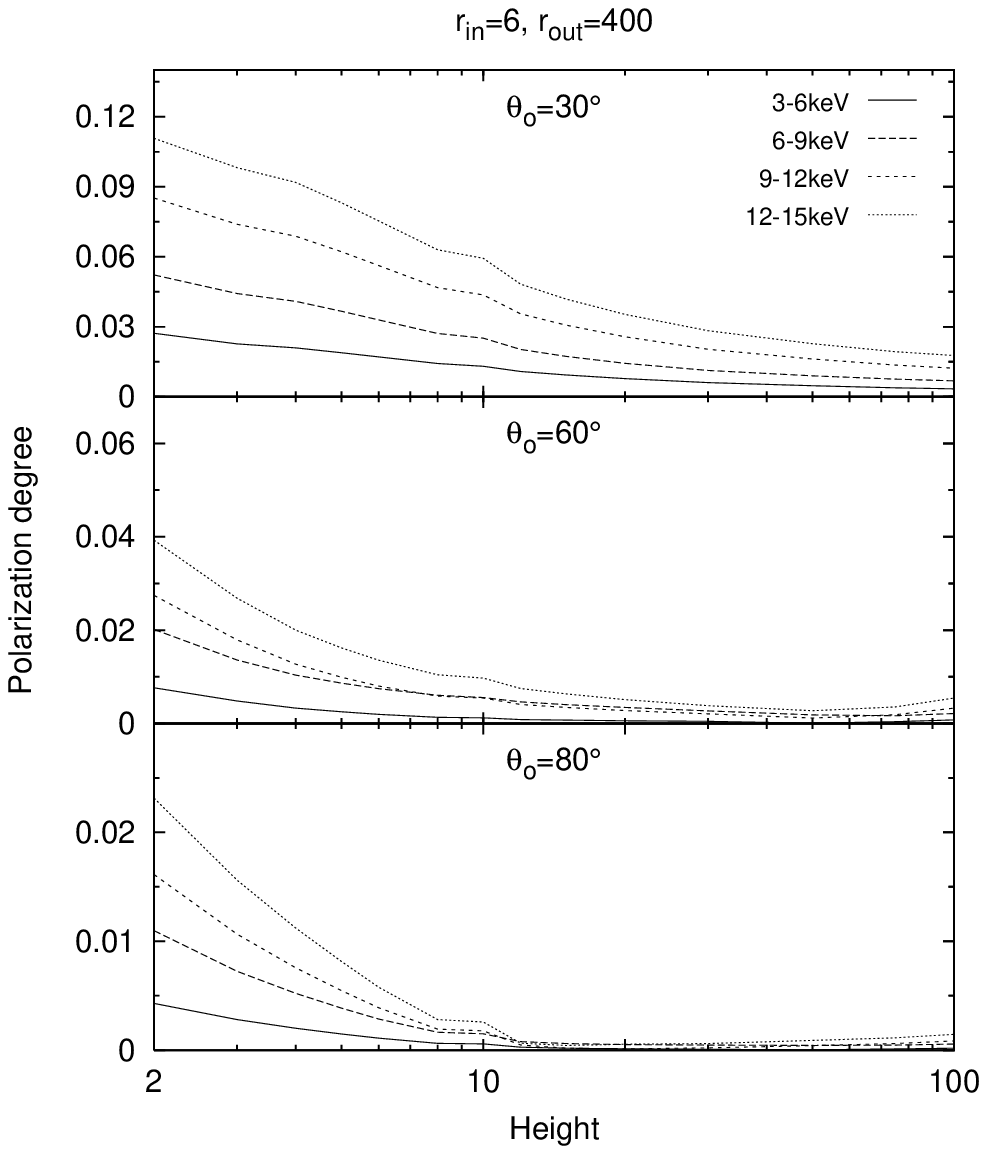}
\hfill
\includegraphics[width=0.48\textwidth]{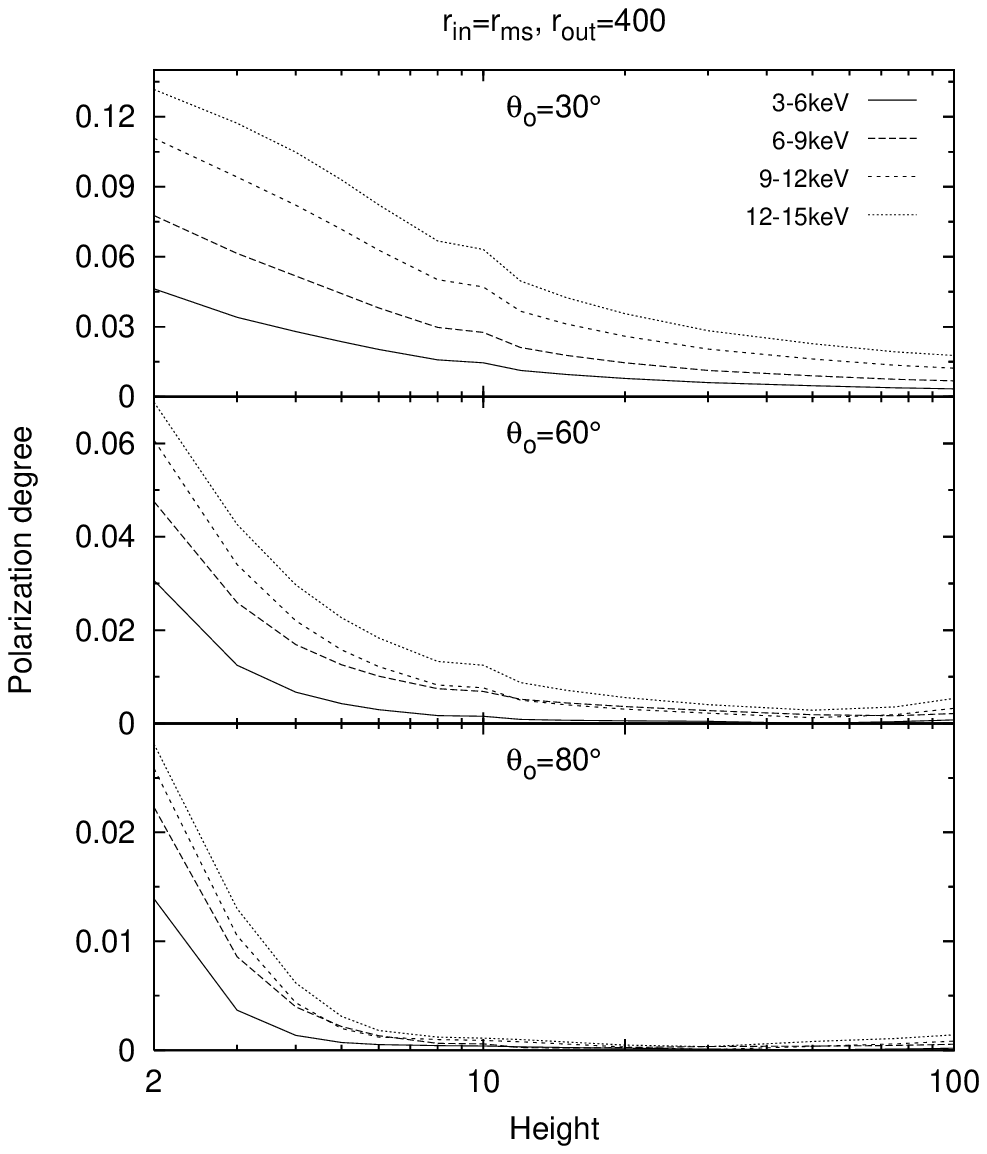}
\mycaption{Net polarization degree of the total (primary
plus reflected) signal as a function of $h$.
Left panel: $r_{\rm{}in}=6$; right panel: $r_{\rm{}in}=1.20$.
The curves are parametrized by the corresponding energy range.}
\end{figure}

\begin{figure}
\dummycaption\label{poldeg2}
\includegraphics[width=0.48\textwidth]{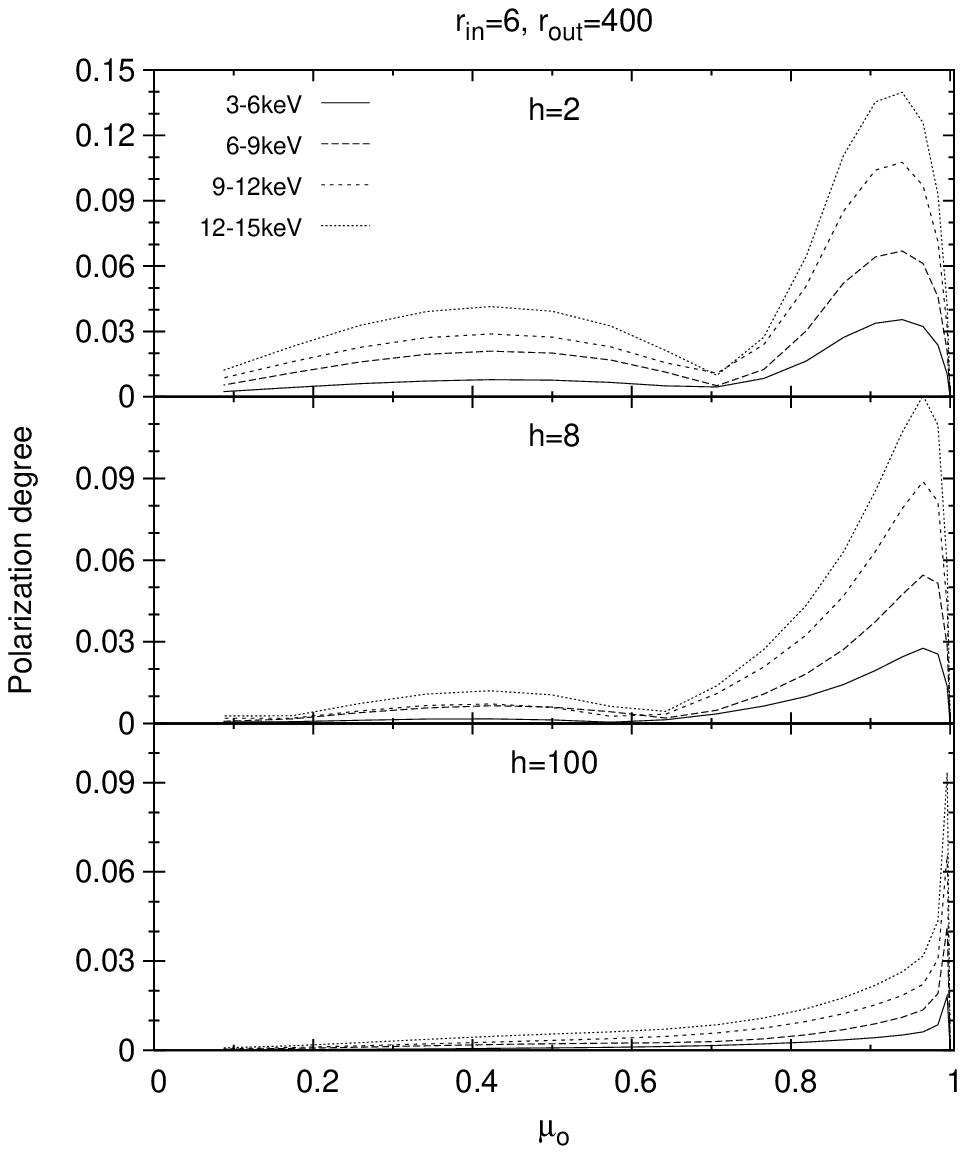}
\hfill
\includegraphics[width=0.48\textwidth]{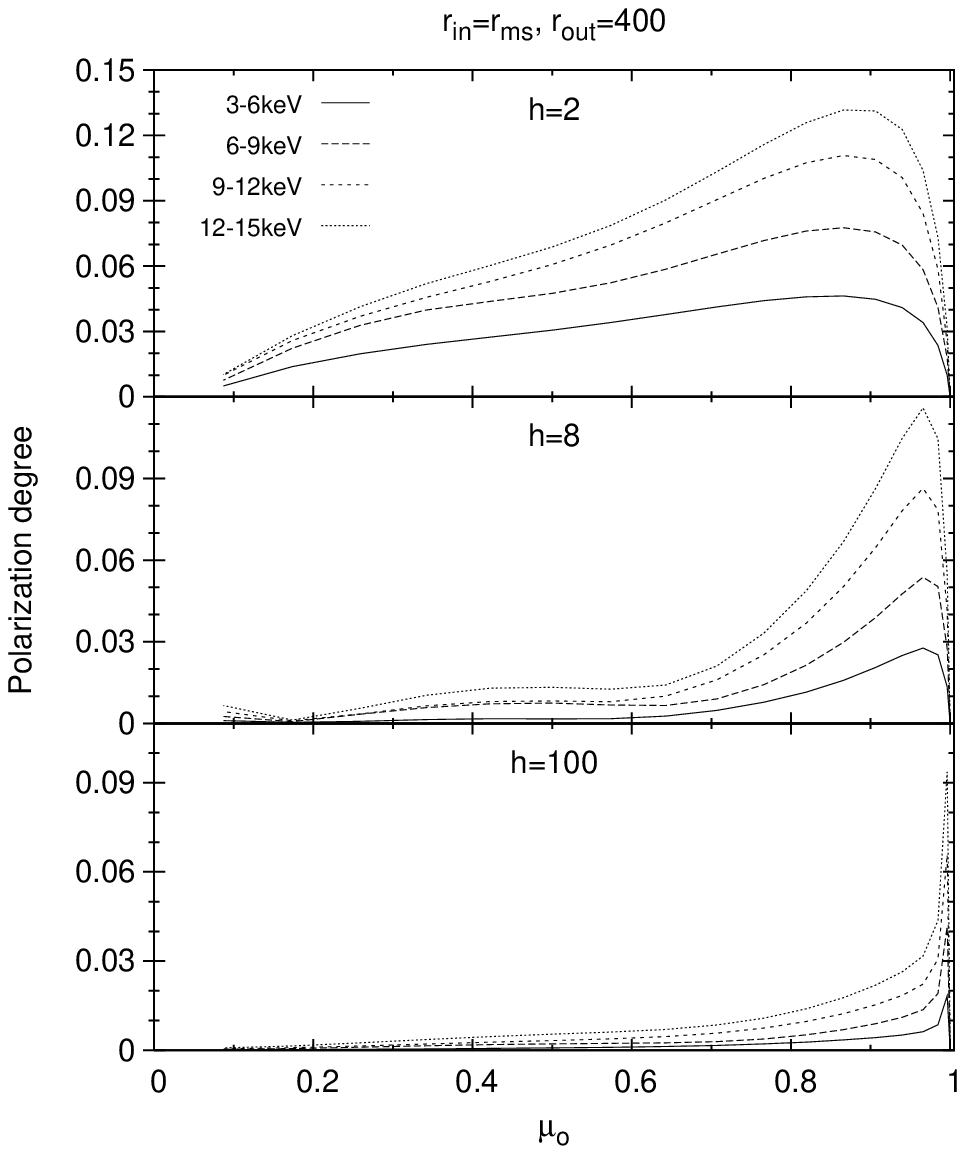}
\mycaption{Net polarization degree of the total (primary
plus reflected) signal as a function of $\mu_{\rm{}o}$.
The same model is shown as in the previous figure.}
\end{figure}

In order to compute observable characteristics one has to
combine the primary power-law continuum with the reflected component.
The polarization
degree of the resulting signal depends on the mutual proportion of the two
components and also on the energy range of an observation. The overall degree of
polarization increases with energy (see bottom panels in
Figs.~\ref{pol}--\ref{pol1}) due to the fact that the intensity of radiation
from the primary source decreases exponentially, the intensity of the reflected
radiation increases with energy (in the energy range $3-15\,$keV) and
the polarization of the reflected light alone is more or less constant.
In our computations we assumed that the irradiating source
emits isotropically and its light is affected only by gravitational redshift
and lensing, according to the source location at $z=h$ on axis. This
results in a dilution of primary light by factor
${\sim}g_{\rm{}h}^2(h,\theta_{\rm{}o})\,l_{\rm{}h}(h,\theta_{\rm{}o})$,
where $g_{\rm{}h}=\sqrt{1-2h/(a^2+h^2)}$ is the redshift of primary
photons reaching directly the
observer, $l_{\rm{}h}$ is the corresponding lensing factor. Here, the redshift
is the dominant relativistic term, while lensing of primary photons is a few
percent at most and it can be safely ignored. Anisotropy of primary  radiation
may further attenuate or amplify the polarization degree of the final signal,
while the polarization angle is rather independent of this influence as long
as the primary light is itself unpolarized.

The polarization of scattered light is also shown in Fig.~\ref{poldeg},
where we plot the polarization degree and the change of the polarization
angle as functions of $h$. Notice that in the Newtonian case only
polarization angles of $0^{\circ}$ or $90^{\circ}$ would be expected
for reasons of symmetry.  The change in angle is due to gravitation for
which we assumed a rapidly rotating black hole.
The two panels in the figure correspond to different locations of
the inner disc edge: $r_{\rm{}in}=6$ and  $r_{\rm{}in}=1.20$,
respectively. The curves are strongly sensitive to $r_{\rm{}in}$ and
$h$, while the dependence on $r_{\rm{}out}$  is weak for a large disc
(here $r_{\rm{}out}=400$). Sensitivity to $r_{\rm{}in}$ is particularly
appealing if one remembers the practical difficulties in estimating
$r_{\rm{}in}$ by fitting spectra. The effect is clearly visible even
for $h\sim20$. Graphs
corresponding to $r_{\rm{}in}=6$ and $a=0.9987$, resemble, in essence quite
closely, the non-rotating case ($a=0$) because dragging effects are most
prominent near the horizon.

Fig.~\ref{polangle1} shows the polarization
degree and angle as functions of the observer's inclination. Again, by
comparing the two cases of different $r_{\rm{}in}$ one can clearly
recognize that the polarization is sensitive to details of the flow near
the inner disc boundary.

The dependence of the polarization degree of overall radiation (primary plus
reflected) on the height of the primary source and the observer inclination
in different energy ranges is shown in Figs.~\ref{poldeg1}--\ref{poldeg2}.

In this section we examined the polarimetric properties of X-ray illuminated
accretion discs in the lamp-post model. From the figures shown it is clear
that observed values of polarization angle and degree are
rather sensitive to the model parameters. The approach adopted
provides additional information with respect to traditional
spectroscopy and so it has great potential for discriminating between
different models. It offers an improved way of measuring rotation of the
black hole because the radiation properties of the inner disc region
most likely reflect the value of the black-hole angular momentum.

While our calculations have been
performed assuming a stationary situation, in reality it is likely that
the height of the illuminating source changes with time, and indeed such
variations have been invoked by \cite{miniutti2003} to explain the
primary and reflected variability patterns of MCG--6-30-15. A complete
time-resolved analysis (including all consequences of the light travel
time in curved space-time) is beyond the scope of this section and we defer
it to future work, assuming that the primary source varies on a
time-scale longer than light-crossing time in the system. This is also a
well-substantiated assumption from a practical point of view,
since feasible techniques will anyway require sufficient integration
time (i.e.\ order of several ksec). Once full temporal resolution is
possible, the analysis described above can be readily extended. Here,
it suffices to note that a variation of $h$ implies a variation of the
observed polarization angle of the reflected radiation. As it is hard to
imagine a physical and/or geometrical effect giving rise to the same
effect, time variability of the polarization angle can be considered
(independently of the details) a very strong signature of strong-field
general relativity effects at work.

New generation photoelectric polarimeters \citep{costa2001} in the
focal plane of large area optics (such as those foreseen for
{\it{}Xeus}) can probe the polarization degree of the order of one percent
in bright AGNs, making polarimetry, along with timing and spectroscopy,
a tool for exploring the properties of the accretion flows in the
vicinity of black holes.

\clearpage
\phantomsection
\addcontentsline{toc}{chapter}{Summary and future prospects}
\chapter*{Summary and future prospects}
 \thispagestyle{empty}
 \enlargethispage*{\baselineskip}
\vspace*{-1em}
In this thesis we have developed a computational tool for modelling
spectral features of X-ray sources in a strong gravitational field.

The following six transfer functions have been computed for light rays
emitted from the equatorial plane of the Kerr
black hole and received by an observer at infinity:
the gravitational and Doppler shift, lensing, emission angle, relative
time delay, change of the polarization angle and azimuthal
emission angle.
The values of these functions for different angular momenta of the
black hole and inclination angles of the observer in Kerr ingoing
coordinates have been stored in the form of tables in a FITS file.
In computations, where parameters of the motion of the matter were needed, we
have assumed a co-rotating
Keplerian disc above the marginally stable orbit and freely falling matter below
it. Graphical representations of the results have been made in the form of an
atlas of contour figures of these functions.

For the modelling of the spectra of an accretion disc, several general
relativistic codes have been developed. Some of them can be used inside a more
general spectral fitting package {\sc xspec} for routine fitting of the data
measured by X-ray satellites. These are the models for the relativistic emission
lines ({\sc kygline}, {\sc kyg1line} and {\sc kyf1ll}), Compton reflection
({\sc kyl1cr} and {\sc kyh1refl}) and for the general use as a relativistic
extension of existing
models (the convolution models {\sc kyconv} and {\sc kyc1onv}). Other components
of the tool have extended features for studying the
non-stationary emission from accretion discs ({\sc kygspot}) and polarimetry
({\sc kyl1cr}).

The newly developed codes have been employed in several applications.
Firstly, we used our new {\sc ky} models inside {\sc xspec} and fitted the
data from {\it XMM-Newton}\/ observations of Seyfert galaxy MCG--6-30-15.
Then we simulated emission from an X-ray illuminated spot orbiting near
a black hole as an application on non-stationary emission from the disc.
And finally, the polarization from an accretion disc illuminated
by a primary source located above the black hole was computed.

There are several possibilities of how one can proceed further.
We would like to:
\begin{itemize} \itemsep -0.3em
\vspace*{-0.4em}
 \item enhance the code so that it would be possible to investigate the
 self-irradiation of the disc,
 \item add more models to the {\sc ky} package, first of all a model for the
 thermal emission,
 \item expand the code so that non-Keplerian discs can be studied as well,
 \item perform more polarization studies with the non-stationary lamp-post
 model,
 \item perform timing analysis of the emission from an orbiting spot in
 different energy bands,
 \item fit more data of X-ray sources acquired by X-ray satellites.
\end{itemize}
Further it would be possible to investigate applications of our code to
quasi-periodic oscillations. Although our code is a two-dimensional one it might
be interesting to compute how oscillations of the disc in the equatorial plane
can affect the observed spectra at infinity.


\appendix
\chapter{Kerr space-time}
\label{kerr}
 \thispagestyle{empty}
 In this Appendix as well as everywhere else in this thesis we use units where
$G\mbh=c=1$ ($\mbh$ is the mass of the central black hole) and we assume that
the angular momentum $a$ of the black hole is positive.

\section{Kerr metric}
The Kerr metric in the Boyer-Lindquist coordinates $(t,r,\theta,\varphi)$ is
\begin{equation}
g_{\mu\nu} = \mat{-(1-\frac{2r}{\rho^2}) & 0 & 0
  & -\frac{2ar\sin^{2}{\! \theta}}{\rho^2} \\[2mm]
  0 & \frac{\rho^2}{\Delta} & 0 & 0 \\[2mm] 0 & 0 & \rho^2 & 0 \\[2mm]
  -\frac{2ar\sin^{2}{\!\theta}}{\rho^2} & 0 & 0
	& \frac{{\cal A}\sin^{2}{\!\theta}}{\rho^2}}\, ,
\end{equation}
where $\rho^2 \equiv r^2+a^2\cos^2\!\theta$, $\Delta \equiv r^2-2r+a^2$ and
${\cal A} \equiv (r^2+a^2)^2-\Delta a^2\sin^2\!\theta$.

%

Let's define special Kerr ingoing coordinates $(\tk,\uk,\muk,\phik)$ by
the following tetrad vectors
\begin{equation}
\begin{array}{cclccl}
\fr{\partial}{\partial \tk} & = & \fr{\partial}{\partial t}\ , &
\dd \tk & = & \dd t + \fr{r^{2}+a^{2}}{\Delta}\dd r\ ,
\\[4mm]
\fr{\partial}{\partial \uk} & = & -r^2\fr{\partial}{\partial r}+
r^2\fr{r^{2}+a^{2}}{\Delta}\left(\fr{\partial}{\partial t}+
\Omega^{\rm H}\fr{\partial}{\partial\varphi}\right)\ ,  \hspace{0.7cm} &
\dd \uk & = & -\fr{1}{r^2}\dd r\ ,
\\[4mm]
\fr{\partial}{\partial \muk} & = & -\fr{1}{\sin\theta}\fr{\partial}
{\partial \theta}\ , & \dd \muk & = & -\sin\theta \dd \theta\ ,\\[4mm]
\fr{\partial}{\partial \phik} & = & \fr{\partial}{\partial \varphi}\ , &
\dd \phik & = & \dd \varphi + \fr{a}{\Delta} \dd r\ ,
\end{array}
\end{equation}
with $\ \Omega^{\rm H}=\fr{a}{r^{2}+a^{2}}$.

The advantage of these coordinates, where $\uk\equiv r^{-1}$ and
$\muk\equiv\cos{\theta}$, is that the Kerr metric is not singular on the
horizon of the black hole and spatial infinity
($r\rightarrow\infty$) is brought to a finite value ($\uk\rightarrow 0$).
Another advantage for numerical computations is that we get rid of the
cosine function.

The relationship between the Boyer-Lindquist coordinate $\varphi$ and the Kerr
ingoing coordinate $\varphi_{\rm K}$, for geodesics coming from infinity
to the equatorial plane, can be expressed in the following way:
\begin{myeqnarray}
\varphi & = & \varphi_{\rm K}+\frac{a}{r_+-r_-}\ln{\frac{r-r_+}{r-r_-}} & &
{\rm for} \ \ a<1\, ,\\
\varphi & = & \varphi_{\rm K}-\frac{1}{r-1} & & {\rm for}\ \ a=1\, ,
\end{myeqnarray}
with $r_\pm=1\pm\sqrt{1-a^2}$ being inner ($-$) and outer ($+$) horizon of the
black hole.

Another useful way to express the transformation between the Boyer-Lindquist
and special Kerr ingoing coordinates is by matrices of transformation
\begin{myeqnarray}
\fr{\partial x^{\mu}}{\partial \hat{x}^{\nu}} & = & \mat{1 & \frac{r^2(r^2+a^2)}
{\Delta}  & 0 & 0 \\[2mm]
 0 & -r^2 & 0 & 0 \\[2mm]
 0 & 0    & -\frac{1}{\sin\theta} & 0 \\[2mm]
 0 & \frac{r^2a}{\Delta}   & 0 & 1} & = &
\mat{1 & \frac{(1+a^2\uk^2)}{\uk^2\Deltak} & 0 & 0 \\[2mm]
     0 & -\frac{1}{\uk^2}  & 0 & 0 \\[2mm]
     0 & 0                 & -\frac{1}{\sqrt{1-\muk^2}} & 0 \\[2mm]
     0 & \frac{a}{\Deltak} & 0 & 1}\, ,\\[4mm]
\fr{\partial \hat{x}^{\mu}}{\partial x^{\nu}}  & = & \hspace*{2mm} \mat{1 &
\frac{r^{2}+a^{2}}{\Delta} & 0 & 0 \\[2mm]
  0 & -\frac{1}{r^2}       & 0 & 0 \\[2mm]
  0 & 0                    & -\sin\theta & 0 \\[2mm]
  0 & \frac{a}{\Delta}     & 0 & 1} & = &
\mat{1 & \frac{1+a^2\uk^2}{\Deltak} & 0 & 0 \\[2mm]
     0 & -\uk^2                     & 0 & 0 \\[2mm]
     0 & 0                          & -\sqrt{1-\muk^2} & 0 \\[2mm]
     0 & \frac{a\uk^2}{\Deltak}     & 0 & 1}\, .\hspace*{9mm}
\end{myeqnarray}

The Kerr metric in special Kerr ingoing coordinates is
\begin{eqnarray}
g_{\hat{\mu}\hat{\nu}} = \mat{-(1-\frac{2\uk}{\rhok^2}) & -\frac{1}{\uk^2} & 0 &
  -\frac{2a\uk(1-\muk^2)}{\rhok^2} \\[2mm]
  -\frac{1}{\uk^2} & 0 & 0 & \frac{a(1-\muk^2)}{\uk^2} \\[2mm] 0 & 0 &
	\frac{\rhok^2}{\uk^2(1-\muk^2)} & 0 \\[2mm]
  -\frac{2a\uk(1-\muk^2)}{\rhok^2} & \frac{a(1-\muk^2)}{\uk^2} & 0 &
	\frac{\Ak(1-\muk^2)}{\uk^2\rhok^2}}\, ,
\end{eqnarray}
with $\rhok^2 \equiv \rho^2/r^2=1+a^2\uk^2\muk^2$,
$\Deltak \equiv \Delta/r^2=1-2\uk+a^2\uk^2$ and
$ \Ak \equiv {\cal A}/r^4=(1+a^2\uk^2)^2-a^2\uk^2\Deltak(1-\muk^2)$.

\section{Light rays in Kerr space-time}
The four-momentum $p^\mu\equiv\frac{{\rm d}x^{\mu}}{{\rm d}\lambda'}$
of photons travelling in Kerr space-time in the
Boyer-Lindquist coordinates is (see e.g.\ \citealt{carter1968} or
\citealt{misner1973})
\begin{eqnarray}
\label{carter1}
p^t & \equiv & \frac{\dd t}{\dd \lambda'} = [a(l-a\sin^2\!\theta)+(r^2+a^2)
(r^2+a^2-al)/\Delta]/\rho^2\, ,\\
\label{carter2}
p^r & \equiv & \frac{\dd r}{\dd \lambda'} = R_{\rm sgn} \{ (r^2+a^2-al)^2 -
\Delta [(l-a)^2+q^2] \}^{1/2}/\rho^2\, ,\\
\label{carter3}
p^\theta & \equiv & \frac{\dd \theta}{\dd \lambda'} =
\Theta_{\rm sgn}[ q^2-\cot^2\!\theta(l^2-a^2\sin^2\!\theta)]^{1/2}/\rho^2\, ,\\
\label{carter4}
p^\varphi & \equiv & \frac{\dd \varphi}{\dd \lambda'} = [ l/\sin^2\!\theta - a +
a(r^2+a^2-al)/\Delta ]/\rho^2\, ,
\end{eqnarray}
where $l=\alpha(1-\mu_\obs^2)^{1/2}=\alpha\sin\theta_\obs$ and
$q^2=\beta^2+\mu_\obs^2(\alpha^2-a^2)$ are Carter's constants of motion with
$\alpha$ and $\beta$ being impact parameters measured perpendicular and
parallel, respectively, to the spin axis of the
black hole projected onto the observer's sky. Here we define $\alpha$ to
be positive when a photon travels in the direction of the four-vector
$\frac{\partial}{\partial \varphi}$ at infinity, and $\beta$ to be positive if
it travels in the direction of $-\frac{\partial}{\partial\theta}$ at infinity.
The parameter $\theta_{\rm o}$ (and $\mu_{\rm o}=\cos{\theta_{\rm o}}$) defines
a point at infinity through which the light ray passes (we consider only light
rays coming to or from the observer at infinity).
Furthermore, we have denoted the sign of the radial component of the momentum
by ${\rm R}_{\rm sgn}$ and the sign of the $\theta$-component of the momentum
by ${\Theta}_{\rm sgn}$. We have chosen an affine parameter $\lambda'$ along
light geodesics in such a way that the conserved energy is normalized to
$-p_t=1$.

The four-momentum transformed into special Kerr ingoing coordinates is
\begin{eqnarray}
\label{ptk}
p^{\tk} \equiv \frac{\dd\tk}{\dd\lambda'} & = & \frac{1}{\rhok^2}
\left[1+a\uk^2(l+a\muk^2)-(1+a^2\uk^2)\frac{1+2\uk-(l^2+q^2)\uk^2}
{\uk(2-a\uk l)+U_{\rm sgn}\sqrt{U}}\right]\, , \\
p^{\uk} \equiv \frac{\dd\uk}{\dd\lambda'} & = &
U_{\rm sgn}\frac{\uk^2\sqrt{U}}{\rhok^2}\, ,\\
p^{\muk} \equiv\frac{\dd\muk}{\dd\lambda'} & = &
M_{\rm sgn}\frac{\uk^2\sqrt{M}}{\rhok^2}\, , \\
p^{\phik}\equiv\frac{\dd\phik}{\dd\lambda'} & = &
\label{pphik}
\frac{\uk^2}{\rhok^2}\left[\frac{l}{1-\muk^2}-a\frac{1+2\uk-(l^2+q^2)\uk^2}
{\uk(2-a\uk l)+U_{\rm sgn}\sqrt{U}}\right]\, ,\hspace*{12mm}
\end{eqnarray}
with $U\equiv 1+(a^2-l^2-q^2)\uk^2+2[(a-l)^2+q^2]\uk^3-a^2q^2\uk^4$ and
$M\equiv q^2+(a^2-l^2-q^2)\muk^2-a^2\muk^4$. The sign of the $\uk$- and
$\muk$-components of the four-momentum is denoted by $U_{\rm sgn}$ and
$M_{\rm sgn}$, respectively. Note that in these coordinates the
four-momentum is not singular on the horizon as opposed to the expressions
in the Boyer-Lindquist coordinates.

\section{Keplerian disc in Kerr space-time}
Matter moves along free stable circular orbits
with a rotational velocity
(see e.g.\ Novikov \& Thorne \citeyear{novikov1973})
\begin{equation}
\omega=\frac{{\rm d} \varphi}{{\rm d}t}=\frac{1}{r^{3/2}+a}\, .
\end{equation}
in a Keplerian disc. The disc resides in the equatorial plane of the
black hole.
Individual components of the four-velocity of the Keplerian disc in the
Boyer-Lindquist coordinates are
\begin{eqnarray}
U^t & = & \frac{r^2+a\sqrt{r}}{r\sqrt{r^2-3r+2a\sqrt{r}}}\, ,\\
U^r & = & 0\, ,\\
U^\theta & = & 0\, ,\\
U^\phi & = & \frac{1}{\sqrt{r(r^2-3r+2a\sqrt{r})}}\, .
\end{eqnarray}

There is no free stable circular orbit below the marginally stable orbit defined by
\begin{eqnarray}
r_{\rm ms} & = & 3+Z_2-\sqrt{(3-Z_1)(3+Z_1+2Z_2)]}
\end{eqnarray}
with $Z_1 = 1+(1-a^2)^{1/3}[(1+a)^{1/3} + (1-a)^{1/3}]$ and
$Z_2 = \sqrt{3a^2+Z_1^2}$.
We suppose that below this orbit the matter is in a free fall down to the
horizon. Thus the matter conserves its specific energy $-U_t$ and its specific
angular momentum $U_\varphi$
\begin{myeqnarray}
-U_t(r<r_\mso) & \equiv & E_\mso \ \equiv\ -U_t(r_\mso) & = &
\frac{r^2_\mso-2r_\mso+a\sqrt{r_\mso}}{r_\mso
\sqrt{r^2_\mso-3r_\mso+2a\sqrt{r_\mso}}}\, , \\
\ \ U_\varphi(r<r_\mso) & \equiv & L_\mso\ \equiv\ \ \ U_\phi(r_\mso) & = &
\frac{r_\mso^2+a^2-2a\sqrt{r_\mso}}
{\sqrt{r_\mso(r^2_\mso-3r_\mso+2a\sqrt{r_\mso})}}\, .
\end{myeqnarray}
When we also consider the normalization condition for the four-velocity,
$U^\mu\,U_\mu=-1$,
then we get the following expressions for its contravariant components
\begin{eqnarray}
U^t(r<r_\mso) & = & \frac{1}{r\Delta}\{[r(r^2+a^2)+2a^2]E_\mso-2aL_\mso\}\, ,\\
U^r(r<r_\mso) & = & -\frac{1}{r\sqrt{r}}\sqrt{[r(r^2+a^2)+2a^2]E^2_\mso-4aE_\mso
L_\mso-(r-2)L^2_\mso-r\Delta}\, ,\hspace*{14mm}\\
U^\theta(r<r_\mso) & = & 0\, ,\\
U^\varphi(r<r_\mso) & = & \frac{1}{r\Delta}[2aE_\mso+(r-2)L_\mso]\, .
\end{eqnarray}

In our calculations we use the following local orthonormal tetrad connected
with the matter in the disc
\begin{eqnarray}
e_{(t)\,\mu} & \equiv & U_\mu\, ,\\
e_{(r)\,\mu} & \equiv & \frac{r}{\sqrt{\Delta(1+U^rU_r)}}\left[\ U^r( U_t,\,
U_r,\,0,\,U_\varphi)+(0,\, 1,\, 0,\, 0)\ \right]\, ,\\
e_{(\theta)\,\mu} & \equiv & (0,\,0,\,r,\,0)\, ,\\
e_{(\varphi)\,\mu} & \equiv & \sqrt{\frac{\Delta}{1+U^rU_r}}\left (-U^\varphi,\,
0,\,0,\,U^t\right )\, .
\end{eqnarray}

\section{Initial conditions of integration}
\label{initial}
In this part of the Appendix we will sum up the initial conditions for
integration of the equation of the geodesic
\begin{equation}
\frac{\DD p^\mu}{\dd \lambda'} = \frac{\dd p^\mu}{\dd \lambda'} +
\Gamma^\mu_{\sigma\tau}p^\sigma p^\tau\,,\quad\mu=\tk,\,\uk,\,\muk,\,\phik
\end{equation}
and equation of the geodesic deviation
\begin{equation}
\frac{\dd^2Y_{\rm j}^\mu}{\dd\lambda'^2}+2\Gamma^\mu_{\sigma\gamma}p^\sigma
\frac{\dd Y_{\rm j}^\gamma}{\dd\lambda'}+\Gamma^\mu_{\sigma\tau,\gamma}p^\sigma
p^\tau Y_{\rm j}^\gamma=0\, ,\quad \mu=\tk,\,\uk,\,\muk,\,\phik\, ,
\quad{\rm j}=1,\,2\,
\end{equation}
which we solve in order to compute transfer functions over the accretion disc
(see Chapter~\ref{chapter1} for details).
Here $p^{\mu}\equiv\frac{{\rm d} x^{\mu}}{{\rm d}\lambda'}$ is a four-momentum
of light, $\lambda'$ is an affine parameter for which the conserved energy
$-p_t=1$, $Y_{\rm j}^\mu$ are two vectors characterizing the distance between
nearby geodesics and $\Gamma_{\mu\nu}^\sigma$ are Christoffel symbols for the
Kerr metric. We solve these equations in special Kerr ingoing coordinates
$\tk,\,\uk,\,\muk,\,\phik$ and we start integrating from the following initial
point:
\begin{equation}
\tk_\init = 0\, ,\quad \uk_\init = 10^{-11}\, ,\quad
\muk_\init  = \mu_\obs = \cos\theta_\obs\, ,\quad \phik_\init = 0\, ,
\end{equation}
where $\theta_{\obs}$ is the inclination angle of the observer at infinity.
The initial values of the four-momentum of light were defined by this point and
eqs.~(\ref{ptk})--(\ref{pphik}) with $U_{\rm sgn}=+1$, $M_{\rm sgn}=0$ for
$\beta=0$ and $M_{\rm sgn}={\rm sign}(\beta)$ for $\beta\neq 0$.

Four-vectors $Y_1^{\mu}$ and $Y_2^{\mu}$ have the initial values
\begin{eqnarray}
\label{Y}
\begin{array}{cclccl}
Y_{1\init}^{\tk} &=& \beta\uk_\init\, , &
 Y_{2\init}^{\tk} &=& \alpha\uk_\init\, ,\\[2mm]
Y_{1\init}^{\uk} &=& -\beta\uk_\init^3\, , &
 Y_{2\init}^{\uk} &=& -\alpha\uk_\init^3\, ,\\[2mm]
Y_{1\init}^{\muk} &=& \uk_\init\sqrt{1-\muk_\init^2}\, , &
 Y_{2\init}^{\muk} &=& -\alpha\uk_\init^2\muk_\init\, ,\\[2mm]
Y_{1\init}^{\phik} &=& \alpha\uk_\init^2\muk_\init/(1-\muk_\init^2)\, ,
 \hspace{6em} &
 Y_{2\init}^{\phik} &=& \uk_\init/\sqrt{1-\muk_\init^2}\, ,
\end{array}
\end{eqnarray}
and their derivatives have the initial values ($\dd Y_{\rm j}^{\hat{\nu}}/
\dd \lambda'=p^{\hat{\sigma}}\,\partial Y_{\rm j}^{\hat{\nu}}/\partial
x^{\hat{\sigma}}$)
\begin{eqnarray}
\label{dY}
\begin{array}{cclccl}
\displaystyle\frac{\dd Y_{1\init}^{\tk}}{\dd\lambda'} &=& \beta\uk_\init^2\, , &
 \displaystyle\frac{\dd Y_{2\init}^{\tk}}{\dd\lambda'} &=& \alpha\uk_\init^2\, ,
 \\[4mm]
\displaystyle\frac{\dd Y_{1\init}^{\uk}}{\dd\lambda'} &=& -3\beta\uk_\init^4\, ,
 & \displaystyle\frac{\dd Y_{2\init}^{\uk}}{\dd\lambda'} &=& -3\alpha\uk_\init^4
 \, ,\\[4mm]
\displaystyle\frac{\dd Y_{1\init}^{\muk}}{\dd\lambda'} &=& \uk_\init^2\sqrt{1-
\muk_\init^2}\, , &
 \displaystyle\frac{\dd Y_{2\init}^{\muk}}{\dd\lambda'} &=&
 -2\alpha\uk_\init^3\muk_\init\, ,\\[4mm]
\displaystyle\frac{\dd Y_{1\init}^{\phik}}{\dd\lambda'} &=& 2\alpha\uk_\init^3\muk_\init/(1-\muk_\init^2)\, , \hspace{4.7em} &
 \displaystyle\frac{\dd Y_{2\init}^{\phik}}{\dd\lambda'} &=& \uk_\init^2/
 \sqrt{1-\muk_\init^2}\, .\hspace{0.6em}
\end{array}
\end{eqnarray}
The initial vectors $Y_{1\init}^{\mu}$ and $Y_{2\init}^{\mu}$ were chosen in such a way that
in the initial point, where we suppose the metric to be a flat-space Minkowski
one, they are perpendicular to each other and to the four-momentum of light,
are space-like and have unit length. We kept only the largest terms in
$\uk_{\rm i}$ in eqs.~(\ref{Y}) and (\ref{dY}).

 \section{Christoffel symbols}
   \enlargethispage*{\baselineskip}
Here we show the Christoffel symbols and their first derivatives
for the Kerr metric in special Kerr ingoing coordinates. They are needed
for the numerical integration of the equation of the geodesic and geodesic
deviation.
\begin{myeqnarray}
\nonumber
\Gamma^{\tk}_{\tk\tk} & = & \uk^2 (1+a^2 \uk^2)(1-a^2 \uk^2 \muk^2)/\rhok^6
 \hspace*{-20cm} &  & \\[0.9mm]
\nonumber
\Gamma^{\tk}_{\tk \uk} & = & \Gamma^{\tk}_{\uk\tk}
& = & 0 \hspace*{10.6cm} \\[0.9mm]
\nonumber
\Gamma^{\tk}_{\tk\muk} & = & \Gamma^{\tk}_{\muk \tk} & = &
2a^2 \uk^3 \muk/\rhok^4 \\[0.9mm]
\nonumber
\Gamma^{\tk}_{\tk\phik} & = & \Gamma^{\tk}_{\phik \tk} & = & -a(1-\muk^2)
\Gamma^{\tk}_{\tk\tk} \\[0.9mm]
\nonumber
\Gamma^{\tk}_{\uk\uk} & = & 0  \hspace*{-20cm} &  & \\[0.9mm]
\nonumber
\Gamma^{\tk}_{\uk\muk} & = & \Gamma^{\tk}_{\muk \uk} & = &
-a^2 \muk/\rhok^2 \\[0.9mm]
\nonumber
\Gamma^{\tk}_{\uk\phik} & = & \Gamma^{\tk}_{\phik \uk} & = &
-a(1-\muk^2)/(\uk\rhok^2) \\[0.9mm]
\nonumber
\Gamma^{\tk}_{\muk\muk} & = & -(1+a^2 \uk^2)/[\uk(1-\muk^2)\rhok^2]
 \hspace*{-20cm} &  & \\[0.9mm]
\nonumber
\Gamma^{\tk}_{\muk\phik} & = & \Gamma^{\tk}_{\phik\muk} & = & -a(1-\muk^2)
\Gamma^{\tk}_{\tk\muk} \\[0.9mm]
\nonumber
\Gamma^{\tk}_{\phik\phik} & = & (1-\muk^2)(1+a^2 \uk^2)[a^2 \uk^3(1-\muk^2)
(1-a^2 \uk^2 \muk^2) - \rhok^4]/(\uk\rhok^6)  \hspace*{-20cm} &  & \\[0.9mm]
\nonumber
\Gamma^{\uk}_{\tk\tk} & = & -\uk^4(1-a^2 \uk^2\muk^2)\Deltak/\rhok^6
 \hspace*{-20cm} &  & \\[0.9mm]
\nonumber
\Gamma^{\uk}_{\tk \uk} & = & \Gamma^{\uk}_{\uk\tk} & = &
-\uk^2(1-a^2 \uk^2\muk^2)/\rhok^4 \\[0.9mm]
\nonumber
\Gamma^{\uk}_{\tk\muk} & = & \Gamma^{\uk}_{\muk \tk} & = & 0 \\[0.9mm]
\nonumber
\Gamma^{\uk}_{\tk\phik} & = & \Gamma^{\uk}_{\phik \tk} & = & -a(1-\muk^2)
\Gamma^{\uk}_{\tk\tk} \\[0.9mm]
\nonumber
\Gamma^{\uk}_{\uk\uk} & = & -2/\uk  \hspace*{-20cm} &  & \\[0.9mm]
\nonumber
\Gamma^{\uk}_{\uk\muk} & = & \Gamma^{\uk}_{\muk \uk} & = &
a^2 \uk^2 \muk/\rhok^2 \\[0.9mm]
\nonumber
\Gamma^{\uk}_{\uk\phik} & = & \Gamma^{\uk}_{\phik \uk} & = & a\uk(1-\muk^2)
[\uk(1-a^2 \uk^2 \muk^2) + \rhok^2]/\rhok^4 \\[0.9mm]
\nonumber
\Gamma^{\uk}_{\muk\muk} & = & \uk\Deltak/[(1-\muk^2)\rhok^2]
\hspace*{-20cm} &  & \\[0.9mm]
\nonumber
\Gamma^{\uk}_{\muk\phik} & = & \Gamma^{\uk}_{\phik\muk} & = & 0 \\[0.9mm]
\nonumber
\Gamma^{\uk}_{\phik\phik} & = & \uk(1-\muk^2)\Deltak[\rhok^4 - a^2 \uk^3
(1-\muk^2)(1-a^2\uk^2\muk^2)]/\rhok^6  \hspace*{-20cm} &  & \\[0.9mm]
\nonumber
\Gamma^{\muk}_{\tk\tk} & = & 2a^2 \uk^5\muk(1-\muk^2)/\rhok^6
\hspace*{-20cm} &  & \\[0.9mm]
\nonumber
\Gamma^{\muk}_{\tk \uk} & = & \Gamma^{\muk}_{\uk\tk} & = & 0 \\[0.9mm]
\nonumber
\Gamma^{\muk}_{\tk\muk} & = & \Gamma^{\muk}_{\muk \tk} & = & 0 \\[0.9mm]
\nonumber
\Gamma^{\muk}_{\tk\phik} & = & \Gamma^{\muk}_{\phik \tk} & = & -2a\uk^3\muk
(1-\muk^2)(1+a^2 \uk^2)/\rhok^6 \\[0.9mm]
\nonumber
\Gamma^{\muk}_{\uk\uk} & = & 0  \hspace*{-20cm} &  & \\[0.9mm]
\nonumber
\Gamma^{\muk}_{\uk\muk} & = & \Gamma^{\muk}_{\muk \uk} & = &
-1/(\uk\rhok^2) \\[0.9mm]
\nonumber
\Gamma^{\muk}_{\uk\phik} & = & \Gamma^{\muk}_{\phik \uk} & = &
a\muk(1-\muk^2)/\rhok^2 \\[0.9mm]
\nonumber
\Gamma^{\muk}_{\muk\muk} & = & \muk(1+a^2 \uk^2)/[(1-\muk^2)\rhok^2]
 \hspace*{-20cm} &  & \\[0.9mm]
\nonumber
\Gamma^{\muk}_{\muk\phik} & = & \Gamma^{\muk}_{\phik\muk} & = & 0 \\[0.9mm]
\nonumber
\Gamma^{\muk}_{\phik\phik} & = & \muk(1+a^2 \uk^2)(1-\muk^2)/\rhok^2 + 2a^2
\uk^3\muk(1-\muk^2)^2(1+a^2 \uk^2+\rhok^2)/\rhok^6  \hspace*{-20cm} &  & \\[0.9mm]
\nonumber
\Gamma^{\phik}_{\tk\tk} & = & a\uk^4(1-a^2 \uk^2 \muk^2)/\rhok^6
\hspace*{-20cm} &  & \\[0.9mm]
\nonumber
\Gamma^{\phik}_{\tk \uk} & = & \Gamma^{\phik}_{\uk\tk} & = & 0 \\[0.9mm]
\nonumber
\Gamma^{\phik}_{\tk\muk} & = & \Gamma^{\phik}_{\muk \tk} & = & 2a\uk^3 \muk/
[(1-\muk^2)\rhok^4] \\[0.9mm]
\nonumber
\Gamma^{\phik}_{\tk\phik} & = & \Gamma^{\phik}_{\phik \tk} & = & -a(1-\muk^2)
\Gamma^{\phik}_{\tk\tk} \\[0.9mm]
\nonumber
\Gamma^{\phik}_{\uk\uk} & = & 0  \hspace*{-20cm} &  & \\[0.9mm]
\nonumber
\Gamma^{\phik}_{\uk\muk} & = & \Gamma^{\phik}_{\muk \uk} & = & -a\muk/
[(1-\muk^2)\rhok^2] \\[0.9mm]
\nonumber
\Gamma^{\phik}_{\uk\phik} & = & \Gamma^{\phik}_{\phik \uk} & = &
\Gamma^{\muk}_{\uk \muk} \\[0.9mm]
\nonumber
\Gamma^{\phik}_{\muk\muk} & = & -a\uk/[(1-\muk^2)\rhok^2]
\hspace*{-20cm} &  & \\[0.9mm]
\nonumber
\Gamma^{\phik}_{\muk\phik} & = & \Gamma^{\phik}_{\phik\muk} & = &
-\muk/(1-\muk^2) - 2a^2 \uk^3\muk/\rhok^4 \\[0.9mm]
\nonumber
\Gamma^{\phik}_{\phik\phik} & = & a\uk(1-\muk^2)[a^2 \uk^3(1-\muk^2)
(1-a^2 \uk^2 \muk^2) - \rhok^4]/\rhok^6 \hspace*{-20cm} &  & \\[\smallskipamount]
\nonumber
\Gamma^{\tk}_{\tk\tk,\uk} & = & 2\uk[\rhok^4 + 2a^2 \uk^2(1-\muk^2) - 4a^2 \uk^2
\muk^2(1 + a^2 \uk^2)]/\rhok^8 \hspace*{-20cm} &  & \\[0.9mm]
\nonumber
\Gamma^{\tk}_{\tk \uk,\uk} & = & \Gamma^{\tk}_{\uk\tk,\uk} & = & 0 \\[0.9mm]
\nonumber
\Gamma^{\tk}_{\tk\muk,\uk} & = & \Gamma^{\tk}_{\muk \tk,\uk} & = &
-2a^2 \uk^2\muk(a^2 \uk^2\muk^2-3)/\rhok^6 \\[0.9mm]
\nonumber
\Gamma^{\tk}_{\tk\phik,\uk} & = & \Gamma^{\tk}_{\phik \tk,\uk} & = &
-a(1-\muk^2)\Gamma^{\tk}_{\tk\tk,\uk} \\[0.9mm]
\nonumber
\Gamma^{\tk}_{\uk\uk,\uk} & = & 0 \hspace*{-20cm} &  & \\[0.9mm]
\nonumber
\Gamma^{\tk}_{\uk\muk,\uk} & = & \Gamma^{\tk}_{\muk \uk,\uk} & = &
2a^4 \uk \muk^3/\rhok^4 \\[0.9mm]
\nonumber
\Gamma^{\tk}_{\uk\phik,\uk} & = & \Gamma^{\tk}_{\phik \uk,\uk} & = &
a(1-\muk^2)(1+3a^2 \uk^2 \muk^2)/(\uk^2\rhok^4) \\[0.9mm]
\nonumber
\Gamma^{\tk}_{\muk\muk,\uk} & = & [1+3a^2 \uk^2 \muk^2-a^2 \uk^2(1 - a^2 \uk^2
\muk^2)]/[\uk^2(1-\muk^2)\rhok^4] \hspace*{-20cm} &  & \\[0.9mm]
\nonumber
\Gamma^{\tk}_{\muk\phik,\uk} & = & \Gamma^{\tk}_{\phik\muk,\uk} & = &
-a(1-\muk^2)\Gamma^{\tk}_{\tk\muk,\uk} \\[0.9mm]
\nonumber
\Gamma^{\tk}_{\phik\phik,\uk} & = & (1-\muk^2)\{\rhok^4[1+3a^2\uk^2\muk^2-a^2
\uk^2(1-a^2\uk^2\muk^2)] \hspace*{-20cm} &  & \\
\nonumber
& & + 2a^2\uk^3(1-\muk^2)[\rhok^4+2a^2\uk^2(1-2a^2\uk^2\muk^2
-3\muk^2)]\}/(\uk^2\rhok^8) \hspace*{-20cm} &  & \\[0.9mm]
\nonumber
\Gamma^{\uk}_{\tk\tk,\uk} & = &
2\uk^3\{\uk[6(1-a^2\uk^2\muk^2)-\rhok^4]-2(1+a^2\uk^2)(1-a^2\uk^2\muk^2)
\hspace*{-20cm} &  & \\
\nonumber
& & + a^2\uk^2[\rhok^4-2(1-\muk^2)]\}/\rhok^8 \hspace*{-20cm} &  & \\[0.9mm]
\nonumber
\Gamma^{\uk}_{\tk \uk,\uk} & = & \Gamma^{\uk}_{\uk\tk,\uk} & = &
-2\uk(1-3a^2\uk^2\muk^2)/\rhok^6 \\[0.9mm]
\nonumber
\Gamma^{\uk}_{\tk\muk,\uk} & = & \Gamma^{\uk}_{\muk \tk,\uk} & = & 0 \\[0.9mm]
\nonumber
\Gamma^{\uk}_{\tk\phik,\uk} & = & \Gamma^{\uk}_{\phik \tk,\uk} & = &
-a(1-\muk^2)\Gamma^{\uk}_{\tk\tk,\uk} \\[0.9mm]
\nonumber
\Gamma^{\uk}_{\uk\uk,\uk} & = & 2/\uk^2 \hspace*{-20cm} &  & \\[0.9mm]
\nonumber
\Gamma^{\uk}_{\uk\muk,\uk} & = & \Gamma^{\uk}_{\muk \uk,\uk} & = &
2a^2\uk\muk/\rhok^4 \\[0.9mm]
\nonumber
\Gamma^{\uk}_{\uk\phik,\uk} & = & \Gamma^{\uk}_{\phik \uk,\uk} & = &
a(1-\muk^2)[2\uk(1-3a^2\uk^2\muk^2)+\rhok^2(1-a^2\uk^2\muk^2)]/\rhok^6 \\[0.9mm]
\nonumber
\Gamma^{\uk}_{\muk\muk,\uk} & = & [2\Deltak-\rhok^2(1-a^2\uk^2)]/[(1-\muk^2)
\rhok^4] \hspace*{-20cm} &  & \\[0.9mm]
\nonumber
\Gamma^{\uk}_{\muk\phik,\uk} & = & \Gamma^{\uk}_{\phik\muk,\uk} & = & 0 \\[0.9mm]
\nonumber
\Gamma^{\uk}_{\phik\phik,\uk} & = &
-(1-\muk^2)\{\rhok^6(1-a^2\uk^2)-2\rhok^4(1+a^2\uk^2)+2a^2\uk^4(1-\muk^2)
[\rhok^4-6(1-a^2\uk^2\muk^2)]\hspace*{-20cm} &  & \\
\nonumber
& & + 2\uk\llbracket\rhok^4+(1+a^2\uk^2)^2
+a^4\uk^4(1-\muk^2)[5(1-\muk^2)-\rhok^2(3+a^2\uk^2\muk^2)]\rrbracket\}/
\rhok^8 \hspace*{-20cm} &  & \\[0.9mm]
\nonumber
\Gamma^{\muk}_{\tk\tk,\uk} & = & -2a^2\uk^4\muk(1-\muk^2)(a^2\uk^2\muk^2-5)/
\rhok^8 \hspace*{-20cm} &  & \\[0.9mm]
\nonumber
\Gamma^{\muk}_{\tk \uk,\uk} & = & \Gamma^{\muk}_{\uk\tk,\uk} & = & 0 \\[0.9mm]
\nonumber
\Gamma^{\muk}_{\tk\muk,\uk} & = & \Gamma^{\muk}_{\muk \tk,\uk} & = & 0 \\[0.9mm]
\nonumber
\Gamma^{\muk}_{\tk\phik,\uk} & = & \Gamma^{\muk}_{\phik \tk,\uk} & = &
2a\uk^2\muk(1-\muk^2)[a^2\uk^2(a^2\uk^2\muk^2-5)-3(1-a^2\uk^2\muk^2)]/
\rhok^8 \\[0.9mm]
\nonumber
\Gamma^{\muk}_{\uk\uk,\uk} & = & 0 \hspace*{-20cm} &  & \\[0.9mm]
\nonumber
\Gamma^{\muk}_{\uk\muk,\uk} & = & \Gamma^{\muk}_{\muk \uk,\uk} & = &
(1+3a^2\uk^2\muk^2)/(\uk^2\rhok^4) \\[0.9mm]
\nonumber
\Gamma^{\muk}_{\uk\phik,\uk} & = & \Gamma^{\muk}_{\phik \uk,\uk} & = &
-2a^3\uk\muk^3(1-\muk^2)/\rhok^4 \\[0.9mm]
\nonumber
\Gamma^{\muk}_{\muk\muk,\uk} & = & 2a^2\uk\muk/\rhok^4
\hspace*{-20cm} &  & \\[0.9mm]
\nonumber
\Gamma^{\muk}_{\muk\phik,\uk} & = & \Gamma^{\muk}_{\phik\muk,\uk} & = & 0
\\[0.9mm]
\nonumber
\Gamma^{\muk}_{\phik\phik,\uk} & = &
2a^2\uk\muk(1-\muk^2)^2\{\rhok^4+\uk[6(1+a^2\uk^2)-a^2\uk^2(1+\muk^2)\rhok^2]\}/
\rhok^8 \hspace*{-20cm} &  & \\[0.9mm]
\nonumber
\Gamma^{\phik}_{\tk\tk,\uk} & = & 4a\uk^3(1-2a^2\uk^2\muk^2)/\rhok^8
\hspace*{-20cm} &  & \\[0.9mm]
\nonumber
\Gamma^{\phik}_{\tk \uk,\uk} & = & \Gamma^{\phik}_{\uk\tk,\uk} & = & 0 \\[0.9mm]
\nonumber
\Gamma^{\phik}_{\tk\muk,\uk} & = & \Gamma^{\phik}_{\muk \tk,\uk} & = &
-2a\uk^2\muk(a^2\uk^2\muk^2-3)/[(1-\muk^2)\rhok^6] \\[0.9mm]
\nonumber
\Gamma^{\phik}_{\tk\phik,\uk} & = & \Gamma^{\phik}_{\phik \tk,\uk} & = &
-a(1-\muk^2)\Gamma^{\phik}_{\tk\tk,\uk} \\[0.9mm]
\nonumber
\Gamma^{\phik}_{\uk\uk,\uk} & = & 0 \hspace*{-20cm} &  & \\[0.9mm]
\nonumber
\Gamma^{\phik}_{\uk\muk,\uk} & = & \Gamma^{\phik}_{\muk \uk,\uk} & = &
2a^3\uk\muk^3/[(1-\muk^2)\rhok^4] \\[0.9mm]
\nonumber
\Gamma^{\phik}_{\uk\phik,\uk} & = & \Gamma^{\phik}_{\phik \uk,\uk} & = &
\Gamma^{\muk}_{\uk\muk,\uk} \\[0.9mm]
\nonumber
\Gamma^{\phik}_{\muk\muk,\uk} & = & -a(1-a^2\uk^2\muk^2)/[(1-\muk^2)
\rhok^4] \hspace*{-20cm} &  & \\[0.9mm]
\nonumber
\Gamma^{\phik}_{\muk\phik,\uk} & = & \Gamma^{\phik}_{\phik\muk,\uk} & = &
2a^2\uk^2\muk(a^2\uk^2\muk^2-3)/\rhok^6 \\[0.9mm]
\nonumber
\Gamma^{\phik}_{\phik\phik,\uk} & = &
a(1-\muk^2)[\rhok^4(a^2\uk^2\muk^2-1)+4a^2\uk^3(1-\muk^2)(1-2a^2\uk^2\muk^2)]/
\rhok^8 \hspace*{-20cm} &  & \\[\smallskipamount]
\nonumber
\Gamma^{\tk}_{\tk\tk,\muk} & = & 4a^2\uk^4\muk(1+a^2\uk^2)(a^2\uk^2\muk^2-2)/
\rhok^8 \hspace*{-20cm} &  & \\[0.9mm]
\nonumber
\Gamma^{\tk}_{\tk \uk,\muk} & = & \Gamma^{\tk}_{\uk\tk,\muk} & = & 0 \\[0.9mm]
\nonumber
\Gamma^{\tk}_{\tk\muk,\muk} & = & \Gamma^{\tk}_{\muk \tk,\muk} & = &
2a^2\uk^3(1-3a^2\uk^2\muk^2)/\rhok^6 \\[0.9mm]
\nonumber
\Gamma^{\tk}_{\tk\phik,\muk} & = & \Gamma^{\tk}_{\phik \tk,\muk} & = &
2a\muk\Gamma^{\tk}_{\tk\tk}-a(1-\muk^2)\Gamma^{\tk}_{\tk\tk,\muk} \\[0.9mm]
\nonumber
\Gamma^{\tk}_{\uk\uk,\muk} & = & 0 \hspace*{-20cm} &  & \\[0.9mm]
\nonumber
\Gamma^{\tk}_{\uk\muk,\muk} & = & \Gamma^{\tk}_{\muk \uk,\muk} & = &
-a^2(1-a^2\uk^2\muk^2)/\rhok^4 \\[0.9mm]
\nonumber
\Gamma^{\tk}_{\uk\phik,\muk} & = & \Gamma^{\tk}_{\phik \uk,\muk} & = &
2a\muk(1+a^2\uk^2)/(\uk\rhok^4) \\[0.9mm]
\nonumber
\Gamma^{\tk}_{\muk\muk,\muk} & = & 2\muk(1+a^2\uk^2)[a^2\uk^2(1-\muk^2)
-\rhok^2]/[\uk(1-\muk^2)^2\rhok^4] \hspace*{-20cm} &  & \\[0.9mm]
\nonumber
\Gamma^{\tk}_{\muk\phik,\muk} & = & \Gamma^{\tk}_{\phik\muk,\muk} & = &
2a\muk\Gamma^{\tk}_{\tk\muk}-a(1-\muk^2)\Gamma^{\tk}_{\tk\muk,\muk} \\[0.9mm]
\nonumber
\Gamma^{\tk}_{\phik\phik,\muk} & = &
-2\muk(1+a^2\uk^2)\{2a^2\uk^3(1-\muk^2)[(1+a^2\uk^2)(1-a^2\uk^2\muk^2)
+a^2\uk^2(1-\muk^2)]\hspace*{-20cm} &  & \\
\nonumber
& & - (1+a^2\uk^2)\rhok^4\}/(\uk\rhok^8) \hspace*{-20cm} &  & \\[0.9mm]
\nonumber
\Gamma^{\uk}_{\tk\tk,\muk} & = & 4a^2\uk^6\muk\Deltak(2-a^2\uk^2\muk^2)/
\rhok^8 \hspace*{-20cm} &  & \\[0.9mm]
\nonumber
\Gamma^{\uk}_{\tk \uk,\muk} & = & \Gamma^{\uk}_{\uk\tk,\muk} & = &
-2a^2\uk^4\muk(a^2\uk^2\muk^2-3)/\rhok^6 \\[0.9mm]
\nonumber
\Gamma^{\uk}_{\tk\muk,\muk} & = & \Gamma^{\uk}_{\muk \tk,\muk} & = & 0 \\[0.9mm]
\nonumber
\Gamma^{\uk}_{\tk\phik,\muk} & = & \Gamma^{\uk}_{\phik \tk,\muk} & = &
2a\muk\Gamma^{\uk}_{\tk\tk}-a(1-\muk^2)\Gamma^{\uk}_{\tk\tk,\muk} \\[0.9mm]
\nonumber
\Gamma^{\uk}_{\uk\uk,\muk} & = & 0 \hspace*{-20cm} &  & \\[0.9mm]
\nonumber
\Gamma^{\uk}_{\uk\muk,\muk} & = & \Gamma^{\uk}_{\muk \uk,\muk} & = &
a^2\uk^2(1-a^2\uk^2\muk^2)/\rhok^4 \\[0.9mm]
\nonumber
\Gamma^{\uk}_{\uk\phik,\muk} & = & \Gamma^{\uk}_{\phik \uk,\muk} & = &
-2a\uk\muk\{\rhok^2(1+a^2\uk^2)+\uk[(1+a^2\uk^2)(1-a^2\uk^2\muk^2)\\
\nonumber
& & & & + 2a^2\uk^2(1-\muk^2)]\}/\rhok^6 \\[0.9mm]
\nonumber
\Gamma^{\uk}_{\muk\muk,\muk} & = & 2\uk\muk\Deltak[\rhok^2-a^2\uk^2(1-\muk^2)]
/[(1-\muk^2)^2\rhok^4] \hspace*{-20cm} &  & \\[0.9mm]
\nonumber
\Gamma^{\uk}_{\muk\phik,\muk} & = & \Gamma^{\uk}_{\phik\muk,\muk} & = & 0
\\[0.9mm]
\nonumber
\Gamma^{\uk}_{\phik\phik,\muk} & = &
2\uk\muk\Deltak\{-(1+a^2\uk^2)\rhok^4+2a^2\uk^3(1-\muk^2)[(1+a^2\uk^2)
(1-a^2\uk^2\muk^2)\hspace*{-20cm} &  & \\
\nonumber
& & + a^2\uk^2(1-\muk^2)]\}/\rhok^8 \hspace*{-20cm} &  & \\[0.9mm]
\nonumber
\Gamma^{\muk}_{\tk\tk,\muk} & = & 2a^2\uk^5[(1-3a^2\uk^2\muk^2)(1-\muk^2)
-2\muk^2(1+a^2\uk^2)]/\rhok^8 \hspace*{-20cm} &  & \\[0.9mm]
\nonumber
\Gamma^{\muk}_{\tk \uk,\muk} & = & \Gamma^{\muk}_{\uk\tk,\muk} & = & 0 \\[0.9mm]
\nonumber
\Gamma^{\muk}_{\tk\muk,\muk} & = & \Gamma^{\muk}_{\muk \tk,\muk} & = & 0 \\[0.9mm]
\nonumber
\Gamma^{\muk}_{\tk\phik,\muk} & = & \Gamma^{\muk}_{\phik \tk,\muk} & = &
-2a\uk^3(1+a^2\uk^2)[(1-3a^2\uk^2\muk^2)(1-\muk^2)-2\muk^2(1+a^2\uk^2)]/
\rhok^8 \\[0.9mm]
\nonumber
\Gamma^{\muk}_{\uk\uk,\muk} & = & 0 \hspace*{-20cm} &  & \\[0.9mm]
\nonumber
\Gamma^{\muk}_{\uk\muk,\muk} & = & \Gamma^{\muk}_{\muk \uk,\muk} & = &
 2a^2\uk\muk/\rhok^4 \\[0.9mm]
\nonumber
\Gamma^{\muk}_{\uk\phik,\muk} & = & \Gamma^{\muk}_{\phik \uk,\muk} & = &
a[\rhok^2(1-\muk^2)-2\muk^2(1+a^2\uk^2)]/\rhok^4 \\[0.9mm]
\nonumber
\Gamma^{\muk}_{\muk\muk,\muk} & = &
(1+a^2\uk^2)[(1-3a^2\uk^2\muk^2)(1-\muk^2)+2\muk^2(1+a^2\uk^2)]/
[(1-\muk^2)^2\rhok^4] \hspace*{-20cm} &  & \\[0.9mm]
\nonumber
\Gamma^{\muk}_{\muk\phik,\muk} & = & \Gamma^{\muk}_{\phik\muk,\muk} & = & 0
\\[0.9mm]
\nonumber
\Gamma^{\muk}_{\phik\phik,\muk} & = &
-1+[(a^2\uk^2+2)(1-a^2\uk^2\muk^2)-\muk^2(3+a^2\uk^2\muk^2)]/\rhok^4
\hspace*{-20cm} &  & \\
\nonumber
& & + 2a^2\uk^3(1-\muk^2)\{(a^2\uk^2+2)[6-\rhok^2(2\muk^2+5)]
-\muk^2(6-\rhok^2+\rhok^4)\}/\rhok^8 \hspace*{-20cm} &  & \\[0.9mm]
\nonumber
\Gamma^{\phik}_{\tk\tk,\muk} & = & 4a^3\uk^6\muk(a^2\uk^2\muk^2-2)/\rhok^8
\hspace*{-20cm} &  & \\[0.9mm]
\nonumber
\Gamma^{\phik}_{\tk \uk,\muk} & = & \Gamma^{\phik}_{\uk\tk,\muk} & = & 0 \\[0.9mm]
\nonumber
\Gamma^{\phik}_{\tk\muk,\muk} & = & \Gamma^{\phik}_{\muk \tk,\muk} & = &
2a\uk^3[(1-3a^2\uk^2\muk^2)(1-\muk^2)+2\muk^2\rhok^2]/[(1-\muk^2)^2\rhok^6]
\\[0.9mm]
\nonumber
\Gamma^{\phik}_{\tk\phik,\muk} & = & \Gamma^{\phik}_{\phik \tk,\muk} & = &
2a\muk\Gamma^{\phik}_{\tk\tk}-a(1-\muk^2)\Gamma^{\phik}_{\tk\tk,\muk} \\[0.9mm]
\nonumber
\Gamma^{\phik}_{\uk\uk,\muk} & = & 0 \hspace*{-20cm} &  & \\[0.9mm]
\nonumber
\Gamma^{\phik}_{\uk\muk,\muk} & = & \Gamma^{\phik}_{\muk \uk,\muk} & = &
-a[(1-a^2\uk^2\muk^2)(1-\muk^2)+2\muk^2\rhok^2]/[(1-\muk^2)^2\rhok^4] \\[0.9mm]
\nonumber
\Gamma^{\phik}_{\uk\phik,\muk} & = & \Gamma^{\phik}_{\phik \uk,\muk} & = &
\Gamma^{\muk}_{\uk\muk,\muk} \\[0.9mm]
\nonumber
\Gamma^{\phik}_{\muk\muk,\muk} & = & -2a\uk\muk[\rhok^2-a^2\uk^2(1-\muk^2)]/
[(1-\muk^2)^2\rhok^4] \hspace*{-20cm} &  & \\[0.9mm]
\nonumber
\Gamma^{\phik}_{\muk\phik,\muk} & = & \Gamma^{\phik}_{\phik\muk,\muk} & = &
-(1+\muk^2)/(1-\muk^2)^2-2a^2\uk^3(1-3a^2\uk^2\muk^2)/\rhok^6] \\[0.9mm]
\nonumber
\Gamma^{\phik}_{\phik\phik,\muk} & = &
2a\uk\muk\{\rhok^4(1+a^2\uk^2)-2a^2\uk^3(1-\muk^2)[a^2\uk^2(1-\muk^2)
\hspace*{-20cm} &  & \\
\nonumber
& & + (1+a^2\uk^2)(1-a^2\uk^2\muk^2)]\}/\rhok^8\hspace*{-20cm} &  &
\end{myeqnarray}

\chapter{Description of FITS files}
\label{fits}
 \thispagestyle{empty}
   \section{Transfer functions in
{\fontfamily{pcr}\fontshape{tt}\selectfont KBHtablesNN.fits}}
\label{appendix3a}

The transfer functions are stored in the file {\tt KBHtablesNN.fits} as binary
extensions and parametrized by the value of the observer inclination angle
$\theta_{\rm o}$ and the horizon of the black hole $r_{\rm h}$. We found
parametrization by  $r_{\rm h}$ more convenient than using the rotational
parameter $a$, although the latter choice may be more common. Each
extension provides values of a particular transfer function for different
radii, which are given in terms of $r-r_{\rm h}$, and for the Kerr
ingoing axial coordinates $\varphi_{\rm K}$. Values of the horizon
$r_{\rm h}$, inclination $\theta_{\rm o}$, radius $r-r_{\rm h}$ and
angle $\varphi_{\rm K}$, at which the functions are evaluated, are defined  as
vectors at the beginning of the FITS file.

The definition of the file {\tt KBHtablesNN.fits}:%
\begin{enumerate} \itemsep -0.4em
\vspace*{-1.5em}
 \item[0.] All of the extensions defined below are binary.
 \item The first extension contains six integers defining which of the
   functions is present in the tables. The integers correspond to the delay,
   $g$-factor, cosine of the local emission angle, lensing, change of the
   polarization angle
   and azimuthal emission angle, respectively. Value $0$
   means that the function is not present in the tables, value $1$ means it is.
 \item The second extension contains a vector of the horizon values in $GM/c^2$
   ($1.00 \le r_{\rm h} \le 2.00$).
 \item The third extension contains a vector of the values of the observer's
   inclination angle $\theta_{\rm o}$ in degrees
   ($0^\circ \le \theta_{\rm o} \le 90^\circ$, $0^\circ$ -- axis, $90^\circ$
   -- equatorial plane).
 \item The fourth extension contains a vector of the values of the radius
 relative to the horizon $r-r_{\rm h}$ in $GM/c^2$.
 \item The fifth extension contains a vector of the values of the azimuthal
 angle $\varphi_{\rm K}$ in radians ($0 \le \varphi_{\rm K} \le 2\pi$).
 Note that $\varphi_{\rm K}$ is a Kerr ingoing axial coordinate, not the
 Boyer-Lindquist one!
 \item All the previous vectors have to have values sorted in an increasing
 order.
 \item In the following extensions the transfer functions are defined, each
 extension is for a particular value of $r_{\rm h}$ and $\theta_{\rm o}$.
 The values of $r_{\rm h}$ and $\theta_{\rm o}$ are changing with each
 extension in the following order:\\ \hspace*{-2.6em}
    \parbox{\textwidth}{
    {\begin{centering}
    $r_{\rm h}[1] \times \theta_{\rm o}[1]$,\\
    $r_{\rm h}[1] \times \theta_{\rm o}[2]$,\\
    $r_{\rm h}[1] \times \theta_{\rm o}[3]$,\\
    \dots \\
    \dots \\
    $r_{\rm h}[2] \times \theta_{\rm o}[1]$,\\
    $r_{\rm h}[2] \times \theta_{\rm o}[2]$,\\
    $r_{\rm h}[2] \times \theta_{\rm o}[3]$,\\
    \dots \\
    \dots \\
    \end{centering}}}\\
 \item Each of these extensions has the same number of columns (up to six).
 In each column, a particular transfer function is stored -- the delay,
 $g$-factor, cosine of the local emission angle, lensing, change of the
 polarization angle and azimuthal emission angle, respectively.
 The order of the functions is important
 but some of the functions may be missing as defined in the first extension
 (see 1.\ above). The functions are:
   \begin{description} \itemsep -2pt
	 \vspace*{-0.2em}
    \item[\rm delay] -- the Boyer-Lindquist time in $GM/c^3$ that elapses
		between the emission of a photon from the disc and absorption of the
		photon by the observer's eye at infinity plus a constant,
    \item[\rm $g$-factor] -- the ratio of the energy of a photon received by the
    observer at infinity to the local energy of the same photon when emitted
              from an accretion disc,
    \item[\rm cosine of the emission angle] -- the cosine of the local emission
		angle between the emitted light ray and local disc normal,
    \item[\rm lensing] -- the ratio of the area at infinity perpendicular to the
    light rays through which photons come to the proper area at the disc
		perpendicular to the light rays and corresponding to the same flux tube,
    \item[\rm change of the polarization angle in radians] --  if the light
		emitted     from the disc is linearly polarized then the direction of
		polarization will be changed by this angle at infinity --
    counter-clockwise if positive, clockwise if negative (we are looking
    towards the coming emitted beam); on the disc we measure the angle of
    polarization with respect to the ``up'' direction perpendicular to the
    disc with respect to the local rest frame; at infinity we also measure the
    angle of polarization with respect to the ``up'' direction perpendicular to
    the disc -- the change of polarization angle is the difference between
    these two angles,
    \item[\rm azimuthal emission angle in radians] -- the angle between the
    projection of the three-momentum of an emitted photon into the disc (in the
    local rest frame co-moving with the disc) and the radial tetrad vector.
   \end{description}
	 For mathematical formulae defining the
functions see eqs.~(\ref{gfac})--(\ref{lensing}),
(\ref{delay1})--(\ref{polar}) and (\ref{azim_angle}) in
Chapter~\ref{transfer_functions}.
 \item Each row corresponds to a particular value of $r-r_{\rm h}$
 (see 4.\ above).
 \item Each element corresponding to a particular column and row is a vector.
 Each element of this vector corresponds to a particular value of
 $\varphi_{\rm K}$ (see 5.\ above).
\end{enumerate}

We have pre-calculated three sets of tables -- {\tt KBHtables00.fits},
{\tt KBHtables50.fits} and {\tt KBHtables99.fits}.
All of these tables were computed for an accretion disc near a Kerr black
hole with no disc corona present. Therefore, ray-tracing in the vacuum Kerr
space-time could be used for calculating the transfer functions.
When computing the transfer functions, it was supposed that the matter in
the disc rotates on stable circular (free) orbits above the marginally stable
orbit. The matter below this orbit is freely falling and has the same energy and
angular momentum as the matter which is on the marginally stable orbit.

The observer is placed in the direction $\varphi = \pi/2$. The black hole
rotates counter-clockwise. All six functions are present in these tables.

Tables are calculated for these values of the black-hole horizon:\\
-- {\tt KBHtables00.fits}: 1.00, 1.05, 1.10, 1.15, \dots, 1.90, 1.95, 2.00
(21 elements),\\
-- {\tt KBHtables50.fits}: 1.00, 1.10, 1.20, \dots, 1.90, 2.00 (11 elements),\\
-- {\tt KBHtables99.fits}: 1.05 (1 element),\\
and for these values of the observer's inclination:\\
-- {\tt KBHtables00.fits}: 0.1, 1, 5, 10, 15, 20, \dots, 80, 85, 89
(20 elements),\\
-- {\tt KBHtables50.fits}: 0.1, 1, 10, 20, \dots, 80, 89 (11 elements),\\
-- {\tt KBHtables99.fits}: 0.1, 1, 5, 10, 15, 20, \dots, 80, 85, 89
(20 elements).

The radii and azimuths at which the functions are evaluated are same for all
three tables:\\
-- radii $r-r_{\rm h}$ are exponentially increasing from 0 to 999
(150 elements),\\
-- values of the azimuthal angle $\varphi_{\rm K}$ are equidistantly spread
from 0 to $2\pi$ radians with a much denser cover ``behind'' the black hole,
i.e.\ near $\varphi_{\rm K} = 1.5\pi$
(because some of the functions are changing heavily in this area for higher
inclination angles, $\theta_{\rm o} > 70^\circ$) (200 elements).

\section{Tables in
{\fontfamily{pcr}\fontshape{tt}\selectfont KBHlineNN.fits}}
\label{appendix3b}
Pre-calculated functions ${\rm d}F(g)\equiv{\rm d}g\,F(g)$ defined in
the eq.~(\ref{conv_function}) are stored in FITS files
{\tt KBHlineNN.fits}. These functions are used by all axisymmetric models.
They are stored as binary extensions and they are parametrized by the value of
the observer inclination angle
$\theta_{\rm o}$ and the horizon of the black hole $r_{\rm h}$. Each
extension provides values for different
radii, which are given in terms of $r-r_{\rm h}$, and for different
$g$-factors. Values of the $g$-factor, radius $r-r_{\rm h}$, horizon
$r_{\rm h}$, and inclination $\theta_{\rm o}$, at which the functions are
evaluated, are defined  as vectors at the beginning of the FITS file.

The definition of the file {\tt KBHlineNN.fits}:%
\begin{enumerate} \itemsep -2pt
\vspace*{-0.5em}
 \item[0.] All of the extensions defined below are binary.
 \item The first extension contains one row with three columns that define
 bins in the $g$-factor:
  \begin{list}{--}{\setlength{\topsep}{-2pt}\setlength{\itemsep}{-2pt}
	                 \setlength{\leftmargin}{1em}}
	 \item integer in the first column defines the width of the bins
	 (0 -- constant, 1 -- exponentially growing),
   \item real number in the second column defines the lower boundary of the
	 first bin (minimum of the $g$-factor),
   \item real number in the third column defines the upper boundary of the
	 last bin (maximum of the $g$-factor).
	\end{list}
 \item The second extension contains a vector of the values of the radius
 relative to the horizon $r-r_{\rm h}$ in $GM/c^2$.
 \item The third extension contains a vector of the horizon values in $GM/c^2$
   ($1.00 \le r_{\rm h} \le 2.00$).
 \item The fourth extension contains a vector of the values of the observer's
   inclination angle $\theta_{\rm o}$ in degrees
   ($0^\circ \le \theta_{\rm o} \le 90^\circ$, $0^\circ$ -- axis, $90^\circ$
   -- equatorial plane).
 \item All the previous vectors have to have values sorted in an increasing
 order.
 \item In the following extensions the functions ${\rm d}F(g)$ are defined, each
 extension is for a particular value of $r_{\rm h}$ and $\theta_{\rm o}$.
 The values of $r_{\rm h}$ and $\theta_{\rm o}$ are changing with each
 extension in the same order as in tables in the {\tt KBHtablesNN.fits} file
 (see the previous section, point 7.). Each extension has one column.
 \item Each row corresponds to a particular value of $r-r_{\rm h}$
 (see 2.\ above).
 \item Each element corresponding to a particular column and row is a vector.
 Each element of this vector corresponds to a value of the function
 for a particular bin in the $g$-factor.
 This bin can be calculated from number of elements of the vector and data
 from the first extension (see 1.\ above).
\end{enumerate}

We have pre-calculated several sets of tables for different limb
darkening/brightening laws and with different resolutions. All of them were
calculated from tables in the {\tt KBHtables00.fits} file (see the previous
section for details) and therefore these tables are calculated for the same
values of the black-hole horizon and observer's inclination. All of these tables
have equidistant bins in the $g$-factor which fall in the interval
$\langle0.001,1.7 \rangle$. Several sets of tables are available:\\
-- {\tt KBHline00.fits} for isotropic emission, see eq.~(\ref{isotropic}),\\
-- {\tt KBHline01.fits} for Laor's limb darkening, see eq.~(\ref{laor}),\\
-- {\tt KBHline02.fits} for Haardt's limb brightening, see eq.~(\ref{haardt}).\\
All of these tables have 300 bins in the $g$-factor and 500 values of the radius
$r-r_{\rm h}$ which are exponentially increasing from 0 to 999.
We have produced also tables
with a lower resolution -- {\tt KBHline50.fits}, {\tt KBHline51.fits}, and
{\tt KBHline52.fits} with 200 bins in the $g$-factor and 300 values of the
radius.

\section{Lamp-post tables in
{\fontfamily{pcr}\fontshape{tt}\selectfont lamp.fits}}
\label{appendix3c}

This file contains pre-calculated values of the functions needed for the
lamp-post model. It is supposed that a primary source of emission is placed
on the axis at a height $h$ above
the Kerr black hole. The matter in the disc rotates on stable circular (free)
orbits above the marginally stable orbit and it is freely falling below this
orbit where it has the same energy and angular momentum as the matter which
is on the marginally stable orbit.
It is assumed that the corona between the source and the disc is optically
thin, therefore ray-tracing in the vacuum Kerr space-time could be used for
computing the functions.

There are five functions stored in the {\tt lamp.fits} file as binary
extensions. They are parametrized by the value of the horizon of the black hole
$r_{\rm h}$, and height $h$, which are defined  as vectors at the beginning of
the FITS file. Currently only tables for $r_{\rm h}=1.05$
(i.e.\ $a\doteq0.9987492$)
and $h=2,\,3,\,4,\,5,\,6,\,8,\,10,\,12,\,15,\,20,\,30,\,50,\,75$ and $100$ are
available.
The functions included are:
 \begin{list}{}{\setlength{\topsep}{-2pt}\setlength{\itemsep}{-2pt}}
  \item[\rm -- angle of emission in degrees] -- the angle under which a photon
	is emitted from a primary source placed at a height $h$ on the axis above the
  black hole measured by a local stationary observer ($0^\circ$ -- a photon is
	emitted downwards, $180^\circ$
	-- a photon is emitted upwards),
  \item[\rm -- radius] -- the radius in $GM/c^2$ at which a photon strikes the
	disc,
  \item[\rm -- $g$-factor] -- the ratio of the energy of a photon hitting the
	disc to the energy of the same photon when emitted from a primary source,
  \item[\rm -- cosine of the incident angle] -- an absolute value of the cosine
	of the local incident angle between the incident light ray and local disc
	normal,
  \item[\rm -- azimuthal incident angle in radians] -- the angle between the
  projection of the three-mo\-mentum of the incident photon into the disc (in
	the local rest frame co-moving with the disc) and the radial tetrad vector.
 \end{list}
For mathematical formulae defining the functions see
eqs.~(\ref{gfac_lamp})--(\ref{azim_angle_inc}) in Section~\ref{lamp-post}.

The definition of the file {\tt lamp.fits}:
\begin{enumerate} \itemsep -0.1em
\vspace*{-0.5em}
 \item[0.] All of the extensions defined below are binary.
 \item The first extension contains a vector of the horizon values in $GM/c^2$,
   though currently only FITS files with tables for one value of the black-hole
	 horizon are accepted ($1.00 \le r_{\rm h} \le 2.00$).
 \item The second extension contains a vector of the values of heights $h$
 of a primary source in $GM/c^2$.
 \item In the following extensions the functions are defined, each extension is
 for a particular value of $r_{\rm h}$ and $h$. The values of $r_{\rm h}$ and
 $h$ are changing with each extension in the following order:\\[1em] \hspace*{-2.6em}
    \parbox{\textwidth}{
    {\begin{centering}
    $r_{\rm h}[1] \times h[1]$,\\
    $r_{\rm h}[1] \times h[2]$,\\
    $r_{\rm h}[1] \times h[3]$,\\
    \dots \\
    \dots \\
    $r_{\rm h}[2] \times h[1]$,\\
    $r_{\rm h}[2] \times h[2]$,\\
    $r_{\rm h}[2] \times h[3]$,\\
    \dots \\
    \dots \\
    \end{centering}}}\\[0.3em]
 \item Each of these extensions has five columns.
 In each column, a particular function is stored -- the angle of emission,
 radius, $g$-factor, cosine of the local incident angle and azimuthal incident
 angle, respectively. The extensions may have a different number of rows.
\end{enumerate}

\section{Coefficient of reflection in
{\fontfamily{pcr}\fontshape{tt}\selectfont
fluorescent\_line.fits}}
\label{appendix3d}
Values of the coefficient of reflection $f(\mu_{\rm i},\mu_{\rm e})$ for
a fluorescent
line are stored for different incident and reflection angles in this file.
For details on the model of scattering used for computations see
\cite{matt1991}.
It is assumed that the incident radiation is a power law with the photon index
$\Gamma=1.7$. The coefficient does not change its angular dependences for other
photon indices, only its normalization changes (see Fig.~14 in\break
\citealt{george1991}). The FITS file consists of three binary extensions:
\begin{list}{}{\setlength{\topsep}{-2pt}\setlength{\itemsep}{-2pt}}
 \item[--] the first extension contains absolute values of the cosine of incident
 angles,
 \item[--] the second extension contains values of the cosine of reflection
 angles,
 \item[--] the third extension contains one column with vector elements, here
 values of the coefficient of reflection are stored for different incident angles
 (rows) and for different reflection angles (elements of a vector).
\end{list}

\section{Tables in
{\fontfamily{pcr}\fontshape{tt}\selectfont refspectra.fits}}
\label{appendix3e}
The function $f(E_\loc;\mu_{\rm i},\mu_{\rm e})$ which gives dependence of
a locally emitted spectrum on the angle of incidence and angle of emission
is stored in this FITS file. The emission is induced by a power-law incident
radiation. Values of this function were
computed by the Monte Carlo simulations of Compton scattering, for details see
\cite{matt1991}. The reflected radiation depends on the photon index $\Gamma$
of the incident radiation.
There are several binary extensions in this fits file:
\begin{list}{}{\setlength{\topsep}{-2pt}\setlength{\itemsep}{-2pt}}
 \item[--] the first extension contains energy values in keV where
 the function $f(E_\loc;\mu_{\rm i},\mu_{\rm e})$ is computed, currently the
 interval from 2 to 300~keV is covered,
 \item[--] the second extension contains the absolute values of the cosine of
 the incident angles,
 \item[--] the third extension contains the values of the cosine of the
 emission angles,
 \item[--] the fourth extension contains the values of the photon indices
 $\Gamma$ of the incident power law,
 currently tables for $\Gamma=1.5,\,1.6,\,\dots,\,2.9$ and $3.0$ are computed,
 \item[--] in the following extensions the function
 $f(E_\loc;\mu_{\rm i},\mu_{\rm e})$ is defined, each extension is
 for a particular value of $\Gamma$; here values of the function are stored as a
 vector for different incident angles (rows) and for different angles of
 emission (columns), each element of this vector corresponds to a value of
 the function for a certain value of energy.
\end{list}

\chapter{Description of the integration routines}
\label{ide}
 \thispagestyle{empty}
	 Here we describe the technical details about the integration routines,
which act as a common driver performing the ray-tracing for various
models of the local emission. The description of non-axisymmetric and
axisymmetric versions are both provided.
An appropriate choice depends on the form of intrinsic emissivity.
Obviously, non-axisymmetric tasks are computationally more demanding.

\section{Non-axisymmetric integration routine
{\fontfamily{pcr}\fontshape{tt}\selectfont ide}}
\label{appendix4a}
This subroutine integrates the local emission and local Stokes parameters
for (partially) polarized emission of the accretion disc near a rotating
(Kerr) black hole (characterized by the angular momentum $a$) for an observer
with an inclination angle $\theta_{\rm o}$.
The subroutine has to be called with ten parameters:
\begin{center}
 {\tt ide(ear,ne,nt,far,qar,uar,var,ide\_param,emissivity,ne\_loc)}
\end{center}
\begin{description} \itemsep -2pt
 \item[{\tt ear}] -- real array of energy bins (same as {\tt ear} for local
 models in {\sc xspec}),
 \item[{\tt ne}] -- integer, number of energy bins (same as {\tt ne} for local
 models in {\sc xspec}),
 \item[{\tt nt}] -- integer, number of grid points in time (${\tt nt}=1$ means
 stationary model),
 \item[{\tt far(ne,nt)}] -- real array of photon flux per bin
             (same as {\tt photar} for local models in {\sc xspec} but with
						 the time resolution),
 \item[{\tt qar(ne,nt)}] -- real array of the Stokes parameter Q divided by the
 energy,
 \item[{\tt uar(ne,nt)}] -- real array of the Stokes parameter U divided by the
 energy,
 \item[{\tt var(ne,nt)}] -- real array of the Stokes parameter V divided by the
 energy,
 \item[{\tt ide\_param}] -- twenty more parameters needed for the integration
 (explained below),
 \item[{\tt emissivity}] -- name of the external emissivity subroutine, where
 the local emission of the disc is defined (explained in detail below),
 \item[{\tt ne\_loc}] -- number of points (in energies) where local photon flux
         (per keV) in the emissivity subroutine is defined.
\end{description}

The description of the {\tt ide\_param} parameters follows:
\begin{description} \itemsep -2pt
 \item[{\tt ide\_param(1)}] -- {\tt a/M} -- the black-hole angular momentum
 ($0 \le {\tt a/M} \le 1$),
 \item[{\tt ide\_param(2)}] -- {\tt theta\_o} -- the observer inclination in
degrees ($0^\circ$ -- pole, $90^\circ$ -- equatorial plane),
 \item[{\tt ide\_param(3)}] -- {\tt rin-rh} -- the inner edge of the non-zero
 disc emissivity relative to the black-hole horizon (in $GM/c^2$),
 \item[{\tt ide\_param(4)}] -- {\tt ms} -- determines whether we also integrate
emission below the marginally stable orbit; if its value is set to zero and
the inner radius of the disc is below the marginally stable orbit then the
emission below this orbit is taken into account, if set to unity it is not,
 \item[{\tt ide\_param(5)}] -- {\tt rout-rh} -- the outer edge of the non-zero
 disc emissivity relative to the black-hole horizon (in $GM/c^2$),
 \item[{\tt ide\_param(6)}] -- {\tt phi} -- the position angle of the axial
 sector of the disc with non-zero emissivity in degrees,
 \item[{\tt ide\_param(7)}] -- {\tt dphi} -- the inner angle of the axial sector
 of the disc with non-zero emissivity in degrees (${\tt dphi} \le 360^\circ$),
 \item[{\tt ide\_param(8)}] -- {\tt nrad} -- the radial resolution of the grid,
 \item[{\tt ide\_param(9)}] -- {\tt division} -- the switch for the spacing of
 the radial grid ($0$ -- equidistant, $1$ -- exponential),
 \item[{\tt ide\_param(10)}] -- {\tt nphi} -- the axial resolution of the grid,
 \item[{\tt ide\_param(11)}] -- {\tt smooth} -- the switch for performing simple
 smoothing ($0$ -- no, $1$ -- yes),
 \item[{\tt ide\_param(12)}] -- {\tt normal} -- the switch for normalizing of
 the final spectrum,\\
   if $=$ 0 -- total flux is unity (usually used for the line),\\
   if $>$ 0 -- flux is unity at the energy = {\tt normal} keV (usually used for
   the continuum),\\
   if $<$ 0 -- final spectrum is not normalized,
 \item[{\tt ide\_param(13)}] -- {\tt zshift} -- the overall redshift of the
 object,
 \item[{\tt ide\_param(14)}] -- {\tt ntable} -- tables to be used, it defines
 a double-digit number {\tt NN} in the name of the FITS file
 {\tt KBHtablesNN.fits} containing the tables ($0 \le {\tt ntable} \le 99$),
 \item[{\tt ide\_param(15)}] -- {\tt edivision} -- the switch for spacing the
 grid in local energies (0 -- equidistant, 1 -- exponential),
 \item[{\tt ide\_param(16)}] -- {\tt periodic} -- if set to unity then local
 emissivity is periodic if set to zero it is not (need not to be set if
 ${\tt nt}=1$),
 \item[{\tt ide\_param(17)}] -- {\tt dt} -- the time step (need not to be set if
 ${\tt nt}=1$),
 \item[{\tt ide\_param(18)}] -- {\tt polar} -- whether the change of the
 polarization angle and/or azimuthal emission angle will be read from FITS
 tables (0 -- no, 1 -- yes),
 \item[{\tt ide\_param(19)}] -- {\tt r0-rh} and
 \item[{\tt ide\_param(20)}] -- {\tt phi0} -- in dynamical computations the
 initial time will be set to the time when photons emitted from the point
 [{\tt r0}, {\tt phi0}] on the disc (in the Boyer-Lindquist coordinates) reach
 the observer.
\end{description}

The subroutine {\tt ide} needs an external emissivity subroutine in which the
local emission and local Stokes parameters are defined. This subroutine has
twelve parameters:\\[3mm]
\hspace*{4mm}{\tt emissivity(\parbox[t]{\textwidth}
{ear\_loc,ne\_loc,nt,far\_loc,qar\_loc,uar\_loc,var\_loc,r,phi,cosine,\\
 phiphoton,first\_emis)}}
\begin{description} \itemsep -2pt
 \item[{\tt ear\_loc(0:ne\_loc)}] -- real array of the local energies where
 the local photon flux {\tt far\_loc} is defined, with special meaning of
 {\tt ear\_loc(0)}
  -- if its value is larger than zero then the local emissivity consists of two
	energy regions where the flux is non-zero; the flux between these
	regions is zero and {\tt ear\_loc(0)} defines the number of points in local
	energies with the zero local flux,
 \item[{\tt ne\_loc}] -- integer, the number of points (in energies) where the
 local photon flux (per keV) is defined,
 \item[{\tt nt}] -- integer, the number of grid points in time (${\tt nt}=1$
 means stationary model),
 \item[{\tt far\_loc(0:ne\_loc,nt)}] -- real array of the local photon flux
 (per keV)
    -- if the local emissivity consists of two separate non-zero regions
    (i.e.\ ${\tt ear\_loc(0)} > 0$) then {\tt far\_loc(0,it)} is the index of
    the last point of the first non-zero local energy region,
\item[{\tt qar\_loc(ne\_loc,nt)}] -- real array of the local Stokes parameter Q
divided by the local energy,
\item[{\tt uar\_loc(ne\_loc,nt)}] -- real array of the local Stokes parameter U
divided by the local energy,
\item[{\tt var\_loc(ne\_loc,nt)}] -- real array of the local Stokes parameter V
divided by the local energy,
\item[{\tt r}] -- the radius in $GM/c^2$ where the local photon flux
{\tt far\_loc} at the local energies {\tt ear\_loc} is wanted
\item[{\tt phi}] -- the azimuth (the Boyer-Lindquist coordinate $\varphi$) where
the local photon flux {\tt far\_loc} at the local energies {\tt ear\_loc} is
wanted,
\item[{\tt cosine}] -- the cosine of the local angle between the emitted ray
 and local disc normal,
 \item[{\tt phiphoton}] -- the angle between the emitted ray projected onto the
 plane of the disc (in the local frame of the moving disc) and the radial
 component of the local tetrad (in radians),
 \item[{\tt first\_emis}] -- boolean, TRUE if we enter the emissivity subroutine
 from the subroutine {\tt ide} for the first time, FALSE if we have already been
 in this subroutine (this is convenient if we want to calculate some initial
 values when we are in the emissivity subroutine for the first time, e.g.\
 trajectory of the falling spot).
\end{description}

\section{Axisymmetric integration routine
{\fontfamily{pcr}\fontshape{tt}\selectfont idre}}
\label{appendix4b}
This subroutine integrates the local axisymmetric emission of an accretion disc
near a rotating (Kerr) black hole (characterized by the angular momentum $a$)
for an observer with an inclination angle~$\theta_{\rm o}$.
The subroutine has to be called with eight parameters:
\begin{center}
 {\tt idre(ear,ne,photar,idre\_param,cmodel,ne\_loc,ear\_loc,far\_loc)}
\end{center}
\begin{description} \itemsep -2pt
 \item[{\tt ear}] -- real array of energy bins (same as {\tt ear} for local
 models in {\sc xspec}),
 \item[{\tt ne}] -- integer, the number of energy bins (same as {\tt ne} for local
 models in {\sc xspec}),
 \item[{\tt photar}] -- real array of the photon flux per bin
             (same as {\tt photar} for local models in {\sc xspec}),
 \item[{\tt idre\_param}] -- ten more parameters needed for the integration
 (explained below),
 \item[{\tt cmodel}] -- 32-byte string with a base name of a FITS file with
 tables for axisymmetric emission (e.g.\ ``{\tt KBHline}'' for
 {\tt KBHlineNN.fits}),
 \item[{\tt ne\_loc}] -- the number of points (in energies) where the local
 photon flux (per keV) is defined in the emissivity subroutine,
 \item[{\tt ear\_loc}] -- array of the local energies where the local
 photon flux {\tt far\_loc} is defined,
 \item[{\tt far\_loc}] -- array of the local photon flux (per keV).
\end{description}

The description of the {\tt idre\_param} parameters follows:
\begin{description} \itemsep -2pt
 \item[{\tt idre\_param(1)}] -- {\tt a/M} -- the black-hole angular momentum
 ($0 \le {\tt a/M} \le 1$),
 \item[{\tt idre\_param(2)}] -- {\tt theta\_o} -- the observer inclination in
degrees ($0^\circ$ -- pole, $90^\circ$ -- equatorial plane),
 \item[{\tt idre\_param(3)}] -- {\tt rin-rh} -- the inner edge of the non-zero
 disc emissivity relative to the black-hole horizon (in $GM/c^2$),
 \item[{\tt idre\_param(4)}] -- {\tt ms} -- determines whether we also integrate
emission below the marginally stable orbit; if its value is set to zero and
the inner radius of the disc is below the marginally stable orbit then the
emission below this orbit is taken into account, if set to unity it is not,
 \item[{\tt idre\_param(5)}] -- {\tt rout-rh} -- the outer edge of the non-zero
 disc emissivity relative to the black-hole horizon (in $GM/c^2$),
 \item[{\tt idre\_param(6)}] -- {\tt smooth} -- the switch for performing simple
 smoothing ($0$ -- no, $1$ -- yes),
 \item[{\tt idre\_param(7)}] -- {\tt normal} -- the switch for normalizing the
 final spectrum,\\
   if $=$ 0 -- total flux is unity (usually used for the line),\\
   if $>$ 0 -- flux is unity at the energy = {\tt normal} keV (usually used for
   the continuum),\\
   if $<$ 0 -- final spectrum is not normalized,
 \item[{\tt idre\_param(8)}] -- {\tt zshift} -- the overall redshift of the
 object,
 \item[{\tt idre\_param(9)}] -- {\tt ntable} -- tables to be used, it defines
 a double-digit number {\tt NN} in the name of the FITS file
 (e.g.\ in {\tt KBHlineNN.fits}) containing the tables
 ($0 \le {\tt ntable} \le 99$),
 \item[{\tt idre\_param(10)}] -- {\tt alpha} -- the radial power-law index.
\end{description}

The subroutine {\tt idre} does not need any external emissivity subroutine.

\chapter{Atlas of transfer functions}
\label{atlas}
 \thispagestyle{empty}
	 In this Appendix we provide a graphical representation of the transfer
functions that are needed for computations of relativistic spectral
profiles. For a mathematical definition of these functions and details of the
adopted notation see Chapter~\ref{chapter1}. Although graphical representation
is not necessary in the process of computations and actual data fitting, we
find it extremely useful and practical for quick order-of-magnitude
estimates. Often, expected values of various quantities can be seen almost
instantly. For example, the magnitude and the range of the energy shift, the
effect of gravitational lensing, the importance of the relative time delay
etc.\ can be estimated from these graphs, assuming only basic parameters of
the light emitting region.

\begin{figure}[tb]
\vspace*{-0.9em}
\dummycaption\label{gfac_00}
\phantomsection\addcontentsline{toc}{section}{g-factor}
\phantomsection\addcontentsline{toc}{subsection}{Inclination 0.1\r{ }}
\includegraphics[width=15cm]{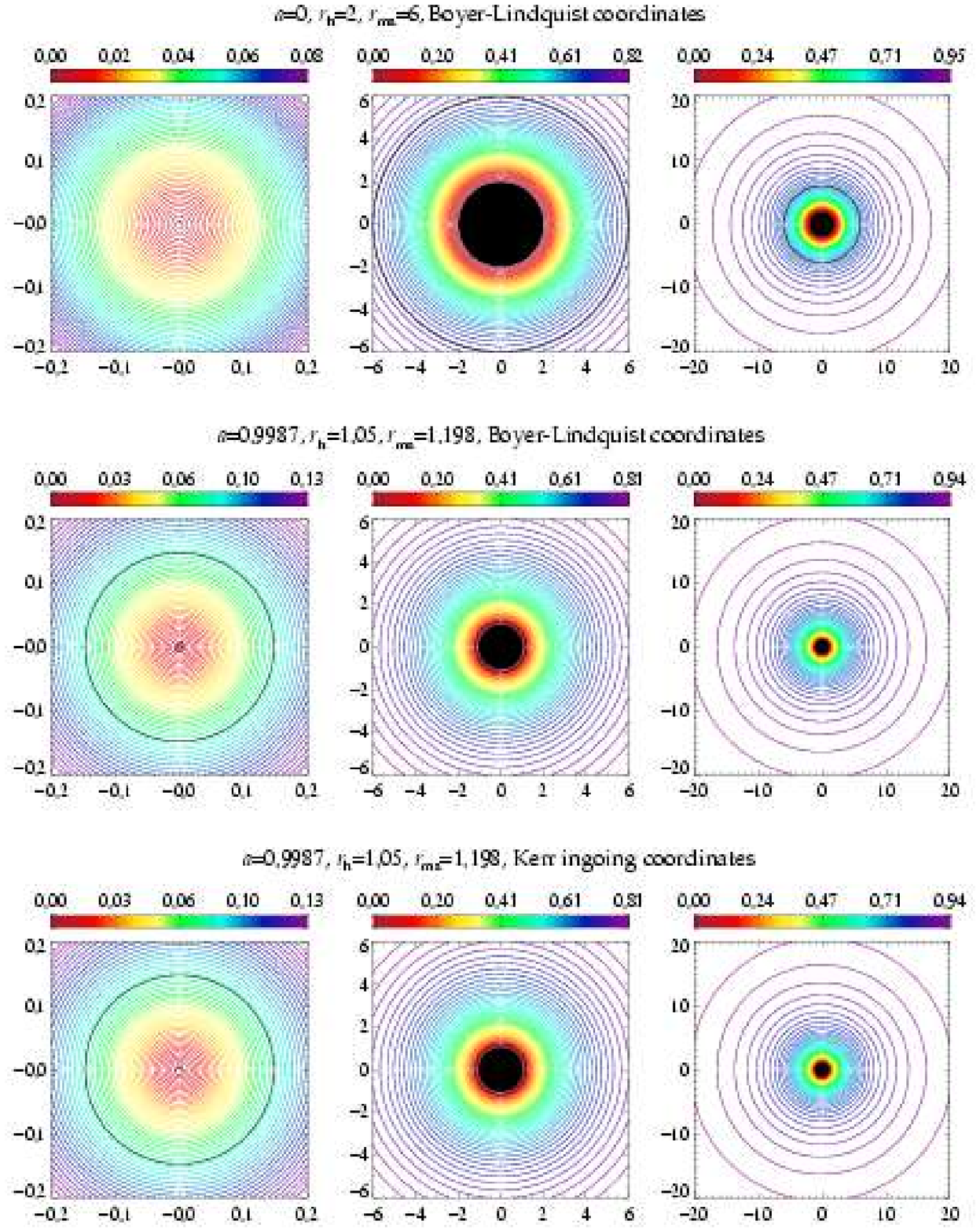}
\vspace{1em}
\mycaption{$g$-factor for the inclination $\theta_{\rm o}=0.1^\circ$.}
\end{figure}

\begin{figure}[tb]
\vspace*{-0.9em}
\dummycaption\label{gfac_45}
\phantomsection\addcontentsline{toc}{subsection}{Inclination 45\r{ }}
\includegraphics[width=15cm]{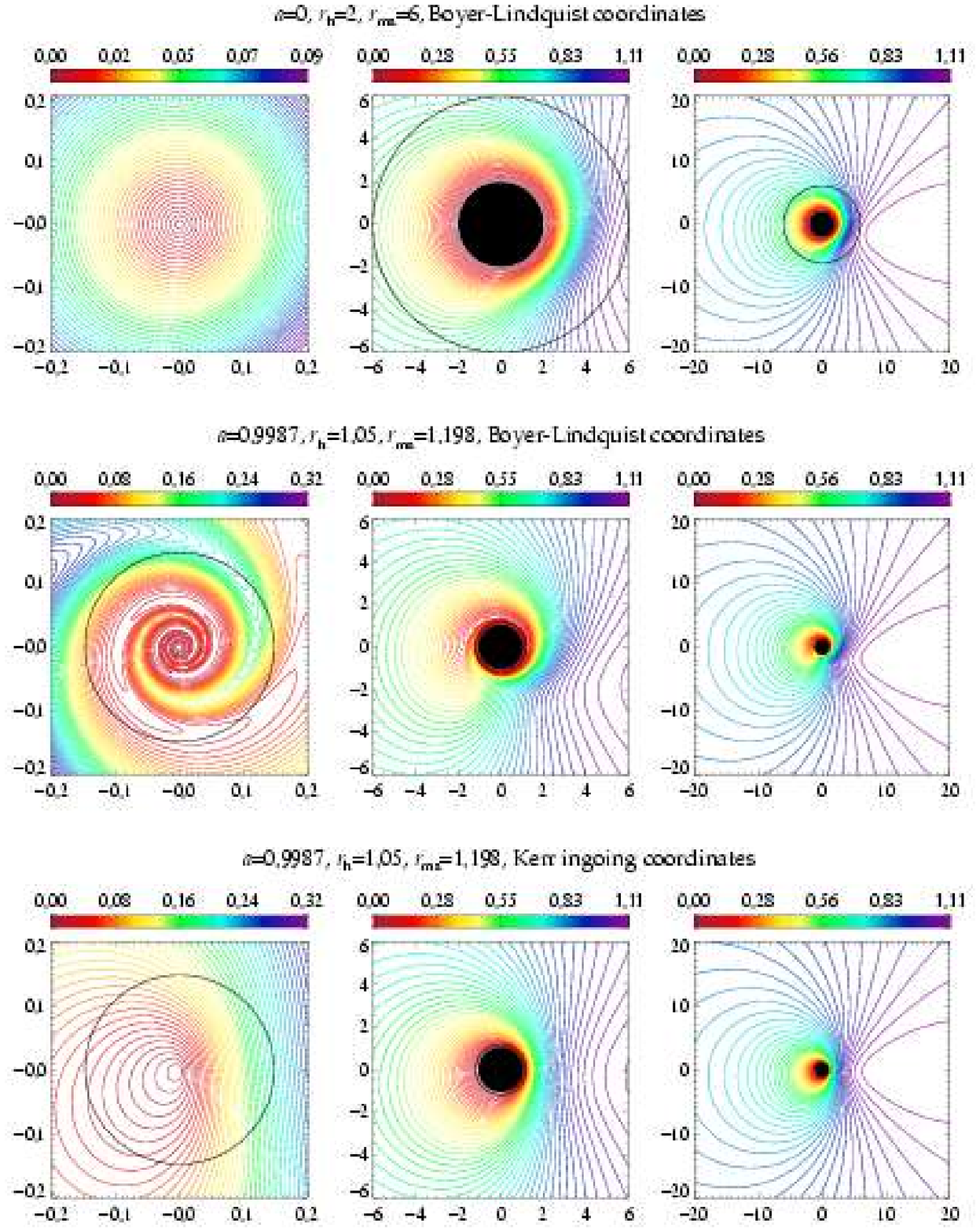}
\vspace{1em}
\mycaption{$g$-factor for the inclination $\theta_{\rm o}=45^\circ$.}
\end{figure}

\begin{figure}[tb]
\vspace*{-0.9em}
\dummycaption\label{gfac_85}
\phantomsection\addcontentsline{toc}{subsection}{Inclination 85\r{ }}
\includegraphics[width=15cm]{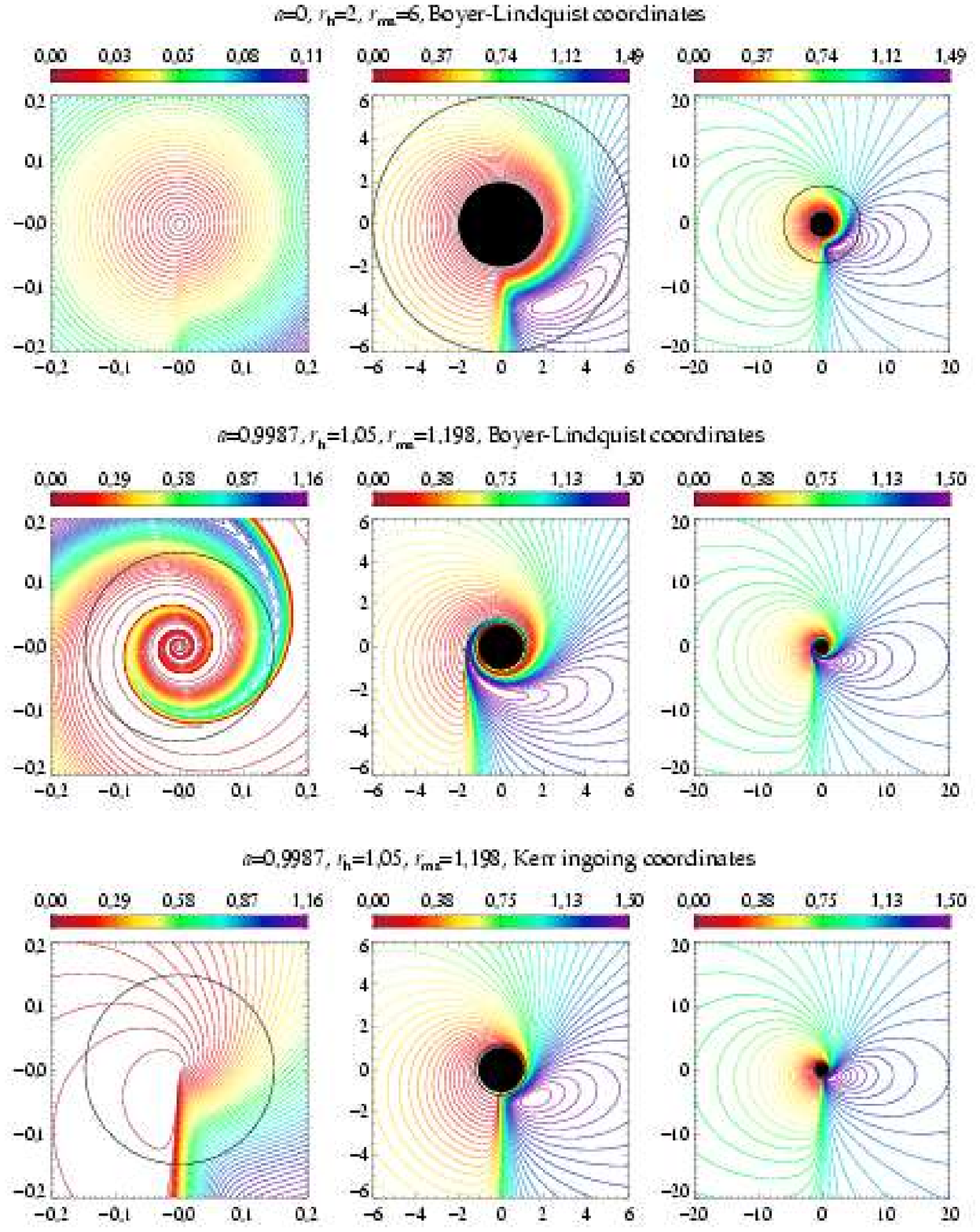}
\vspace{1em}
\mycaption{$g$-factor for the inclination $\theta_{\rm o}=85^\circ$.}
\end{figure}

\begin{figure}[tb]
\vspace*{-0.9em}
\dummycaption\label{Angle_00}
\phantomsection\addcontentsline{toc}{section}{Emission angle}
\phantomsection\addcontentsline{toc}{subsection}{Inclination 0.1\r{ }}
\includegraphics[width=15cm]{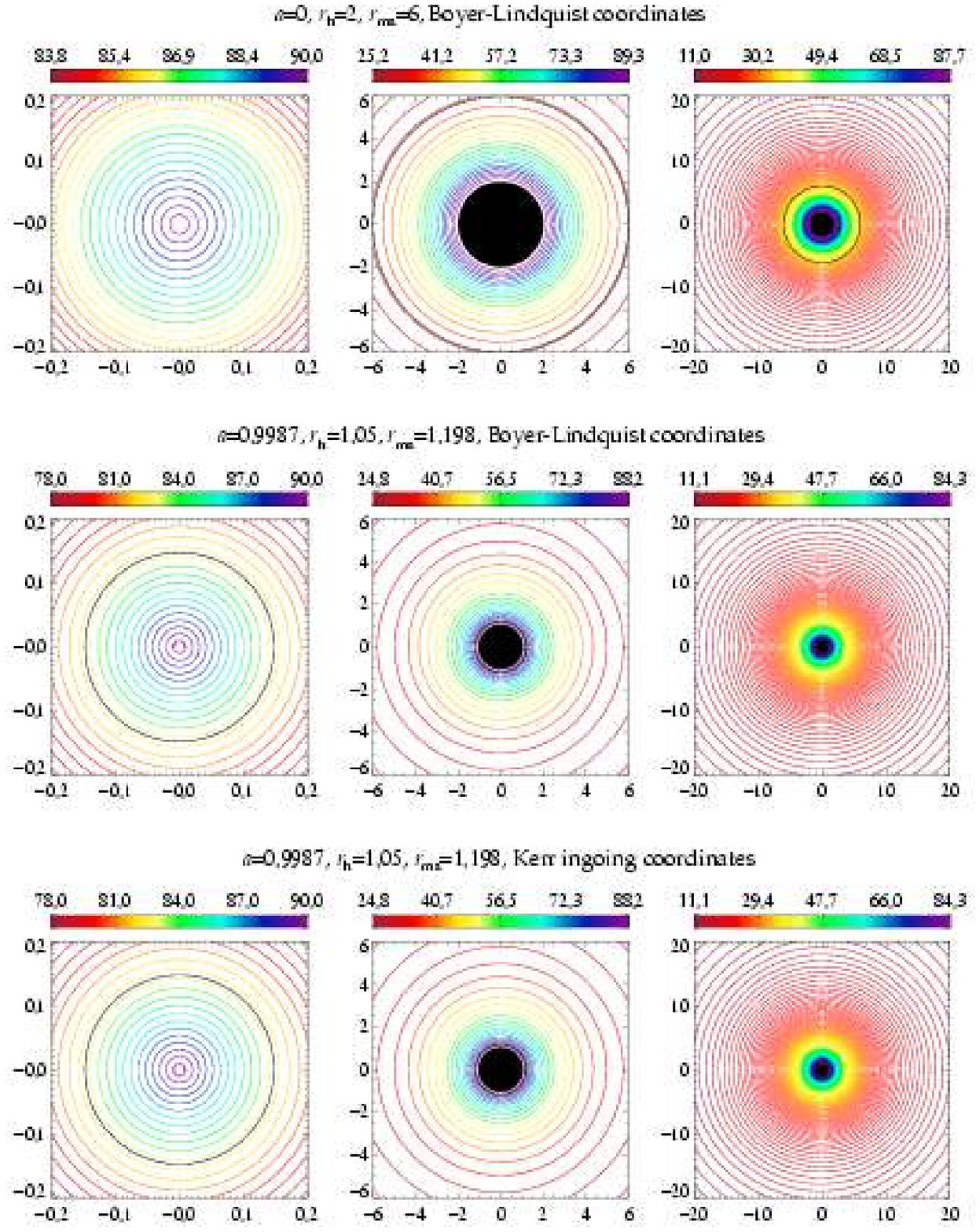}
\vspace{1em}
\mycaption{Emission angle for the inclination $\theta_{\rm o}=0.1^\circ$.}
\end{figure}

\begin{figure}[tb]
\vspace*{-0.9em}
\dummycaption\label{Angle_45}
\phantomsection\addcontentsline{toc}{subsection}{Inclination 45\r{ }}
\includegraphics[width=15cm]{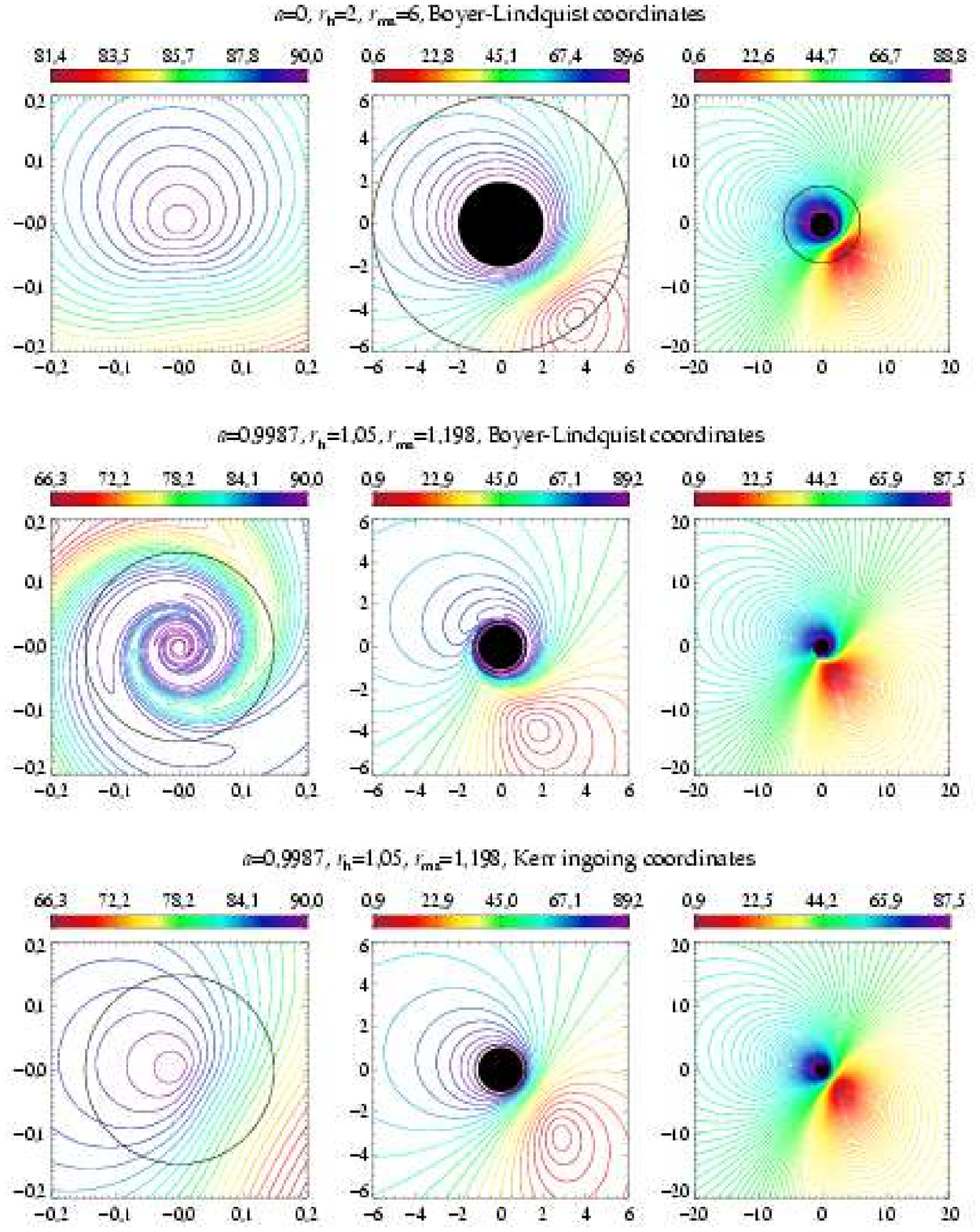}
\vspace{1em}
\mycaption{Emission angle for the inclination $\theta_{\rm o}=45^\circ$.}
\end{figure}

\begin{figure}[tb]
\vspace*{-0.9em}
\dummycaption\label{Angle_85}
\phantomsection\addcontentsline{toc}{subsection}{Inclination 85\r{ }}
\includegraphics[width=15cm]{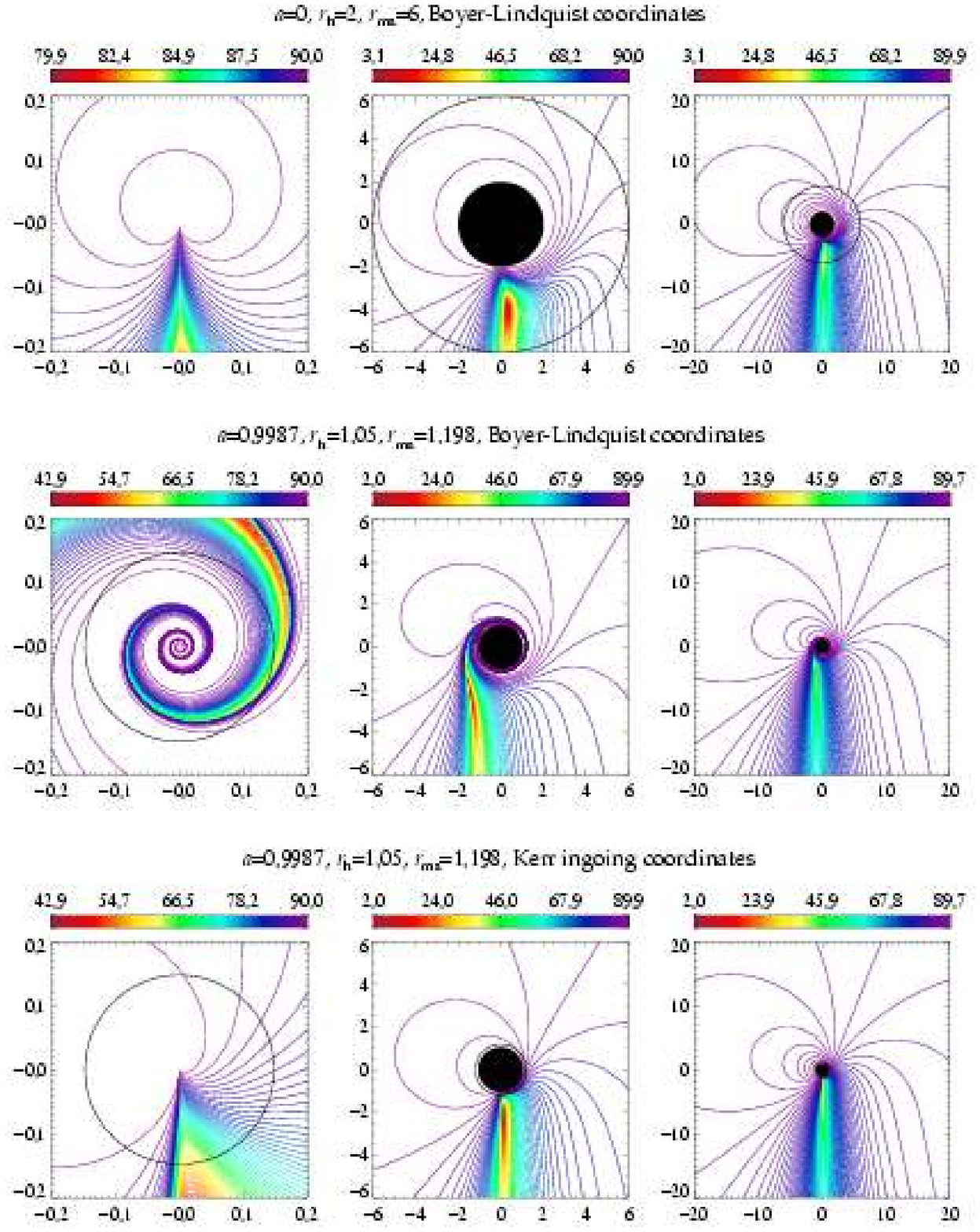}
\vspace{1em}
\mycaption{Emission angle for the inclination $\theta_{\rm o}=85^\circ$.}
\end{figure}

\begin{figure}[tb]
\vspace*{-0.9em}
\dummycaption\label{lens_00}
\phantomsection\addcontentsline{toc}{section}{Lensing}
\phantomsection\addcontentsline{toc}{subsection}{Inclination 0.1\r{ }}
\includegraphics[width=15cm]{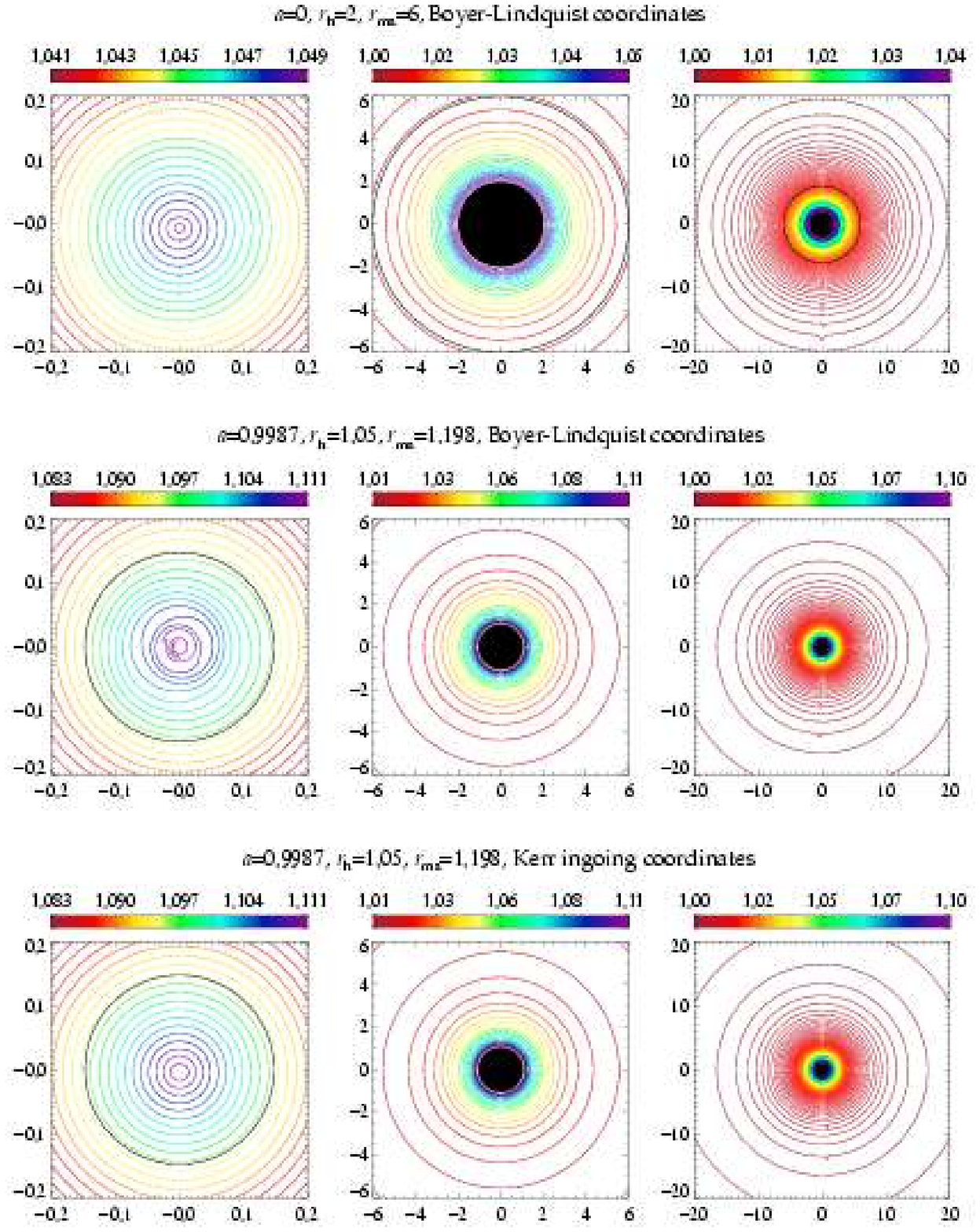}
\vspace{1em}
\mycaption{Lensing for the inclination $\theta_{\rm o}=0.1^\circ$.}
\end{figure}

\begin{figure}[tb]
\vspace*{-0.9em}
\dummycaption\label{lens_45}
\phantomsection\addcontentsline{toc}{subsection}{Inclination 45\r{ }}
\includegraphics[width=15cm]{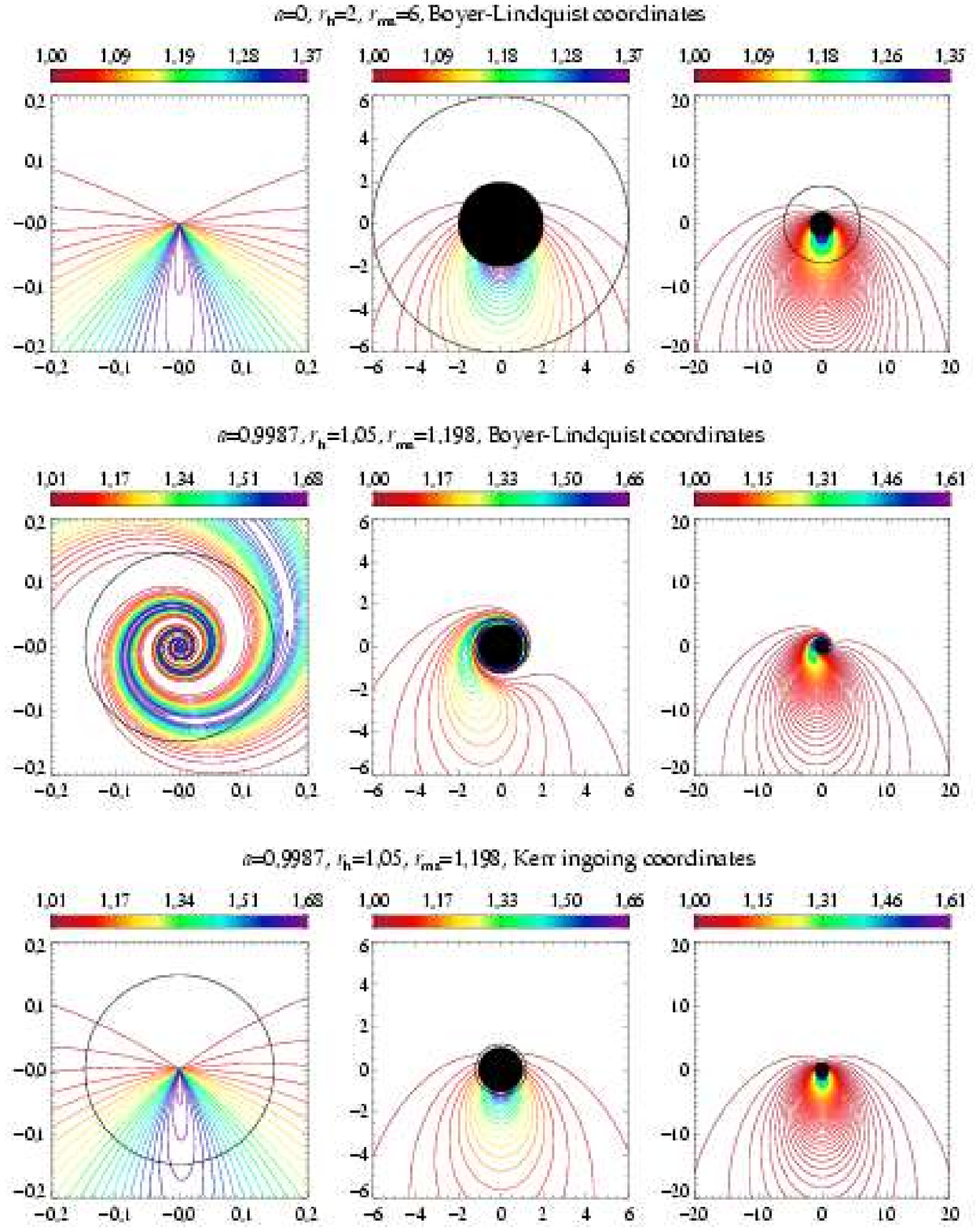}
\vspace{1em}
\mycaption{Lensing for the inclination $\theta_{\rm o}=45^\circ$.}
\end{figure}

\begin{figure}[tb]
\vspace*{-0.9em}
\dummycaption\label{lens_85}
\phantomsection\addcontentsline{toc}{subsection}{Inclination 85\r{ }}
\includegraphics[width=15cm]{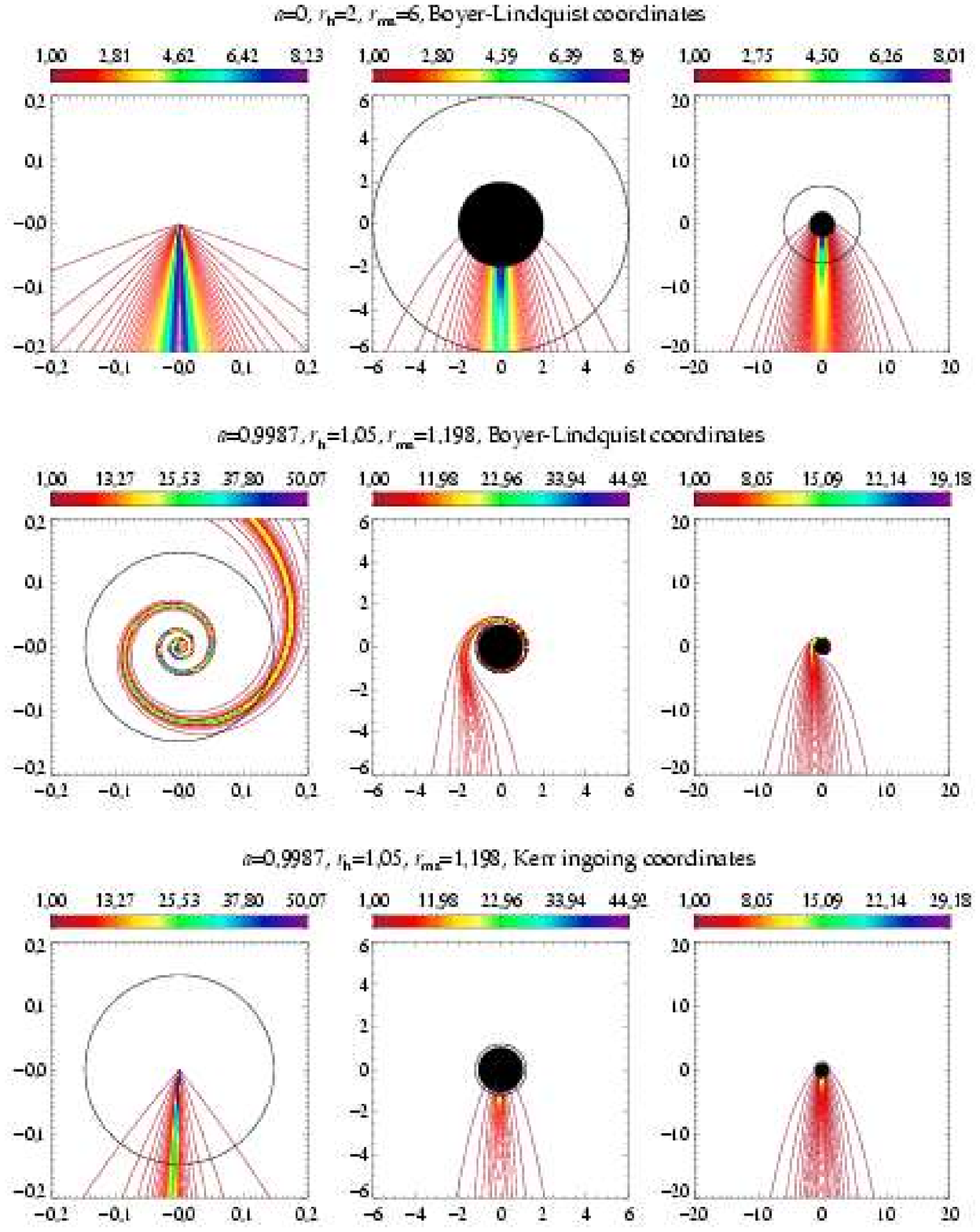}
\vspace{1em}
\mycaption{Lensing for the inclination $\theta_{\rm o}=85^\circ$.}
\end{figure}

\begin{figure}[tb]
\vspace*{-0.9em}
\dummycaption\label{del_00}
\phantomsection\addcontentsline{toc}{section}{Relative time delay}
\phantomsection\addcontentsline{toc}{subsection}{Inclination 0.1\r{ }}
\includegraphics[width=15cm]{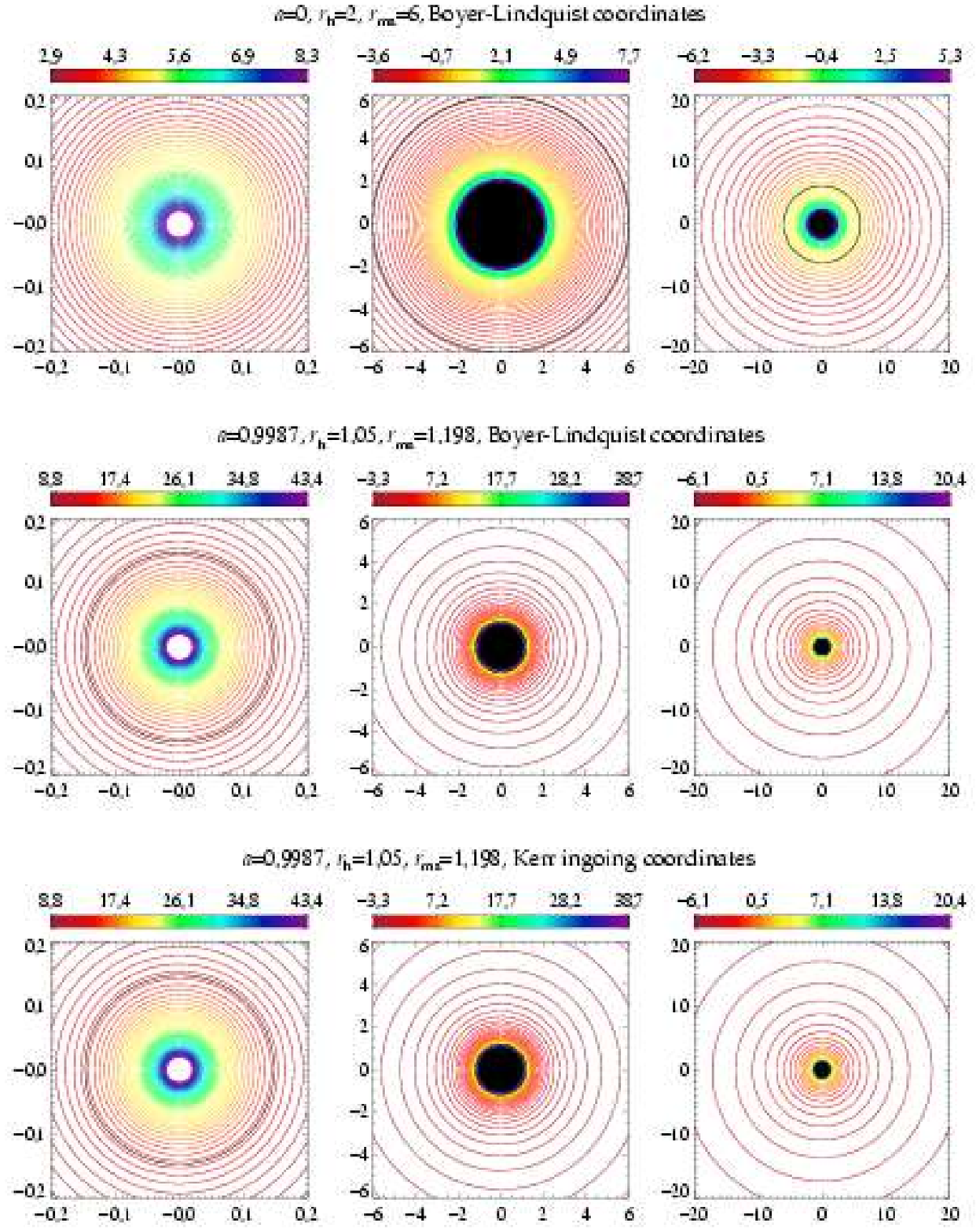}
\vspace{1em}
\mycaption{Relative time delay for the inclination $\theta_{\rm o}=0.1^\circ$.}
\end{figure}

\begin{figure}[tb]
\vspace*{-0.9em}
\dummycaption\label{del_45}
\phantomsection\addcontentsline{toc}{subsection}{Inclination 45\r{ }}
\includegraphics[width=15cm]{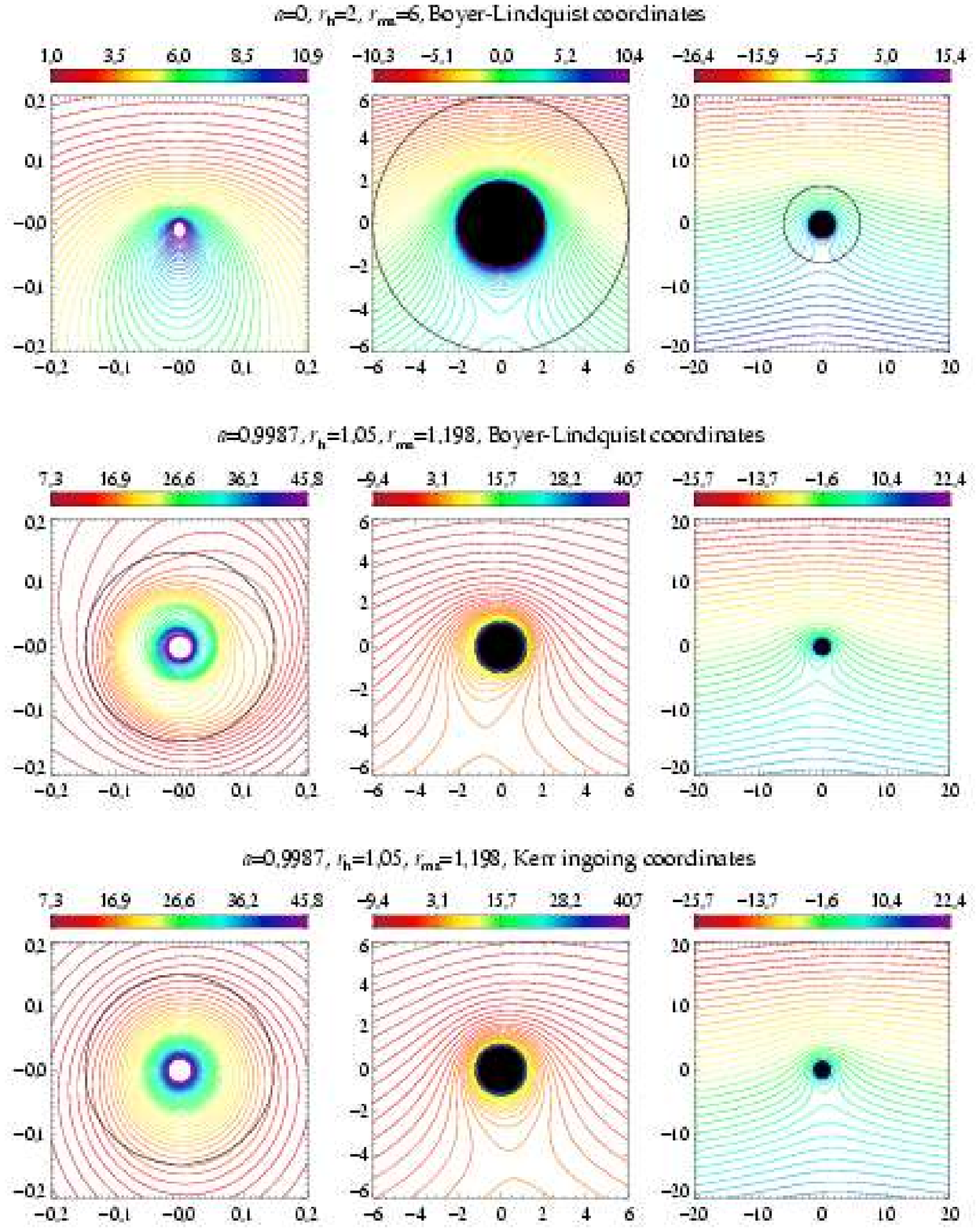}
\vspace{1em}
\mycaption{Relative time delay for the inclination $\theta_{\rm o}=45^\circ$.}
\end{figure}

\begin{figure}[tb]
\vspace*{-0.9em}
\dummycaption\label{del_85}
\phantomsection\addcontentsline{toc}{subsection}{Inclination 85\r{ }}
\includegraphics[width=15cm]{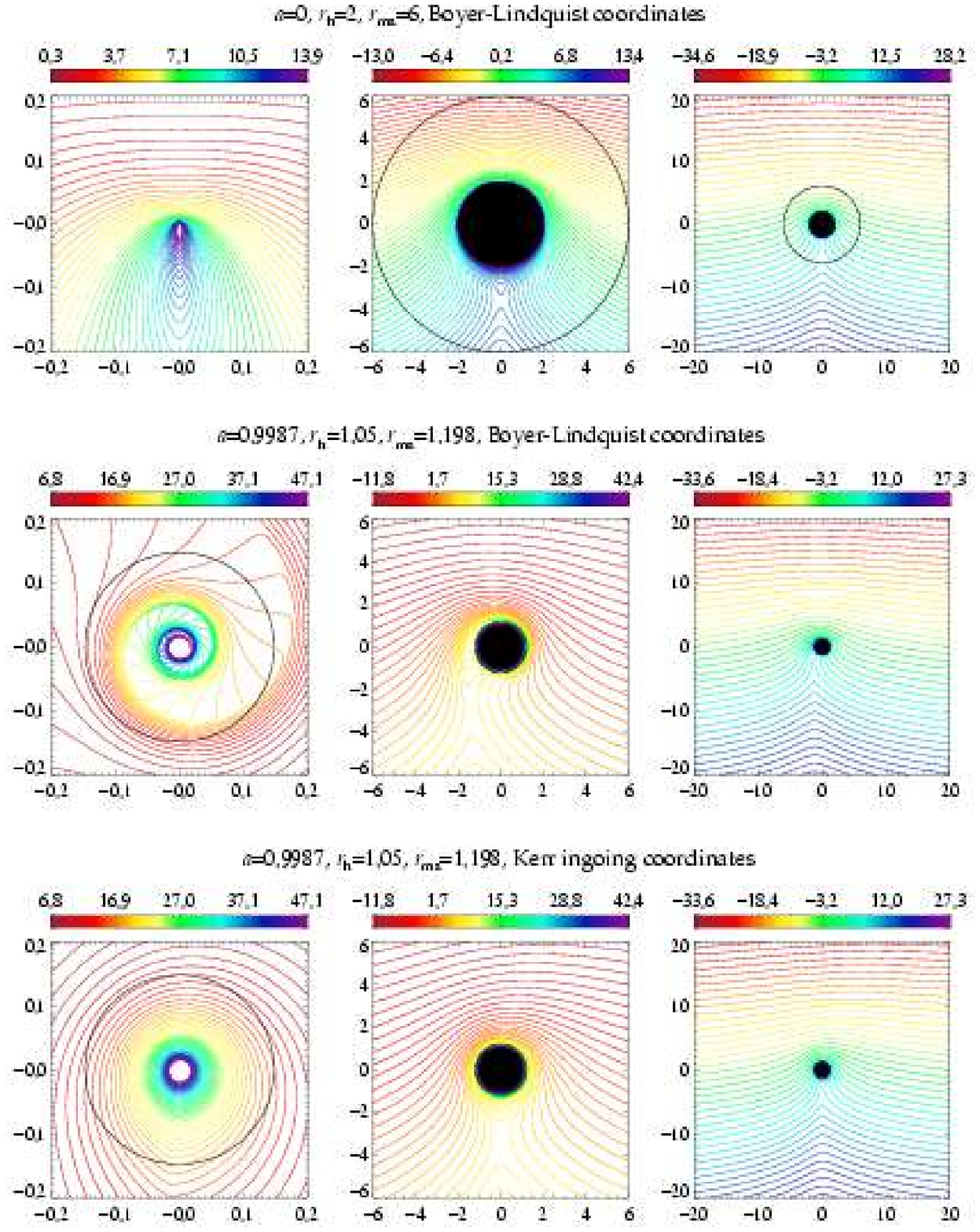}
\vspace{1em}
\mycaption{Relative time delay for the inclination $\theta_{\rm o}=85^\circ$.}
\end{figure}

\begin{figure}[tb]
\vspace*{-0.9em}
\dummycaption\label{pol_00}
\phantomsection\addcontentsline{toc}{section}{Change of polarization angle}
\phantomsection\addcontentsline{toc}{subsection}{Inclination 0.1\r{ }}
\includegraphics[width=15cm]{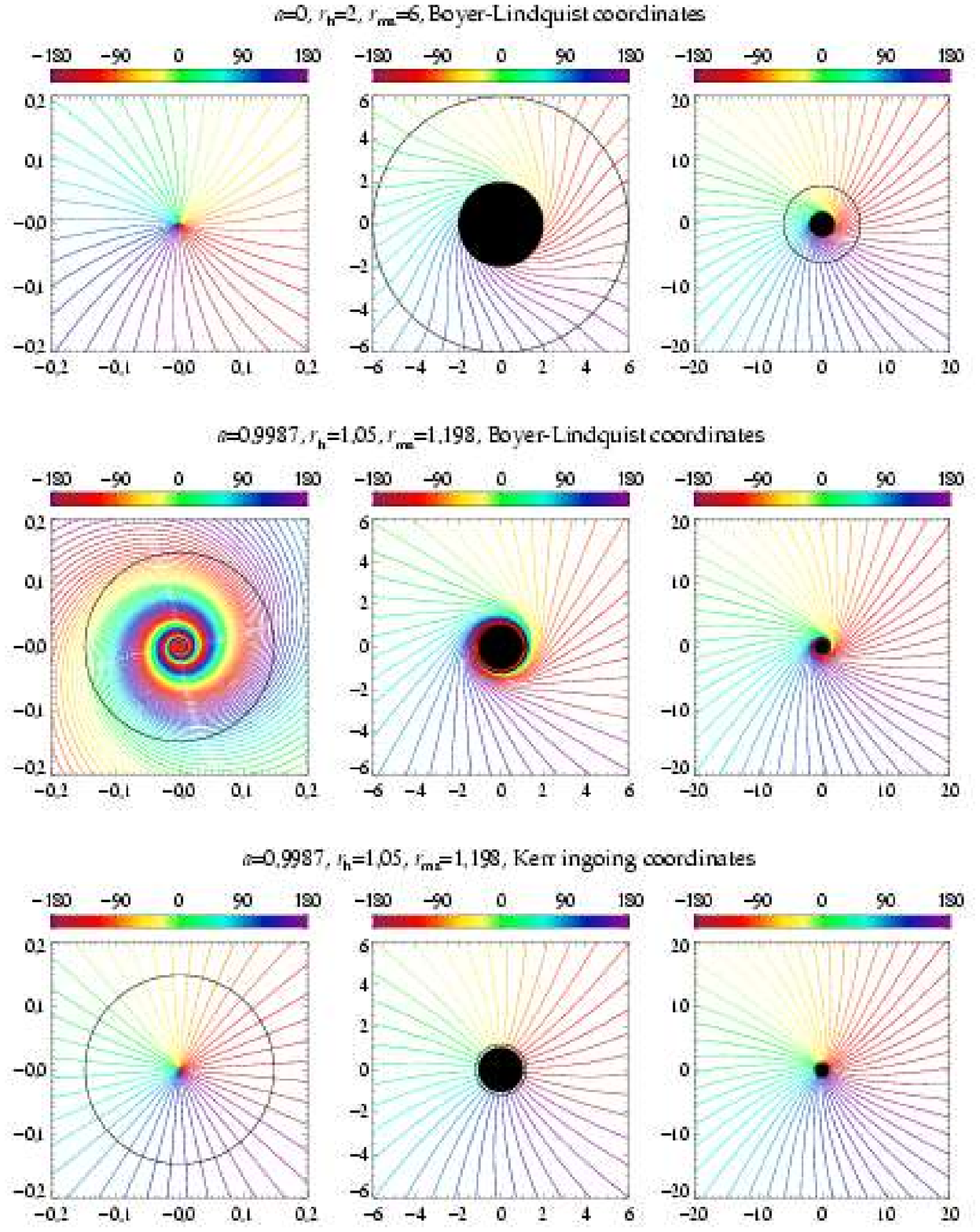}
\vspace{1em}
\mycaption{Change of the polarization angle for the inclination
$\theta_{\rm o}=0.1^\circ$.}
\end{figure}

\begin{figure}[tb]
\vspace*{-0.9em}
\dummycaption\label{pol_45}
\phantomsection\addcontentsline{toc}{subsection}{Inclination 45\r{ }}
\includegraphics[width=15cm]{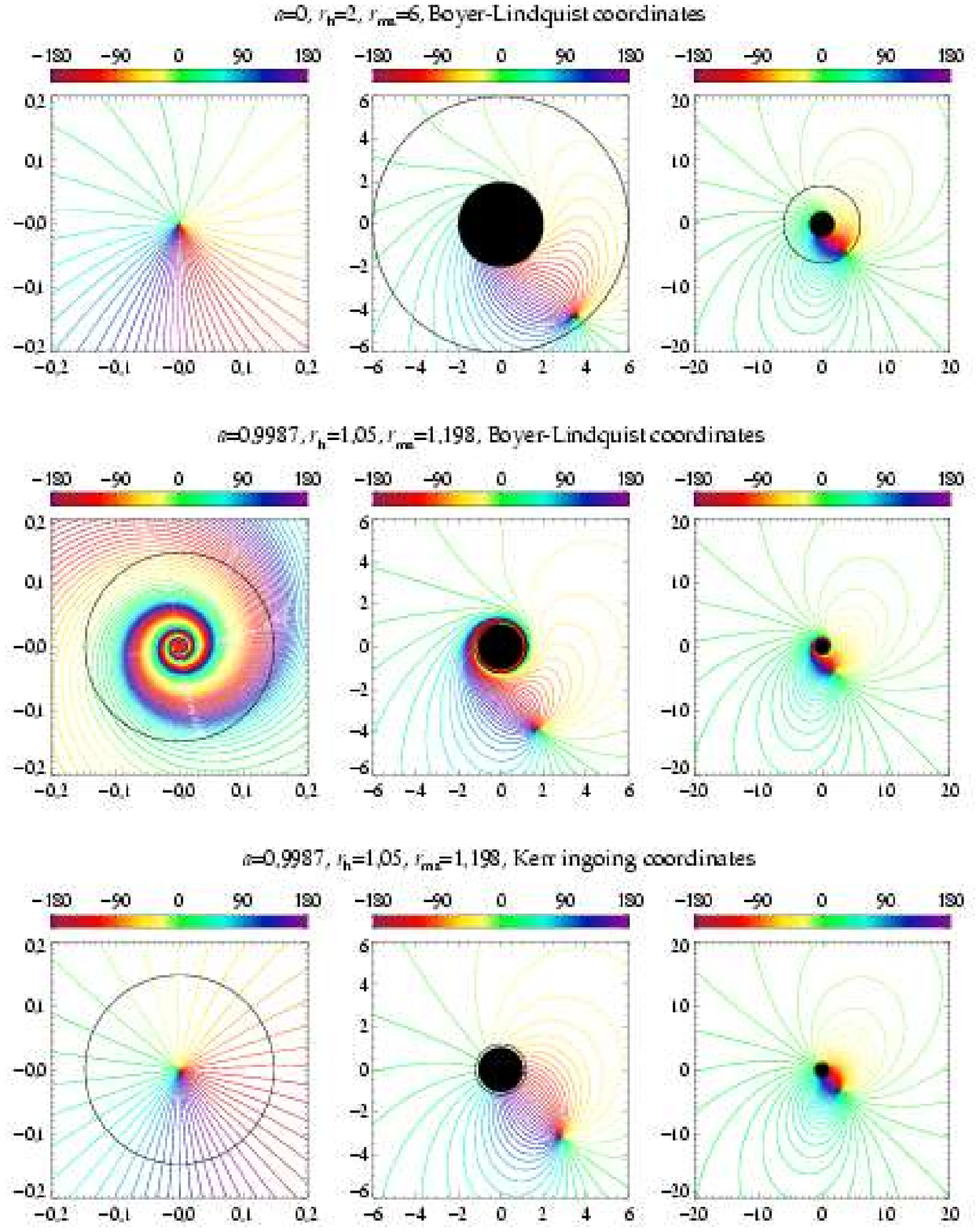}
\vspace{1em}
\mycaption{Change of the polarization angle for the inclination
$\theta_{\rm o}=45^\circ$.}
\end{figure}

\begin{figure}[tb]
\vspace*{-0.9em}
\dummycaption\label{pol_85}
\phantomsection\addcontentsline{toc}{subsection}{Inclination 85\r{ }}
\includegraphics[width=15cm]{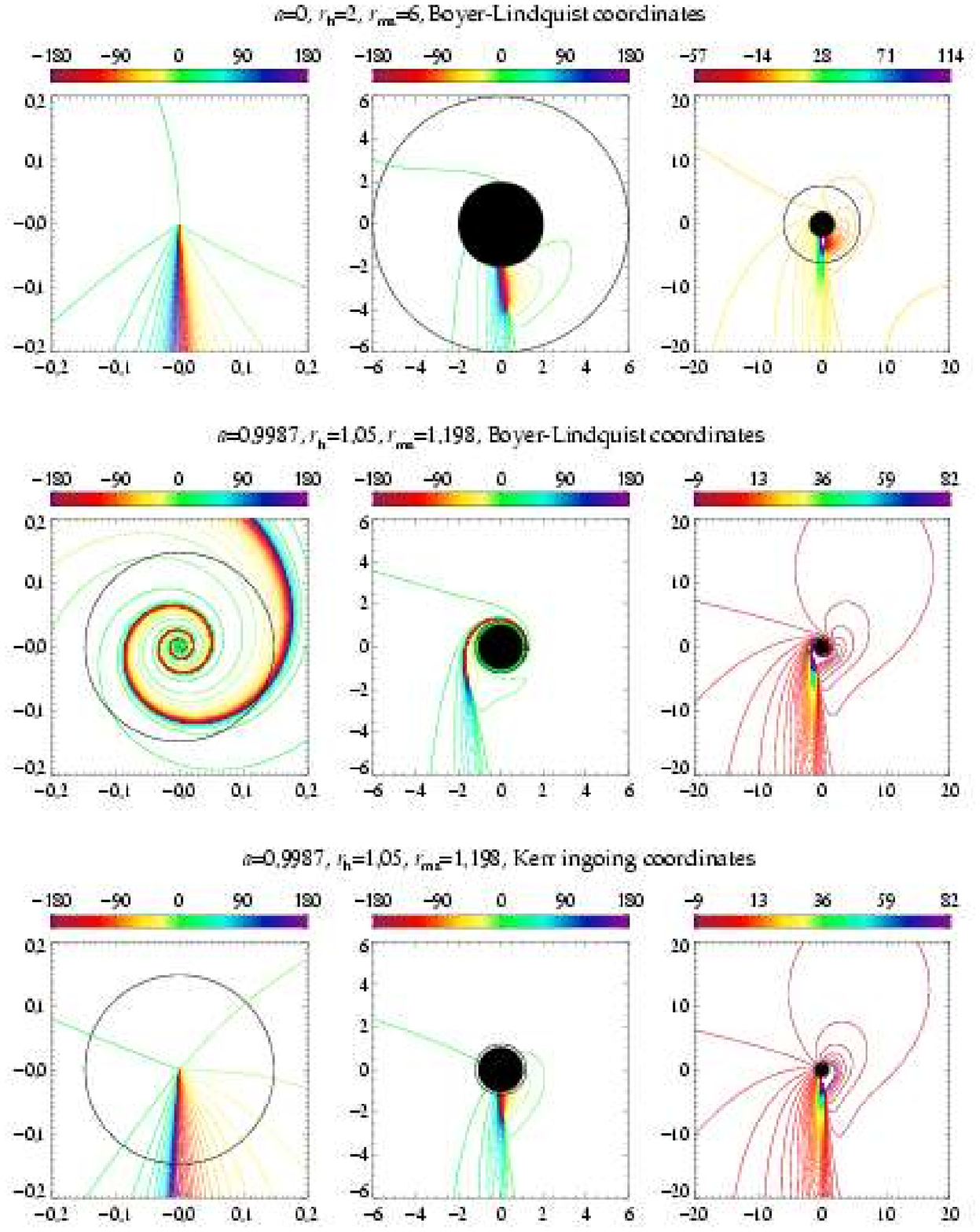}
\vspace{1em}
\mycaption{Change of the polarization angle for the inclination
$\theta_{\rm o}=85^\circ$.}
\end{figure}

\begin{figure}[tb]
\vspace*{-0.9em}
\dummycaption\label{phiph_00}
\phantomsection\addcontentsline{toc}{section}{Azimuthal emission angle}
\phantomsection\addcontentsline{toc}{subsection}{Inclination 0.1\r{ }}
\includegraphics[width=15cm]{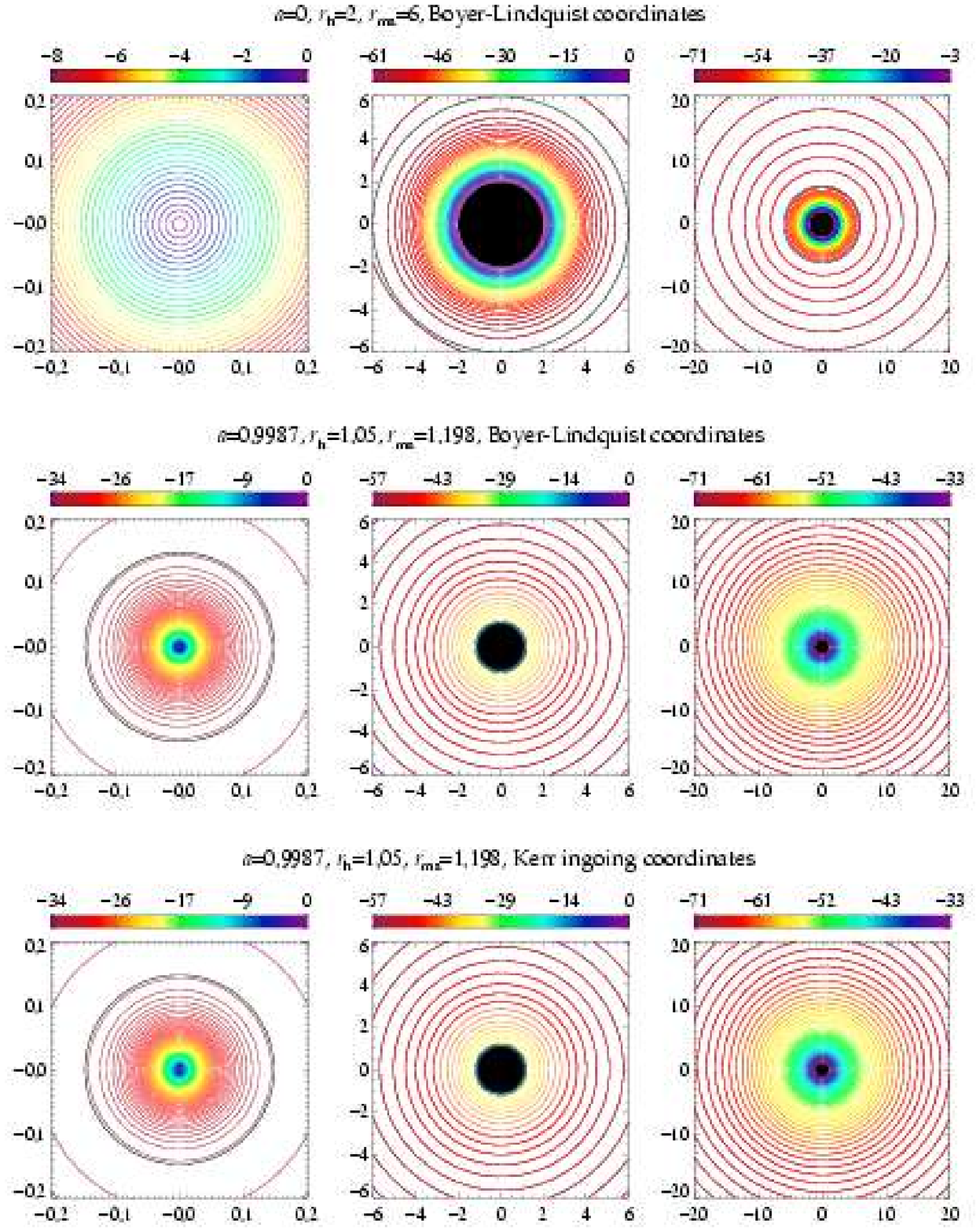}
\vspace{1em}
\mycaption{Azimuthal emission angle for the inclination $\theta_{\rm o}=0.1^\circ$.}
\end{figure}

\begin{figure}[tb]
\vspace*{-0.9em}
\dummycaption\label{phiph_45}
\phantomsection\addcontentsline{toc}{subsection}{Inclination 45\r{ }}
\includegraphics[width=15cm]{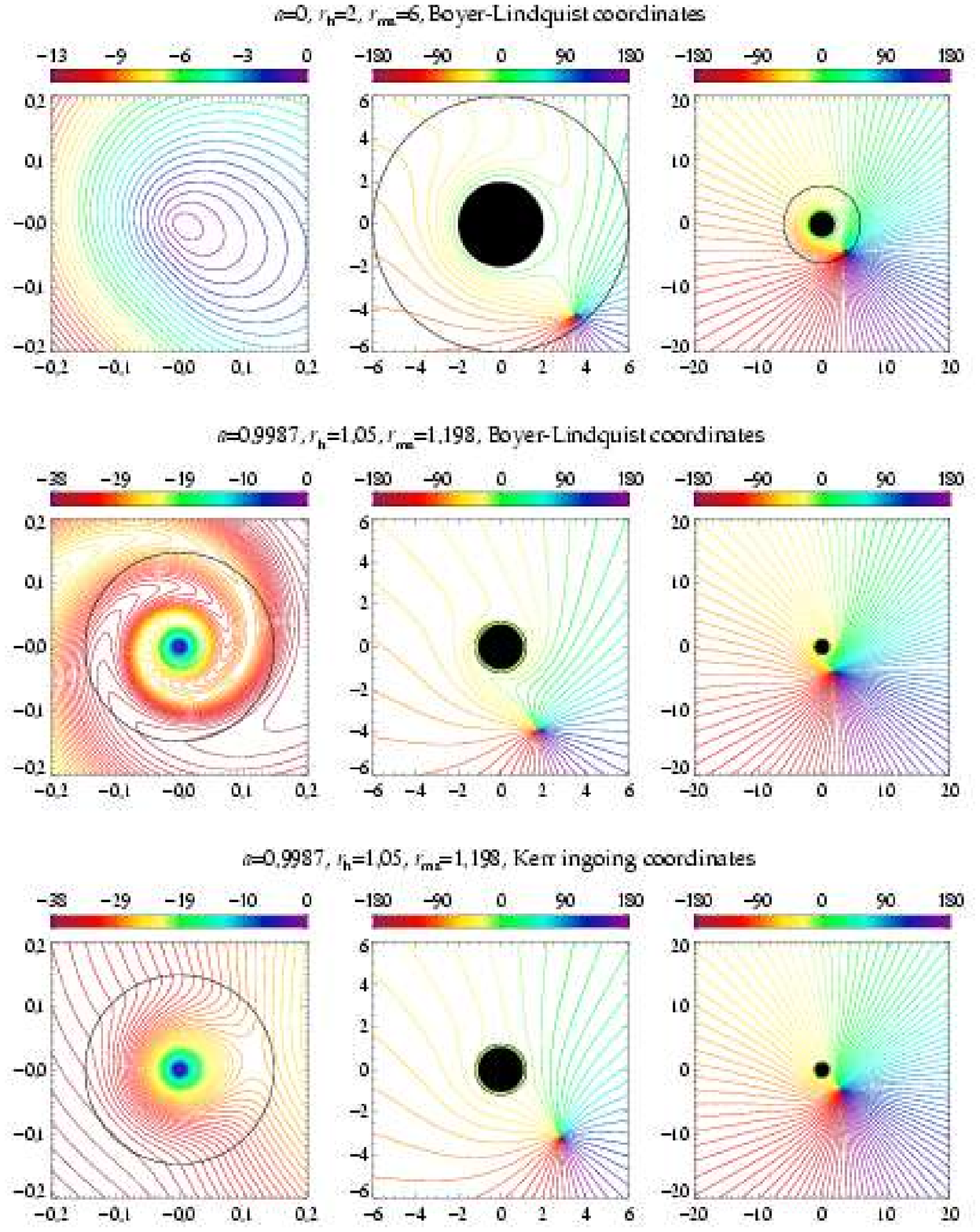}
\vspace{1em}
\mycaption{Azimuthal emission angle for the inclination $\theta_{\rm o}=45^\circ$.}
\end{figure}

\begin{figure}[tb]
\vspace*{-0.9em}
\dummycaption\label{phiph_85}
\phantomsection\addcontentsline{toc}{subsection}{Inclination 85\r{ }}
\includegraphics[width=15cm]{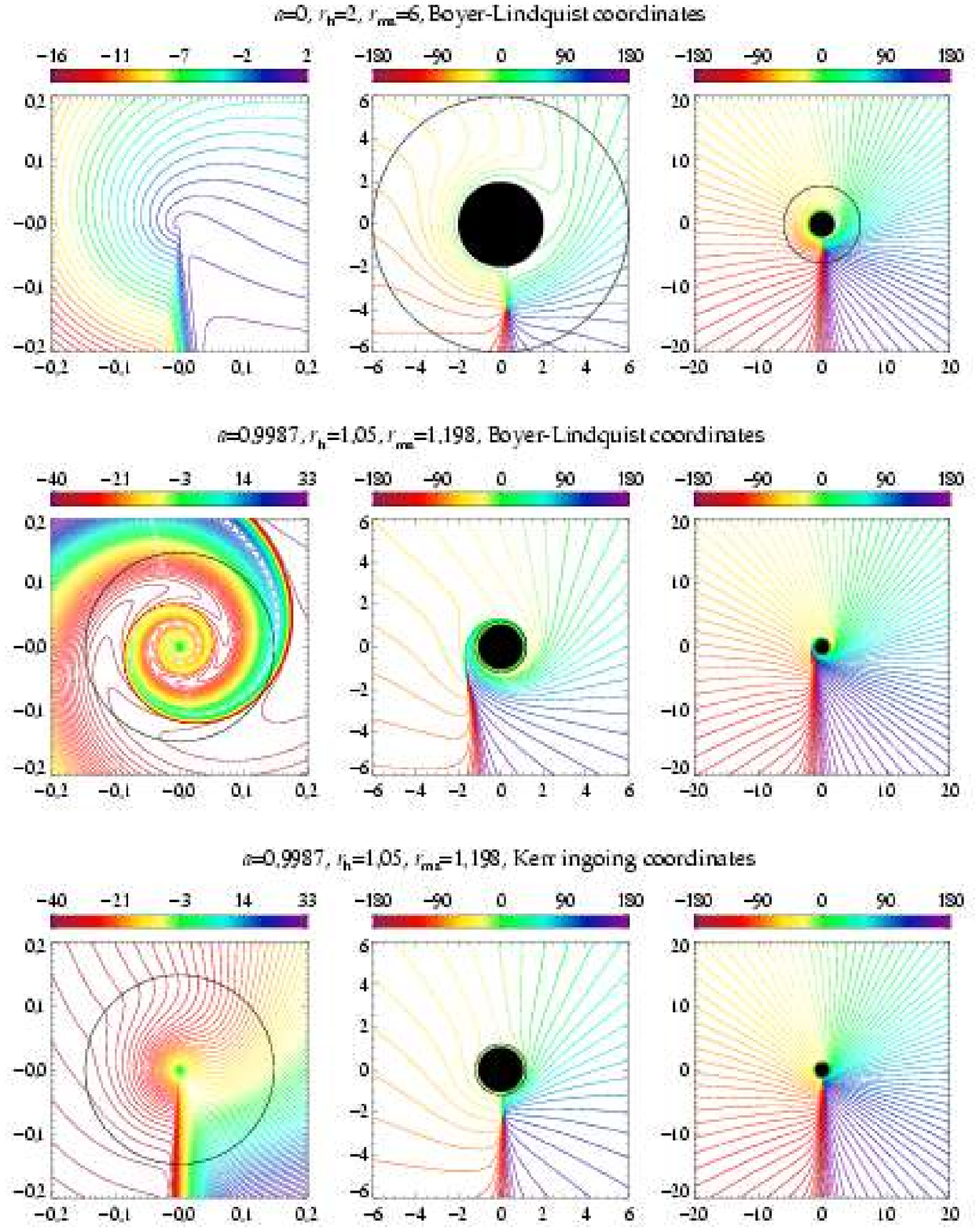}
\vspace{1em}
\mycaption{Azimuthal emission angle for the inclination $\theta_{\rm o}=85^\circ$.}
\end{figure}

\begin{figure}[tb]
\vspace*{-0.9em}
\dummycaption\label{transf_00}
\phantomsection\addcontentsline{toc}{section}{Overall transfer function}
\phantomsection\addcontentsline{toc}{subsection}{Inclination 0.1\r{ }}
\includegraphics[width=15cm]{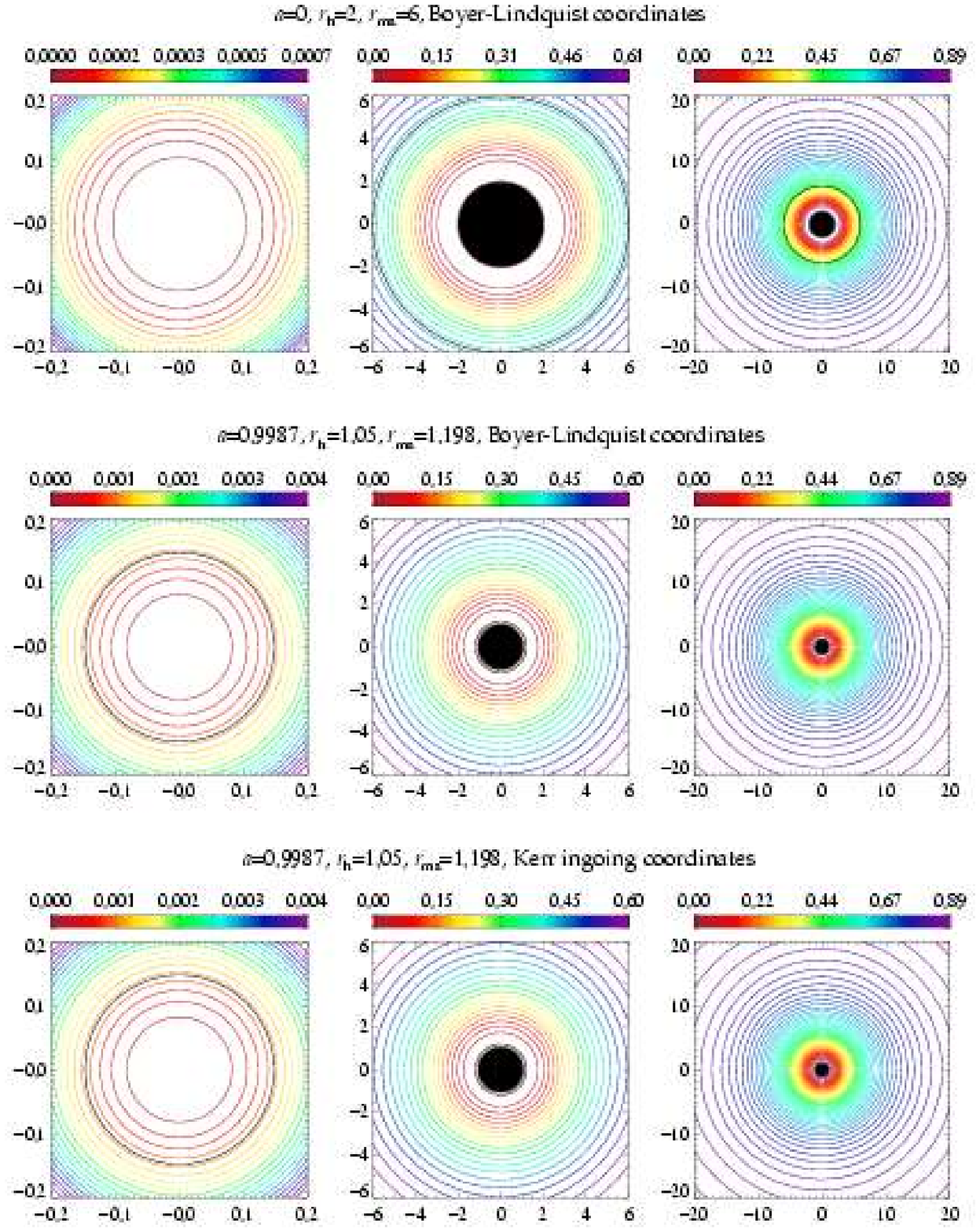}
\vspace{1em}
\mycaption{Overall transfer function for the inclination $\theta_{\rm o}=0.1^\circ$.}
\end{figure}

\begin{figure}[tb]
\vspace*{-0.9em}
\dummycaption\label{transf_45}
\phantomsection\addcontentsline{toc}{subsection}{Inclination 45\r{ }}
\includegraphics[width=15cm]{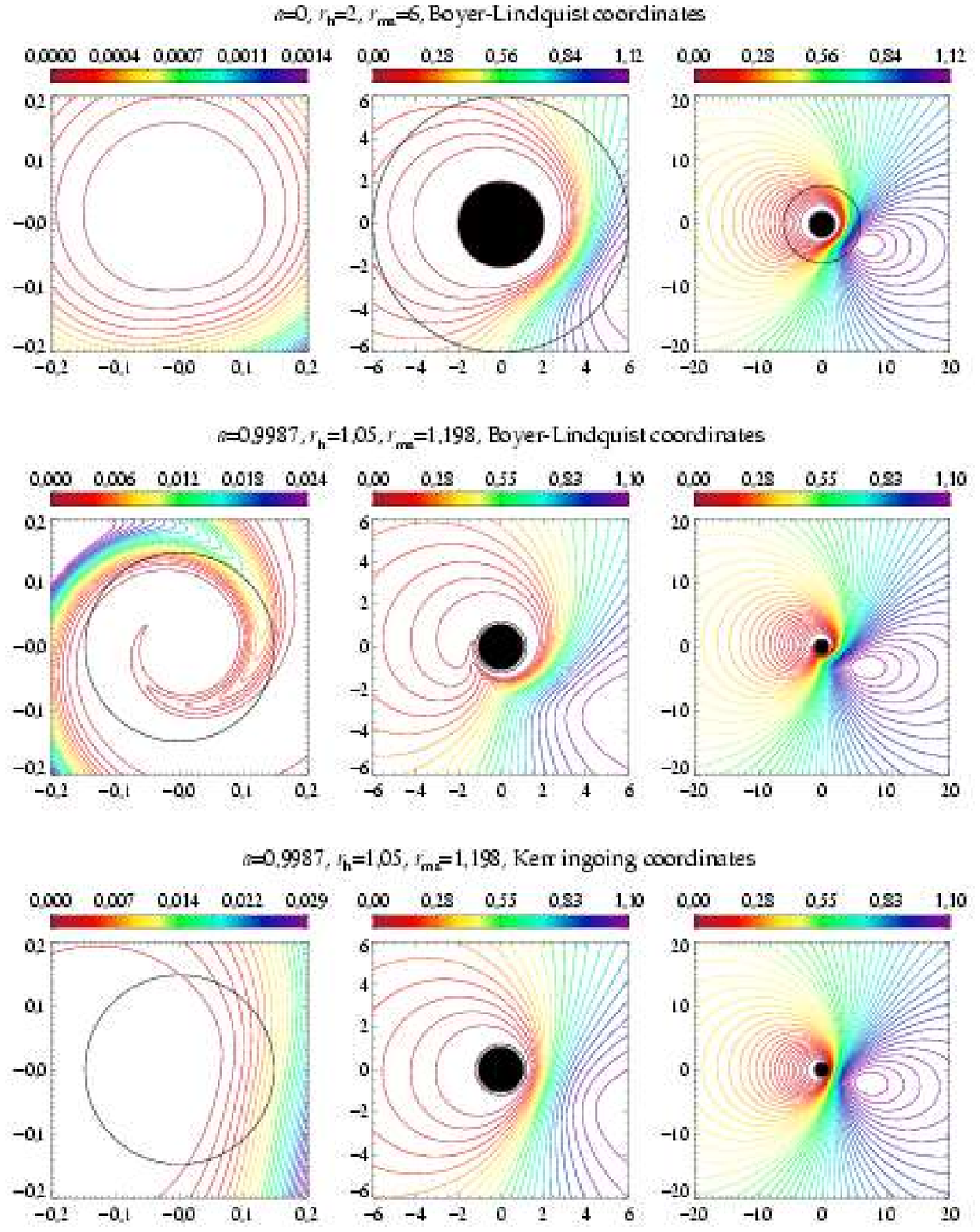}
\vspace{1em}
\mycaption{Overall transfer function for the inclination $\theta_{\rm o}=45^\circ$.}
\end{figure}

\begin{figure}[tb]
\vspace*{-0.9em}
\dummycaption\label{transf_85}
\phantomsection\addcontentsline{toc}{subsection}{Inclination 85\r{ }}
\includegraphics[width=15cm]{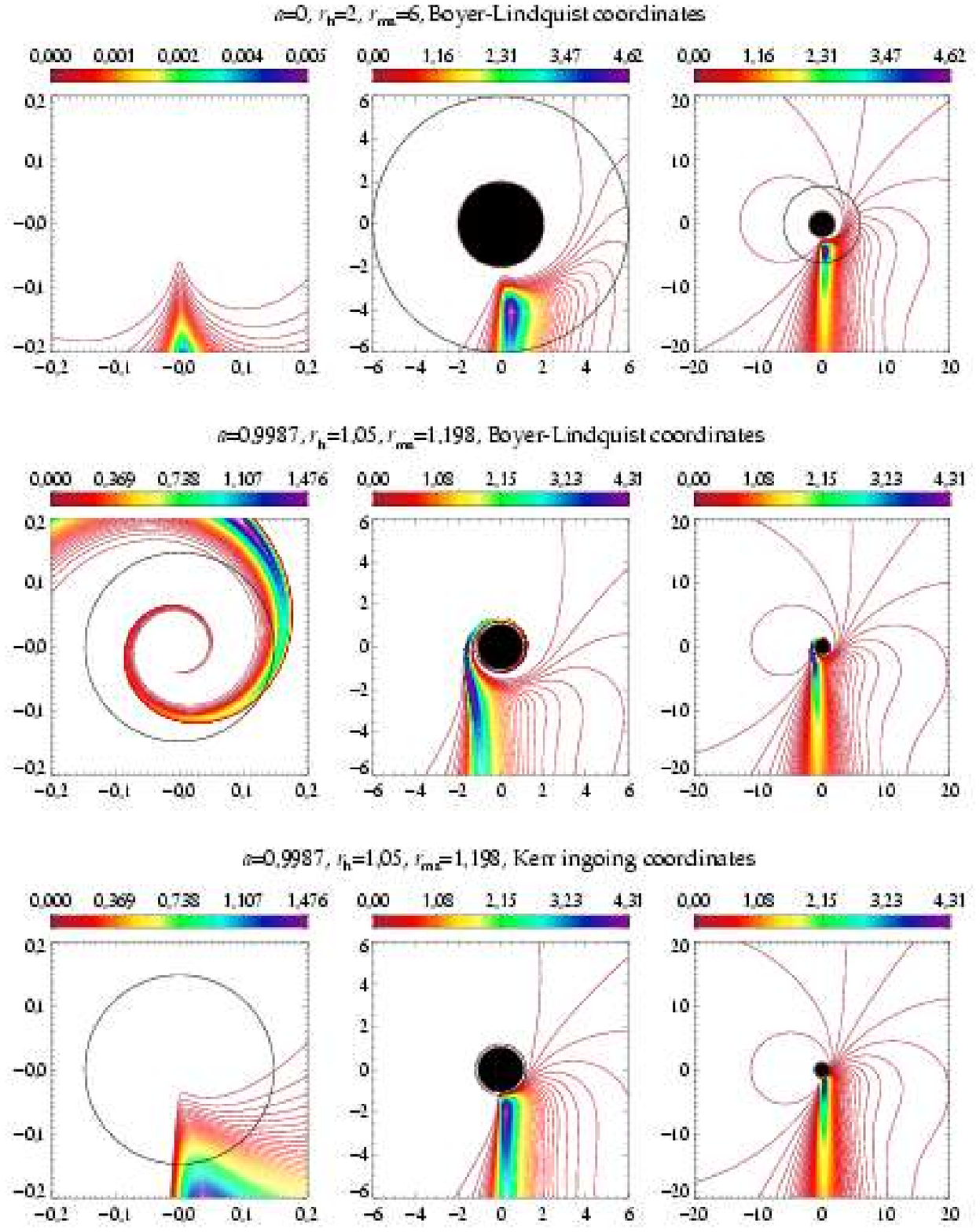}
\vspace{1em}
\mycaption{Overall transfer function for the inclination $\theta_{\rm o}=85^\circ$.}
\end{figure}

\begin{figure}[tb]
\vspace*{-0.9em}
\dummycaption\label{functions1_1000}
\phantomsection\addcontentsline{toc}{section}{Scale 1000x1000}
\phantomsection\addcontentsline{toc}{subsection}{g-factor, emission angle and lensing}
\includegraphics[width=15cm]{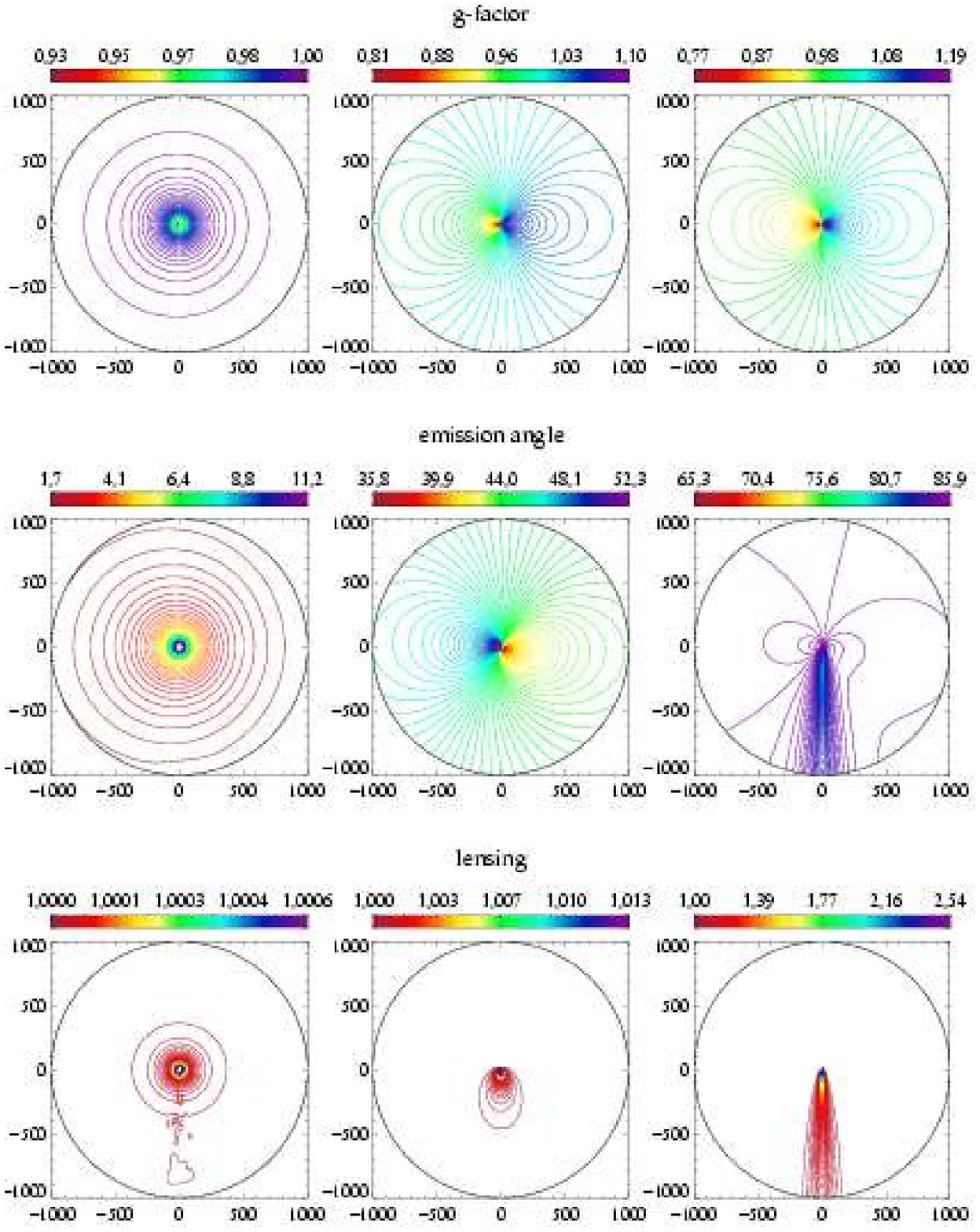}
\vspace{0.2em}
\mycaption{$g$-factor, emission angle and lensing for the inclinations
$\theta_{\rm o}=0.1^\circ,\,45^\circ$ and $85^\circ$ (from left to right).
The strange features in the graph for lensing with the observer inclination
$0.1^\circ$ are due to numerical errors which arise when integrating geodesics
passing close to the axis in spherical coordinates. Notice, though, that the
errors are not significantly large.}
\end{figure}

\begin{figure}[tb]
\vspace*{-0.9em}
\dummycaption\label{functions2_1000}
\phantomsection\addcontentsline{toc}{subsection}{Rel. time delay, 
change of polar. angle and azimuthal emission angle}
\includegraphics[width=15cm]{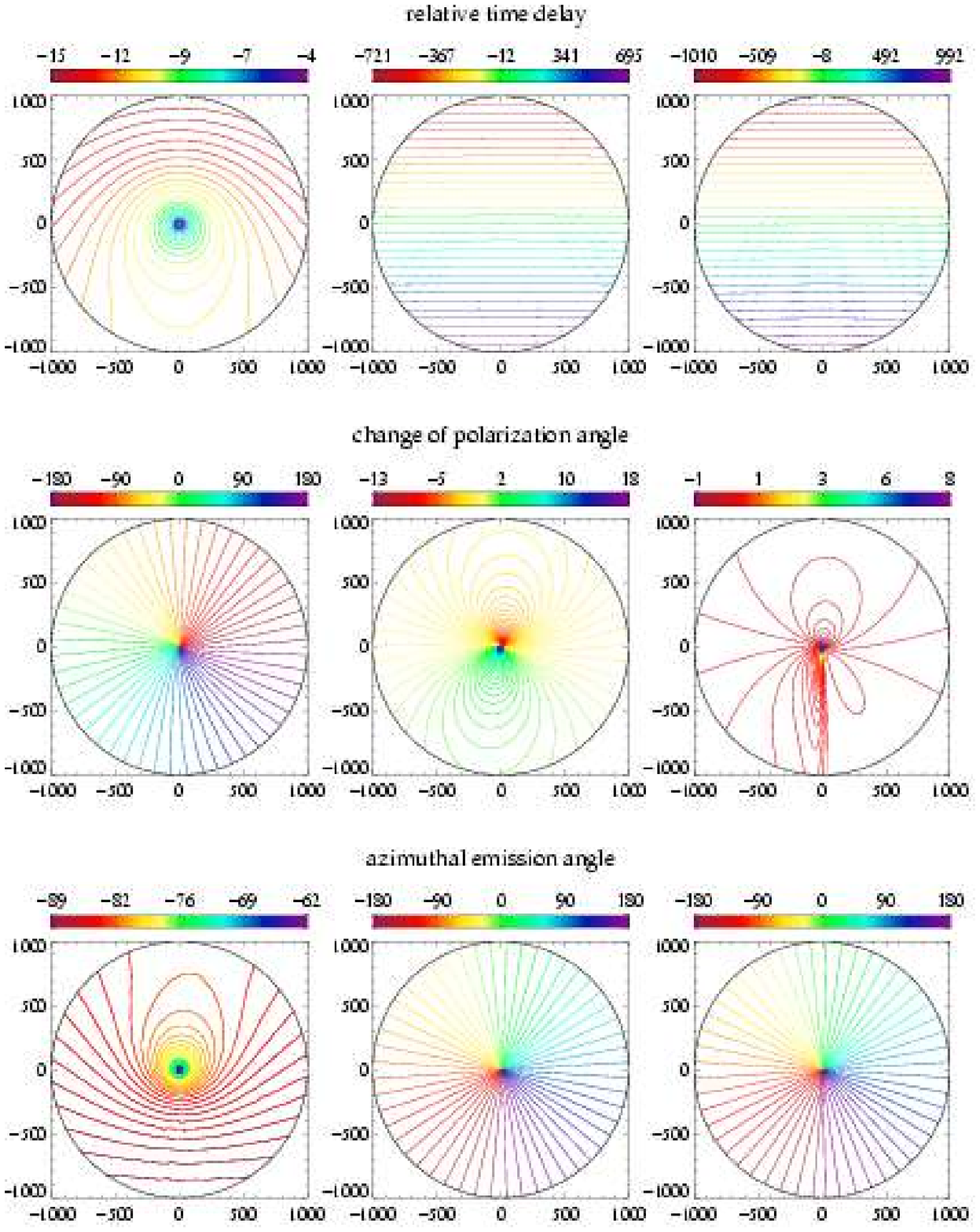}
\vspace{1em}
\mycaption{Relative time delay, change of the polarization angle and azimuthal
emission angle for the inclinations $\theta_{\rm o}=0.1^\circ,\,45^\circ$ and
$85^\circ$ (from left to right).}
\end{figure}

The following contour graphs of the functions represent a top view of the
equatorial plane for three different observer's inclination angles
($\theta_{\rm o}=0.1^\circ,\, 45^\circ$ and $85^\circ$) in four different
spatial scales (three columns in Figs.~\ref{gfac_00}--\ref{transf_85} and
the coarsest scale in Figs.~\ref{functions1_1000}--\ref{functions2_1000}).
In the finest scale (the first column) we define the radial coordinate
in a different way to that in the other columns. Specifically, we use
${r^{\prime}}^2\equiv{x^{\prime}}^2+{y^{\prime}}^2=(r-r_{\rm h})^2$.
The use of $r^{\prime}$ brings the horizon $r_{\rm h}$ to the origin so that
the region just outside the black hole is well resolved in these plots.
We show figures for two values of an angular
momentum $a$ of the black hole (rows in Figs.~\ref{gfac_00}--\ref{transf_85}),
particularly for the Schwarzschild black hole
($a=0$, the horizon is at $r_{\rm h}=2$ and the marginally stable orbit is at
$r_{\rm ms}=6$) and for an almost extreme Kerr black hole ($a\doteq 0.9987$,
the horizon is at $r_{\rm h}=1.05$ and the marginally stable orbit is at
$r_{\rm ms}\doteq 1.198$). The clock-wise distortion of the contours visible in
the second row is due to the frame-dragging near a rapidly
rotating Kerr black hole, which is clearly visible in the Boyer-Lindquist
coordinates. This effect is largely eliminated by the transformation
from the Boyer-Lindquist to the Kerr ingoing coordinates (the last row of panels
in Figs.~\ref{gfac_00}--\ref{transf_85}).

An observer is located to the top of the graphs. The black hole rotates
counter-clockwise. The values of the functions are encoded by a colour scale,
as indicated above each panel. The marginally stable orbit is also shown
(drawn as a circle) where relevant.

The atlas shown here is only a part of a larger set that contains figures for
more values of the observer's inclination and angular momentum of the black hole
(including an extreme Kerr black hole). An interested reader can ask for the
whole atlas by contacting the author at 
\href{mailto:dovciak@mbox.troja.mff.cuni.cz}{{\tt dovciak@mbox.troja.mff.cuni.cz}}.

\clearpage

\clearpage
\clearpage

\clearpage
\clearpage
\clearpage

\clearpage
\clearpage
\clearpage

\clearpage
\clearpage
\clearpage

\clearpage
\clearpage
\clearpage

\clearpage
\clearpage
\clearpage

\clearpage
\clearpage
\clearpage

\clearpage
\clearpage


\clearpage
\phantomsection\addcontentsline{toc}{chapter}{Bibliography}
\setlength{\bibsep}{0.2em}
\bibliographystyle{mynatbib}
\bibliography{references}

\providecommand{\noopsort}[1]{}
\begin{thebibliography}{} \thispagestyle{empty}

\bibitem[{Abramowicz} {\em et~al.}(1991){Abramowicz}, {Bao}, {Lanza} \&
  {Zhang}]{abramowicz1991}
{Abramowicz} M.~A., {Bao} G., {Lanza} A.  \& {Zhang} X.-H. (1991).
\newblock {X-ray variability power spectra of active galactic nuclei}.
\newblock {\em \aap\/}, {\bf 245}, 454{.
\newblock}\ads
  {http://adsabs.harvard.edu/cgi-bin/nph-bib_query?bibcode=1991A\%26A...245..4%
54A&amp;db_key=AST}.

\bibitem[{Abramowicz} {\em et~al.}(1996){Abramowicz}, {Beloborodov}, {Chen} \&
  {Igumenshchev}]{abramowicz1996}
{Abramowicz} M.~A., {Beloborodov} A.~M., {Chen} X.-M.  \& {Igumenshchev} I.~V.
  (1996).
\newblock {Special relativity and the pseudo-Newtonian potential.}
\newblock {\em \aap\/}, {\bf 313}, 334{.
\newblock}\ads
  {http://adsabs.harvard.edu/cgi-bin/nph-bib_query?bibcode=1996A\%26A...313..3%
34A&amp;db_key=AST}.

\bibitem[{Agol}(1997){Agol}]{agol1997}
{Agol} E. (1997).
\newblock {The effects of magnetic fields, absorption, and relativity on the
  polarization of accretion disks around supermassive black holes}.
\newblock {\em Ph.D.~Thesis\/}{.
\newblock}\ads
  {http://adsabs.harvard.edu/cgi-bin/nph-bib_query?bibcode=1997PhDT.........2A%
&amp;db_key=AST}.

\bibitem[{Arnaud}(1996){Arnaud}]{arnaud1996}
{Arnaud} K.~A. (1996).
\newblock {XSPEC: The first ten years}.
\newblock In {\em {Astronomical data analysis software and systems V}\/},
  volume 101, page~17. Jacoby~G. \& Barnes~J., ASP Conf.\ Series{.
\newblock}\ads
  {http://adsabs.harvard.edu/cgi-bin/nph-bib_query?bibcode=1996adass...5...17A%
&amp;db_key=AST}.

\bibitem[{Artemova}, {Bj\"ornsson} \& {Novikov}(1996){Artemova}, {Bj\"ornsson}
  \& {Novikov}]{artemova1996}
{Artemova} I.~V., {Bj\"ornsson} G.  \& {Novikov} I.~D. (1996).
\newblock {Modified Newtonian potentials for the description of relativistic
  effects in accretion disks around black holes}.
\newblock {\em \apj\/}, {\bf 461}, 565{.
\newblock}\ads
  {http://adsabs.harvard.edu/cgi-bin/nph-bib_query?bibcode=1996ApJ...461..565A%
&amp;db_key=AST}.

\bibitem[{Asaoka}(1989){Asaoka}]{asaoka1989}
{Asaoka} I. (1989).
\newblock {X-ray spectra at infinity from a relativistic accretion disk around
  a Kerr black hole}.
\newblock {\em \pasj\/}, {\bf 41}, 763{.
\newblock}\ads
  {http://adsabs.harvard.edu/cgi-bin/nph-bib_query?bibcode=1989PASJ...41..763A%
&amp;db_key=AST}.

\bibitem[{Ballantyne}, {Ross} \& {Fabian}(2001){Ballantyne}, {Ross} \&
  {Fabian}]{ballantyne2001}
{Ballantyne} D.~R., {Ross} R.~R.  \& {Fabian} A.~C. (2001).
\newblock {X-ray reflection by photoionized accretion discs}.
\newblock {\em \mnras\/}, {\bf 327}, 10{.
\newblock}\ads
  {http://adsabs.harvard.edu/cgi-bin/nph-bib_query?bibcode=2001MNRAS.327...10B%
&amp;db_key=AST}.

\bibitem[{Ballantyne}, {Vaughan} \& {Fabian}(2003){Ballantyne}, {Vaughan} \&
  {Fabian}]{ballantyne2003}
{Ballantyne} D.~R., {Vaughan} S.  \& {Fabian} A.~C. (2003).
\newblock {A two-component ionized reflection model of MCG--6-30-15}.
\newblock {\em \mnras\/}, {\bf 342}, 239{.
\newblock}\ads
  {http://adsabs.harvard.edu/cgi-bin/nph-bib_query?bibcode=2003MNRAS.342..239B%
&amp;db_key=AST}.

\bibitem[{Bao}, {Hadrava} \& {\O stgaard}(1994){Bao}, {Hadrava} \& {\O
  stgaard}]{bao1994}
{Bao} G., {Hadrava} P.  \& {\O stgaard} E. (1994).
\newblock {Emission-line profiles from a relativistic accretion disk and the
  role of its multiple images}.
\newblock {\em \apj\/}, {\bf 435}, 55{.
\newblock}\ads
  {http://adsabs.harvard.edu/cgi-bin/nph-bib_query?bibcode=1994ApJ...435...55B%
&amp;db_key=AST}.

\bibitem[{Bao} {\em et~al.}(1997){Bao}, {Hadrava}, {Wiita} \& {Xiong}]{bao1997}
{Bao} G., {Hadrava} P., {Wiita} P.  \& {Xiong} Y. (1997).
\newblock {Polarization variability of active galactic nuclei and X-ray
  binaries}.
\newblock {\em \apj\/}, {\bf 487}, 142{.
\newblock}\ads
  {http://adsabs.harvard.edu/cgi-bin/nph-bib_query?bibcode=1997ApJ...487..142B%
&amp;db_key=AST}.

\bibitem[{\noopsort{Baoa}}{Bao}, {Wiita} \&
  {Hadrava}(1996){\noopsort{Baoa}}{Bao}, {Wiita} \& {Hadrava}]{bao1996}
{\noopsort{Baoa}}{Bao} G., {Wiita} P.~J.  \& {Hadrava} P. (1996).
\newblock {Energy-dependent polarization variability as a black hole
  signature}.
\newblock {\em \prl\/}, {\bf 77}, 12{.
\newblock}\ads
  {http://adsabs.harvard.edu/cgi-bin/nph-bib_query?bibcode=1996PhRvL..77...12B%
&amp;db_key=AST}.

\bibitem[{Bardeen}(1973){Bardeen}]{bardeen1973}
{Bardeen} J.~M. (1973).
\newblock {Rapidly rotating stars, disks, and black holes}.
\newblock In {\em {Black holes}\/}, page 241. C.~de~Witt and B.~S.~de~Witt,
  Gordon \& Breach, New York.

\bibitem[{Bardeen}, {Press} \& {Teukolsky}(1972){Bardeen}, {Press} \&
  {Teukolsky}]{bardeen1972}
{Bardeen} J.~M., {Press} W.~H.  \& {Teukolsky} S.~A. (1972).
\newblock {Rotating black holes: locally nonrotating frames, energy extraction,
  and scalar synchrotron radiation}.
\newblock {\em \apj\/}, {\bf 178}, 347{.
\newblock}\ads
  {http://adsabs.harvard.edu/cgi-bin/nph-bib_query?bibcode=1972ApJ...178..347B%
&amp;db_key=AST}.

\bibitem[{Beckwith} \& {Done}(2004){Beckwith} \& {Done}]{beckwith2004}
{Beckwith} K.  \& {Done} C. (2004).
\newblock {Iron line profiles in strong gravity}.
\newblock {\em \mnras\/}.
\newblock In press. \href{http://arxiv.org/abs/astro-ph/0402199}
  {[astro-ph/0402199]}.

\bibitem[{Bianchi} {\em et~al.}(2004){Bianchi}, {Matt}, {Balestra}, {Guainazzi}
  \& {Perola}]{bianchi2004}
{Bianchi} S., {Matt} G., {Balestra} I., {Guainazzi} M.  \& {Perola} G.~C.
  (2004).
\newblock {X-ray reprocessing in Seyfert galaxies: simultaneous
  XMM-Newton/BeppoSAX observations}.
\newblock {\em \aap\/}.
\newblock Submitted. \href{http://arxiv.org/abs/astro-ph/0404308}
  {[astro-ph/0404308]}.

\bibitem[{Bi\v{c}\'{a}k}, {Semer\'{a}k} \& {Hadrava}(1993){Bi\v{c}\'{a}k},
  {Semer\'{a}k} \& {Hadrava}]{bicak1993}
{Bi\v{c}\'{a}k} J., {Semer\'{a}k} O.  \& {Hadrava} P. (1993).
\newblock {Collimation effects of the Kerr field}.
\newblock {\em \mnras\/}, {\bf 263}, 545{.
\newblock}\ads
  {http://adsabs.harvard.edu/cgi-bin/nph-bib_query?bibcode=1993MNRAS.263..545B%
&amp;db_key=AST}.

\bibitem[{Blandford} \& {Znajek}(1977){Blandford} \& {Znajek}]{blandford1977}
{Blandford} R.~D.  \& {Znajek} R.~L. (1977).
\newblock {Electromagnetic extraction of energy from Kerr black holes}.
\newblock {\em \mnras\/}, {\bf 179}, 433{.
\newblock}\ads
  {http://adsabs.harvard.edu/cgi-bin/nph-bib_query?bibcode=1977MNRAS.179..433B%
&amp;db_key=AST}.

\bibitem[{Boller} {\em et~al.}(2001){Boller}, {Keil}, {Tr{\" u}mper},
  {O'Brien}, {Reeves} \& {Page}]{boller2001}
{Boller} T., {Keil} R., {Tr{\" u}mper} J., {O'Brien} P.~T., {Reeves} J.  \&
  {Page} M. (2001).
\newblock {Detection of an X-ray periodicity in the narrow-line Seyfert 1
  galaxy Mrk 766 with XMM-Newton}.
\newblock {\em \aap\/}, {\bf 365}, L146{.
\newblock}\ads
  {http://adsabs.harvard.edu/cgi-bin/nph-bib_query?bibcode=2001A\%26A...365L.1%
46B&amp;db_key=AST}.

\bibitem[{Bondi} \& {Hoyle}(1944){Bondi} \& {Hoyle}]{bondi1944}
{Bondi} H.  \& {Hoyle} F. (1944).
\newblock {On the mechanism of accretion by stars}.
\newblock {\em \mnras\/}, {\bf 104}, 273{.
\newblock}\ads
  {http://adsabs.harvard.edu/cgi-bin/nph-bib_query?bibcode=1944MNRAS.104..273B%
&amp;db_key=AST}.

\bibitem[{Bowyer} {\em et~al.}(1965){Bowyer}, {Byram}, {Chubb} \&
  {Friedman}]{bowyer1965}
{Bowyer} S., {Byram} E.~T., {Chubb} T.~A.  \& {Friedman} H. (1965).
\newblock {Cosmic X-ray sources}.
\newblock {\em Science\/}, {\bf 147}, 394{.
\newblock}\ads
  {http://adsabs.harvard.edu/cgi-bin/nph-bib_query?bibcode=1965Sci...147..394B%
&amp;db_key=AST}.

\bibitem[{Bromley}, {Chen} \& {Miller}(1997){Bromley}, {Chen} \&
  {Miller}]{bromley1997}
{Bromley} B.~C., {Chen} K.  \& {Miller} W.~A. (1997).
\newblock {Line emission from an accretion disk around a rotating black hole:
  toward a measurement of frame dragging}.
\newblock {\em \apj\/}, {\bf 475}, 57{.
\newblock}\ads
  {http://adsabs.harvard.edu/cgi-bin/nph-bib_query?bibcode=1997ApJ...475...57B%
&amp;db_key=AST}.

\bibitem[{\noopsort{Cadez}}{{\v C}ade{\v z}}, {Calvani} \&
  {Fanton}(2003){\noopsort{Cadez}}{{\v C}ade{\v z}}, {Calvani} \&
  {Fanton}]{cadez2003}
{\noopsort{Cadez}}{{\v C}ade{\v z}} A., {Calvani} M.  \& {Fanton} C. (2003).
\newblock {X-ray iron line profiles from warped accretion discs}.
\newblock {\em Memorie della Societa Astronomica Italiana\/}, {\bf 74}, 446{.
\newblock}\ads
  {http://adsabs.harvard.edu/cgi-bin/nph-bib_query?bibcode=2003MmSAI..74..446C%
&amp;db_key=AST}.

\bibitem[{Carter}(1968){Carter}]{carter1968}
{Carter} B. (1968).
\newblock {Global structure of the Kerr family of gravitational fields}.
\newblock {\em Phys. Rev.}, {\bf 174}, 1559{.
\newblock}\ads
  {http://adsabs.harvard.edu/cgi-bin/nph-bib_query?bibcode=1968PhRv..174.1559C%
&amp;db_key=PHY}.

\bibitem[{Chandrasekhar}(1931){Chandrasekhar}]{chandrasekhar1931}
{Chandrasekhar} S. (1931).
\newblock {The maximum mass of ideal white dwarfs}.
\newblock {\em \apj\/}, {\bf 74}, 81{.
\newblock}\ads
  {http://adsabs.harvard.edu/cgi-bin/nph-bib_query?bibcode=1931ApJ....74...81C%
&amp;db_key=AST}.

\bibitem[{Chandrasekhar}(1960){Chandrasekhar}]{chandrasekhar1960}
{Chandrasekhar} S. (1960).
\newblock {\em {Radiative transfer}\/}.
\newblock Dover publications, New York{.
\newblock}\ads
  {http://adsabs.harvard.edu/cgi-bin/nph-bib_query?bibcode=1960ratr.book.....C%
&amp;db_key=AST}.

\bibitem[{Chandrasekhar}(1983){Chandrasekhar}]{chandrasekhar1983}
{Chandrasekhar} S. (1983).
\newblock {\em {The mathematical theory of black holes}\/}.
\newblock Clarendon Press/Oxford University Press (International Series of
  Monographs on Physics.~Volume 69), p.~663{.
\newblock}\ads
  {http://adsabs.harvard.edu/cgi-bin/nph-bib_query?bibcode=1983mtbh.book.....C%
&amp;db_key=AST}.

\bibitem[{Chen} \& {Eardley}(1991){Chen} \& {Eardley}]{chen1991}
{Chen} K.  \& {Eardley} D.~M. (1991).
\newblock {Polarization properties of emission lines from relativistic
  accretion disks}.
\newblock {\em \apj\/}, {\bf 382}, 125{.
\newblock}\ads
  {http://adsabs.harvard.edu/cgi-bin/nph-bib_query?bibcode=1991ApJ...382..125C%
&amp;db_key=AST}.

\bibitem[{\noopsort{Connorsa}}{Connors}, {Piran} \&
  {Stark}(1980){\noopsort{Connorsa}}{Connors}, {Piran} \& {Stark}]{connors1980}
{\noopsort{Connorsa}}{Connors} P.~A., {Piran} T.  \& {Stark} R.~F. (1980).
\newblock {Polarization features of X-ray radiation emitted near black holes}.
\newblock {\em \apj\/}, {\bf 235}, 224{.
\newblock}\ads
  {http://adsabs.harvard.edu/cgi-bin/nph-bib_query?bibcode=1980ApJ...235..224C%
&amp;db_key=AST}.

\bibitem[{\noopsort{Connorsb}}{Connors} \&
  {Stark}(1977){\noopsort{Connorsb}}{Connors} \& {Stark}]{connors1977}
{\noopsort{Connorsb}}{Connors} P.~A.  \& {Stark} R.~F. (1977).
\newblock {Observable gravitational effects on polarised radiation coming from
  near a black hole}.
\newblock {\em Nature\/}, {\bf 269}, 128{.
\newblock}\ads
  {http://adsabs.harvard.edu/cgi-bin/nph-bib_query?bibcode=1977Natur.269..128C%
&amp;db_key=AST}.

\bibitem[{Costa} {\em et~al.}(2001){Costa}, {Soffitta}, {Bellazzini}, {Brez},
  {Lumb} \& {Spandre}]{costa2001}
{Costa} E., {Soffitta} P., {Bellazzini} R., {Brez} A., {Lumb} N.  \& {Spandre}
  G. (2001).
\newblock {An efficient photoelectric X-ray polarimeter for the study of black
  holes and neutron stars}.
\newblock {\em \nat\/}, {\bf 411}, 662{.
\newblock}\ads
  {http://adsabs.harvard.edu/cgi-bin/nph-bib_query?bibcode=2001Natur.411..662C%
&amp;db_key=AST}.

\bibitem[{Cunningham}(1975){Cunningham}]{cunningham1975}
{Cunningham} C.~T. (1975).
\newblock {The effects of redshifts and focusing on the spectrum of an
  accretion disk around a Kerr black hole}.
\newblock {\em \apj\/}, {\bf 202}, 788{.
\newblock}\ads
  {http://adsabs.harvard.edu/cgi-bin/nph-bib_query?bibcode=1975ApJ...202..788C%
&amp;db_key=AST}.

\bibitem[{Cunningham}(1976){Cunningham}]{cunningham1976}
{Cunningham} C.~T. (1976).
\newblock {Returning radiation in accretion disks around black holes.}
\newblock {\em \apj\/}, {\bf 208}, 534{.
\newblock}\ads
  {http://adsabs.harvard.edu/cgi-bin/nph-bib_query?bibcode=1976ApJ...208..534C%
&amp;db_key=AST}.

\bibitem[{Curtis}(1918){Curtis}]{curtis1918}
{Curtis} H.~D. (1918).
\newblock {\em Publications of Lick Observatory\/}, {\bf 13}, 11{.
\newblock}\ads
  {http://adsabs.harvard.edu/cgi-bin/nph-bib_query?bibcode=1918PLicO..13.0000C%
&amp;db_key=AST}.

\bibitem[{Czerny} {\em et~al.}(2004){Czerny}, {R{\' o}{\. z}a{\' n}ska},
  {Dov{\v c}iak}, {Karas} \& {Dumont}]{czerny2004}
{Czerny} B., {R{\' o}{\. z}a{\' n}ska} A., {Dov{\v c}iak} M., {Karas} V.  \&
  {Dumont} A.-M. (2004).
\newblock {The structure and radiation spectra of illuminated accretion disks
  in AGN. II. Flare/spot model of X-ray variability}.
\newblock {\em \aap\/}, {\bf 420}, 1{.
\newblock}\ads
  {http://adsabs.harvard.edu/cgi-bin/nph-bib_query?bibcode=2004A\%26A...420...%
.1C&amp;db_key=AST}.

\bibitem[{Dabrowski} {\em et~al.}(1997){Dabrowski}, {Fabian}, {Iwasawa},
  {Lasenby} \& {Reynolds}]{dabrowski1997}
{Dabrowski} Y., {Fabian} A.~C., {Iwasawa} K., {Lasenby} A.~N.  \& {Reynolds}
  C.~S. (1997).
\newblock {The profile and equivalent width of the X-ray iron emission line
  from a disc around a Kerr black hole}.
\newblock {\em \mnras\/}, {\bf 288}, L11{.
\newblock}\ads
  {http://adsabs.harvard.edu/cgi-bin/nph-bib_query?bibcode=1997MNRAS.288L..11D%
&amp;db_key=AST}.

\bibitem[{Damour}(1980){Damour}]{damour1980}
{Damour} T. (1980).
\newblock {Mechanical, electrodynamical and thermodynamical properties of black
  holes}.
\newblock In {\em Gravitational Radiation, Collapsed Objects, and Exact
  Solutions\/}, page 454{.
\newblock}\ads
  {http://adsabs.harvard.edu/cgi-bin/nph-bib_query?bibcode=1980grco.symp..454D%
&amp;db_key=AST}.

\bibitem[{Damour} {\em et~al.}(1978){Damour}, {Ruffini}, {Hanni} \&
  {Wilson}]{damour1978}
{Damour} T., {Ruffini} R., {Hanni} R.~S.  \& {Wilson} J.~R. (1978).
\newblock {Regions of magnetic support of a plasma around a black hole}.
\newblock {\em \prd\/}, {\bf 17}, 1518{.
\newblock}\ads
  {http://adsabs.harvard.edu/cgi-bin/nph-bib_query?bibcode=1978PhRvD..17.1518D%
&amp;db_key=AST}.

\bibitem[{\noopsort{Dovciaka}}{Dov{\v c}iak} {\em
  et~al.}(2004a){\noopsort{Dovciaka}}{Dov{\v c}iak}, {Bianchi}, {Guainazzi},
  {Karas} \& {Matt}]{dovciak2004a}
{\noopsort{Dovciaka}}{Dov{\v c}iak} M., {Bianchi} S., {Guainazzi} M., {Karas}
  V.  \& {Matt} G. (\protect{2004a}).
\newblock {Relativistic spectral features from X-ray-illuminated spots and the
  measure of the black hole mass in active galactic nuclei}.
\newblock {\em \mnras\/}, {\bf 350}, 745{.
\newblock}\ads
  {http://adsabs.harvard.edu/cgi-bin/nph-bib_query?bibcode=2004MNRAS.350..745D%
&amp;db_key=AST}.

\bibitem[{\noopsort{Dovciakb}}{Dov\v{c}iak} {\em
  et~al.}(2004b){\noopsort{Dovciakb}}{Dov\v{c}iak}, {Karas}, {Martocchia},
  {Matt} \& {Yaqoob}]{dovciak2004b}
{\noopsort{Dovciakb}}{Dov\v{c}iak} M., {Karas} V., {Martocchia} A., {Matt} G.
  \& {Yaqoob} T. (\protect{2004b}).
\newblock {XSPEC model to explore spectral features from black hole sources}.
\newblock In {\em Proc. of the workshop on processes in the vicinity of black
  holes and neutron stars\/}. S.~Hled\'{\i}k \& Z.~Stuchl\'{\i}k, Opava.
\newblock In press{.
\newblock}\ads
  {http://adsabs.harvard.edu/cgi-bin/nph-bib_query?bibcode=2004astro.ph..7330D%
&amp;db_key=PRE&amp;high=4190a5aca304944}.

\bibitem[{\noopsort{Dovciakc}}{Dov{\v c}iak}, {Karas} \&
  {Matt}(2004){\noopsort{Dovciakc}}{Dov{\v c}iak}, {Karas} \&
  {Matt}]{dovciak2004c}
{\noopsort{Dovciakc}}{Dov{\v c}iak} M., {Karas} V.  \& {Matt} G. (2004).
\newblock {Polarization signatures of strong gravity in AGN accretion discs}.
\newblock {\em \mnras\/}.
\newblock In press{.
\newblock}\ads
  {http://adsabs.harvard.edu/cgi-bin/nph-bib_query?doi=10.1111\%2Fj.1365-2966.%
2004.08396.x&amp;db_key=AST&amp;high=4190a5aca304944}.

\bibitem[{\noopsort{Dovciakd}}{Dov\v{c}iak}, {Karas} \&
  {Yaqoob}(2004){\noopsort{Dovciakd}}{Dov\v{c}iak}, {Karas} \&
  {Yaqoob}]{dovciak2004}
{\noopsort{Dovciakd}}{Dov\v{c}iak} M., {Karas} V.  \& {Yaqoob} T. (2004).
\newblock {An extended scheme for fitting X-ray data with accretion disk
  spectra in the strong gravity regime}.
\newblock {\em \apjs\/}, {\bf 153}, 205{.
\newblock}\ads
  {http://adsabs.harvard.edu/cgi-bin/nph-bib_query?bibcode=2004ApJS..153..205D%
&amp;db_key=AST}.

\bibitem[{Einstein}(1916){Einstein}]{einstein1916}
{Einstein} A. (1916).
\newblock {Die Grundlage ger allgemeinen Relativit\"{a}tstheorie}.
\newblock {\em Ann. Phys. {\rm (Leipzig)}\/}, {\bf 49}, 769.

\bibitem[{\noopsort{Fabiana}}{Fabian} {\em
  et~al.}(2000){\noopsort{Fabiana}}{Fabian}, {Iwasawa}, {Reynolds} \&
  {Young}]{fabian2000}
{\noopsort{Fabiana}}{Fabian} A.~C., {Iwasawa} K., {Reynolds} C.~S.  \& {Young}
  A.~J. (2000).
\newblock {Broad iron lines in active galactic nuclei}.
\newblock {\em \pasp\/}, {\bf 112}, 1145{.
\newblock}\ads
  {http://adsabs.harvard.edu/cgi-bin/nph-bib_query?bibcode=2000PASP..112.1145F%
&amp;db_key=AST}.

\bibitem[{\noopsort{Fabianb}}{Fabian} {\em
  et~al.}(1989){\noopsort{Fabianb}}{Fabian}, {Rees}, {Stella} \&
  {White}]{fabian1989}
{\noopsort{Fabianb}}{Fabian} A.~C., {Rees} M.~J., {Stella} L.  \& {White} N.~E.
  (1989).
\newblock {X-ray fluorescence from the inner disc in Cygnus X-1}.
\newblock {\em \mnras\/}, {\bf 238}, 729{.
\newblock}\ads
  {http://adsabs.harvard.edu/cgi-bin/nph-bib_query?bibcode=1989MNRAS.238..729F%
&amp;db_key=AST}.

\bibitem[{\noopsort{Fabianc}}{Fabian} \&
  {Vaughan}(2003){\noopsort{Fabianc}}{Fabian} \& {Vaughan}]{fabian2003}
{\noopsort{Fabianc}}{Fabian} A.~C.  \& {Vaughan} S. (2003).
\newblock {The iron line in MCG--6-30-15 from XMM-Newton: evidence for
  gravitational light bending?}
\newblock {\em \mnras\/}, {\bf 340}, L28{.
\newblock}\ads
  {http://adsabs.harvard.edu/cgi-bin/nph-bib_query?bibcode=2003MNRAS.340L..28F%
&amp;db_key=AST}.

\bibitem[{\noopsort{Fabiand}}{Fabian} {\em
  et~al.}(2002){\noopsort{Fabiand}}{Fabian}, {Vaughan}, {Nandra}, {Iwasawa},
  {Ballantyne}, {Lee}, {de Rosa}, {Turner} \& {Young}]{fabian2002}
{\noopsort{Fabiand}}{Fabian} A.~C., {Vaughan} S., {Nandra} K., {Iwasawa} K.,
  {Ballantyne} D.~R., {Lee} J.~C., {de Rosa} A., {Turner} A.  \& {Young} A.~J.
  (2002).
\newblock {A long hard look at MCG--6-30-15 with XMM-Newton}.
\newblock {\em \mnras\/}, {\bf 335}, L1{.
\newblock}\ads
  {http://adsabs.harvard.edu/cgi-bin/nph-bib_query?bibcode=2002MNRAS.335L...1F%
&amp;db_key=AST}.

\bibitem[{\noopsort{Felice}}{de Felice}, {Nobili} \&
  {Calvani}(1974){\noopsort{Felice}}{de Felice}, {Nobili} \&
  {Calvani}]{felice1974}
{\noopsort{Felice}}{de Felice} F., {Nobili} L.  \& {Calvani} M. (1974).
\newblock {Blackhole physics}.
\newblock {\em \aap\/}, {\bf 30}, 111{.
\newblock}\ads
  {http://adsabs.harvard.edu/cgi-bin/nph-bib_query?bibcode=1974A\%26A....30..1%
11D&amp;db_key=AST}.

\bibitem[{Fishbone} \& {Moncrief}(1976){Fishbone} \& {Moncrief}]{fishbone1976}
{Fishbone} L.~G.  \& {Moncrief} V. (1976).
\newblock {Relativistic fluid disks in orbit around Kerr black holes}.
\newblock {\em \apj\/}, {\bf 207}, 962{.
\newblock}\ads
  {http://adsabs.harvard.edu/cgi-bin/nph-bib_query?bibcode=1976ApJ...207..962F%
&amp;db_key=AST}.

\bibitem[{Frank}, {King} \& {Raine}(2002){Frank}, {King} \& {Raine}]{frank2002}
{Frank} J., {King} A.  \& {Raine} D.~J. (2002).
\newblock {\em {Accretion Power in Astrophysics: Third Edition}\/}.
\newblock Cambridge University Press{.
\newblock}\ads
  {http://adsabs.harvard.edu/cgi-bin/nph-bib_query?bibcode=2002apa..book.....F%
&amp;db_key=AST}.

\bibitem[{George} \& {Fabian}(1991){George} \& {Fabian}]{george1991}
{George} I.~M.  \& {Fabian} A.~C. (1991).
\newblock {X-ray reflection from cold matter in active galactic nuclei and
  X-ray binaries}.
\newblock {\em \mnras\/}, {\bf 249}, 352{.
\newblock}\ads
  {http://adsabs.harvard.edu/cgi-bin/nph-bib_query?bibcode=1991MNRAS.249..352G%
&amp;db_key=AST}.

\bibitem[{Gerbal} \& {Pelat}(1981){Gerbal} \& {Pelat}]{gerbal1981}
{Gerbal} D.  \& {Pelat} D. (1981).
\newblock {Profile of a line emitted by an accretion disk -- influence of the
  geometry upon its shape parameters}.
\newblock {\em \aap\/}, {\bf 95}, 18{.
\newblock}\ads
  {http://adsabs.harvard.edu/cgi-bin/nph-bib_query?bibcode=1981A\%26A....95...%
18G&amp;db_key=AST}.

\bibitem[{Ghisellini}, {Haardt} \& {Matt}(1994){Ghisellini}, {Haardt} \&
  {Matt}]{ghisellini1994}
{Ghisellini} G., {Haardt} F.  \& {Matt} G. (1994).
\newblock {The contribution of the obscuring torus to the X-ray spectrum of
  Seyfert galaxies -- a test for the unification model}.
\newblock {\em \mnras\/}, {\bf 267}, 743{.
\newblock}\ads
  {http://adsabs.harvard.edu/cgi-bin/nph-bib_query?bibcode=1994MNRAS.267..743G%
&amp;db_key=AST}.

\bibitem[{Giacconi} {\em et~al.}(1962){Giacconi}, {Gursky}, {Paolini} \&
  {Rossi}]{giacconi1962}
{Giacconi} R., {Gursky} H., {Paolini} F.~R.  \& {Rossi} B.~B. (1962).
\newblock {Evidence for X-rays from sources outside the Solar system}.
\newblock {\em \prl\/}, {\bf 9}, 439{.
\newblock}\ads
  {http://adsabs.harvard.edu/cgi-bin/nph-bib_query?bibcode=1962PhRvL...9..439G%
&amp;db_key=PHY}.

\bibitem[{Gierli\'{n}ski}, {Maciolek-Nied\'{z}wiecki} \&
  {Ebisawa}(2001){Gierli\'{n}ski}, {Maciolek-Nied\'{z}wiecki} \&
  {Ebisawa}]{gierlinski2001}
{Gierli\'{n}ski} M., {Maciolek-Nied\'{z}wiecki} A.  \& {Ebisawa} K. (2001).
\newblock {Application of a relativistic accretion disc model to X-ray spectra
  of LMC X-1 and GRO J1655-40}.
\newblock {\em \mnras\/}, {\bf 325}, 1253{.
\newblock}\ads
  {http://adsabs.harvard.edu/cgi-bin/nph-bib_query?bibcode=2001MNRAS.325.1253G%
&amp;db_key=AST}.

\bibitem[{Goyder} \& {Lasenby}(2004){Goyder} \& {Lasenby}]{goyder2004}
{Goyder} R.  \& {Lasenby} A.~N. (2004).
\newblock {Inferring the coronal flaring patterns in AGN from reverberation
  maps}.
\newblock {\em \mnras\/}.
\newblock In press. \href{http://arxiv.org/abs/astro-ph/0309518}
  {[astro-ph/0309518]}.

\bibitem[{Guainazzi}(2003){Guainazzi}]{guainazzi2003}
{Guainazzi} M. (2003).
\newblock {The history of the iron K{$\alpha$} line profile in the Piccinotti
  AGN ESO 198-G24}.
\newblock {\em \aap\/}, {\bf 401}, 903{.
\newblock}\ads
  {http://adsabs.harvard.edu/cgi-bin/nph-bib_query?bibcode=2003A\%26A...401..9%
03G&amp;db_key=AST}.

\bibitem[{Guainazzi} {\em et~al.}(1999){Guainazzi}, {Matt}, {Molendi}, {Orr},
  {Fiore}, {Grandi}, {Matteuzzi}, {Mineo}, {Perola}, {Parmar} \&
  {Piro}]{guainazzi1999}
{Guainazzi} M., {Matt} G., {Molendi} S., {Orr} A., {Fiore} F., {Grandi} P.,
  {Matteuzzi} A., {Mineo} T., {Perola} G.~C., {Parmar} A.~N.  \& {Piro} L.
  (1999).
\newblock {BeppoSAX confirms extreme relativistic effects in the X-ray spectrum
  of MCG--6-30-15}.
\newblock {\em \aap\/}, {\bf 341}, L27{.
\newblock}\ads
  {http://adsabs.harvard.edu/cgi-bin/nph-bib_query?bibcode=1999A\%26A...341L..%
27G&amp;db_key=AST}.

\bibitem[{Haardt}(1993){Haardt}]{haardt1993a}
{Haardt} F. (1993).
\newblock {Anisotropic Comptonization in thermal plasmas -- spectral
  distribution in plane-parallel geometry}.
\newblock {\em \apj\/}, {\bf 413}, 680{.
\newblock}\ads
  {http://adsabs.harvard.edu/cgi-bin/nph-bib_query?bibcode=1993ApJ...413..680H%
&amp;db_key=AST}.

\bibitem[{Haardt} \& {Maraschi}(1993){Haardt} \& {Maraschi}]{haardt1993}
{Haardt} F.  \& {Maraschi} L. (1993).
\newblock {X-ray spectra from two-phase accretion disks}.
\newblock {\em \apj\/}, {\bf 413}, 507{.
\newblock}\ads
  {http://adsabs.harvard.edu/cgi-bin/nph-bib_query?bibcode=1993ApJ...413..507H%
&amp;db_key=AST}.

\bibitem[{Haardt}, {Maraschi} \& {Ghisellini}(1994){Haardt}, {Maraschi} \&
  {Ghisellini}]{haardt1994}
{Haardt} F., {Maraschi} L.  \& {Ghisellini} G. (1994).
\newblock {A model for the X-ray and ultraviolet emission from Seyfert galaxies
  and galactic black holes}.
\newblock {\em \apjl\/}, {\bf 432}, L95{.
\newblock}\ads
  {http://adsabs.harvard.edu/cgi-bin/nph-bib_query?bibcode=1994ApJ...432L..95H%
&amp;db_key=AST}.

\bibitem[{Hameury}, {Marck} \& {Pelat}(1994){Hameury}, {Marck} \&
  {Pelat}]{hameury1994}
{Hameury} J.-M., {Marck} J.-A.  \& {Pelat} D. (1994).
\newblock {e$^+$ -- e$^-$ annihilation lines from accretion discs around Kerr
  black holes}.
\newblock {\em \aap\/}, {\bf 287}, 795{.
\newblock}\ads
  {http://adsabs.harvard.edu/cgi-bin/nph-bib_query?bibcode=1994A\%26A...287..7%
95H&amp;db_key=AST}.

\bibitem[{Hanisch} {\em et~al.}(2001){Hanisch}, {Farris}, {Greisen}, {Pence},
  {Schlesinger}, {Teuben}, {Thompson} \& {Warnock}]{hanisch2001}
{Hanisch} R.~J., {Farris} A., {Greisen} E.~W., {Pence} W.~D., {Schlesinger}
  B.~M., {Teuben} P.~J., {Thompson} R.~W.  \& {Warnock} A. (2001).
\newblock {Definition of the flexible image transport system (FITS)}.
\newblock {\em \aap\/}, {\bf 376}, 359{.
\newblock}\ads
  {http://adsabs.harvard.edu/cgi-bin/nph-bib_query?bibcode=2001A\%26A...376..3%
59H&amp;db_key=AST}.

\bibitem[{Hawking}(1971){Hawking}]{hawking1971}
{Hawking} S.~W. (1971).
\newblock {Gravitationally collapsed objects of very low mass}.
\newblock {\em \mnras\/}, {\bf 152}, 75{.
\newblock}\ads
  {http://adsabs.harvard.edu/cgi-bin/nph-bib_query?bibcode=1971MNRAS.152...75H%
&amp;db_key=AST}.

\bibitem[{Hawking}(1975a){Hawking}]{hawking1975}
{Hawking} S.~W. (1975a).
\newblock {Particle creation by black holes}.
\newblock {\em Commun. Math. Phys.}, {\bf 43}, 199.
\newblock See also in \cite{hawking1975a}.

\bibitem[{Hawking}(1975b){Hawking}]{hawking1975a}
{Hawking} S.~W. (1975b).
\newblock {Particle creation by black holes}.
\newblock In {\em Quantum gravity; Proceedings of the Oxford Symposium,
  Harwell, Berks., England, February 15, 16, 1974\/}, page 219. Oxford,
  Clarendon Press{.
\newblock}\ads
  {http://adsabs.harvard.edu/cgi-bin/nph-bib_query?bibcode=1975qugr.symp..219H%
&amp;db_key=AST}.

\bibitem[{Hawking} \& {Ellis}(1973){Hawking} \& {Ellis}]{hawking1973}
{Hawking} S.~W.  \& {Ellis} G.~F.~R. (1973).
\newblock {\em {The large scale structure of space-time}\/}.
\newblock Cambridge Monographs on Mathematical Physics, London: Cambridge
  University Press{.
\newblock}\ads
  {http://adsabs.harvard.edu/cgi-bin/nph-bib_query?bibcode=1973lsss.book.....H%
&amp;db_key=AST}.

\bibitem[{Hawley}, {Gammie} \& {Balbus}(1995){Hawley}, {Gammie} \&
  {Balbus}]{hawley1995}
{Hawley} J.~F., {Gammie} C.~F.  \& {Balbus} S.~A. (1995).
\newblock {Local three-dimensional magnetohydrodynamic simulations of accretion
  disks}.
\newblock {\em \apj\/}, {\bf 440}, 742{.
\newblock}\ads
  {http://adsabs.harvard.edu/cgi-bin/nph-bib_query?bibcode=1995ApJ...440..742H%
&amp;db_key=AST}.

\bibitem[{\noopsort{Iwasawaa}}{Iwasawa} {\em
  et~al.}(1998){\noopsort{Iwasawaa}}{Iwasawa}, {Fabian}, {Brandt}, {Kunieda},
  {Misaki}, {Terashima} \& {Reynolds}]{iwasawa1998}
{\noopsort{Iwasawaa}}{Iwasawa} K., {Fabian} A.~C., {Brandt} W.~N., {Kunieda}
  H., {Misaki} K., {Terashima} Y.  \& {Reynolds} C.~S. (1998).
\newblock {Detection of an X-ray periodicity in the Seyfert galaxy IRAS
  18325-5926}.
\newblock {\em \mnras\/}, {\bf 295}, L20{.
\newblock}\ads
  {http://adsabs.harvard.edu/cgi-bin/nph-bib_query?bibcode=1998MNRAS.295L..20I%
&amp;db_key=AST}.

\bibitem[{\noopsort{Iwasawab}}{Iwasawa} {\em
  et~al.}(1996){\noopsort{Iwasawab}}{Iwasawa}, {Fabian}, {Reynolds}, {Nandra},
  {Otani}, {Inoue}, {Hayashida}, {Brandt}, {Dotani}, {Kunieda}, {Matsuoka} \&
  {Tanaka}]{iwasawa1996}
{\noopsort{Iwasawab}}{Iwasawa} K., {Fabian} A.~C., {Reynolds} C.~S., {Nandra}
  K., {Otani} C., {Inoue} H., {Hayashida} K., {Brandt} W.~N., {Dotani} T.,
  {Kunieda} H., {Matsuoka} M.  \& {Tanaka} Y. (1996).
\newblock {The variable iron K emission line in MCG--6-30-15}.
\newblock {\em \mnras\/}, {\bf 282}, 1038{.
\newblock}\ads
  {http://adsabs.harvard.edu/cgi-bin/nph-bib_query?bibcode=1996MNRAS.282.1038I%
&amp;db_key=AST}.

\bibitem[{\noopsort{Karasa}}{Karas}(1996){\noopsort{Karasa}}{Karas}]{karas1996}
{\noopsort{Karasa}}{Karas} V. (1996).
\newblock {Light curve of a source orbiting a black hole: a fitting formula}.
\newblock {\em \apj\/}, {\bf 470}, 743{.
\newblock}\ads
  {http://adsabs.harvard.edu/cgi-bin/nph-bib_query?bibcode=1996ApJ...470..743K%
&amp;db_key=AST}.

\bibitem[{\noopsort{Karasb}}{Karas}, {Hur{\' e}} \& {Semer{\'
  a}k}(2004){\noopsort{Karasb}}{Karas}, {Hur{\' e}} \& {Semer{\'
  a}k}]{karas2004}
{\noopsort{Karasb}}{Karas} V., {Hur{\' e}} J.-M.  \& {Semer{\' a}k} O. (2004).
\newblock {Topical review: gravitating discs around black holes}.
\newblock {\em Classical and Quantum Gravity\/}, {\bf 21}, 1{.
\newblock}\ads
  {http://adsabs.harvard.edu/cgi-bin/nph-bib_query?bibcode=2004CQGra..21R...1K%
&amp;db_key=PHY}.

\bibitem[{\noopsort{Karasc}}{Karas}, {Lanza} \&
  {Vokrouhlick\'{y}}(1995){\noopsort{Karasc}}{Karas}, {Lanza} \&
  {Vokrouhlick\'{y}}]{karas1995}
{\noopsort{Karasc}}{Karas} V., {Lanza} A.  \& {Vokrouhlick\'{y}} D. (1995).
\newblock {Emission-line profiles from self-gravitating thin disks}.
\newblock {\em \apj\/}, {\bf 440}, 108{.
\newblock}\ads
  {http://adsabs.harvard.edu/cgi-bin/nph-bib_query?bibcode=1995ApJ...440..108K%
&amp;db_key=AST}.

\bibitem[{\noopsort{Karasd}}{Karas} \& {Vokrouhlick{\'
  y}}(1991){\noopsort{Karasd}}{Karas} \& {Vokrouhlick{\' y}}]{karas1991}
{\noopsort{Karasd}}{Karas} V.  \& {Vokrouhlick{\' y}} D. (1991).
\newblock {Dynamics of charged particles near a black hole in a magnetic
  field}.
\newblock {\em Journal de Physique I\/}, {\bf 1}, 1005{.
\newblock}\ads
  {http://adsabs.harvard.edu/cgi-bin/nph-bib_query?bibcode=1991JPhy1...1.1005K%
&amp;db_key=PHY}.

\bibitem[{\noopsort{Karase}}{Karas}, {Vokrouhlick\'{y}} \&
  {Polnarev}(1992){\noopsort{Karase}}{Karas}, {Vokrouhlick\'{y}} \&
  {Polnarev}]{karas1992}
{\noopsort{Karase}}{Karas} V., {Vokrouhlick\'{y}} D.  \& {Polnarev} A.~G.
  (1992).
\newblock {In the vicinity of a rotating black hole -- a fast numerical code
  for computing observational effects}.
\newblock {\em \mnras\/}, {\bf 259}, 569{.
\newblock}\ads
  {http://adsabs.harvard.edu/cgi-bin/nph-bib_query?bibcode=1992MNRAS.259..569K%
&amp;db_key=AST}.

\bibitem[{Kato}, {Fukue} \& {Mineshige}(1998){Kato}, {Fukue} \&
  {Mineshige}]{kato1998}
{Kato} S., {Fukue} J.  \& {Mineshige} S. (1998).
\newblock {\em {Black-hole accretion disks}\/}.
\newblock Kyoto, Japan: Kyoto University Press{.
\newblock}\ads
  {http://adsabs.harvard.edu/cgi-bin/nph-bib_query?bibcode=1998bhad.conf.....K%
&amp;db_key=AST}.

\bibitem[{Kerr}(1963){Kerr}]{Kerr1963}
{Kerr} R.~P. (1963).
\newblock {Gravitational field of a spinning mass as an example of
  algebraically special metrics}.
\newblock {\em \prl\/}, {\bf 11}, 237{.
\newblock}\ads
  {http://adsabs.harvard.edu/cgi-bin/nph-bib_query?bibcode=1963PhRvL..11..237K%
&amp;db_key=PHY}.

\bibitem[{Kojima}(1991){Kojima}]{kojima1991}
{Kojima} Y. (1991).
\newblock {The effects of black hole rotation on line profiles from accretion
  discs}.
\newblock {\em \mnras\/}, {\bf 250}, 629{.
\newblock}\ads
  {http://adsabs.harvard.edu/cgi-bin/nph-bib_query?bibcode=1991MNRAS.250..629K%
&amp;db_key=AST}.

\bibitem[{Kormendy} \& {Richstone}(1995){Kormendy} \&
  {Richstone}]{kormendy1995}
{Kormendy} J.  \& {Richstone} D. (1995).
\newblock {Inward bound -- the search for supermassive black holes in galactic
  nuclei}.
\newblock {\em \araa\/}, {\bf 33}, 581{.
\newblock}\ads
  {http://adsabs.harvard.edu/cgi-bin/nph-bib_query?bibcode=1995ARA\%26A..33..5%
81K&amp;db_key=AST}.

\bibitem[{Krolik} \& {Hawley}(2002){Krolik} \& {Hawley}]{krolik2002}
{Krolik} J.  \& {Hawley} J.~F. (2002).
\newblock {Where is the inner edge of an accretion disk around a black hole?}
\newblock {\em \apj\/}, {\bf 573}, 754{.
\newblock}\ads
  {http://adsabs.harvard.edu/cgi-bin/nph-bib_query?bibcode=2002ApJ...573..754K%
&amp;db_key=AST}.

\bibitem[{Laor}(1991){Laor}]{laor1991}
{Laor} A. (1991).
\newblock {Line profiles from a disc around a rotating black hole}.
\newblock {\em \apj\/}, {\bf 376}, 90{.
\newblock}\ads
  {http://adsabs.harvard.edu/cgi-bin/nph-bib_query?bibcode=1991ApJ...376...90L%
&amp;db_key=AST}.

\bibitem[{Laor}, {Netzer} \& {Piran}(1990){Laor}, {Netzer} \&
  {Piran}]{laor1990}
{Laor} A., {Netzer} H.  \& {Piran} T. (1990).
\newblock {Massive thin accretion discs. II. Polarization}.
\newblock {\em \mnras\/}, {\bf 242}, 560{.
\newblock}\ads
  {http://adsabs.harvard.edu/cgi-bin/nph-bib_query?bibcode=1990MNRAS.242..560L%
&amp;db_key=AST}.

\bibitem[{Laplace}(1796){Laplace}]{laplace1796}
{Laplace} P.~S. (1796).
\newblock {\em Le Syst\`{e}me du Monde Vol.~II. Des Mouvements R\'{e}els des
  Corps C\'{e}lestes {\rm (Paris: Duprat)}\/}.

\bibitem[{Laplace}(1799){Laplace}]{laplace1799}
{Laplace} P.~S. (1799).
\newblock {Beweis des Satzes, dass die anziehende Kraft bey einem
  Weltk\"{o}rper so gross seyn k\"{o}nne, dass das Licht davon nicht
  ausstr\"{o}men kann}.
\newblock {\em Allgemeine Geographische Ephemeriden {\rm (Weimar)}\/}, {\bf
  4}(1).
\newblock Engl. trans. in \cite{hawking1973} or \cite{stephani2003}.

\bibitem[{Lee} {\em et~al.}(2000){Lee}, {Fabian}, {Reynolds}, {Brandt} \&
  {Iwasawa}]{lee2000}
{Lee} J.~C., {Fabian} A.~C., {Reynolds} C.~S., {Brandt} W.~N.  \& {Iwasawa} K.
  (2000).
\newblock {The X-ray variability of the Seyfert 1 galaxy MCG--6-30-15 from long
  ASCA and RXTE observations}.
\newblock {\em \mnras\/}, {\bf 318}, 857{.
\newblock}\ads
  {http://adsabs.harvard.edu/cgi-bin/nph-bib_query?bibcode=2000MNRAS.318..857L%
&amp;db_key=AST}.

\bibitem[{Lee} {\em et~al.}(2001){Lee}, {Ogle}, {Canizares}, {Marshall},
  {Schulz}, {Morales}, {Fabian} \& {Iwasawa}]{lee2001}
{Lee} J.~C., {Ogle} P.~M., {Canizares} C.~R., {Marshall} H.~L., {Schulz} N.~S.,
  {Morales} R., {Fabian} A.~C.  \& {Iwasawa} K. (2001).
\newblock {Revealing the dusty warm absorber in MCG--6-30-15 with the Chandra
  high-energy transmission grating}.
\newblock {\em \apjl\/}, {\bf 554}, L13{.
\newblock}\ads
  {http://adsabs.harvard.edu/cgi-bin/nph-bib_query?bibcode=2001ApJ...554L..13L%
&amp;db_key=AST}.

\bibitem[{Lynden-Bell}(1969){Lynden-Bell}]{lynden-bell1969}
{Lynden-Bell} D. (1969).
\newblock {Galactic nuclei as collapsed old quasars}.
\newblock {\em \nat\/}, {\bf 223}, 690{.
\newblock}\ads
  {http://adsabs.harvard.edu/cgi-bin/nph-bib_query?bibcode=1969Natur.223..690L%
&amp;db_key=AST}.

\bibitem[{Madejski} {\em et~al.}(1993){Madejski}, {Done}, {Turner},
  {Mushotzky}, {Serlemitsos}, {Fiore}, {Sikora} \& {Begelman}]{madejski1993}
{Madejski} G.~M., {Done} C., {Turner} T.~J., {Mushotzky} R.~F., {Serlemitsos}
  P., {Fiore} F., {Sikora} M.  \& {Begelman} M.~C. (1993).
\newblock {Solving the mystery of the X-ray periodicity in the Seyfert galaxy
  NGC 6814}.
\newblock {\em \nat\/}, {\bf 365}, 626{.
\newblock}\ads
  {http://adsabs.harvard.edu/cgi-bin/nph-bib_query?bibcode=1993Natur.365..626M%
&amp;db_key=AST}.

\bibitem[{Magdziarz} \& {Zdziarski}(1995){Magdziarz} \&
  {Zdziarski}]{magdziarz1995}
{Magdziarz} P.  \& {Zdziarski} A.~A. (1995).
\newblock {Angle-dependent Compton reflection of X-rays and gamma-rays}.
\newblock {\em \mnras\/}, {\bf 273}, 837{.
\newblock}\ads
  {http://adsabs.harvard.edu/cgi-bin/nph-bib_query?bibcode=1995MNRAS.273..837M%
&amp;db_key=AST}.

\bibitem[{\noopsort{Martocchiaa}}{Martocchia}, {Karas} \&
  {Matt}(2000){\noopsort{Martocchiaa}}{Martocchia}, {Karas} \&
  {Matt}]{martocchia2000}
{\noopsort{Martocchiaa}}{Martocchia} A., {Karas} V.  \& {Matt} G. (2000).
\newblock {Effects of Kerr space-time on spectral features from X-ray
  illuminated accretion discs}.
\newblock {\em \mnras\/}, {\bf 312}, 817{.
\newblock}\ads
  {http://adsabs.harvard.edu/cgi-bin/nph-bib_query?bibcode=2000MNRAS.312..817M%
&amp;db_key=AST}.

\bibitem[{\noopsort{Martocchiab}}{Martocchia} \&
  {Matt}(1996){\noopsort{Martocchiab}}{Martocchia} \& {Matt}]{martocchia1996}
{\noopsort{Martocchiab}}{Martocchia} A.  \& {Matt} G. (1996).
\newblock {Iron K$\alpha$ line intensity from accretion discs around rotating
  black holes}.
\newblock {\em \mnras\/}, {\bf 282}, L53{.
\newblock}\ads
  {http://adsabs.harvard.edu/cgi-bin/nph-bib_query?bibcode=1996MNRAS.282L..53M%
&amp;db_key=AST}.

\bibitem[{\noopsort{Martocchiac}}{Martocchia}, {Matt} \&
  {Karas}(2002){\noopsort{Martocchiac}}{Martocchia}, {Matt} \&
  {Karas}]{martocchia2002a}
{\noopsort{Martocchiac}}{Martocchia} A., {Matt} G.  \& {Karas} V. (2002).
\newblock {On the origin of the broad, relativistic iron line of MCG--6-30-15
  observed by XMM-Newton}.
\newblock {\em \aap\/}, {\bf 383}, L23{.
\newblock}\ads
  {http://adsabs.harvard.edu/cgi-bin/nph-bib_query?bibcode=2002A\%26A...383L..%
23M&amp;db_key=AST}.

\bibitem[{\noopsort{Martocchiad}}{Martocchia} {\em
  et~al.}(2002){\noopsort{Martocchiad}}{Martocchia}, {Matt}, {Karas}, {Belloni}
  \& {Feroci}]{martocchia2002b}
{\noopsort{Martocchiad}}{Martocchia} A., {Matt} G., {Karas} V., {Belloni} T.
  \& {Feroci} M. (2002).
\newblock {Evidence for a relativistic iron line in GRS 1915+105}.
\newblock {\em \aap\/}, {\bf 387}, 215{.
\newblock}\ads
  {http://adsabs.harvard.edu/cgi-bin/nph-bib_query?bibcode=2002A\%26A...387..2%
15M&amp;db_key=AST}.

\bibitem[{Mason} {\em et~al.}(2003){Mason}, {Branduardi-Raymont}, {Ogle},
  {Page}, {Puchnarewicz}, {Behar}, {C{\' o}rdova}, {Davis}, {Maraschi},
  {McHardy}, {O'Brien}, {Priedhorsky} \& {Sasseen}]{mason2003}
{Mason} K.~O., {Branduardi-Raymont} G., {Ogle} P.~M., {Page} M.~J.,
  {Puchnarewicz} E.~M., {Behar} E., {C{\' o}rdova} F.~A., {Davis} S.,
  {Maraschi} L., {McHardy} I.~M., {O'Brien} P.~T., {Priedhorsky} W.~C.  \&
  {Sasseen} T.~P. (2003).
\newblock {The X-ray spectrum of the Seyfert I galaxy Markarian 766: dusty warm
  absorber or relativistic emission lines?}
\newblock {\em \apj\/}, {\bf 582}, 95{.
\newblock}\ads
  {http://adsabs.harvard.edu/cgi-bin/nph-bib_query?bibcode=2003ApJ...582...95M%
&amp;db_key=AST}.

\bibitem[{\noopsort{Matta}}{Matt}(1993){\noopsort{Matta}}{Matt}]{matt1993b}
{\noopsort{Matta}}{Matt} G. (1993).
\newblock {X-ray polarization properties of a centrally illuminated accretion
  disc}.
\newblock {\em \mnras\/}, {\bf 260}, 663{.
\newblock}\ads
  {http://adsabs.harvard.edu/cgi-bin/nph-bib_query?bibcode=1993MNRAS.260..663M%
&amp;db_key=AST}.

\bibitem[{\noopsort{Mattb}}{Matt}, {Fabian} \&
  {Ross}(1993){\noopsort{Mattb}}{Matt}, {Fabian} \& {Ross}]{matt1993}
{\noopsort{Mattb}}{Matt} G., {Fabian} A.~C.  \& {Ross} R.~R. (1993).
\newblock {X-ray photoionized accretion discs -- ultraviolet and X-ray
  continuum spectra and polarization}.
\newblock {\em \mnras\/}, {\bf 264}, 839{.
\newblock}\ads
  {http://adsabs.harvard.edu/cgi-bin/nph-bib_query?bibcode=1993MNRAS.264..839M%
&amp;db_key=AST}.

\bibitem[{\noopsort{Mattc}}{Matt} \& {Perola}(1992){\noopsort{Mattc}}{Matt} \&
  {Perola}]{matt1992a}
{\noopsort{Mattc}}{Matt} G.  \& {Perola} G.~C. (1992).
\newblock {The iron K$\alpha$ response in an X-ray illuminated relativistic
  disc and a black hole mass estimate}.
\newblock {\em \mnras\/}, {\bf 259}, 433{.
\newblock}\ads
  {http://adsabs.harvard.edu/cgi-bin/nph-bib_query?bibcode=1992MNRAS.259..433M%
&amp;db_key=AST}.

\bibitem[{\noopsort{Mattd}}{Matt}, {Perola} \&
  {Piro}(1991){\noopsort{Mattd}}{Matt}, {Perola} \& {Piro}]{matt1991}
{\noopsort{Mattd}}{Matt} G., {Perola} G.~C.  \& {Piro} L. (1991).
\newblock {The iron line and high energy bump as X-ray signatures of cold
  matter in Seyfert 1 galaxies}.
\newblock {\em \aap\/}, {\bf 247}, 25{.
\newblock}\ads
  {http://adsabs.harvard.edu/cgi-bin/nph-bib_query?bibcode=1991A\%26A...247...%
25M&amp;db_key=AST}.

\bibitem[{\noopsort{Matte}}{Matt} {\em et~al.}(1992){\noopsort{Matte}}{Matt},
  {Perola}, {Piro} \& {Stella}]{matt1992}
{\noopsort{Matte}}{Matt} G., {Perola} G.~C., {Piro} L.  \& {Stella} L. (1992).
\newblock {Iron K$\alpha$ line from X-ray illuminated relativistic discs}.
\newblock {\em \aap\/}, {\bf 257}, 63.
\newblock {Ibid.} (1992), {\bf 263}, 453{.
\newblock}\ads
  {http://adsabs.harvard.edu/cgi-bin/nph-bib_query?bibcode=1992A\%26A...257...%
63M&amp;db_key=AST}.

\bibitem[{\noopsort{Mattf}}{Matt}, {Perola} \&
  {Stella}(1993){\noopsort{Mattf}}{Matt}, {Perola} \& {Stella}]{matt1993a}
{\noopsort{Mattf}}{Matt} G., {Perola} G.~C.  \& {Stella} L. (1993).
\newblock {Multiple-peaked line profiles from relativistic disks at high
  inclination angles}.
\newblock {\em \aap\/}, {\bf 267}, 643{.
\newblock}\ads
  {http://adsabs.harvard.edu/cgi-bin/nph-bib_query?bibcode=1993A\%26A...267..6%
43M&amp;db_key=AST}.

\bibitem[{\noopsort{Matthewsa}}{Matthews} {\em
  et~al.}(1960){\noopsort{Matthewsa}}{Matthews}, {Bolton}, {Greenstein},
  {M\"{u}nch} \& {Sandage}]{matthews1960}
{\noopsort{Matthewsa}}{Matthews} T.~A., {Bolton} J.~G., {Greenstein} J.~L.,
  {M\"{u}nch} G.  \& {Sandage} A.~R. (1960).
\newblock {107th Meeting Amer. Astr. Soc., Dec.~31, 1960 (unpublished)}.
\newblock {\bf 21}, 148.
\newblock See {\em Sky and Tel.} (1961).

\bibitem[{\noopsort{Matthewsb}}{Matthews} \&
  {Sandage}(1962){\noopsort{Matthewsb}}{Matthews} \& {Sandage}]{matthews1962}
{\noopsort{Matthewsb}}{Matthews} T.~A.  \& {Sandage} A. (1962).
\newblock {3C 196 as a second radio star}.
\newblock {\em \pasp\/}, {\bf 74}, 406{.
\newblock}\ads
  {http://adsabs.harvard.edu/cgi-bin/nph-bib_query?bibcode=1962PASP...74R.406M%
&amp;db_key=AST}.

\bibitem[{McClintock} \& {Remillard}(2003){McClintock} \&
  {Remillard}]{mcclintock2003}
{McClintock} J.~E.  \& {Remillard} R.~A. (2003).
\newblock {Black hole binaries}.
\newblock In {\em {Compact stellar X-ray sources}\/}. W.~H.~G. Lewin \& M. van
  der Klis, Cambridge University Press.
\newblock \href{http://arxiv.org/abs/astro-ph/0306213}
  {\mbox{[astro-ph/0306213]}}.

\bibitem[{Merloni} \& {Fabian}(2001){Merloni} \& {Fabian}]{merloni2001}
{Merloni} A.  \& {Fabian} A.~C. (2001).
\newblock {Thunderclouds and accretion discs: a model for the spectral and
  temporal variability of Seyfert 1 galaxies}.
\newblock {\em \mnras\/}, {\bf 328}, 958{.
\newblock}\ads
  {http://adsabs.harvard.edu/cgi-bin/nph-bib_query?bibcode=2001MNRAS.328..958M%
&amp;db_key=AST}.

\bibitem[{Michell}(1784){Michell}]{michell1784}
{Michell} J. (1784).
\newblock {On the means of discovering the distance, magnitude, etc.\ of the
  fixed stars, in consequence of the Diminution of the Velocity of their Light,
  in case such a Diminution should be found to take place in any of them, and
  such other Data should be procured from Observations, as would be farther
  necessary for that Purpose}.
\newblock {\em Philosophical Transactions of the Royal Society of London\/},
  {\bf 74}, 35.

\bibitem[{\noopsort{Millera}}{Miller} {\em
  et~al.}(2002){\noopsort{Millera}}{Miller}, {Fabian}, {Wijnands}, {Remillard},
  {Wojdowski}, {Schulz}, {Di Matteo}, {Marshall}, {Canizares}, {Pooley} \&
  {Lewin}]{miller2002a}
{\noopsort{Millera}}{Miller} J.~M., {Fabian} A.~C., {Wijnands} R., {Remillard}
  R.~A., {Wojdowski} P., {Schulz} N.~S., {Di Matteo} T., {Marshall} H.~L.,
  {Canizares} C.~R., {Pooley} D.  \& {Lewin} W.~H.~G. (2002).
\newblock {Resolving the composite Fe K{$\alpha$} emission line in the Galactic
  black hole Cygnus X-1 with Chandra}.
\newblock {\em \apj\/}, {\bf 578}, 348{.
\newblock}\ads
  {http://adsabs.harvard.edu/cgi-bin/nph-bib_query?bibcode=2002ApJ...578..348M%
&amp;db_key=AST}.

\bibitem[{\noopsort{Millerb}}{Miller} {\em
  et~al.}(2002){\noopsort{Millerb}}{Miller}, {Fabian}, {in't Zand}, {Reynolds},
  {Wijnands}, {Nowak} \& {Lewin}]{miller2002b}
{\noopsort{Millerb}}{Miller} J.~M., {Fabian} A.~C., {in't Zand} J.~J.~M.,
  {Reynolds} C.~S., {Wijnands} R., {Nowak} M.~A.  \& {Lewin} W.~H.~G. (2002).
\newblock {A relativistic Fe K{$\alpha$} emission line in the
  intermediate-luminosity {\it BeppoSAX} spectrum of the Galactic microquasar
  V4641 Sgr}.
\newblock {\em \apjl\/}, {\bf 577}, L15{.
\newblock}\ads
  {http://adsabs.harvard.edu/cgi-bin/nph-bib_query?bibcode=2002ApJ...577L..15M%
&amp;db_key=AST}.

\bibitem[{Miniutti} {\em et~al.}(2003){Miniutti}, {Fabian}, {Goyder} \&
  {Lasenby}]{miniutti2003}
{Miniutti} G., {Fabian} A.~C., {Goyder} R.  \& {Lasenby} A.~N. (2003).
\newblock {The lack of variability of the iron line in MCG--6-30-15: general
  relativistic effects}.
\newblock {\em \mnras\/}, {\bf 344}, L22{.
\newblock}\ads
  {http://adsabs.harvard.edu/cgi-bin/nph-bib_query?bibcode=2003MNRAS.344L..22M%
&amp;db_key=AST}.

\bibitem[{Miniutti}, {Fabian} \& {Miller}(2004){Miniutti}, {Fabian} \&
  {Miller}]{miniutti2004}
{Miniutti} G., {Fabian} A.~C.  \& {Miller} J.~M. (2004).
\newblock {The relativistic Fe emission line in XTE J1650-500 with BeppoSAX:
  evidence for black hole spin and light-bending effects?}
\newblock {\em \mnras\/}, {\bf 351}, 466{.
\newblock}\ads
  {http://adsabs.harvard.edu/cgi-bin/nph-bib_query?bibcode=2004MNRAS.351..466M%
&amp;db_key=AST}.

\bibitem[{Misner}, {Thorne} \& {Wheeler}(1973){Misner}, {Thorne} \&
  {Wheeler}]{misner1973}
{Misner} C.~W., {Thorne} K.~S.  \& {Wheeler} J.~A. (1973).
\newblock {\em {Gravitation}\/}.
\newblock W.~H.~Freedman \& Co., San Fransisco{.
\newblock}\ads
  {http://adsabs.harvard.edu/cgi-bin/nph-bib_query?bibcode=1973grav.book.....M%
&amp;db_key=AST}.

\bibitem[{Nandra} {\em et~al.}(1997){Nandra}, {George}, {Mushotzky}, {Turner}
  \& {Yaqoob}]{nandra1997}
{Nandra} K., {George} I.~M., {Mushotzky} R.~F., {Turner} T.~J.  \& {Yaqoob} T.
  (1997).
\newblock {ASCA observations of Seyfert 1 galaxies. II. Relativistic iron
  K$\alpha$ emission}.
\newblock {\em \apj\/}, {\bf 477}, 602{.
\newblock}\ads
  {http://adsabs.harvard.edu/cgi-bin/nph-bib_query?bibcode=1997ApJ...477..602N%
&amp;db_key=AST}.

\bibitem[{Nayakshin} \& {Kazanas}(2002){Nayakshin} \& {Kazanas}]{nayakshin2002}
{Nayakshin} S.  \& {Kazanas} D. (2002).
\newblock {On time-dependent X-ray reflection by photoionized accretion disks:
  implications for Fe K{$\alpha$} line reverberation studies of active galactic
  nuclei}.
\newblock {\em \apj\/}, {\bf 567}, 85{.
\newblock}\ads
  {http://adsabs.harvard.edu/cgi-bin/nph-bib_query?bibcode=2002ApJ...567...85N%
&amp;db_key=AST}.

\bibitem[{Newman} {\em et~al.}(1965){Newman}, {Couch}, {Chinnapared}, {Exton},
  {Prakash} \& {Torrence}]{newman1965}
{Newman} E.~T., {Couch} R., {Chinnapared} K., {Exton} A., {Prakash} A.  \&
  {Torrence} R. (1965).
\newblock {Metric of a rotating, charged mass}.
\newblock {\em J. Math. Phys.}, {\bf 6}, 918.

\bibitem[{Nishikawa} {\em et~al.}(2001){Nishikawa}, {Koide}, {Shibata},
  {Kudoh}, {Sol} \& {Hughes}]{nishikawa2001}
{Nishikawa} K.-I., {Koide} S., {Shibata} K., {Kudoh} T., {Sol} H.  \& {Hughes}
  J.~P. (2001).
\newblock {3-D general relativistic MHD simulations of generating jets}.
\newblock {\em Bulletin of the American Astronomical Society\/}, {\bf 33},
  1498{.
\newblock}\ads
  {http://adsabs.harvard.edu/cgi-bin/nph-bib_query?bibcode=2001AAS...19913204N%
&amp;db_key=AST}.

\bibitem[{Nordstr\"{o}m}(1918){Nordstr\"{o}m}]{nordstrom1918}
{Nordstr\"{o}m} G. (1918).
\newblock {On the energy of the gravitational field in Einstein's theory}.
\newblock {\em Proc. Kon. Ned. Akad. Wet.}, {\bf 20}, 1238.

\bibitem[{Novikov} \& {Thorne}(1973){Novikov} \& {Thorne}]{novikov1973}
{Novikov} I.~D.  \& {Thorne} K.~S. (1973).
\newblock {Astrophysics of black holes}.
\newblock In {\em {Black holes}\/}, page 343. C.~de~Witt and B.~S.~de~Witt,
  Gordon \& Breach, New York.

\bibitem[{Ogura}, {Ohuo} \& {Kojima}(2000){Ogura}, {Ohuo} \&
  {Kojima}]{ogura2000}
{Ogura} J., {Ohuo} N.  \& {Kojima} Y. (2000).
\newblock {Profiles and polarization properties of emission lines from
  relativistic disks}.
\newblock {\em \pasj\/}, {\bf 52}, 841{.
\newblock}\ads
  {http://esoads.eso.org/cgi-bin/nph-bib_query?bibcode=2000PASJ...52..841O&amp%
;db_key=AST}.

\bibitem[{Oppenheimer} \& {Volkoff}(1939){Oppenheimer} \&
  {Volkoff}]{oppenheimer1939}
{Oppenheimer} J.~R.  \& {Volkoff} G.~M. (1939).
\newblock {On massive neutron cores}.
\newblock {\em Phys. Rev.}, {\bf 55}, 374{.
\newblock}\ads
  {http://adsabs.harvard.edu/cgi-bin/nph-bib_query?bibcode=1939PhRv...55..374O%
&amp;db_key=PHY}.

\bibitem[{Ostriker}(1964){Ostriker}]{ostriker1964}
{Ostriker} J. (1964).
\newblock {The equilibrium of self-gravitating rings}.
\newblock {\em \apj\/}, {\bf 140}, 1067{.
\newblock}\ads
  {http://adsabs.harvard.edu/cgi-bin/nph-bib_query?bibcode=1964ApJ...140.1067O%
&amp;db_key=AST}.

\bibitem[{Paczy\'{n}ski}(1978a){Paczy\'{n}ski}]{paczynski1978a}
{Paczy\'{n}ski} B. (1978a).
\newblock {A model of self-gravitating accretion disk}.
\newblock {\em Acta Astronomica\/}, {\bf 28}, 91{.
\newblock}\ads
  {http://adsabs.harvard.edu/cgi-bin/nph-bib_query?bibcode=1978AcA....28...91P%
&amp;db_key=AST}.

\bibitem[{Paczy\'{n}ski}(1978b){Paczy\'{n}ski}]{paczynski1978}
{Paczy\'{n}ski} B. (1978b).
\newblock {A model of self-gravitating accretion disk with a hot corona}.
\newblock {\em Acta Astronomica\/}, {\bf 28}, 241{.
\newblock}\ads
  {http://adsabs.harvard.edu/cgi-bin/nph-bib_query?bibcode=1978AcA....28..241P%
&amp;db_key=AST}.

\bibitem[{Pariev} \& {Bromley}(1998){Pariev} \& {Bromley}]{pariev1998}
{Pariev} V.~I.  \& {Bromley} B.~C. (1998).
\newblock {Line emission from an accretion disk around a black hole: effects of
  disk structure}.
\newblock {\em \apj\/}, {\bf 508}, 590{.
\newblock}\ads
  {http://adsabs.harvard.edu/cgi-bin/nph-bib_query?bibcode=1998ApJ...508..590P%
&amp;db_key=AST}.

\bibitem[{Pariev}, {Bromley} \& {Miller}(2001){Pariev}, {Bromley} \&
  {Miller}]{pariev2001}
{Pariev} V.~I., {Bromley} B.~C.  \& {Miller} W.~A. (2001).
\newblock {Estimation of relativistic accretion disk parameters from iron line
  emission}.
\newblock {\em \apj\/}, {\bf 547}, 649{.
\newblock}\ads
  {http://adsabs.harvard.edu/cgi-bin/nph-bib_query?bibcode=2001ApJ...547..649P%
&amp;db_key=AST}.

\bibitem[{Penrose} \& {Floyd}(1971){Penrose} \& {Floyd}]{penrose1971}
{Penrose} R.  \& {Floyd} G.~R. (1971).
\newblock {Black holes -- extraction of rotational energy}.
\newblock {\em Nature Physical Science\/}, {\bf 229}, 177{.
\newblock}\ads
  {http://adsabs.harvard.edu/cgi-bin/nph-bib_query?bibcode=1971NPhS..229..177P%
&amp;db_key=AST}.

\bibitem[{Petrucci} \& {Henri}(1997){Petrucci} \& {Henri}]{petrucci1997}
{Petrucci} P.~O.  \& {Henri} G. (1997).
\newblock {Anisotropic illumination of AGN's accretion disk by a non thermal
  source. II. General relativistic effects.}
\newblock {\em \aap\/}, {\bf 326}, 99{.
\newblock}\ads
  {http://adsabs.harvard.edu/cgi-bin/nph-bib_query?bibcode=1997A\%26A...326...%
99P&amp;db_key=AST}.

\bibitem[{Phillips} \& {M\'{e}sz\'{a}ros}(1986){Phillips} \&
  {M\'{e}sz\'{a}ros}]{phillips1986}
{Phillips} K.~C.  \& {M\'{e}sz\'{a}ros} P. (1986).
\newblock {Polarization and beaming of accretion disc radiation.}
\newblock {\em \apj\/}, {\bf 310}, 284{.
\newblock}\ads
  {http://adsabs.harvard.edu/cgi-bin/nph-bib_query?bibcode=1986ApJ...310..284P%
&amp;db_key=AST}.

\bibitem[{Poutanen} \& {Fabian}(1999){Poutanen} \& {Fabian}]{poutanen1999}
{Poutanen} J.  \& {Fabian} A.~C. (1999).
\newblock {Spectral evolution of magnetic flares and time lags in accreting
  black hole sources}.
\newblock {\em \mnras\/}, {\bf 306}, L31{.
\newblock}\ads
  {http://adsabs.harvard.edu/cgi-bin/nph-bib_query?bibcode=1999MNRAS.306L..31P%
&amp;db_key=AST}.

\bibitem[{Pozdnyakov}, {Sobo\v{l}} \& {Sunyaev}(1979){Pozdnyakov}, {Sobo\v{l}}
  \& {Sunyaev}]{pozdnyakov1979}
{Pozdnyakov} L.~A., {Sobo\v{l}} I.~M.  \& {Sunyaev} R.~A. (1979).
\newblock {The profile evolution of X-ray spectral lines due to Comptonization
  -- Monte Carlo computations}.
\newblock {\em \aap\/}, {\bf 75}, 214{.
\newblock}\ads
  {http://adsabs.harvard.edu/cgi-bin/nph-bib_query?bibcode=1979A\%26A....75..2%
14P&amp;db_key=AST}.

\bibitem[{Price} \& {Thorne}(1986){Price} \& {Thorne}]{price1986}
{Price} R.~H.  \& {Thorne} K.~S. (1986).
\newblock {Membrane viewpoint on black holes: properties and evolution of the
  stretched horizon}.
\newblock {\em \prd\/}, {\bf 33}, 915{.
\newblock}\ads
  {http://adsabs.harvard.edu/cgi-bin/nph-bib_query?bibcode=1986PhRvD..33..915P%
&amp;db_key=AST}.

\bibitem[{Pringle} \& {Rees}(1972){Pringle} \& {Rees}]{pringle1972}
{Pringle} J.~E.  \& {Rees} M.~J. (1972).
\newblock {Accretion disc models for compact X-ray sources}.
\newblock {\em \aap\/}, {\bf 21}, 1{.
\newblock}\ads
  {http://adsabs.harvard.edu/cgi-bin/nph-bib_query?bibcode=1972A\%26A....21...%
.1P&amp;db_key=AST}.

\bibitem[{Rauch} \& {Blandford}(1994){Rauch} \& {Blandford}]{rauch1994}
{Rauch} K.~P.  \& {Blandford} R.~D. (1994).
\newblock {Optical caustics in a Kerr space-time and the origin of rapid X-ray
  variability in active galactic nuclei}.
\newblock {\em \apj\/}, {\bf 421}, 46{.
\newblock}\ads
  {http://adsabs.harvard.edu/cgi-bin/nph-bib_query?bibcode=1994ApJ...421...46R%
&amp;db_key=AST}.

\bibitem[{Rees}(1975){Rees}]{rees1975}
{Rees} M.~J. (1975).
\newblock {Expected polarization properties of binary X-ray sources}.
\newblock {\em \mnras\/}, {\bf 171}, 457{.
\newblock}\ads
  {http://adsabs.harvard.edu/cgi-bin/nph-bib_query?bibcode=1975MNRAS.171..457R%
&amp;db_key=AST}.

\bibitem[{Reissner}(1916){Reissner}]{reissner1916}
{Reissner} H. (1916).
\newblock {\"{U}ber die Eigengravitation des elektrischen Feldes nach
  Einsteinschen Theorie}.
\newblock {\em Ann. Phys. {\rm (Leipzig)}\/}, {\bf 59}, 106.

\bibitem[{Reynolds} \& {Begelman}(1997){Reynolds} \& {Begelman}]{reynolds1997}
{Reynolds} C.~S.  \& {Begelman} M.~C. (1997).
\newblock {Iron fluorescence from within the innermost stable orbit of black
  hole accretion disks}.
\newblock {\em \apj\/}, {\bf 488}, 109{.
\newblock}\ads
  {http://adsabs.harvard.edu/cgi-bin/nph-bib_query?bibcode=1997ApJ...488..109R%
&amp;db_key=AST}.

\bibitem[{Reynolds} \& {Nowak}(2003){Reynolds} \& {Nowak}]{reynolds2003}
{Reynolds} C.~S.  \& {Nowak} M.~A. (2003).
\newblock {Fluorescent iron lines as a probe of astrophysical black hole
  systems}.
\newblock {\em Phys. Rep.}, {\bf 377}, 389{.
\newblock}\ads
  {http://adsabs.harvard.edu/cgi-bin/nph-bib_query?bibcode=2003PhR...377..389R%
&amp;db_key=AST}.

\bibitem[{Reynolds} {\em et~al.}(1999){Reynolds}, {Young}, {Begelman} \&
  {Fabian}]{reynolds1999}
{Reynolds} C.~S., {Young} A.~J., {Begelman} M.~C.  \& {Fabian} A.~C. (1999).
\newblock {X-ray iron line reverberation from black hole accretion disks}.
\newblock {\em \apj\/}, {\bf 514}, 164{.
\newblock}\ads
  {http://adsabs.harvard.edu/cgi-bin/nph-bib_query?bibcode=1999ApJ...514..164R%
&amp;db_key=AST}.

\bibitem[{Rhoades} \& {Ruffini}(1974){Rhoades} \& {Ruffini}]{rhoades1974}
{Rhoades} C.~E.  \& {Ruffini} R. (1974).
\newblock {Maximum mass of a neutron star}.
\newblock {\em \prl\/}, {\bf 32}, 324{.
\newblock}\ads
  {http://adsabs.harvard.edu/cgi-bin/nph-bib_query?bibcode=1974PhRvL..32..324R%
&amp;db_key=PHY}.

\bibitem[{\noopsort{Rozanska}}{R{\' o}{\. z}a{\' n}ska} {\em
  et~al.}(2002){\noopsort{Rozanska}}{R{\' o}{\. z}a{\' n}ska}, {Dumont},
  {Czerny} \& {Collin}]{rozanska2002}
{\noopsort{Rozanska}}{R{\' o}{\. z}a{\' n}ska} A., {Dumont} A.-M., {Czerny} B.
  \& {Collin} S. (2002).
\newblock {The structure and radiation spectra of illuminated accretion discs
  in active galactic nuclei. I. Moderate illumination}.
\newblock {\em \mnras\/}, {\bf 332}, 799{.
\newblock}\ads
  {http://adsabs.harvard.edu/cgi-bin/nph-bib_query?bibcode=2002MNRAS.332..799R%
&amp;db_key=AST}.

\bibitem[{Ruszkowski}(2000){Ruszkowski}]{ruszkowski2000}
{Ruszkowski} M. (2000).
\newblock {X-ray iron line variability for the model of an orbiting flare above
  a black hole accretion disc}.
\newblock {\em \mnras\/}, {\bf 315}, 1{.
\newblock}\ads
  {http://adsabs.harvard.edu/cgi-bin/nph-bib_query?bibcode=2000MNRAS.315....1R%
&amp;db_key=AST}.

\bibitem[{Salpeter}(1964){Salpeter}]{salpeter1964}
{Salpeter} E.~E. (1964).
\newblock {Accretion of interstellar matter by massive objects}.
\newblock {\em \apj\/}, {\bf 140}, 796{.
\newblock}\ads
  {http://adsabs.harvard.edu/cgi-bin/nph-bib_query?bibcode=1964ApJ...140..796S%
&amp;db_key=AST}.

\bibitem[{Schmidt}(1963){Schmidt}]{schmidt1963}
{Schmidt} M. (1963).
\newblock {3C 273: a star-like object with large red-shift}.
\newblock {\em \nat\/}, {\bf 197}, 1040{.
\newblock}\ads
  {http://adsabs.harvard.edu/cgi-bin/nph-bib_query?bibcode=1963Natur.197.1040S%
&amp;db_key=AST}.

\bibitem[{Schneider}, {Ehlers} \& {Falco}(1992){Schneider}, {Ehlers} \&
  {Falco}]{schneider1992}
{Schneider} P., {Ehlers} J.  \& {Falco} E.~E. (1992).
\newblock {\em {Gravitational lenses}\/}.
\newblock Springer-Verlag Berlin Heidelberg New York{.
\newblock}\ads
  {http://adsabs.harvard.edu/cgi-bin/nph-bib_query?bibcode=1992grle.book.....S%
&amp;db_key=AST}.

\bibitem[{Schnittman} \& {Bertschinger}(2004){Schnittman} \&
  {Bertschinger}]{schnittman2004}
{Schnittman} J.~D.  \& {Bertschinger} E. (2004).
\newblock {The harmonic structure of high-frequency quasi-periodic oscillations
  in accreting black holes}.
\newblock {\em \apj\/}, {\bf 606}, 1098{.
\newblock}\ads
  {http://adsabs.harvard.edu/cgi-bin/nph-bib_query?bibcode=2004ApJ...606.1098S%
&amp;db_key=AST}.

\bibitem[{Schwarzschild}(1916){Schwarzschild}]{schwarzschild1916}
{Schwarzschild} K. (1916).
\newblock {From the observatory}.
\newblock {\em \pasp\/}, {\bf 28}, 269{.
\newblock}\ads
  {http://adsabs.harvard.edu/cgi-bin/nph-bib_query?bibcode=1916PASP...28..269S%
&amp;db_key=AST}.

\bibitem[{Semer{\' a}k} \& {Karas}(1999){Semer{\' a}k} \& {Karas}]{semerak1999}
{Semer{\' a}k} O.  \& {Karas} V. (1999).
\newblock {Pseudo-Newtonian models of a rotating black hole field}.
\newblock {\em \aap\/}, {\bf 343}, 325{.
\newblock}\ads
  {http://adsabs.harvard.edu/cgi-bin/nph-bib_query?bibcode=1999A\%26A...343..3%
25S&amp;db_key=AST}.

\bibitem[{Seyfert}(1943){Seyfert}]{seyfert1943}
{Seyfert} C.~K. (1943).
\newblock {Nuclear emission in spiral nebulae.}
\newblock {\em \apj\/}, {\bf 97}, 28{.
\newblock}\ads
  {http://adsabs.harvard.edu/cgi-bin/nph-bib_query?bibcode=1943ApJ....97...28S%
&amp;db_key=AST}.

\bibitem[{Shakura} \& {Sunyaev}(1973){Shakura} \& {Sunyaev}]{shakura1973}
{Shakura} N.~I.  \& {Sunyaev} R.~A. (1973).
\newblock {Black holes in binary systems. Observational appearance.}
\newblock {\em \aap\/}, {\bf 24}, 337{.
\newblock}\ads
  {http://adsabs.harvard.edu/cgi-bin/nph-bib_query?bibcode=1973A\%26A....24..3%
37S&amp;db_key=AST}.

\bibitem[{Shapiro}, {Lightman} \& {Eardley}(1976){Shapiro}, {Lightman} \&
  {Eardley}]{shapiro1976}
{Shapiro} S.~L., {Lightman} A.~P.  \& {Eardley} D.~M. (1976).
\newblock {A two-temperature accretion disk model for Cygnus X-1 -- structure
  and spectrum}.
\newblock {\em \apj\/}, {\bf 204}, 187{.
\newblock}\ads
  {http://adsabs.harvard.edu/cgi-bin/nph-bib_query?bibcode=1976ApJ...204..187S%
&amp;db_key=AST}.

\bibitem[{Sharp}(1979){Sharp}]{sharp1979}
{Sharp} N.~A. (1979).
\newblock {Geodesics in black hole space-times}.
\newblock {\em General Relativity and Gravitation\/}, {\bf 10}, 659{.
\newblock}\ads
  {http://adsabs.harvard.edu/cgi-bin/nph-bib_query?bibcode=1979GReGr..10..659S%
&amp;db_key=AST}.

\bibitem[{Speith}, {Riffert} \& {Ruder}(1995){Speith}, {Riffert} \&
  {Ruder}]{speith1995}
{Speith} R., {Riffert} H.  \& {Ruder} H. (1995).
\newblock {The photon transfer function for accretion disks around a Kerr black
  hole}.
\newblock {\em Computer Physics Communications\/}, {\bf 88}, 109{.
\newblock}\ads
  {http://adsabs.harvard.edu/cgi-bin/nph-bib_query?bibcode=1995CoPhC..88..109S%
&amp;db_key=PHY}.

\bibitem[{Stella}(1990){Stella}]{stella1990}
{Stella} L. (1990).
\newblock {Measuring black hole mass through variable line profiles from
  accretion disks}.
\newblock {\em Nature\/}, {\bf 344}, 747{.
\newblock}\ads
  {http://adsabs.harvard.edu/cgi-bin/nph-bib_query?bibcode=1990Natur.344..747S%
&amp;db_key=AST}.

\bibitem[{Stephani}(2003){Stephani}]{stephani2003}
{Stephani} H. (2003).
\newblock {Laplace, Weimar, Schiller and the birth of black hole theory}.
\newblock {\em ArXiv General Relativity and Quantum Cosmology e-prints\/}.
\newblock \href{http://arxiv.org/abs/gr-qc/0304087}{[gr-qc/0304087]}.

\bibitem[{Sunyaev} \& {Titarchuk}(1985){Sunyaev} \& {Titarchuk}]{sunyaev1985}
{Sunyaev} R.~A.  \& {Titarchuk} L.~G. (1985).
\newblock {Comptonization of low-frequency radiation in accretion disks.
  Angular distribution and polarization of hard radiation}.
\newblock {\em \aap\/}, {\bf 143}, 374{.
\newblock}\ads
  {http://adsabs.harvard.edu/cgi-bin/nph-bib_query?bibcode=1985A\%26A...143..3%
74S&amp;db_key=AST}.

\bibitem[{Tanaka} {\em et~al.}(1995){Tanaka}, {Nandra}, {Fabian}, {Inoue},
  {Otani}, {Dotani}, {Hayashida}, {Iwasawa}, {Kii}, {Kunieda}, {Makino} \&
  {Matsuoka}]{tanaka1995}
{Tanaka} Y., {Nandra} K., {Fabian} A.~C., {Inoue} H., {Otani} C., {Dotani} T.,
  {Hayashida} K., {Iwasawa} K., {Kii} T., {Kunieda} H., {Makino} F.  \&
  {Matsuoka} M. (1995).
\newblock {Gravitationally redshifted emission implying an accretion disk and
  massive black hole in the active galaxy MCG--6-30-15}.
\newblock {\em \nat\/}, {\bf 375}, 659{.
\newblock}\ads
  {http://adsabs.harvard.edu/cgi-bin/nph-bib_query?bibcode=1995Natur.375..659T%
&amp;db_key=AST}.

\bibitem[{Thorne}(1974){Thorne}]{thorne1974}
{Thorne} K.~S. (1974).
\newblock {Disk-accretion onto a black hole. II. Evolution of the hole}.
\newblock {\em \apj\/}, {\bf 191}, 507{.
\newblock}\ads
  {http://adsabs.harvard.edu/cgi-bin/nph-bib_query?bibcode=1974ApJ...191..507T%
&amp;db_key=AST}.

\bibitem[{Thorne}, {Price} \& {MacDonald}(1986){Thorne}, {Price} \&
  {MacDonald}]{thorne1986}
{Thorne} K.~S., {Price} R.~H.  \& {MacDonald} D.~A. (1986).
\newblock {\em Black holes: the membrane paradigm\/}.
\newblock New Haven: Yale University Press{.
\newblock}\ads
  {http://adsabs.harvard.edu/cgi-bin/nph-bib_query?bibcode=1986bhmp.book.....T%
&amp;db_key=AST}.

\bibitem[{\noopsort{Turnera}}{Turner}, {Kraemer} \&
  {Reeves}(2004){\noopsort{Turnera}}{Turner}, {Kraemer} \&
  {Reeves}]{turner2004}
{\noopsort{Turnera}}{Turner} T.~J., {Kraemer} S.~B.  \& {Reeves} J.~N. (2004).
\newblock {Transient relativistically shifted lines as a probe of black hole
  systems}.
\newblock {\em \apj\/}, {\bf 603}, 62{.
\newblock}\ads
  {http://adsabs.harvard.edu/cgi-bin/nph-bib_query?bibcode=2004ApJ...603...62T%
&amp;db_key=AST}.

\bibitem[{\noopsort{Turnerb}}{Turner} {\em
  et~al.}(2002){\noopsort{Turnerb}}{Turner}, {Mushotzky}, {Yaqoob}, {George},
  {Snowden}, {Netzer}, {Kraemer}, {Nandra} \& {Chelouche}]{turner2002}
{\noopsort{Turnerb}}{Turner} T.~J., {Mushotzky} R.~F., {Yaqoob} T., {George}
  I.~M., {Snowden} S.~L., {Netzer} H., {Kraemer} S.~B., {Nandra} K.  \&
  {Chelouche} D. (2002).
\newblock {Narrow components within the Fe K{$\alpha$} profile of NGC 3516:
  evidence of the importance of general relativistic effects?}
\newblock {\em \apjl\/}, {\bf 574}, L123{.
\newblock}\ads
  {http://adsabs.harvard.edu/cgi-bin/nph-bib_query?bibcode=2002ApJ...574L.123T%
&amp;db_key=AST}.

\bibitem[{Usui}, {Nishida} \& {Eriguchi}(1998){Usui}, {Nishida} \&
  {Eriguchi}]{usui1998}
{Usui} F., {Nishida} S.  \& {Eriguchi} Y. (1998).
\newblock {Emission-line profiles from self-gravitating toroids around black
  holes}.
\newblock {\em \mnras\/}, {\bf 301}, 721{.
\newblock}\ads
  {http://adsabs.harvard.edu/cgi-bin/nph-bib_query?bibcode=1998MNRAS.301..721U%
&amp;db_key=AST}.

\bibitem[{Viergutz}(1993){Viergutz}]{viergutz1993}
{Viergutz} S.~U. (1993).
\newblock {Image generation in Kerr geometry. I. Analytical investigations on
  the stationary emitter-observer problem}.
\newblock {\em \aap\/}, {\bf 272}, 355{.
\newblock}\ads
  {http://adsabs.harvard.edu/cgi-bin/nph-bib_query?bibcode=1993A\%26A...272..3%
55V&amp;db_key=AST}.

\bibitem[{\noopsort{Villiers}}{de Villiers} \&
  {Hawley}(2003){\noopsort{Villiers}}{de Villiers} \& {Hawley}]{villiers2003}
{\noopsort{Villiers}}{de Villiers} J.  \& {Hawley} J.~F. (2003).
\newblock {Global general relativistic magnetohydrodynamic simulations of
  accretion tori}.
\newblock {\em \apj\/}, {\bf 592}, 1060{.
\newblock}\ads
  {http://adsabs.harvard.edu/cgi-bin/nph-bib_query?bibcode=2003ApJ...592.1060D%
&amp;db_key=AST}.

\bibitem[{Wald}(1974){Wald}]{wald1974}
{Wald} R.~M. (1974).
\newblock {Black hole in a uniform magnetic field}.
\newblock {\em \prd\/}, {\bf 10}, 1680{.
\newblock}\ads
  {http://adsabs.harvard.edu/cgi-bin/nph-bib_query?bibcode=1974PhRvD..10.1680W%
&amp;db_key=PHY}.

\bibitem[{Wald}(1998){Wald}]{wald1998}
{Wald} R.~M. (1998).
\newblock {\em Black holes and relativistic stars\/}.
\newblock Chicago: University of Chicago Press{.
\newblock}\ads
  {http://adsabs.harvard.edu/cgi-bin/nph-bib_query?bibcode=1998bhrs.conf.....W%
&amp;db_key=AST}.

\bibitem[{Walker} \& {Penrose}(1970){Walker} \& {Penrose}]{walker1970}
{Walker} M.  \& {Penrose} R. (1970).
\newblock {On quadratic first integrals of the geodesic equations for type
  \{22\} space-times}.
\newblock {\em Commun. Math. Phys.}, {\bf 18}, 265.

\bibitem[{Wilms} {\em et~al.}(2001){Wilms}, {Reynolds}, {Begelman}, {Reeves},
  {Molendi}, {Staubert} \& {Kendziorra}]{wilms2001}
{Wilms} J., {Reynolds} C.~S., {Begelman} M.~C., {Reeves} J., {Molendi} S.,
  {Staubert} R.  \& {Kendziorra} E. (2001).
\newblock {XMM-EPIC observation of MCG--6-30-15: direct evidence for the
  extraction of energy from a spinning black hole?}
\newblock {\em \mnras\/}, {\bf 328}, L27{.
\newblock}\ads
  {http://adsabs.harvard.edu/cgi-bin/nph-bib_query?bibcode=2001MNRAS.328L..27W%
&amp;db_key=AST}.

\bibitem[{\noopsort{Yaqooba}}{Yaqoob} {\em
  et~al.}(2003){\noopsort{Yaqooba}}{Yaqoob}, {George}, {Kallman},
  {Padmanabhan}, {Weaver} \& {Turner}]{yaqoob2003}
{\noopsort{Yaqooba}}{Yaqoob} T., {George} I.~M., {Kallman} T.~R., {Padmanabhan}
  U., {Weaver} K.~A.  \& {Turner} T.~J. (2003).
\newblock {Fe XXV and Fe XXVI diagnostics of the black hole and accretion disk
  in active galaxies: Chandra time-resolved grating spectroscopy of NGC 7314}.
\newblock {\em \apj\/}, {\bf 596}, 85{.
\newblock}\ads
  {http://adsabs.harvard.edu/cgi-bin/nph-bib_query?bibcode=2003ApJ...596...85Y%
&amp;db_key=AST}.

\bibitem[{\noopsort{Yaqoobb}}{Yaqoob} \&
  {Padmanabhan}(2004){\noopsort{Yaqoobb}}{Yaqoob} \& {Padmanabhan}]{yaqoob2004}
{\noopsort{Yaqoobb}}{Yaqoob} T.  \& {Padmanabhan} U. (2004).
\newblock {The cores of the Fe K lines in Seyfert 1 galaxies observed by the
  Chandra high energy grating}.
\newblock {\em \apj\/}, {\bf 604}, 63{.
\newblock}\ads
  {http://adsabs.harvard.edu/cgi-bin/nph-bib_query?bibcode=2004ApJ...604...63Y%
&amp;db_key=AST}.

\bibitem[{Zakharov}(1994){Zakharov}]{zakharov1994}
{Zakharov} A.~F. (1994).
\newblock {On the hotspot near a Kerr black-hole -- Monte-Carlo simulations}.
\newblock {\em \mnras\/}, {\bf 269}, 283{.
\newblock}\ads
  {http://adsabs.harvard.edu/cgi-bin/nph-bib_query?bibcode=1994MNRAS.269..283Z%
&amp;db_key=AST}.

\bibitem[{Zeldovich} \& {Novikov}(1964){Zeldovich} \& {Novikov}]{zeldovich1964}
{Zeldovich} Y.~B.  \& {Novikov} I.~D. (1964).
\newblock {\em Dokl. Acad. Nauk. SSSR\/}, {\bf 158}, 811.

\bibitem[{\.{Z}ycki} \& {Czerny}(1994){\.{Z}ycki} \& {Czerny}]{zycki1994}
{\.{Z}ycki} P.  \& {Czerny} B. (1994).
\newblock {The iron K$\alpha$ line from a partially ionized reflecting medium
  in an active galactic nucleus}.
\newblock {\em \mnras\/}, {\bf 266}, 653{.
\newblock}\ads
  {http://adsabs.harvard.edu/cgi-bin/nph-bib_query?bibcode=1994MNRAS.266..653Z%
&amp;db_key=AST}.

\end{thebibliography}
\end{document}